%% file: main.tex
\pdfoutput=1
\documentclass[
11pt,
oneside,
english,
singlespacing,
parskip, 
headsepline,
]{MastersDoctoralThesis} % The class file specifying the document structure

\usepackage[utf8]{inputenc}
\usepackage[T1]{fontenc}
\usepackage{mathpazo} 
\usepackage{tikz-cd}
\usepackage[sorting=none,style=phys,biblabel=brackets,]{biblatex}
\DeclareFieldFormat{titlecase}{#1}
\DeclareFieldFormat[article,book]{title}{#1}
\bibliography{Literature}
\usepackage{caption}
\usepackage{ragged2e}
\usepackage{amssymb}
\usepackage{amsmath, nccmath}
\usepackage{bbm}
\usepackage{amscd}
\usepackage{latexsym}
\usepackage{graphicx}
\usepackage{ifthen}
\usepackage{epsfig}
\usepackage{setspace}
\usepackage{subcaption}
\usepackage{hyperref}
\usepackage{dsfont}
\usepackage{mathrsfs}
\usepackage{color}
\usepackage{cancel}
\usepackage{mathtools}
\usepackage[shortlabels]{enumitem}
\usepackage[toc,page]{appendix}
\usepackage{float}
\usepackage{adjustbox}
\usepackage{titlesec}
\hypersetup{
 colorlinks=true,
 linkcolor=blue,
 citecolor=magenta,
}
\usepackage{array}
\newcolumntype{P}[1]{>{\centering\arraybackslash}p{#1}}
\newcolumntype{L}[1]{>{\raggedright\arraybackslash}p{#1}}
\usepackage{makecell}

\catcode`@=11
\renewcommand{\section}{\@startsection{section}{1}{0pt}{\medskipamount}
{\medskipamount}{\large\bf}}
\numberwithin{equation}{section}
\catcode`@=12

\newcommand{\mathsym}[1]{{}}
\newcommand{\unicode}[1]{{}}

\newcommand{\N}{\mathds N}
\newcommand{\R}{\mathds R}
\newcommand{\Acal}{{\cal A}}
\newcommand{\Fcal}{{\cal F}}
\newcommand{\Ecal}{{\cal E}}
\newcommand{\Bcal}{{\cal B}}
\newcommand{\Dcal}{{\cal D}}
\newcommand{\Pcal}{{\cal P}}
\newcommand{\Kcal}{{\cal K}}
\newcommand{\Mcal}{{\cal M}}
\newcommand{\Ical}{{\cal I}}
\newcommand{\Jcal}{{\cal J}}
\newcommand{\Lcal}{{\cal L}}
\newcommand{\tY}{{\widetilde{Y}}}

\newcommand{\cn}{{\mathrm{cn}\bigl(\sfrac{\tau}{\epsilon},k\bigr)}}
\newcommand{\sn}{{\mathrm{sn}\bigl(\sfrac{\tau}{\epsilon},k\bigr)}}
\newcommand{\dn}{{\mathrm{dn}\bigl(\sfrac{\tau}{\epsilon},k\bigr)}}

\def\im{\mathrm{i}}
\def\ep{\mathrm{e}}
\def\pa{\mbox{$\partial$}}
\def\diff{\mathrm{d}}
\def\tr{\mathrm{tr}}
\def\sfrac#1#2{{\textstyle\frac{#1}{#2}}}
\def\>{\rangle}
\def\<{\langle}
\def\+{\dagger}
\def\we{{\wedge}}
\def\={\ =\ }
\def\und{\quad\textrm{and}\quad}
\def\with{\quad\textrm{with}\quad}
\def\for{\quad\textrm{for}\quad}
\newcommand{\unity}{\mathbbm{1}}
\def\3j#1#2#3#4#5#6{\begin{pmatrix} #1&#2&#3\\#4&#5&#6 \end{pmatrix}}
\def\s3j#1#2{\begin{pmatrix} #1\\#2 \end{pmatrix}}
\def\rep#1#2{\bigl[\begin{smallmatrix}#1\\#2\end{smallmatrix}\bigr]}
\DeclareMathOperator{\sech}{sech}

\thesistitle{Solutions of Yang--Mills theory in \\ four-dimensional de Sitter space} % Your thesis title, this is used in the title and abstract, print it elsewhere with \ttitle
\supervisor{Prof. Olaf \textsc{Lechtenfeld}} % Your supervisor's name, this is used in the title page, print it elsewhere with \supname
\examiner{} % Your examiner's name, this is not currently used anywhere in the template, print it elsewhere with \examname
\degree{Doctor of Philosophy} % Your degree name, this is used in the title page and abstract, print it elsewhere with \degreename
\author{Kaushlendra \textsc{Kumar}} % Your name, this is used in the title page and abstract, print it elsewhere with \authorname
\addresses{} % Your address, this is not currently used anywhere in the template, print it elsewhere with \addressname

\subject{Theoretical Physics} % Your subject area, this is not currently used anywhere in the template, print it elsewhere with \subjectname
\keywords{} % Keywords for your thesis, this is not currently used anywhere in the template, print it elsewhere with \keywordnames
\university{\href{https://www.uni-hannover.de/en/}{Leibniz Universität Hannover}} % Your university's name and URL, this is used in the title page and abstract, print it elsewhere with \univname
\faculty{\href{https://www.maphy.uni-hannover.de/en/}{Fakultät f\"{u}r Mathematik and Physik}} % Your faculty's name and URL, this is used in the title page and abstract, print it elsewhere with \facname

\AtBeginDocument{
\hypersetup{pdftitle=\ttitle} % Set the PDF's title to your title
\hypersetup{pdfauthor=\authorname} % Set the PDF's author to your name
\hypersetup{pdfkeywords=\keywordnames} % Set the PDF's keywords to your keywords
}

\begin{document}

\frontmatter % Use roman page numbering style (i, ii, iii, iv...) for the pre-content pages

\pagestyle{plain} % Default to the plain heading style until the thesis style is called for the body content

%----------------------------------------------------------------------------------------
%	TITLE PAGE
%----------------------------------------------------------------------------------------

\begin{titlepage}
\begin{center}

\huge{\bf \ttitle}\\

\vfill

\large {Von der Fakultät für Mathematik und Physik} \\[0.3cm]
{der Gottfried Wilhelm Leibniz Universität Hannover} \\[1cm]

{zur Erlangung des akademischen Grades} \\[0.3cm]
{Doktor der Naturwissenschaften} \\[0.3cm]
{Dr. rer. nat.} \\[1cm]

{genehmigte Dissertation von} \\[1.5cm]
\Large {\bf M.Sc. Kaushlendra Kumar} \\[3cm]

\Large {2022}

\end{center}
\end{titlepage}

\pagebreak
~
\vfill 
\RaggedRight
{\bf Referent:} Prof. Dr. Olaf Lechtenfeld \\[0.3cm]
{\bf Korreferent:} Prof. Dr. Domenico Giulini \\[0.3cm]
{\bf Korreferent:} Prof. Dr. Gleb Arutyunov \\[0.3cm]
{\bf Tag der Promotion:} 04.05.2022 \\[6cm]

Printed and/or published with the support of the German Academic Exchange Service.

\clearpage
\addchaptertocentry{Resources} 
\vspace*{1cm}

\begin{center}
    {\huge Resources}
\end{center}

\vspace{10pt}

{\Large This thesis is based on following published research articles:}
\begin{enumerate}
\justifying
    \item Kaushlendra Kumar, Olaf Lechtenfeld, and Gabriel Picanço Costa, \\ 
    Trajectories of charged particles in knotted electromagnetic fields, \\
    \href{https://doi.org/10.1088/1751-8121/ac7c49}{{\it Journal of Physics A: Mathematical \& Theoretical} {\bf 55} (2022) 315401}.
    \item Kaushlendra Kumar, Gabriel Picanço Costa, and Lukas Hantzko, \\
    Conserved charges for rational electromagnetic knots, \\
    \href{https://doi.org/10.1140/epjp/s13360-022-02563-4}{{\it European Physical Journal Plus} {\bf 137} (2022) 407}.
    \item Kaushlendra Kumar, Olaf Lechtenfeld, and Gabriel Picanço Costa, \\
    Instability of cosmic Yang-Mills fields, \\
    \href{https://doi.org/10.1016/j.nuclphysb.2021.115583}{{\it Nuclear Physics B} {\bf 973} (2021) 115583}.
    \item Kaushlendra Kumar and Olaf Lechtenfeld, \\
    On rational electromagnetic fields,\\
    \href{https://doi.org/10.1016/j.physleta.2020.126445}{{\it Physics Letters A} {\bf 384} (2020) 126445}.
\end{enumerate}

\vspace{10pt}
{\Large Following notebooks validate the data presented in this thesis:}

\begin{itemize}
\justifying
    \item Kaushlendra Kumar and Gabriel Picanço Costa,\\
    ``Yang-Mills Theory in $4$-Dimensional de Sitter Space" \\
    from the Notebook Archive (2022), \href{https://notebookarchive.org/2022-04-dbzrbqg}{https://notebookarchive.org/2022-04-dbzrbqg}.
    \item Kaushlendra Kumar, Olaf Lechtenfeld and Gabriel Picanco Costa,\\
    ``Trajectories of Charged Particles in Knotted Electromagnetic Fields"\\
    from the Notebook Archive (2022), \href{https://notebookarchive.org/2022-05-7es6sj9}{https://notebookarchive.org/2022-05-7es6sj9}.
\end{itemize}

\clearpage
%----------------------------------------------------------------------------------------
%	Zusammenfassung
%----------------------------------------------------------------------------------------
\clearpage
\addchaptertocentry{Zusammenfassung} 
\vspace*{1cm}

\begin{center}
    {\huge \it Zusammenfassung}
\end{center}

\begin{center}
\justifying
Diese Doktorarbeit befasst sich mit der Analyse einiger Yang-Mills-Lösungen auf vier dimensionalen de Sitter Raum $\diff{S}_4$. Die konforme Äquivalenz dieses Raums mit einem endlichen Lorentz-Zylinder über der $3$-Sphäre $S^3$ und auch mit Teilen des Minkowski-Raums---zusammen mit der Tatsache, dass die Yang-Mills-Theorie in der 4-dimensionalen Raumzeit konform invariant ist---hat kürzlich zur Entdeckung einer Familie rational verknoteter elektromagnetischer Feldkonfigurationen geführt. Diese „Basisknoten“-Lösungen der Maxwell-Gleichungen, auch bekannt als Abelsche Yang-Mills-Theorie, sind mit den hypersphärischen Harmonischen $Y_{j,m,n}$ auf der $S^3$ gekennzeichnet und haben nette Eigenschaften wie endliche Energie, endliche Wirkung und das Vorhandensein einer konservierten topologischen Größe namens Helizität. Ihre Feldlinien bilden geschlossene Knotenschleifen im dreidimensionalen Euklidischen Raum. Diese könnten in der Kosmologie des frühen Universums eine Rolle spielen, um das symmetrische Higgs-Vakuum zu stabilisieren.

Wir untersuchen Symmetrieaspekte dieser elektromagnetischen Knotenkonfigurationen und berechnen alle erhaltenen Ladungen für die konforme Gruppe $SO(2,4)$ im Zusammenhang mit dem $4$-dimensionalen Minkowski-Raum für eine komplexe Linearkombination dieser Basislösungen für ein festes $j$. Die Berücksichtigung einer solchen komplexen linearen Kombination ist wichtig, da diese rationalen Basislösungen verwendet werden können, um auf diese Weise jede endliche Energiefeldkonfiguration zu erzeugen; wir demonstrieren diese Tatsache mit einigen bekannten Ergebnissen für bestimmte modifizierte Hopf--Ran\~{a}da-Knoten. Wir finden, dass die skalaren Ladungen entweder verschwinden oder proportional zur Energie sind. Für die nicht verschwindenden Vektorladungen finden wir eine schöne geometrische Struktur, die auch die Berechnung ihrer sphärischen Komponenten erleichtert. Wir finden auch, dass die Helizität mit der Energie zusammenhängt. Darüber hinaus charakterisieren wir den Unterraum von Nullfeldern und präsentieren einen Ausdruck für den elektromagnetischen Fluss bei null unendlich, der mit der Gesamtenergie übereinstimmt, wodurch die Energieerhaltung validiert wird. Schließlich untersuchen wir die Trajektorien von Punktladungen im Hintergrund solcher Basisknotenkonfigurationen. Dazu finden wir je nach Feldkonfiguration und verwendetem Parametersatz unterschiedliche Verhaltensweisen. Dazu gehören eine Beschleunigung von Teilchen durch das elektromagnetische Feld aus dem Ruhezustand auf ultrarelativistische Geschwindigkeiten, eine schnelle Konvergenz ihrer Flugbahnen in wenige schmale Kegel, asymptotisch für einen ausreichend hohen Wert der Kopplung, und ein ausgeprägtes Verdrehen und Wenden der Bahnen in kohärenter Weise.

Wir analysieren die lineare Stabilität der „kosmischen Yang-Mills-Felder“ der $SU(2)$ gegen-über allgemeinen Eichfeldstörungen, während wir die Metrik eingefroren halten, indem wir den (zeitabhängigen) Yang-Mills-Fluktuationsoperator um sie herum diagonalisieren und die Floquet-Theorie auf ihre Eigenfrequenzen und normale Modi anwenden. Mit Ausnahme der exakt lösbaren $SO(4)$-Singulett-Perturbation, die linear marginal stabil, aber nichtlinear begrenzt ist, wachsen generische Normalmoden aufgrund von Resonanzeffekten häufig exponentiell an. Selbst bei sehr hohen Energien werden alle kosmischen Yang-Mills-Hintergründe linear instabil gemacht.

\vspace{10pt}
{\bf Schlüsselwörter:} Yang--Mills-Theorie, vierdimensionaler de Sitter-Raum, elektromagnetische Knoten, kosmische Eichfelder.
\end{center}

\clearpage
%----------------------------------------------------------------------------------------
%	ABSTRACT PAGE
%----------------------------------------------------------------------------------------

\begin{abstract}
\addchaptertocentry{\abstractname} 
\justifying
This doctoral work deals with the analysis of some Yang--Mills solutions on $4$-dimensional de Sitter space $\diff{S}_4$. The conformal equivalence of this space with a finite Lorentzian cylinder over the $3$-sphere $S^3$ and also with parts of Minkowski space---together with the fact that Yang--Mills theory is conformally invariant in $4$-dimensional spacetime---has recently led to the discovery of a family of rational knotted electromagnetic field configurations. These ``basis-knot" solutions of the Maxwell equations, aka Abelian Yang--Mills theory, are labelled with the hyperspherical harmonics $Y_{j,m,n}$ of the $S^3$ and have nice properties such as finite-energy, finite-action and presence of a conserved topological quantity called helicity. Their field lines form closed knotted loops in the $3$-dimensional Euclidean space. Moreover, in the non-Abelian case of the gauge group $SU(2)$ there exist time-dependent solutions of Yang--Mills equation on $\diff{S}_4$ in terms of Jacobi elliptic functions that are of cosmological significance. These might play a role in early-universe cosmology for stabilizing the symmetric Higgs vacuum.

We study symmetry aspects of these electromagnetic knot configurations and compute all conserved charges for the conformal group $SO(2,4)$ associated with the $4$-dimensional Minkowski space for a complex linear combination of these basis solutions for a fixed $j$. Consideration of such complex linear combination is important because these rational basis solutions can be used to generate any finite energy field configuration in this way; we demonstrate this fact with some known results for certain modified Hopf--Ran\~{a}da knots. We find that the scalar charges either vanish or are proportional to the energy. For the non-vanishing vector charges we find a nice geometric structure that facilitates computation of their spherical components as well. We also find that helicity is related to the energy. Moreover, we characterize the subspace of null fields and present an expression for the electromagnetic flux at null infinity that matches with the total energy, thus validating the energy conservation. Finally, we investigate the trajectories of point charges in the background of such basis-knot configurations. To this end, we find a variety of behaviors depending on the
field configuration and the parameter set used. This includes an acceleration of particles by the electromagnetic field from rest to ultrarelativistic speeds, a quick convergence of their trajectories into a few narrow cones asymptotically for
sufficiently high value of the coupling, and a pronounced twisting and turning of trajectories in a coherent fashion.

We analyze the linear stability of the $SU(2)$ ``cosmic Yang--Mills fields” against general gauge-field perturbations while keeping the metric frozen, by diagonalizing the (time-dependent) Yang–Mills fluctuation operator around them and applying Floquet theory to its eigenfrequencies and normal modes. Except for the exactly solvable SO(4) singlet perturbation, which is found to be marginally stable linearly but bounded nonlinearly, generic normal modes often grow exponentially due to resonance effects. Even at very high energies, all cosmic Yang–Mills backgrounds are rendered linearly unstable.

\vspace{10pt}

\end{abstract}

\cleardoublepage
%----------------------------------------------------------------------------------------
%	ACKNOWLEDGEMENTS
%----------------------------------------------------------------------------------------
\justifying

\begin{acknowledgements}
\addchaptertocentry{\acknowledgementname} 
First, I would like to thank my supervisor, Olaf Lechtenfeld, for his constant support and encouragements at all stages of my PhD in his research group. He was always available for discussions and clarifications and, over the years, has propelled me to think critically for myself on a given research question. He has also very kindly helped me to improve various manuscripts not just related to our joint projects but also otherwise. I also thank him for suggesting the wonderful ``Saalburg Summer Schools" where I learned a lot about the state of the art in various research areas. 

I would also like to thank Biswajit Chakraborty who helped me at a critical juncture of my career when I made a transition to theoretical physics and who introduced me to the beautiful world of noncommutative geometry. I am also thankful to many teachers like Indresh Parihar, Diwakar Nath Jha, Anand Kumar (Super 30), Anand Dasgupta and Sunandan Gangopadhyay for their wonderful mentoring at various stages of learning and making physics/mathematics so interesting for me. 

I am grateful to Domenico Giulini and Gleb Arutyunov for revieing my thesis and suggesting valuable improvements. I thank Rolf Haug for agreeing to be the head of my PhD committee. 

Furthermore, I thank my colleagues Gleb Zhilin, Gabriel Picanço Costa, Joshua Cork and Till Bargheer for several interesting and valuable discussions (both academic and otherwise) at ITP (Institute for Theoretical Physics), Leibniz University Hannover (LUH).

I am grateful  to the DAAD (Deutscher Akademischer Austauschdienst) for the doctoral research grant 57381412 that enabled me to pursue my PhD research work at LUH. I am also grateful to the ITP for providing excellent working environment and equipments. I am thankful to Tanja Wießner and Brigitte Weskamp for bureaucratic support. 

Last, but not the least, I thank my family and friends for all their love and support. The culmination of this thesis is also a testament to their patience and hard work. I would attribute my keen interest in mathematics from an early age to the teaching environment at home created by my father (being a mathematics teacher).

\end{acknowledgements}

\cleardoublepage
%----------------------------------------------------------------------------------------
%	LIST OF CONTENTS/FIGURES/TABLES PAGES
%----------------------------------------------------------------------------------------

{\hypersetup{linkcolor=violet} \tableofcontents} % Prints the main table of contents

\mainmatter % Begin numeric (1,2,3...) page numbering

\pagestyle{thesis} 

\include{Chapters/Chapter1}

\include{Chapters/Chapter2}

\include{Chapters/Chapter3}

\include{Chapters/Chapter4}

\include{Chapters/Chapter5}

\include{Chapters/Chapter6}

\include{Chapters/Chapter7} 

%----------------------------------------------------------------------------------------
%	THESIS CONTENT - APPENDICES
%----------------------------------------------------------------------------------------

\appendix 

\include{Appendices/AppendixA}
\include{Appendices/AppendixB}
\include{Appendices/AppendixC}
%\include{Appendices/AppendixD}
%----------------------------------------------------------------------------------------
%	BIBLIOGRAPHY
%----------------------------------------------------------------------------------------

\printbibliography
\clearpage
%----------------------------------------------------------------------------------------

% CURRICULUM VITAE

%----------------------------------------------------------------------------------------
\pagestyle{empty}

\begin{center}
    {\Large \bf Curriculum Vitae}
\end{center}

\vspace{1cm}

1. {\bf Personal}

\begin{tabular}{ll}
Name & Kaushlendra Kumar \\
Date of Birth & 06/05/1993 \\
Place of Birth & Varanasi, India \\
Nationality & Indian \\
Email & kaushal.kumar224@gmail.com \\
Website & \href{https://k-kumar.netlify.app/}{https://k-kumar.netlify.app/}
\end{tabular}

\vspace{5pt}
2. {\bf Education.}

\begin{tabular}{L{4cm}L{11cm}}
    10/2018 -- 09/2022 & PhD physics, Institute for Theoretical Physics, Leibniz University Hannover.   Examination date: 04.05.2022 \\[0.6cm]
    07/2015 -- 05/2018 & MSc Physics, Indian Institute of Science Education and Research, Kolkata.   Examination date: 20.07.2017 \\[0.6cm]
    07/2010 -- 05/2015 & BTech \& MTech Biochemical Engineering, Indian Institute of Technology, BHU, Varanasi. Examination date: 27.05.2015 \\[0.6cm]
    05/2010 & Senior School Certificate, science stream (Physics, Chemistry, Mathematics), Jesus \& Mary Academy, Darbhanga. \\[0.6cm]
    05/2008 & Secondary School Certificate, Jawahar Navodaya Vidayalaya, Piprakothi, East Champaran.
\end{tabular}

\vspace{5pt}
3. {\bf List of publications.} \\[4pt]
3.1  {\it Refereed.} 
\begin{enumerate}
    \item K. Kumar, O. Lechtenfeld and G. Pican{\c c}o Costa, Trajectories of charged particles in knotted electromagnetic fields, {\it J. Phys. A: Math. Theor.} {\bf 55} (2022) 315401.
    \item L. Hantzko, K. Kumar and G. Pican{\c c}o Costa, Conserved charges for rational electromagnetic knots, {\it Eur. Phys. J. Plus} {\bf 137} (2022) 407.
    \item K. Kumar, O. Lechtenfeld and G. Pican{\c c}o Costa, Instability of cosmic Yang--Mills fields, {\it Nucl. Phys. B} {\bf 973} (2021) 115583.
    \item K. Kumar and O. Lechtenfeld, On rational electromagnetic fields, {\it Phys. Lett. A} {\bf 384} (2020) 126445.
    \item K. Kumar and B. Chakraborty, Spectral distances on the doubled Moyal plane using Dirac eigenspinors,
    {\it Phys. Rev. D} {\bf 97} (2018) 086019.
    \item Y.C. Devi, K. Kumar, B. Chakraborty and F.G. Scholtz, Revisiting Connes' finite spectral distance on noncommutative spaces: Moyal plane and fuzzy sphere, {\it Int. J. Geom. Methods Mod. Phys.} {\bf 15} (2018) 1850204.
    \item K. Kumar, S. Prajapat and B. Chakraborty, On the role of Schwinger's SU(2) generators for simple harmonic oscillator in 2D Moyal plane, {\it Eur. Phys. J. Plus} {\bf 130} (2015) 120.
\end{enumerate}

3.2 {\it Preprint.}
\begin{itemize}
    \item K. Kumar, O. Lechtenfeld, G. Pican{\c c}o Costa and J. R{\"o}hrig, Yang--Mills solutions on Minkowski space via non-compact coset spaces, arXiv:2206.12009 [hep-th].
\end{itemize}

\end{document}

%% file: Chapters/Chapter1.tex
\chapter{Introduction} 
\label{Chapter1} 
\def\={\ =\ }
\justifying

The $4$-dimensional de Sitter space $\diff{S}_4$ plays an important role in gravity. It is one of the three (topological) types\footnote{There are $3$ types of FLRW spacetime viz. Minkowski space ($\kappa{=}0$), de Sitter space ($\kappa{=}1$), and Anti-de Sitter space ($\kappa{=}{-1}$) according to the global topology of the backgroud $3$-space (labelled by $\kappa$).} of Friedmann--Lemaître--Robertson--Walker (FLRW) spacetime that can model a homogeneous and isotropic universe like ours (at distance scales of 100 Mpsec). This has a positive global scalar curvature of the underlying $3$-space (viewed as global foliation) that is consistent with the observed positive cosmological constant $\Lambda$ aka {\it dark energy} that is argued to be fueling the accelerated expansion of our universe. It is believed that our universe is asymptotically de Sitter which means in the future, when the dark energy dominates, our universe would become de Sitter.

Gauge theory, in particular Yang--Mills theory, is of central importance in the classical description of fundamental forces of nature like electromagnetism, weak and strong nuclear forces. Classical Yang--Mills theory also has applications in other physics areas such as QCD confinement of high energy physics and spin-orbit interaction in condensed matter physics. Even gravity can be understood as a gauge theory. It is, therefore, natural to seek solutions of Yang--Mills theory on four-dimensional de Sitter space. Furthermore, owing to the conformal relation of $\diff{S}_4$ with Minkowski space $\R^{1,3}$, it turns out that even Abelian Yang--Mills theory aka electromagnetism studied on the former has nice application since the solutions can be pulled back to Minkowski space (of our laboratory) owing to the conformal invariance of the Yang--Mills theory in $4$-dimensions.

\vspace{12pt}
%----------------------------------------------------------------------------------------
\section{Electromagnetic knots}
\vspace{2pt}
Theoretical discovery of electromagnetic knots dates back to 1989 when Rañada \cite{Ranada89} constructed them using the Hopf map. These finite-energy finite-action vacuum solutions of Maxwell's equations are constructed from a pair of complex scalar fields $\phi$ and $\theta$ on the $4$-dimensional spacetime where the $3$-space is compatified to $S^3$ with the addition of a point at infinity. These solutions are thus characterised by a topological quantity called the Hopf index of the following Hopf map ($A=1,2,3,4$):
\begin{equation}
    h:\ S^3 \to S^2\ ,\quad \{ \omega_{_A} \} \mapsto \{x_1,x_2,x_3\}
\end{equation}
where $S^2$, arising from the compactification of the complex plane $\mathds{C}$, has coordinates $x_i$, satisfying $x_1^2 + x_2^2 + x_3^2 = 1$, that are constructed from the $S^3$ coordinates $\omega_{_A}$, satisfying $\omega_1^2 + \omega_2^2 + \omega_3^2 + \omega_4^2 = 1$, as follows
\begin{equation}
    x_1\ :=\ 2(\omega_1\omega_2 + \omega_3\omega_4)\ ,\ x_2\ :=\ 2(\omega_1\omega_4 - \omega_2\omega_3)\ ,\ x_3\ :=\ (\omega_1^2+\omega_3^2) - (\omega_2^2 + \omega_4^2)\ .
\end{equation}
The level curves of these complex-valued functions $\theta$ and $\phi$ are identified with electric and magnetic field lines. Several other approaches to construct such electromagnetic knots have been developed since then such as  Bateman's complex Euler potentials, conformal inversion and Penrose twistors (see \cite{ABT17} for a full review).

Apart from these there exists another way of constructing these knotted electromagnetic fields via the conformal correspondence between de Sitter space $\diff{S}_4$ and Minkowski space $\R^{1,3}$, while passing through a finite Lorentzian $S^3$-cylinder, as was shown in \cite{LZ17}. In this method, one obtains a complete family\footnote{In the sense that any given finite-energy rational Maxwell solution can be expanded in terms of these basis configurations.} of such electromagnetic knotted configurations that are labelled with hyperspherical harmonics $Y_{j;m,n}$ of the $3$-sphere. The de Sitter space enjoys a larger symmetry group in $SO(1,4)$ whose subgroup $SO(4)\cong (SU(2)\times SU(2))/\mathds{Z}_2$ is made use of in this construction by working with the $S^3$-cylinder. Here one employs the right-action of $SU(2)$---the group manifold of $S^3$---on the $3$-sphere to write down the gauge field in terms of the left-invariant one-forms of $S^3$. The resulting Maxwell's equation can be solved analytically and are then pulled back to the Minkowski space using the conformal map. This ``de Sitter" method of construction has advantage over the others because of the $SO(4)$ covariant treatment of Maxwell theory. 

Knotted electromagnetic fields might become important for future applications because of their unique topological properties. It is, therefore, important to seek experimental settings to generate those fields and to study scenarios with them. Irvine and Bouwmeester \cite{Irvine08} discuss the generation of knotted fields using Laguerre--Gaussian beams and predict potential applications in atomic particle trapping, the manipulation of cold atomic ensembles, helicity injection for plasma confinement, and in the generation of soliton-like solutions in a nonlinear medium. Moreover, laser beams with knotted polarization singularities were recently employed to produce some simple knotted field configurations, including the one with figure-$8$ topology in the lab \cite{LSMetal18}.

\vspace{12pt}
%----------------------------------------------------------------------------------------
\section{Cosmic $SU(2)$ Yang--Mills fields}
\vspace{2pt}

Finding analytic solution to the Einstein--Yang--Mills system of equations arising from the following action (without topological term $\sim F\we F$)
\begin{equation}
    S \= S_{YM} + S_{EH} \= \frac{1}{2\kappa}\int_M \diff^4x \sqrt{-g}\,(R+2\Lambda) + \frac{1}{8g^2}\int_M \mathrm{Tr}(F\we *F)\ ,
\end{equation}
where $\Lambda$ is the cosmological term and $\kappa,g$ are coupling constants, is not possible in general. There is, however, one scenario where such a solution can be obtained and that is FLRW cosmology. Here the Yang--Mills equation decouple from Einstein equation due to the conformal invariance of the former in $4$-dimensional spacetime $M$. This means that given a solution of Yang--Mills equation in one of these FLRW spacetime, the corresponding scale factor can be obtained via Friedmann equations. Such a solution with finite energy and action do exist for the de Sitter case together with the $SU(2)$ gauge group \cite{AFF76,Luescher77,Schechter77}. 

A {\it crucial} ingredient of the Standard Model of cosmology called inflation can also be tackled with homogeneous and isotropic non-Abelian Yang--Mills fields in the Minkowski type (spatially flat) FLRW background in theories of gauge-flation or chromo-natural inflation (see \cite{MSS13} for a review). Another more minimalistic approach towards tackling this issue was recently put forth by Daniel Friedan \cite{Friedan}. He considers a coupled Einstein--Yang--Mills--Higgs system where a rapidly oscillating isotropic $SU(2)$ gauge field stabilizes the symmetric Higgs vacuum in a de Sitter type (spatially closed) FLRW spacetime. 

Based on these, it is only natural to analyze the stability behaviour of such ``cosmic Yang--Mills fields" under generic linear perturbation of the Yang--Mills field equation. For the gauge-flation scenario such an analysis has been done before, but for the later scenario there only exits result for spin-$j{=}0$ case of these $SU(2)$ gauge fields \cite{Hosotani}.

In light of this, we present here a complete stability analysis of these $SU(2)$ solutions in a closed FLRW universe. This analysis is in contrast with that of the guage-flation one where conformal invariance is broken; our homogeneous and isotropic gauge field would give rise to inhomogeneous Yang--Mills fields on flat FLRW spacetime. For the sake of simplicity we keep the background metric fixed in this perturbation analysis. While this does mean that our analysis is still partial, we can argue for the relevance of our analysis as follows:
\begin{enumerate}[(a)]
    \item In $4$-dimensional spacetime, the gauge fields decouple with the background metric. Therefore, fluctuation of the latter does not affect the gauge fields of our theory.
    \item The fluctuation of the gauge fields is extremely rapid when compared to the evolution of the background metric and its subsequent fluctuations. The former, therefore, do not experience any significant effect due to such slow metric fluctuations.
\end{enumerate}

The only known family of finite-energy $SU(2)$ Yang-Mills field configuration on FLRW spacetime are obtained, in an efficient manner, by employing the following conformal correspondence between de Sitter space and the cylinder $\Ical\times S^3$ for $\Ical := (0,\pi)$. This conformal map arises via a temporal reparametrization
and Weyl rescaling~\cite{Hosotani,Volkov,ILP17prl,Friedan},
\begin{equation} \label{deSittermetric}
\begin{aligned}
   \diff s_{\textrm{dS}_4}^2 &\= -\diff t^2 + \ell^2\cosh^2\!\sfrac{t}{\ell}\,\diff\Omega_3^2
   \= \sfrac{\ell^2}{\sin^2\!\tau} \bigl(-\diff\tau^2 + \diff\Omega_3^2\bigr) \\
   &\qquad \for t\in(-\infty,+\infty)\ \Leftrightarrow\ \tau \in (0,\pi)\ ,
\end{aligned}
\end{equation}
where $\diff\Omega_3^{\ 2}$ is the round metric on $S^3$, and $\ell$ is the de Sitter radius. We observe that the relation between the conformal time $\tau$ and co-moving time~$t$ in \eqref{deSittermetric} fixes the cosmological constant to~$\Lambda=3/\ell^2$. At this point, we can employ an $S^3$-symmetric ansatz for the gauge field by noting that $SU(2)$ is the group manifold of $S^3$. This yields an ODE for some scalar function $\psi(\tau)$ parametrized by the conformal time, that is nothing but a Newton's equation for a classical point particle under the influence of a double-well potential
\begin{equation} \label{Vpot}
V(\psi) \= \sfrac12 (\psi^2-1)^2\ .
\end{equation}
The solution for these anharmonic oscillators are well known in terms of Jacobi elliptic functions that depend on time. Although these solutions exists for all three FLRW metrics, only spatially closed one admits an isotropic solution under the above conformal transformation. For de Sitter type FLRW spacetime we have
\begin{equation} \label{FLRW}
\begin{aligned}
  \diff s^2 &\= -\diff t^2 + a(t)^2\,\diff\Omega_3^{\ 2}
  \= a(\tau)^2\bigl(-\diff\tau^2 + \diff\Omega_3^{\ 2}\bigr) \\
  &\qquad\for t\in(0,t_{\textrm{max}})\ \Leftrightarrow\ \tau \in \Ical \equiv (0,T')\ ,
\end{aligned}
\end{equation}
where we impose a big-bang initial condition~$a(0){=}0$, so that
\begin{equation} 
\diff\tau \= \frac{\diff t}{a(t)} \qquad\textrm{with}\qquad 
\tau(t{=}0) = 0 \quad\und\quad \tau(t{=}t_{\textrm{max}}) =: T'<\infty\ .
\end{equation}
The lifetime $t_{\textrm{max}}$ of the universe 
can be infinite (big rip, $a(t_{\textrm{max}}){=}\infty$) 
or finite (big crunch, $a(t_{\textrm{max}}){=}0$).
Moreover, bouncing cosmologies as in (\ref{deSittermetric}) are also allowed but will not be pursued here. 

We notice that the contribution of these $SO(4)$-symmetric Yang--Mills fields, and their stress-energy tensor, in the one-way coupling with the background de Sitter FLRW metric via Friedmann equation (as discussed before) is such that it only modifies the scale factor $a(\tau)$. It is well known that the equation of motion governing this scale factor arises as a Newton's equation with the following (cosmological) potential 
\begin{equation} \label{Wpot}
W(a) \= \sfrac12 a^2 - \sfrac{\Lambda}{6} a^4\ ,
\end{equation}
which is another anharmonic oscillator (although inverted). 

The pair of solutions $(\psi,a)$ corresponding to \eqref{Vpot} and \eqref{Wpot} yields an exact classical Einstein--Yang--Mills configuration. The conserved mechanical energy $E$ for $\psi$ fixes the same i.e. $E'$ for $a$ via the Wheeler--DeWitt constraint
\begin{equation}
    E' \= \epsilon E\ ;\quad E\ :=\ \frac{1}{2}\dot{\psi}^2 + V(\psi) \und E'\ :=\ \frac{1}{2}\dot{a}^2 + W(a)\ ,
\end{equation}
where the overdot denotes a derivative with respect to conformal time and $\epsilon$ depends on coupling constants. 

One can also introduce a complex scaler Higgs field $\phi$ in the fundamental SU(2) representation to the Standard Model of cosmology with Higgs potential
\begin{equation}
U(\phi) \= \sfrac12\,\lambda^2\,\bigl(\phi^\+\phi-\sfrac12 v^2\bigr)^2\ ,
\end{equation}
where $v/\sqrt{2}$ is the Higgs vev and $\lambda v$ is the Higgs mass. It turns out that imposing an $SO(4)$-invariance makes the Higgs field $\phi\equiv0$, which provides us with a definite positive cosmological constant of
\begin{equation}
\Lambda \= \kappa\,U(0) \= \sfrac18\,\kappa\,\lambda^2 v^4\ ,
\end{equation}
where $\kappa$ is the gravitational coupling.
The full Einstein--Yang--Mills--Higgs action (in standard notation),
\begin{equation}
S \= \int\!\diff^4 x\ \sqrt{-g}\ \Bigl\{ \sfrac{1}{2\kappa} R + \sfrac{1}{8g^2} \tr F_{\mu\nu}F^{\mu\nu}
-D_\mu\phi^\+ D^\mu\phi - U(\phi) \Bigr\}\ ,
\end{equation}
reduces in the SO(4)-invariant sector to
\begin{equation}
S[a,\psi,\Lambda] \= 12\pi^2 \int_0^{T'}\!\!\diff\tau\ \Bigl\{
\sfrac{1}{\kappa} \bigl(-\sfrac12\dot{a}^2+W(a)\bigr) + 
\sfrac{1}{2g^2} \bigl(\sfrac12\dot{\psi}^2-V(\psi)\bigr) \Bigr\}\ ,
\end{equation}
where $g$ is the gauge coupling.

If the Yang--Mills energy $E$ is large enough it could propel an eternal expansion of the universe that is accompanied by rapid fluctuations of the gauge field. The coupling of this Yang--Mills field with the Higgs field stabilizes the symmetric vacuum $\phi{\equiv}0$ at the local maximum of~$U$ through a parametric resonance effect, as long as $a$ is not too large. Eventually, when $a$ exceeds a critical value of electro-weak symmetry breaking scale $a_{\textrm{EW}}$, the Higgs field will begin to roll down towards a minimum of~$U$, thus breaking the SO(4) symmetry. The corresponding time $t_{\textrm{EW}}$ signifies the electroweak phase transition in the early universe. This is a rather unconventional scenario put forward recently by Friedan~\cite{Friedan}. 

\vspace{12pt}
%----------------------------------------------------------------------------------------
\section{Outline and summary of results}
\vspace{2pt}

In the next two chapters we review the mathematical preliminaries that builds up gauge theory. In Chapter \ref{Chapter2} we present a brief but thorough review of the mathematics of background geometry such as manifold, fibre bundles, etc. and symmetry in physics such as Lie groups, Lie algebra and their representations. Next, in Chapter \ref{Chapter3} we review the construction of gauge theory via principal bundle formalism.

In Chapter \ref{Chapter4} we discuss the geometry of and calculus on the $3$-sphere apart from demonstrating the conformal equivalence of the $S^3$-cylinder with Minkowski space. We then present Yang--Mills equation for a $SO(4)$-asymmetric $SU(2)$ Yang--Mills theory on the $S^3$-cylinder and discuss its two limiting cases where analytic solutions can be obtained. 

We present the construction of electromagnetic knotted field configurations in Chapter \ref{Chapter5} and study their symmetry feature and other properties. We analyze the effect of the de Sitter group $SO(1,4)$, i.e. the isometry group of $\diff{S}_4$, on these solutions. To that end, we demonstrate the emergence of the Poincare group $ISO(1,3)$ of $\R^{1,3}$ from $SO(1,4)$ in the limit $\ell \to \infty$. We observe that only the subgroup $SO(3)$ is common in the two cases. 

We then proceed, also in Chapter \ref{Chapter5}, to compute all the Noether charges associated with the conformal group $SO(2,4)$ viz. energy, momentum, angular momentum, boost, dilatation and special conformal transformations (SCT) for a linear combination---in terms of complex coefficients $\Lambda_{j,m,n}$---of these basis-knot configurations. These conserved charge densities are evaluated on de Sitter space at $\tau{=}0$ where considerable simplifications occur demonstrating the usefulness of this ``de Sitter method". We find that the dilatation vanishes while the scalar SCT charge $V_0$ is proportional to the energy $E$. Furthermore, the boosts $K_i$ vanish and the vector SCT charges $V_i$ are proportional to the momenta $P_i$. Interestingly, for the vector charge densities viz. momenta $p_i$, angular momenta $l_i$ and vector SCT $v_i$ we find that the one-form, e.g.~$p_i\diff x^i$ constructed on the spatial slice $\R^3 \hookrightarrow \R^{1,3}$ is proportional to a similar one on de Sitter space. This correspondence allows us to compute additional charges $(p_r,p_\theta,p_\phi)$ by the action of such one-forms on spherical vector fields $(\pa_r,\pa_\theta,\pa_\phi)$. At $j{=}0$ it turns out that there are only four independent non-zero charges: the energy $E$ and momenta $P_i$. The situation for higher spin $j$ is more complicated, but some of the components of the charges in spherical coordinates are found to vanish for arbitrary $j$. The action of $so(3)$ generators $\mathcal{D}_a$ on the indices of $\Lambda_{j,m,n}$ can easily be obtained for a fixed $j$ owing to the $SO(4)$ isometry. This allows for an action of these generators on the charges. For (the Cartesian components of) the vector charges this action is found to inherit the original $so(3)$ Lie algebraic structure, as expected. We also compute the correct coefficients $\Lambda_{0;0,n}$ corresponding to two interesting generalisations of the Hopfian solution obtained via Bateman's construction in \cite{HSS15}, which allow us to validate our generic formulae of these charges. We also demonstrate the relationship of energy with the conserved helicity.

Furthermore, in Chapter \ref{Chapter5} we characterize the moduli space of null solutions which turns out to be a complete-intersection projective complex variety of complex dimension $2j{+}1$. We also demonstrate how the energy flux is radiated to infinity with an energy profile that is concentrated along the lightcone situated at the origin (selected by these solutions). Finally, we study the trajectories of multiple identical charged particles in the background of these basis knot configurations. We employ several initial conditions for these charged particles and find interesting features of the trajectories like coherent twisting, ultrarelativistic acceleration of particles starting from rest and a quick convergence of their trajectories into a few narrow cones asymptotically for sufficiently high value of the coupling. 

In Chapter \ref{Chapter6} we first review the classical configurations $(A_\mu,g_{\mu\nu})$ 
in terms of Newtonian solutions $(\psi,a)$ for the anharmonic oscillator pair~$(V,W)$. We then investigate arbitrary small perturbations of the gauge field departing from the time-dependent background~$A_\mu$ parametrized by the ``gauge energy'' $E$. Later on we linearize the Yang--Mills equation around it and diagonalize the fluctuation operator to obtain a spectrum of time-dependent natural frequencies. To decide about the linear stability of the cosmic Yang--Mills configurations we have to analyze the long-time behavior of the solutions to Hill's equation for all these normal modes. To that end, we employ Floquet theory to learn that their growth rate is determined 
by the stroboscopic map or monodromy, which is easily computed numerically for any given mode. 
We do so for a number of low-frequency normal modes and find, when varying~$E$, an alternating sequence of stable (bounded) and unstable (exponentially growing) fluctuations. The unstable bands roughly correspond to the parametric resonance frequencies.
With growing ``gauge energy'' the runaway perturbation modes become more prominent,
and some of them persist in the infinite-energy limit, where we detect universal natural frequencies and monodromies. A special role is played by the SO(4)-invariant fluctuation of $j{=}0$ mode, 
which merely shifts the parameter~$E$ of the background. We treat it exactly and beyond the linear regime. This ``singlet'' mode turns out to be marginally stable, i.e. it has a vanishing Lyapunov exponent. Its linear growth, however, gets limited by nonlinear effects of the full fluctuation equation, whose analytic solutions exhibit wave beat behavior.

Finally we present a brief summary of our work in Chapter \ref{Chapter7} and present the future outlook. We also collect several relevant data in various appendices and present a direct map between the cylinder and Minkowski space via Carter--Penrose transformation, avoiding the de Sitter detour, in appendix \ref{appendCPtransf}.

%% file: Chapters/Chapter2.tex
\chapter{Geometry \& Symmetry}
\label{Chapter2}

%----------------------------------------------------------------------------------------

\newcommand{\ex}[1]{{\bf Example #1}}
\newcommand{\defn}[1]{\noindent{\bf Definition #1}}
\newcommand{\vecF}[1]{\mathfrak{X}(#1)}
\newcommand{\rem}[1]{\noindent{\bf Remark #1}}
\newcommand{\thm}[1]{\noindent{\bf Theorem #1}}

%----------------------------------------------------------------------------------------
\justifying
Two of the most important concepts that play a fundamental role in physics are geometry of background space and presence of symmetries. They have well understood mathematical foundation in differential geometry and Lie groups/algebras respectively. Here we present a short exposition of these mathematical topics that is essential towards building the subsequent Gauge theory. The following contents are built upon some basic mathematical structures like vector spaces, groups and topological spaces all of which can be found in \cite{Nakahara}. We refrain from presenting proofs of the statements in this chapter and suggest \cite{Baez&Muniain,Bleecker}, which are the source for most of the contents in this chapter. For details on Lie groups/algebras and their representations we refer to classic texts \cite{wybourne} and \cite{Hall}.

\vspace{12pt}
%----------------------------------------------------------------------------------------
\section{Manifolds}
\vspace{2pt}
\rem{2.1.1} The idea of a Manifold generalizes the notion of differentiation, or more precisely calculus on $\R^n$, in the same way as a Topological space allows one to study the notion of continuity in a more abstract way. In this thesis, we will only be concerned with (finite-dimensional) smooth manifolds, and subsequently, smooth structures on it.

\defn{2.1.1} An n-dimensional {\it manifold} $M$ is a topological space\footnote{Some required technicalities like paracompactness and Haussdorffness have been assumed.} equipped with an atlas consisting of charts $(U_i,\phi_i)$ such that
\begin{itemize}
    \item $U_i$ are open sets and the maps $\phi_i$ are homeomorphism between $U_i$ and some open ball in $\R^n$, and
    \item the {\it transition maps} $\phi_2\circ \phi_1^{-1}:\ \phi_1(U_1\cap U_2) \rightarrow \phi_2(U_1\cap U_2)$ is infinitely differentiable for any two charts $(U_1,\phi_1)$ and $(U_2,\phi_2)$.
\end{itemize}

\ex{2.1.1} A nice example of a smooth manifold is the round $n$-sphere
\begin{equation}
    S^n := \lbrace {\bf x} \in \R^{n+1} ~|~ {\bf x\cdot x} = 1 \rbrace
\end{equation}
embedded in $\R^{n+1}$ that requires two charts: $U_N = S^n - \{(0,0,\ldots,1)\}$ and $U_S := S^n - \{(0,0,\ldots,-1)\}$, along-with their respective stereographic projections:
\begin{equation}
 \begin{aligned}
    \phi_N(\bf{x})\ &:=\ \left( \sfrac{x_1}{1-x^{n+1}},\sfrac{x_2}{1-x_{n+1}},\ldots,\sfrac{x_n}{1-x_{n+1}} \right)\und \\
    \phi_S(\bf{x})\ &:=\ \left( \sfrac{x_1}{1+x_{n+1}},\sfrac{x_2}{1+x_{n+1}},\ldots,\sfrac{x_n}{1+x_{n+1}} \right)
 \end{aligned}
\end{equation}
with $x_{n+1} = 1 - x_1^2 - x_2^2 - \ldots - x_n^2$.

\defn{2.1.2} Given a manifold $M$ a map $f: M \rightarrow \R$ is called {\it smooth} or $C^\infty$ if it is infinitely differentiable. The set of all such smooth maps are denoted as $C^\infty(M)$.

\defn{2.1.3} {\it Diffeomorphism} $f:M \rightarrow N$ between any two manifolds $M$ and $N$ is a bijection such that both $f$ and $f^{-1}$ are smooth. The set of all diffeomorphisms from $M$ to itself forms a group called $\mathrm{Diff}(M)$.

\rem{2.1.2} Notice here that the differentiability of such a map $f$ is decided using local charts: given a chart $(U,\phi)$ in $M$ containing $p\in U$ and another chart $(V,\psi)$ in $N$ containing $f(p)\in V$, one can differentiate the map $\psi\circ f\circ\phi^{\text -1}$ using standard calculus. A crucial notion in differential geometry is that of a tangent vector. It can intrinsically be defined in terms of a {\it curve} on a manifold.

\defn{2.1.4} A {\it curve} $\sigma :\ (-\epsilon,\epsilon) \subset \R \to M$ on a manifold $M$ is defined as a smooth map from some open interval of the real line to $M$.

\defn{2.1.5} The {\it pull-back of a map $f$} by another map $\phi$ is defined as
\begin{equation}
    \phi^*f\ :=\ f\circ\phi\ .
\end{equation}

\defn{2.1.6} A tangent to a curve $\sigma$ at a point $p=\sigma(0)$ on a manifold $M$ is defined by the map
\begin{equation}
    \sigma'(t):\ C^\infty(M) \rightarrow \R\ ,\quad \sigma'(t{=}0)[f]\ :=\ \sfrac{\pa}{\pa t}(\sigma^*f)\big|_{t=0}
\end{equation}
This idea can be generalized as follows.

\defn{2.1.7} A {\it tangent vector} at a point $m\in M$ is a map $v_m:\ C^\infty(M) \rightarrow \R$ that satisfies the following properties \begin{itemize}
    \item $v_m[f+g] \= v_m[f] + v_m[g]\quad  \forall\quad f,g \in C^\infty(M)$\ ,
    \item $v_m[a\,f] \= a\,v_m[f]\quad \forall\quad a\in \R\ \&\ f\in C^\infty(M)$\ , and
    \item $v_m[f\,g] \= f(p)\,v_m[g] + g(p)\,v_m[f] \quad\forall\quad f,g\in C^\infty(M)$.
\end{itemize}

\defn{2.1.8} The set of all such tangent vectors at $m\in M$ is known as the {\it tangent space} at $m$ and is denoted by $T_mM$, which forms a vector space over $\R$. 

\rem{2.1.3} This vector space is spanned by vectors $\big\{\sfrac{\scriptsize\pa}{\scriptsize\pa x^i}\big|_m \equiv \pa_i|_m \big\}$ where $x^i$ are the coordinate functions in some chart $(U, \phi)$ containing $m$, i.e. $\phi(m) = (x^1,x^2,\ldots,x^n)$ and whose actions is defined by 
\begin{equation}
    \sfrac{\scriptsize\pa}{\scriptsize\pa x^i}\big|_m[f] \ :=\ \sfrac{\scriptsize\pa (\phi^* f)}{\scriptsize\pa x^i}\quad \forall \quad f\in C^\infty(N)\ .
\end{equation}
The dimension of $T_mM$ is, therefore, equal to the dimension of the manifold $M$.

\defn{2.1.9} Given a map $\varphi: M \rightarrow N$ between two manifolds $M$ and $N$ and a vector $v\in T_mM$, the {\it push-forward} of $v$ by $\varphi$ is defined by
\begin{equation}
    \varphi_*v[f]\ :=\ v[\phi^*f]\quad \forall \quad f\in C^\infty(N)\ .
\end{equation}

\rem{2.1.4} Note that the push-forward induces a map between following tangent spaces:
\begin{equation}
    \phi_*:\ T_mM \rightarrow T_{h(m)}N
\end{equation}
for manifolds $M$ and $N$.

\defn{2.1.10} The vector space dual to $T_mM$ is called {\it cotangent space} and is denoted by $T^*_mM$.

\rem{2.1.5} The vector space $T^*_mM$ is spanned by covectors $\diff{x}^i|_m$ defined as 
\begin{equation}
    \langle \diff{x}^i|_m\, ,\, v\rangle\ :=\ v[x^i]\quad \forall \quad v \in T_mM
\end{equation}
and has the same dimension as $T_mM$.

\defn{2.1.11} The {\it pull-back} $\varphi^*\omega\in T^*_mM$ of a covector $\omega \in T^*_nN$ by a map $\varphi: M \rightarrow N$ between two manifolds $M$ and $N$ is defined via
\begin{equation}
    \langle\varphi^*\omega\, ,\, v \rangle\ :=\ \langle \omega\, , \, \varphi_*v \rangle
\end{equation}
with $n = \varphi(m)$.

\vspace{12pt}
%----------------------------------------------------------------------------------------
\section{Fibre bundles}
\vspace{2pt}
\rem{2.2.1} The theory of bundles has proved to be the correct way of studying classical gauge theories like general relativity and Yang-Mills theory. Here we shall confine ourselves to bundles constructed on/with manifolds.

\defn{2.2.1} A {\it bundle} is a triple $(E,M,\pi)$ consisting of a manifold $E$, {\it aka} the target space, a manifold $M$, {\it aka} the base space, and a continuous surjection $\pi: E \rightarrow M$, {\it aka} the projection map. It is denoted diagrammatically as 
\begin{tikzcd}
   &E \arrow[d, "\pi"] \\
   &M
\end{tikzcd}

\defn{2.2.2} A bundle $(E,M,\pi)$ is called a {\it fibre bundle} with typical fibre $F$ if the inverse image of all $p \in M$ under $\pi$ is isomorphic to some space $F$, i.e. $\pi^{-1}(p) \cong F$. If $F$ is a vector space then the bundle $(E,M,\pi)$ becomes a {\it vector bundle}.

\defn{2.2.3} A pair of fibre bundles $(E,M,\pi)$ and $(\widetilde{E},\widetilde{M},\widetilde{\pi})$ are called {\it isomorphic} (as bundles) if there exist a pair of diffeomorphisms $\varphi: E \rightarrow \widetilde{E}$ and $\psi: M \rightarrow \widetilde{M}$ such that $\widetilde{\pi}\circ \varphi = \psi\circ\pi$ and $\pi\circ\varphi^{-1} = \psi^{-1}\circ\widetilde{\pi}$. Diagrammatically this means that the following diagram, and its inverse, commutes 
\begin{tikzcd}
   E \arrow[r, "\varphi"] \arrow[d, "\pi"] &\widetilde{E} \arrow[d, "\widetilde{\pi}"] \\
   M \arrow[r, "\psi"] &\widetilde{M}
\end{tikzcd}

\defn{2.2.4} A vector bundle $(E,M,\pi)$ with typical fibre $F$ is called {\it locally trivial} if for any $U \subset M$ the induced bundle $(\pi^{-1}(U),U,\pi|_{\pi^{-1}(U)})$ is isomorphic to the product bundle $(U\times F,U,\pi_1)$, where $\pi_1$ is the projection in the first slot. The set $(U,\varphi)$\footnote{Notice, here, that $\psi=Id_U$.} is known as a {\it local trivialization} of the vector bundle. A {\it trivial} bundle is one where $E=M\times F$ and $\pi = \pi_1$ (projection in the first slot).

\rem{2.2.2} We shall only deal with locally trivial bundles here. A vector bundle $(E,M,\pi)$ will, sometimes, be simply denoted $E$. A classic example of a bundle that is not globally trivial is the following.

\ex{2.2.1} A M\"{o}bius strip $(E,S^1,\pi)$ with fibre $[0,1]$ is locally isomorphic\footnote{Meaning that they share local trivialization $(U,\varphi)$ for any $x\in U$.} to the trivial bundle $(S^1\times [0,1],S^1,\pi_1)$. The former, however, is not trivial as transporting any vector across a loop yields the corresponding vector, albeit inverted.

\defn{2.2.5} Given a fibre bundle $(E,M,\pi)$ with typical fibre $F$ and a pair of local trivializations $(U_i,\varphi_i)$ and $(U_j,\varphi_j)$ with $U_i\cup U_j \neq 0$ on it, we obtain {\it transition functions} $g_{ij}(x)$ with $x\in U_i\cap U_j$ of the vector bundle $(U_i\cap U_j)\times F$ by realizing that $\varphi_j\circ \varphi_i(x,v) = (x,g_{ij}(x)v)$ for any vector $v\in F$. The set of such transition functions $g_{ij}(x)$ forms a group $G \subset End(F)$ called the {\it structure group}.

\defn{2.2.6} A (local) {\it section} $\sigma$ (for some local trivialization $(U,\varphi)$) of a fibre bundle $(E,M,\pi)$ is a smooth map $\sigma: (U) M \rightarrow E$ such that $\pi\circ\sigma = (Id_U) Id_M$. The set of all such (local) global sections are denoted $(\Gamma^\infty(U,E))$ $\Gamma^\infty(M,E)$. The space of global sections is also denoted, more simply, as $\Gamma^\infty(E)$. 

\vspace{12pt}
%----------------------------------------------------------------------------------------
\section{Lie groups}
\vspace{2pt}
\rem{2.3.1} The notion of symmetry in modern physics is analyzed with tools from the theory of Lie groups and Lie algebras. We will only consider finite, matrix Lie groups in what follows.

\defn{2.3.1} A {\it Lie group} ({\it aka} continuous group) $G$ is both a group and a smooth, finite-dimensional manifold where the group operation, $\cdot: G\times G \rightarrow G$, as well as the inversion map, $^{-1}: G \rightarrow G$ satisfying $g^{-1}\cdot g = g\cdot g^{-1}=e~\forall~ g\in G$ with identity element $e$, are smooth.

\rem{2.3.2} We will omit the group multiplication symbol $\cdot$ and use $\mathds{1}$ interchangeably with $e$ for the matrix Lie groups (to be considered below) from now onward.

\defn{2.3.2} A Lie group homomorphism $\varphi: G \rightarrow H$ between Lie groups $G$ and $H$ is a smooth map that is also a group homomorphism. The map $\varphi$ becomes an isomorphism if it is bijective and its inverse $\varphi^{-1}$ is smooth; the Lie groups $G$ and $H$ then becomes isomorphic.

\defn{2.3.3} A {\it positive-definite inner product} is a bilinear map $\<\cdot,\cdot\>: V\times V \xrightarrow{\sim} \R$ for a vector space $V\ni x,y$ that is
\begin{enumerate}[(a)]
    \item symmetric: $\<x,y\>=\<y,x\>$\ ,
    \item positive-definite: $\<x,x\>>0$\ , and
    \item nondegenerate: i.e. if $\<x,y\>=0$ for all $y \implies x=0$\ .
\end{enumerate}
If $\<x,x\>\not>0$ then the inner product is called {\it indefinite}.

\rem{2.3.3} Most of the important Lie groups in physics emerges as matrix Lie groups by considering invertible linear maps $V\xrightarrow{\sim}V$ on some finite-dimensional vector space $V$ that preserves a given inner product on $V$. Such {\it general linear groups} are denoted $GL(V)$. Furthermore, these are Lie subgroups\footnote{A Lie subgroup $H\subset G$ is a subgroup as well as a submanifold of $G$. It turns out to be a Lie group in itself.} of the general Linear groups $GL(n,\R)$ or $GL(n,{\mathds C})$ consisting of invertible linear maps on vector field $\R$ or $\mathds{C}$ respectively. This becomes evident when we consider Riemannian manifolds which contains a metric (see Section \ref{covectorFields}).

\ex{2.3.1} The special linear groups $SL(n,\R)$ (over $\R$) and $SL(n,{\mathds C})$ (over ${\mathds C}$) are $n\times n$ invertible matrices with unit determinant, i.e.
\begin{equation}
    SL(n,\R/{\mathds C}) \= \{A \in GL(n,\R/{\mathds C})~|~ detA = 1 \}\ .
\end{equation}

\ex{2.3.2} The Unitary group $U(n)$ is the set of $n\times n$ complex-valued matrices whose inverse is the same as its conjugate transpose:
\begin{equation}
    U(n) \= \{A \in GL(n,{\mathds C})~ |~ A^{-1} = \bar{A}^T\equiv A^\+ \}\ .
\end{equation}
The special unitary group is defined as
\begin{equation}
    SU(n) \= \{ A \in U(n)~ |~ detA = 1 \}\ .
\end{equation}

\rem{2.3.4} The special unitary groups $SU(n)$ preserves the standard inner product on ${\mathds C}^n \ni {\bf x},{\bf y}$ given by 
\begin{equation}
    \langle {\bf x}, {\bf y} \rangle_{{\mathds C}} \= \sum\limits_{i=1}^n\bar{x}_iy_i
\end{equation}
and also the norm of a given vector (induced from this inner product).

\ex{2.3.3} The orthogonal group $O(k,n)$ are the $(k+n)\times (k+n)$ matrices that preserve the following inner product (${\bf x},{\bf y} \in \R^{k+n} $):
\begin{equation}
    \langle {\bf x}, {\bf y} \rangle_{k,n}\ :=\ -x_1y_1 - \ldots - x_ky_k + x_{k+1}y_{k+1} +\ldots + x_{k+n}y_{k+n}\ .
\end{equation}
One can show that $A\in O(k,n)$ if and only if
\begin{equation}
    A^T\eta^{(k,n)} A \= \eta^{(k,n)}\ ,
\end{equation}
where $\eta^{(k,n)}$ is the diagonal matrix $diag(1,\ldots1,-1,\ldots,-1)$. The special orthogonal group is defined as
\begin{equation}
    SO(k,n)\ :=\ \{ A \in O(k,n)~|~ detA = 1\}\ .
\end{equation}
The orthogonal group $O(0,n)$ is denoted $O(n)$ and is more famously defined as
\begin{equation}
    O(n) := \{ A \in GL(n,\R) ~|~ A^{-1} = A^T \}\ .
\end{equation}
Similarly, the special orthogonal group $SO(0,n)$ is denoted as $SO(n)$.

\rem{2.3.5} The group $SO(n)$ consists of rotations in $n$-dimensions and is thus knows as isometry group of the $n$-sphere $S^n$. The group $O(n)$ consists of rotations as well as reflections. The groups $SO(1,3)$ and $SO(2,4)$ known, respectively, as the {\it Lorentz group} and the {\it Conformal group} are of special significance in physics. Another important group for us in this thesis is the {\it de Sitter group} $SO(1,4)$.

\ex{2.3.4} The {\it Poincar\'{e} group} or the {\it inhomogeneous Lorentz group} $ISO(1,3)$ consists of Lorentz transformation together with translations and is defined as
\begin{equation}
    ISO(1,3) := \{ I_{{\bf x}} A ~|~ A \in SO(1,3) \}\ ,
\end{equation}
where the translations $I_{{\bf x}}$ acts on a given vector ${\bf y} \in \R^{1,3}$ as
\begin{equation}
    I_{{\bf x}}{\bf y} = {\bf x} + {\bf y}\ .
\end{equation}

\rem{2.3.6} It can be shown that the group $SO(4)$ decomposes into two copies of $SU(2)$ and that there exist a $2{-}1$ ($2$ to $1$) homomorphism
\begin{equation}
    SU(2)\times SU(2) \xrightarrow{2{-}1} SO(4)\ .
\end{equation}
This is a rather generic fact that arises when one considers spin-groups (e.g. $Spin(4)$ here), which provide universal cover\footnote{This is a topological term that refers to a connected, simply connected group that projects down to the given group $G$ via a smooth surjection $\pi$ such that any open $U \in G$ lifts to a disjoint union $\pi^{-1}(U)$ whose members are isomorphic (individually) to $U$.} to some group (e.g. $SO(4)$ here). To this end, we note down the following relevant facts here:
\begin{equation}
    Spin(4)\cong SU(2)\times SU(2)\ ,\quad Spin(3)\cong SU(2)\ ,\ \mathrm{and}\quad Spin(1,3)\cong SL(2,{\mathds C})\ .
\end{equation}

\defn{2.3.4} The {\it left action} of a Lie group $G$ (aka left $G$-action) on a set $M$ is defined by the map
\begin{equation}
    \varphi_l :\ G\times M \rightarrow M\ ,\quad (g,m) \mapsto \varphi_l(g,m)\ \equiv\ gm
\end{equation}
that satisfies the following properties:
\begin{itemize}
    \item $em \= m$ for the identity element $e \in G$ and every $m\in M$, and
    \item $g_1(g_2m) \= (g_1g_2)m$ for all $g_1,g_2 \in G$ and $m \in M$.
\end{itemize}
The set $M$ is known as a {\it homogeneous space}, that splits into an {\it orbit space} $M/G$ of equivalence classes of orbits
\begin{equation}
    O_m\ :=\ \{ n \in M ~|~ \exists\ g \in G : n = \varphi_l(g,m) \}\ .
\end{equation}

\defn{2.3.5} The {\it left translations} or {\it left multiplications} of a Lie group $G$ is its diffeomorphism $l_g\in Diff(G)$ defined by
\begin{equation}
    l_g :\ G \rightarrow G\ ,\quad h \mapsto gh\ .
\end{equation}

\rem{2.3.7} An equivalent notion of {\it right action} defined by $\varphi_r(g,m) := mg$ also exits and is of prime importance for the principle $G$-bundles that we will discuss in the next chapter. It is important to note here that one can induce a right-action $\varphi_r$ from a given left-action $\varphi_l$ as follows:
\begin{equation}
    \varphi_r(g,m)\ :=\ \varphi_l(g^{-1},m) \quad \forall\quad m \in M\ . 
\end{equation}
Similarly, we have the notion of right translations $r_g$ for the Lie group $G$ but we will only deal with left translations here.

\defn{2.3.6} The {\it coset space} $G/H$ for a Lie group $G$ and its subgroup $H\subset G$ is defined as
\begin{equation}
    G/H\ :=\ \{ gH ~|~ g \in G \}\quad \mathrm{where}\quad gH\ :=\ \{ gh ~|~ h \in H \}\ .
\end{equation}

\rem{2.3.8} There exist a natural left $G$-action (and hence, a left $H$-action) on $G/H$ defined by
\begin{equation}
    \varphi_l(g',gH) = g'gH\quad \forall\quad g,g' \in G\ .
\end{equation}

\defn{2.3.7} A left $G$-action on $M$ is called {\it free} if, for all $m\in M$, $gm=m$ implies that $g=e$. The action would be {\it transitive} if for all $m,m'\in M$ there exists $g\in G$ such that $m=gm'$.

\rem{2.3.9} If the left action of $G$ on $M$ is free then every orbit $O_m$ is diffeomorphic to the Lie group $G$.

\defn{2.3.8} The {\it stability/isotropy subgroup} $G_m$ of a left $G$-action on $M\ni m$ for a Lie group $G$ is its closed subgroup defined by
\begin{equation}
    G_m\ :=\ \{ g \in G ~|~ gm = m \}\ .
\end{equation}

\thm{2.3.1} For a transitive left $G$-action there exist an isomorphism\footnote{Some technicalities like $G$ be locally compact and $M$ be locally compact and connected are required.} $G/G_m \cong M$ between the coset space $G/G_m$ and the homogeneous space $M$, for any $m\in M$, given by
\begin{equation}
    j_p :\ G/G_p \rightarrow M\ ,\quad gG_p \mapsto gp\ .
\end{equation}

\rem{2.3.10} For a Lie group $G$ and its closed subgroup $H$ (e.g. $G_m$) the homogeneous space $G/H$ can be canonically endowed with the structure of a smooth manifold. Moreover, there exist a canonical projection from $G$ to $G/H$ given by
\begin{equation}\label{CanonicalProj}
    \pi_0 :\ G \rightarrow G/H\ ,\qquad g \mapsto gH\ .
\end{equation}

\ex{2.3.5} The round $n$-sphere $S^n$ is diffeomorphic to the following homogenoeus space:
\begin{equation}
    S^n\ \cong\ SO(n{+}1)/SO(n)\ .
\end{equation}
In particular, we have that $S^3 \cong SO(4)/SO(3)$.

\ex{2.3.6} A $(2n{+}1)$-sphere can also be realized as the following homogeneous space:
\begin{equation}
    S^{2n{+}1}\ \cong\ SU(n{+}1)/SU(n)\ .
\end{equation}
A special case is $S^3\cong SU(2)$, where the group $SU(1)$ is trivial.

\vspace{12pt}
%----------------------------------------------------------------------------------------
\section{Vector fields}
\vspace{2pt}
\defn{2.4.1} The {\it tangent bundle} $(TM,M,\pi)$ over a manifold $M$ is nothing but a union of tangent spaces at all point of the manifold i.e.
\begin{equation}
    TM \= \bigcup\limits_{p\in M} T_p M\ ,
\end{equation}
where the projection $\pi$ just picks out the base point of a given vector in $TM$. 

\rem{2.4.1} The dimension of the tangent bundle for an $n$-dimensional manifold is $2n$, as its members are ($n$-dimensional) vectors of some $T_pM$ labelled by the coordinate ($n$-tuple) of a base point $p\in M$.

\defn{2.4.2} A {\it vector field} $X$ is a smooth section of the tangent bundle $(TM,M,\pi)$. The set of all such vector fields are denoted as $\vecF{M} := \Gamma^\infty(TM)$.

\rem{2.4.2} Given a vector field $X\in \vecF{M}$ and a smooth function $f\in C^\infty(M)$, one can show that $Xf$, defined by
\begin{equation}
    Xf(p)\ :=\ X_p[f] \for X_p \in T_pM\ ,
\end{equation}
is also smooth, i.e. $Xf \in C^\infty(M)$. The set of vector fields $\vecF{M}$ do not form a vector space rather a module over the algebra of smooth functions $C^\infty(M)$\footnote{An algebra is just a vector space $V$ equipped with a multiplication operation $V\times V \rightarrow V$, which for functions is just composition. A module over an algebra is the same thing as a vector space over a field.}. Nevertheless, one can choose a basis of vector fields $\left\{ \sfrac{\pa}{\pa x^i} \equiv \pa_i \right\}$ on some local chart $(U,\varphi)$ with coordinate functions $x^i$ (see Remark 2.1.3) to write $X \in\vecF{M}$ as
\begin{equation}
    X \= \sum\limits_{i=1}^n (Xx^i)\,\pa_i\ .
\end{equation}

\defn{2.4.3} There is a well defined notion of {\it Lie bracket} associated with vector fields defined as
\begin{equation}
    [X,Y]f\ :=\ X(Yf) - Y(Xf)\ ,
\end{equation}
which satisfy the following properties  
\begin{itemize}
    \item bilinearity: $[\cdot,\cdot] :\ \vecF{M}\times \vecF{M} \xrightarrow{\sim} \vecF{M}$,
    \item anti-symmetry: $[X,Y] = -[Y,X]$, and
    \item Jacobi identity: $[X,[Y,Z]] + [Y,[Z,X]] + [Z,[X,Y]] = 0$,
\end{itemize} 
for all $X,Y,Z \in \vecF{M}$.

\rem{2.4.3} In general, it is not possible to induce a push-forward (see Definition 2.1.9) between vector fields $\vecF{M} \ni X$ and $\vecF{N} \ni Y$ of two manifolds $M$ and $N$ with a given map $\varphi: M\rightarrow N$ (it works if $\varphi$ is a diffeomorphism). Nevertheless, it is useful to define the following relation.

\defn{2.4.4} A vector field $X \in \vecF{M}$ is called {\it h-related} to another vector field $Y \in \vecF{M}$ and is written as $Y = \varphi_*X$ if for all $p\in M$ we have that
\begin{equation}
    \varphi_*X_p \= Y_{\varphi(p)}\ .
\end{equation}

\rem{2.4.4} If $X_1$ and $X_2$ is h-related to $Y_1$ and $Y_2$ respectively then $[X_1,X_2]$ is h-related to $[Y_1,Y_2]$ for all $X_1,X_2,Y_1,Y_2 \in \vecF{M}$ i.e.
\begin{equation}
    \varphi_*[X_1,X_2] \= [Y_1,Y_2]\ .
\end{equation}

\defn{2.4.5} Given a vector field $X\in \vecF{M}$ its {\it integral curve} through $p\in M$ is a curve 
\begin{equation}
    \sigma_X :\ (-\epsilon,\epsilon) \rightarrow M\ ,
\end{equation}
such that 
\begin{equation}
    \sigma_X(0) \= p \und \sigma_X'(t) \= X_{\sigma_X(t)}
\end{equation}
for all $t\in (-\epsilon,\epsilon)$.

\defn{2.4.6} If the interval $(-\epsilon,\epsilon)$ can be extended to whole of $\R$ for any $p \in M$ then the vector field, and also the underlying manifold, is called {\it complete}.

\rem{2.4.5} Presence of singularities (e.g. a black hole) on a manifold makes it incomplete. We will only consider complete manifolds in this thesis.

\vspace{12pt}
%----------------------------------------------------------------------------------------
\section{Lie algebra}
\vspace{2pt}
\defn{2.5.1} A {\it left-invariant} vector field $X$ for a Lie group $G$ is defined as $the$ vector field which is $l_g$-related to itself for all $g\in G$:
\begin{equation}
    l_{g*}X \= X \quad \mathrm{or}\quad l_{g*}X_{g'} = X_{gg'}
\end{equation}
for all $g' \in G$.

\defn{2.5.2} The set of all left-invariant vector fields---forming a vector space that is denoted as $L(G)$---together with the Lie bracket $[\cdot,\cdot]$ i.e. $\left(L(G),[\cdot,\cdot]\right)$ is known as the {\it Lie algebra} of $G$\footnote{Remark 2.4.4 is crucial here.}.

\thm{2.5.1} We have a Lie algebra isomorphism\footnote{It is an invertible map that preserves the Lie algebraic structure.} $(T_eG,[\cdot,\cdot]) \cong (L(G),[\cdot,\cdot])$ given by
\begin{equation}
    j:\ T_eG \rightarrow L(G)\ ,\quad A\mapsto j(A)\equiv L^A\ ,
\end{equation}
where the left-invariant vector field $L^A$ is defined by
\begin{equation}
    L^A_g\ :=\ l_{g*}A \in T_gG\ .
\end{equation}

\rem{2.5.1} Note that the Lie bracket on $T_eG$ is induced from the one on $L(G)$ via the map $j$, i.e.
\begin{equation}
    [A,B]\ :=\ j^{-1}[L^A,L^B] \= l_{g{-1}*}[L^A_g,L^B_g]
\end{equation}
for any $g\in G$. This takes the form of usual matrix commutator for matrix Lie groups that we are dealing with. Thus we find that $\textrm{dim}L(G) = \textrm{dim}T_eG = \textrm{dim}G$.

\ex{2.5.1} We note down below in Table \ref{LieTable} Lie algebras $T_eG$ of some Lie groups $G$ that we encountered before.
\begin{table}[!htbp]
    \centering
    \begin{tabular}{|p{0.15\textwidth}|p{0.6\textwidth}|}
    \hline
        $G$ & $T_eG$ \\
    \hline\hline
        $GL(n,\R/\mathds{C})$ & $M(n,\R/\mathds{C}) := $ The set of $n\times n$ real/complex matrices \\
    \hline
        $SO(n)$ & $so(n) := \{ A \in M(n,\R)\ |\ A^T = -A \}$  \\
    \hline
        $SU(n)$ & $su(n) := \{ A \in M(n,\mathds{C})\ |\ \bar{A} = -A\ \&\ \tr{A} = 0 \}$ \\
    \hline
    \end{tabular}
    \caption{A list of Lie groups and their Lie algebras.}
    \label{LieTable}
\end{table}

\thm{2.5.2} A Lie group homomorphism $\varphi : G \rightarrow H$ between Lie groups $G$ and $H$ induces a Lie algebra homomorphism
\begin{equation}
    \diff{\varphi}\equiv \varphi_*:\ T_eG \rightarrow T_eH\ , \quad \mathrm{i.e.}\quad \varphi_*[A,B] \= [\varphi_*A,\varphi_*B]\ .
\end{equation}

\rem{2.5.2} Given a basis $\{ E_1,\ldots, E_n \}$ of $L(G)\cong T_eG$ for an $n$-dimensional Lie group $G$, its structure constants $f_{ij}^{\ \ k}$ are defined by
\begin{equation}
    [E_i,E_j] \= \sum\limits_{k=1}^n f_{ij}^{\ \ k}\,E_k\ .
\end{equation}

\ex{2.5.2} The structure constants for the Lie algebra $su(2)$ are given by $f_{ij}^{\ \ k} = 2\,\varepsilon_{ij}^{\ \ k}$, where $\varepsilon_{ij}^{\ \ k}$ is the $3$-dimensional Levi-Civita symbol.

\thm{2.5.3} A left-invariant vector field $X$ on a Lie group $G$ is complete.

\rem{2.5.3} A consequence of the above theorem is that there exist a unique integral curve
\begin{equation}
    t \mapsto \sigma_{L^A}(t)
\end{equation}
for all $t\in \R$ and left-invariant vector field $L^A$ constructed from a given $A\in T_eG$ of the Lie group $G$.

\defn{2.5.3} The {\it exponential map} for a Lie group $G$ is defined by
\begin{equation}
    \mathrm{exp}:\ T_eG \rightarrow G\ , \quad A \mapsto \mathrm{exp}(A)\ :=\ \sigma_{L^A}(t{=}1)\ .
\end{equation}

\rem{2.5.4} The exponential map is locally diffeomorphic. Furthermore, it lifts $T_eG$ to a connected component of $G$ (connected to $e$) and is a surjection when $G$ is compact.

\defn{2.5.4} A {\it one-parameter subgroup} of a Lie group $G$ is a smooth homomorphism $\mu :\ \R \rightarrow G$ from the additive group $\R$ into $G$ i.e.
\begin{equation}
    \mu(t_1+t_2) \= \mu(t_1)\mu(t_2).
\end{equation}

\thm{2.5.4} If $\mu: \R \rightarrow G$ is a one-parameter subgroup of $G$ then, for all $t\in \R$,
\begin{equation}
    \mu(t) \= \mathrm{exp}(tA) \with  A\ :=\ \mu_*\sfrac{\diff}{\diff{t}}\big|_0\ .
\end{equation}

\rem{2.5.5} The above theorem shows that there exist a one-to-one correspondence between one-parameter subgroups of a Lie group $G$ and its Lie algebra $T_eG$. 

\defn{2.5.5} Given a vector field $X\in \vecF{M}$ one defines the {\it flow generated by X} as the one-parameter group $\{ \varphi_t : t \in \R \}$ where the set of maps $\varphi_t : M \rightarrow M$ are nothing but the integral curves
\begin{equation}
    \varphi_t(m) \= \sigma_X(t) \with \sigma_X(0) \= m\ .
\end{equation}
One can define the {\it Lie derivative of $Y$ along $X$}, for $Y\in \vecF{F}$, by
\begin{equation}
    \Lcal_X Y \= \frac{\diff}{\diff t}(\varphi_t^{-1})_*(Y)\Big|_{t=0}\ .
\end{equation}

\rem{2.5.6} Interestingly, one can show that $\Lcal_XY = [X,Y]$.

\vspace{12pt}
%----------------------------------------------------------------------------------------
\section{Tensor fields}
\vspace{2pt}
\label{covectorFields}
\defn{2.6.1} A {\it (p,q)-tensor} $T^{p,q}(V)$ for a vector space $V$ and its dual $V^*$ is the space of multilinear functions
\begin{equation}
    f:\ V^*\times\overset{p}{\cdots}\times V^*\times V\times\overset{q}{\cdots}\times V \xrightarrow{\sim} \R
\end{equation}
and is, alternatively, denoted using tensor products as $V\otimes\overset{q}{\cdots}\otimes V\otimes V^*\otimes\overset{p}{\cdots}\otimes V^*$.

\rem{2.6.1} We note down have the following results/observations:
\begin{equation}
    T^{0,0}(V):=\R\ ,\quad T^{0,1}(V)\ \cong\ V^* \und T^{1,0}(V) \equiv (V^*)^*\ \cong\ V
\end{equation}
for any finite-dimensional vector space $V$. Elements of $T^{p,q}(V)$ can be easily expanded in terms of a given basis of $V$ and the corresponding dual basis of $V^*$.

\defn{2.6.2} A {\it (p,q)-tensor bundle} $(T^{p,q}M,M,\pi)$ over a manifold $M$ is the following disjoint union of tensors:
\begin{equation}
    T^{p,q}M \= \bigcup\limits_{m\in M}T^{p,q}(T_mM)
\end{equation}
where the projection $\pi$ associates the base point $m\in M$ of a given vector in $T^{p,q}M$.

\rem{2.6.2} The tangent bundle $TM$ is isomorphic to $T^{1,0}M$, while the bundle $T^{0,1}M$ is isomorphic to the so called {\it cotangent bundle} $T^*M$ defined analogously to Definition 2.4.1 before.

\defn{2.6.3} The {\it exterior algebra} over a vector space $V$ -- denoted as $\bigwedge V$ -- is the algebra\footnote{Plainly speaking, it is a linear combination of all possible (finitely many) anti-symmetrized tensor products of vectors in $V$.} generated by the so called {\it wedge product} $\wedge$ satisfying
\begin{equation}
    v_1\wedge v_2 \= - v_2\wedge v_1
\end{equation}
for all $v_1,v_2 \in V$. A subspace of $\bigwedge V$ consisting of linear combinations of $k$--fold wedge products of vectors in $V$ is denoted $\bigwedge^k (V)$.

\subsection{Differential forms}

\defn{2.6.4} A {\it $k$-form} $\omega$ is a $(0,k)$-tensor field $\omega \in \Gamma^\infty(M,T^{0,k}M)$ such that $\omega_m \in \bigwedge^k(T_mM)$ for all $m\in M$.

\rem{2.6.3} The space of $k$-forms is denoted $\Omega^k(M)$. Notice that $\Omega^0(M) \equiv C^\infty(M)$ and the only member of $\Omega^n(M)$ (upto to a scalar multiple), known as the {\it volume form}, is denoted $\mu$ or $\diff{V}$. The space $\Omega^n(M)$ splits into two equivalence classes $\{ [\mu_+], [\mu_-] \}$ with the relation $\mu \sim \mu'$ defined by $\mu = \lambda \mu'$ such that $\mu \in [\mu_+]$ if $\lambda > 0$ and $\mu \in [\mu_-]$ if $\lambda < 0$.

\rem{2.6.4} On a chart $(U,\varphi)$ of $M$ with coordinates $\varphi(m)=(x^1,\ldots,x^n)$ one can expand a $k$--form $\omega$ as 
\begin{equation}
    \omega \= \sfrac{1}{k!}\sum\limits_{i_1,\ldots,i_k = 1}^n\omega_{i_1,\ldots,i_k}\,\diff{x}^{i_1}\wedge\ldots\wedge\diff{x}^{i_k}
\end{equation}
where the coefficients $\omega_{i_1,\ldots,i_k} \in C^\infty(M)$ are totally anti-symmetric in its indices.

\defn{2.6.5} The {\it exterior derivative} $\diff$ is the linear map $\diff : \Omega^k(M) \xrightarrow{\sim} \Omega^{k+1}(M)$ defined, for all $\omega \in \Omega^k(M)$, by
\begin{equation}
\diff{\omega} \= \sfrac{1}{k!}\sum\limits_{i_1,\ldots,i_k = 1}^n\pa_i(\omega_{i_1,\ldots,i_k})\,\diff{x}^i\wedge\diff{x}^{i_1}\wedge\ldots\wedge\diff{x}^{i_k}\ ,
\end{equation}
where the partial derivative $\pa_i$ is taken with respect to coordinate $x^i$. It can also be defined in a coordinate independent way, for all $X_1,\ldots,X_{k+1} \in \vecF{M}$, as
\begin{equation}
 \begin{aligned}
    \diff{\omega}(X_1,\ldots,X_{k+1}) &\= \sum\limits_{i=1}^{k+1} (-1)^{i+1}\, X_i(\omega(X_i,\ldots,\hat{X}_i,\ldots,X_{k+1})) \\
    &\qquad + \sum\limits_{1\leq i < j \leq n} (-1)^{i+j}\, \omega([X_i,X_j],X_1,\ldots,\hat{X}_i,\ldots,\hat{X}_j,\ldots,X_{k+1})\ ,
 \end{aligned}
\end{equation}
where the circumflex means that the symbol beneath it is omitted.

\rem{2.6.5} The action of exterior derivative follow {\it graded Leibniz rule}:
\begin{equation}
    \diff{(\alpha\wedge\beta)} \= \diff{\alpha}\wedge\beta + (-1)^p\,\alpha\wedge\diff{\beta}
\end{equation}
for all $\alpha\in \Omega^p(M)$ and $\beta\in\Omega^q(M)$. Moreover, it can easily be shown (using, e.g.~the first definition above) that
\begin{equation}
    \diff^2\ \equiv\ \diff\circ\diff \= 0\ .
\end{equation}

\defn{2.6.6} For a map $\varphi : M \rightarrow N$ and a $k$-form $\omega \in \Omega^k(N)$ its {\it pull-back} $\varphi^*\omega \in \Omega^k(M)$ is defined via
\begin{equation}
    (\varphi^*\omega)_m(X_1,\ldots,X_k)\ :=\ \omega_{\varphi(m)}(\varphi_{*}X_1,\ldots,\varphi_{*}X_k)
\end{equation}
for all $X_1,\ldots,X_k \in \vecF{M}$. 

\rem{2.6.6} It can be shown that the exterior derivative is {\it natural}, i.e.~it is compatible with the pull-back:
\begin{equation}
    \varphi^*(\diff{\omega}) \= \diff{(\varphi^*\omega)}\ ,
\end{equation}
for any $\omega\in \Omega^k(M)$. Furthermore, one can prove that
\begin{equation}
    \varphi^*(\alpha\wedge\beta) = \varphi^*\alpha\wedge\varphi^*\beta \und (\varphi\circ\tilde{\varphi})^*\omega = \tilde{\varphi}^*(\varphi^*\omega)\ .
\end{equation}

\defn{2.6.7} The {\it Lie derivative of $\omega$ along $X$}, for $\omega\in\Omega^k(M)$ and $X\in \vecF{M}$, can be defined analogous to Definition 2.5.5:
\begin{equation}
    \Lcal_X\omega \= \frac{\diff}{\diff t}(\varphi_t^{-1})^*(\omega)\Big|_{t=0}\ .
\end{equation}

\rem{2.6.7} There exists a nice formula by \'{E}lie Cartan for the Lie derivative of differential forms given, for any $X\in \vecF{M}$, by
\begin{equation}
    \Lcal_X \= \diff\circ\iota_X + \iota_X\circ\diff\ ,
\end{equation}
where the linear map $\iota_X: \Omega^p(M) \xrightarrow{\sim} \Omega^{p-1}(M)$ is the {\it interior product} defined by  
\begin{equation}
    (\iota_X\omega)(X_1,\ldots,X_{p-1}) \= \omega(X,X_1,\ldots,X_{p-1})
\end{equation}
for $\omega \in \Omega^p(M)$ and $X_1,\ldots,X_{p-1} \in \vecF{M}$.

\subsection{Metric}

\defn{2.6.8} A {\it metric} $g$ is a $(0,2)$-tensor field $g \in \Gamma^\infty(T^{0,2}M)$ where $g_m$ for every $m\in M$ defines an inner product on the vector space $T_mM$. If this inner product is positive definite then the manifold is called {\it Riemannian}. If the metric $g$ has an underlying inner product that is indefinite then the manifold is called {\it pseudo-Riemannian}.

\rem{2.6.8} A metric $g$ on an $n$-dimensional manifold $M$ can be written, in local coordinates, as
\begin{equation}
    g \= \sum\limits_{i=1}^n\, g_{ij}\,\diff{x}^i\otimes\diff{x}^j \with g_{ij} := g(\pa_i,\pa_j)
\end{equation}
being the {\it metric components} for the basis vector fields $\pa_i,\pa_j \in \vecF{M}$. Often times, in physics literature, the tensor product sign is ignored. The fact that $g$ is non-degenerate means that the matrix $g$ can be inverted to yield another matrix $g^{-1}$ with components $g^{-1}_{ij} =: g^{ij}$ that, in turn, defines a $(2,0)$-tensor field called the {\it induced metric}. These can be used to raise or lower indices of any tensor component, e.g.
\begin{equation}
    T_{i'}^{\ j'k'} := g_{ii'}g^{jj'}g^{kk'}T^i_{\ jk}\quad \forall \quad T \in \Gamma^\infty(M,T^{1,2}M)\ .
\end{equation}

\ex{2.6.1} A standard example of a Reimannian manifold is $\R^n$ with metric
\begin{equation}
    g \= (\diff{x}^1)^2 + \ldots + (\diff{x}^n)^2\ .
\end{equation}
A prominent example of a psuedo-Riemannian manifold is {\it Minkowski space} $\R^{1,n}$ with metric
\begin{equation}
    g \= -(\diff{x}^1)^2 + (\diff{x}^2)^2 + \ldots + (\diff{x}^n)^2\ .
\end{equation}

\defn{2.6.9} The {\it signature} of a pseudo-Riemannian manifold $(M,g)$ of dimension $n$ is a count of the number of positive and negative eigenvalues of the matrix $g_{ij}$ and is denoted by an $n$-tuple of $+$ and $-$. A $4$-dimensional Lorentzian manifold has one of the signature different from the rest three.

\ex{2.6.2} The signature of the Minkowski space $\R^{1,3}$ as presented above is $(-,+,+,+)$, and is thus a Lorentzian manifold.

\defn{2.6.10} Two (pseudo-)Riemannian manifolds $(M,g_M)$ and $(N,g_N)$ are called {\it conformal} if their metrices are related by some smooth function $\Omega^2 \in C^\infty(N)$: 
\begin{equation}
    g_M \= \Omega^2\,g_N\ .
\end{equation}

\defn{2.6.11} A locally {\it orthonormal basis} of $1$-forms $\{ e^i \}$ (aka coframe) with $i=1,\ldots,n$ for an $n$-dimensional Riemannian manifold $M$ satisfy, for all $e^i,e^j \in \Omega^1(M)$,
\begin{equation}
    g(e^i,e^j) \= g^{ij} = \delta^{ij}\ .
\end{equation}
Similarly, one defines locally orthonormal basis of vector fields (aka frame) $\{ X_1,\ldots,X_n \}$ by demanding that they satisfy
\begin{equation}
    g_{ij} \= g(X_i,X_j) \= \delta_{ij}
\end{equation}
for all $X_i,X_j \in \vecF{M}$.

\subsection{Maurer--Cartan form}
\defn{2.6.12} A $k$-form $\omega \in \Omega^k(G)$ of a Lie group $G$ is said to {\it left-invariant} if, for all $g\in G$,
\begin{equation}
    l_g^*\omega \= \omega\ ,\quad \mathrm{i.e.} \quad l_g^*(\omega_{g'}) \= \omega_{g^{-1}g'}\ \forall\ g' \in G\ .
\end{equation}
The set of all left-invariant $1$-forms on $G$ is denoted $L^*(G)$. 

\rem{2.6.9} From Remark 2.6.6 we see that if $\omega$ is left-invariant then so is its exterior derivative:
\begin{equation}
    l_g^*(\diff{\omega}) \= \diff{l_g^*\omega}\ .
\end{equation}

\rem{2.6.10} Similar to $j$ of Theorem 2.5.1, there exist an isomorphism $\tilde{j}$ between $T^*_eG$ and $L^*(G)$ given by
\begin{equation}
    \tilde{j}:\ T^*_eG \rightarrow L^*(G)\ ,\quad \omega \mapsto \tilde{j}(\omega) \equiv \lambda^\omega\ ,
\end{equation}
where the left-invariant $1$-forms $\lambda^\omega$ is defined as
\begin{equation}
    \lambda^\omega_g\ :=\ l_{g-1}^*(\omega) \in T^*_gG\ .
\end{equation} 
Notice that these $1$-forms are dual to the left-invariant vector fields $L^A$ for $A \in T_eG$:
\begin{equation}
    \< \lambda^\omega, L^A \>_g \= \< \omega, A\>
\end{equation}
for all $g \in G$.

\rem{2.6.11} For a basis $\{ E_1,\ldots,E_n \}$ of $L(G)$ with dim$G=n$ (see Remark 2.5.2) we can define the dual basis $\{ \omega^1,\ldots,\omega^n \}$ of $L^*(G)$ by
\begin{equation}
    \< \omega^i, E_j \> \= \delta^i_j\ ,
\end{equation}
which satisfy the Maurer--Cartan structure equation
\begin{equation}
    \diff{\omega}^i + \frac{1}{2} f_{jk}^{\ \ i}\,\omega^j\wedge\omega^k\ .
\end{equation}

\defn{2.6.13} The Maurer--Cartan $1$-form $\Omega_l$ is {\it the} $L(G)$-valued $1$-form on $G$ that, for any $A \in T_gG$, gives a left-invariant vector field as follows
\begin{equation}
    \< \Omega_l, A \>_{g'}\ :=\ l_{g'*}(l_{g^{-1}*}A) 
\end{equation}
for any $g' \in G$.

\rem{2.6.12} It is more useful to consider the Maurer--Cartan $1$-form to be valued in $T_eG \cong L(G)$, which yields a nice result:
\begin{equation}
    \< \Omega_l, L^A \> \= A\ .
\end{equation}
It can be shown that the Maurer-Cartan $1$-form takes the following explicit form for the matrix Lie groups that we are intereseted in,
\begin{equation}
    \Omega_l^{ij} \= \sum\limits_{k=1}^n (g^{-1})^{ik}\, \diff{g}^{kj}\ ,
\end{equation}
where $g^{ij}$ are the coordinates for the matrix Lie group $G$ on a given chart. 

\vspace{12pt}
%----------------------------------------------------------------------------------------
\section{Integration} 
\vspace{2pt}
\defn{2.7.1} A manifold $M$ is called {\it orientable} if it is equipped with a nowhere vanishing volume form $\mu$. An {\it orientation} of $M$ refers to a choice made for one of the the two equivalence classes in Remark 2.6.3; this is usually chosen to be the positive one.

\ex{2.7.1} The standard volume form on $\R^n$, which is an oriented manifold, is given by 
\begin{equation}
    \mu \= \diff{x}^1\wedge\ldots\wedge\diff{x}^n\ . 
\end{equation}
M\"{o}bius strip provides a classic example of a nonorientable manifold.

\defn{2.7.2} {\it Integration} of a compactly supported\footnote{This simply means that it vanishes outside of a compact subset in $M$} volume form $\mu$ of an oriented manifold $M$ that is covered with charts $(U_1,\phi_1),\ldots,(U_N,\phi_N)$ can be defined, using a partition of unity\footnote{It is a set of smooth functions $f_i \in C^\infty(M)$ that vanish outside of $U_i$, lies within $[0,1]$, and satisfies the condition: $\sum_{i=1}^Nf_i(m) = 1\ \forall\ m\in M$.} $\{ f_i \}$, as 
\begin{equation}
    \int_M \mu \= \sum\limits_{i=1}^N \int_{\phi(U_i)} (\phi_i^{-1})^*(f_i\mu)\ .
\end{equation}

\rem{2.7.1} It can be shown that this definition of integration is independent of the choice of a chart. 

\thm{2.7.1} For an oriented manifold $M$ with boundary $\pa M$ (that inherits an induced orientation from $M$) the integral of a compactly supported $(n{-}1)$-form $\omega \in \Omega^{n{-}1}(M)$ is related to an integral of its exterior derivative $\diff{\omega}$:
\begin{equation}
    \int_M \diff{\omega} \= \int_{\pa M} \omega\ .
\end{equation}

\rem{2.7.2} This is the famous {\it Stockes' theorem}. Notice that if the manifold $M$ has no boundary then the above integral vanishes.

\defn{2.7.3} For an $n$-dimensional, oriented, Riemannian manifold $(M,g)$ the {\it Riemannian volume form} is given by
\begin{equation}
    \mu_g \= \sqrt{|g|}\,\diff{x}^1\wedge\cdots\wedge\diff{x}^n\ ,
\end{equation}
where $|g| := |\mathrm{det}(g_{ij})|$ for metric components $g_{ij}$.

\rem{2.7.3} Integration on Riemannian manifolds are performed using volume form $\mu_g$. 
  
\vspace{12pt}  
%----------------------------------------------------------------------------------------
\section{Hodge duality}
\vspace{2pt}
\rem{2.8.1} On an $n$-dimensional pseudo-Riemannian manifold $(M,g)$ one can induce an inner product on $\Omega^k(M)$---generated by $k$-forms $\{ \diff{x}^{i_1}\wedge\cdots\wedge\diff{x}^{i_k} \}$ with $i_1,\ldots,i_k \in \{ 1,\ldots,n \}$---that is given by
\begin{equation}
    \Big\< \diff{x}^{i_1}\wedge\cdots\wedge\diff{x}^{i_k}, \diff{x}^{j_1}\wedge\cdots\wedge\diff{x}^{j_k} \Big\> \= g^{i_1j_1}\cdots g^{i_kj_k}\ ,
\end{equation}
where $g^{ij} := g(\diff{x}^i,\diff{x}^j)$.

\defn{2.8.1} The {\it Hodge star operator} $\ast : \Omega^k(M) \xrightarrow{\sim} \Omega^{n-k}(M)$ is a linear map that is defined, on an oriented $n$-dimensional pseudo-Riemannian manifold $(M,g)$ with volume form $\mu$, as
\begin{equation}
    \alpha\wedge\ast\beta \= \< \alpha,\beta \>\, \mu
\end{equation}
for all $\alpha,\beta \in \Omega^k(M)$. Here $\ast\beta$ is called the {\it Hodge dual} of $\beta$.

\rem{2.8.2} For an $n$-dimensional oriented pseudo-Riemannian manifold $(M,g)$ with signature $(-,\overset{s}{\ldots},-,+,\overset{n-s}{\ldots},+)$ the following condition holds true on $\Omega^p(M)$:
\begin{equation}
    \ast^2 \= (-1)^{p(n-p)+s}\ .
\end{equation}

\ex{2.8.1} For the Minkowski space $\R^{1,3}$ with signature $(-,+,+,+)$, coordinates $ ~ \\(x^0, x^1, x^2, x^3) =: (t,x,y,z)$ and volume form $\diff{V} = \diff{t}\wedge\diff{x}\wedge\diff{y}\wedge\diff{z}$ we have the following results:
%\begin{equation}
 \begin{align*}
    &\ast\diff{z} \= -\diff{t}\wedge\diff{x}\wedge\diff{y}\ , &
    &\ast\,\diff{t}\wedge\diff{x} \= -\diff{y}\wedge\diff{z}\ , & &\ast\,\diff{t}\wedge\diff{y} \= \diff{x}\wedge\diff{z}\ ,\\
    &\ast\diff{t}\wedge\diff{z} \= -\diff{x}\wedge\diff{y}\ , & &\ast\,\diff{x}\wedge\diff{y} \= -\diff{t}\wedge\diff{z}\ , & &\ast\,\diff{x}\wedge\diff{z} \= -\diff{t}\wedge\diff{y}\ ,\\
    &\ast\diff{y}\wedge\diff{z} \= \diff{t}\wedge\diff{x}\ , & &\ast\,\diff{t}\wedge\diff{x}\wedge\diff{y} \= -\diff{z}\ , & &\ast\,\diff{t}\wedge\diff{x}\wedge\diff{z} \= \diff{y}\ ,\\
    &\ast\diff{t}\wedge\diff{y}\wedge\diff{z} \= -\diff{x}\ , & &\ast\,\diff{x}\wedge\diff{y}\wedge\diff{z} \= -\diff{t}\ , & &\ast\,\diff{t}\wedge\diff{x}\wedge\diff{y}\wedge\diff{z} \= -1\ .
 \end{align*}
%\end{equation}

\vspace{12pt}
%----------------------------------------------------------------------------------------
\section{Representations}
\vspace{2pt}
\defn{2.9.1} A {\it representation of a Lie group} $G$ is a Lie group homomorphism
\begin{equation}
    \Pi :\ G \rightarrow \mathrm{GL}(V)
\end{equation}
for a finite dimensional vector space $V$. 

\defn{2.9.2} A {\it representation of a Lie algebra} $\mathfrak{g}$, for a finite dimensional vector space $V$, is a Lie algebra homomorphism
\begin{equation}
    \pi :\ \mathfrak{g} \rightarrow \mathrm{gl}(V)\ ,
\end{equation}
where $\mathrm{gl}(V) := \mathrm{End}(V)$\footnote{The endomorphism of $V$ denoted $\mathrm{End}(V)$ is the space of linear maps $V\xrightarrow{\sim} V$.}.

\ex{2.9.1} The {\it standard/fundamental representation} of a Lie group $G \ni g$ is $\Pi(g) = g$ and of a Lie algbra $\mathfrak{g} \ni X$ is $\pi(X) = X$. 

\ex{2.9.2} For matrix Lie groups $G$ with Lie algebra $\mathfrak{g}$, its {\it adjoint representation} is the following homomorphism
\begin{equation}
    \mathrm{Ad}:\ G \rightarrow \mathrm{GL}(\mathfrak{g})\ ,\quad g \mapsto \mathrm{Ad}_g\ ,
\end{equation}
where the {\it adjoint map} $\mathrm{Ad}_A$ is defined by
\begin{equation}
    \mathrm{Ad}_g:\ \mathfrak{g} \xrightarrow{\sim} \mathfrak{g}\ ,\quad X \mapsto \mathrm{Ad}_g(X)\ :=\ gXg^{-1}\ .
\end{equation}

\rem{2.9.1} It can be shown that the map $\mathrm{Ad}$ induces (see Theorem 2.5.2) the following lie algebra homomorphism
\begin{equation}
    \mathrm{Ad}_* \equiv \mathrm{ad}:\ \mathfrak{g} \rightarrow \mathrm{gl}(\mathfrak{g})\ ,\quad X \mapsto \mathrm{ad}_X\ ,
\end{equation}
where the action of the map $\mathrm{ad}_X$ can be shown to be the following 
\begin{equation}
    \mathrm{ad}_X:\ \mathfrak{g}\rightarrow \mathfrak{g}\ , \quad Y\mapsto \mathrm{ad}_X(Y)\ :=\ [X,Y]\ .
\end{equation}
This is known as the {\it adjoint representation} of a finite-dimensional Lie algebra $\mathfrak{g}$. 

\vspace{12pt}
%----------------------------------------------------------------------------------------
\section{Maxwell equations}
\vspace{2pt}
\rem{2.10.1} For the nice\footnote{Here it means a connected and simply connected manifold.} manifolds that we are interested in this thesis the Maxwell theory boils down to a choice of the gauge potential $A$, which is a $1$-form on the manifold.

\defn{2.10.1} For Minkowski space $\R^{1,3}$ we choose the gauge potential $A$ as
\begin{equation}
    A \= A_0\,\diff{t} + A_i\,\diff{x}^i\ ,
\end{equation}
and the corresponding field strength $F$ is organized as
\begin{equation}
    F\ :=\ \diff{A} \= E_i\,\diff{x}^i\wedge\diff{t} + \frac{1}{2} B_i\, \varepsilon^{i}_{\ jk}\,\diff{x}^j\wedge\diff{x}^k
\end{equation}
with electric field ${\bf E}$ and magnetic field ${\bf B}$.

\rem{2.10.2} The source free Maxwell equations viz.
\begin{equation}
    \nabla\cdot {\bf B} \= 0 \und \nabla\times{\bf E} + \frac{\pa {\bf B}}{\pa t} \= 0
\end{equation}
are given by 
\begin{equation}
    \diff{F} \= 0\ .
\end{equation}
Notice that, due to Remark 2.6.5, the above condition is trivially satisfied here.

\rem{2.10.3} The other two Maxwell equations with source viz.
\begin{equation}
    \nabla\cdot{\bf E} \= \rho \und \nabla\times{\bf B} - \frac{\pa {\bf E}}{\pa t} \= {\bf J}
\end{equation}
with {\it charge density} $\rho$ and {\it current density} ${\bf J}$ is given by 
\begin{equation}
    \delta F \= j \with j \= -\rho\,\diff{t} + J_i\,\diff{x}^i \in \Omega^1(\R^{1,3})\ , 
\end{equation}
where the {\it codifferential} $\delta := *\diff*$.

\rem{2.10.4} An important feature of the Maxwell theory is that it possesses gauge symmetry, i.e. the transformation 
\begin{equation}
    A \rightarrow A + \diff{f}\ ,
\end{equation}
for all $f\in C^\infty(\R^{1,3})$, leaves the field strength $F$ invariant. There are many ways to fix this redundancy by employing some kind of gauge-fixing. On Minkowski space we can always work in the so called ``temporal gauge" where $A_0 = 0$ (see Chapter 6 of \cite{Baez&Muniain}).

\rem{2.10.5} Noticing that $\delta^2 = \pm *\diff^2* = 0$ and applying $\delta$ on the Maxwell equations with source $j$, we arrive at the following {\it continuity equation}:
\begin{equation}
    0 \= \delta^2 F \= \delta j \quad \implies \quad \frac{\pa \rho}{\pa t} + \nabla\cdot{\bf J} \= 0\ .
\end{equation}

%% file: Chapters/Chapter3.tex
%auto-ignore
\chapter{Gauge theory}
\label{Chapter3}

\justifying

All four known forces of nature viz. gravity, electromagnetism, weak and strong nuclear forces can be described (at least classically) in terms of a gauge theory. The study of modern gauge theory requires the notion of principal bundles and various structures on it. While it is possible to study a lot of physics---including Yang--Mills theory---with just vector bundles alone, e.g.~as in \cite{Baez&Muniain}, a lot of deep physics arising from the underlying topology of the base manifold can not be fully appreciated without following principle bundles approach. We review, in this chapter, the construction of gauge theory from principal bundles without bothering about proof of any statement, all of which can be found in \cite{Isham}. For a quick review of Yang--Mills theory one may refer to \cite{Bleecker} or a nice review article by Daniel and Viallet \cite{DV80}.

\vspace{12pt}
%----------------------------------------------------------------------------------------
\section{Principal $G$-bundles}
\vspace{2pt}
\defn{3.1.1} A bundle $(P,M,\pi)$ is called a {\it principle $G$-bundle} and depicted, diagrammatically, as 
\begin{tikzcd}
   G \arrow[r, "r_g"] &P \arrow[d, "\pi"] \\
                               &M
\end{tikzcd}
if there exist a free right action of the Lie group $G$ on $P$ and, moreover, if there exist a bundle isomorphism between $(P,M,\pi)$ and $(P,P/G,\pi_0)$, with the canonical projection $\pi_0$ \eqref{CanonicalProj}, meaning that the diagram
\begin{tikzcd}
   P \arrow[d, "\pi"] \arrow[r, "\varphi"] &P \arrow[d, "\pi_0"] \\
   M \arrow[r, "\psi"]  &P/G
\end{tikzcd}
commutes.

\rem{3.1.1} Notice that the structure group in this case is $G$. Furthermore, the fibre $\pi^{-1}(\{m\})$, for any $m \in M$, is diffeomorphic to $G$, but it does not have a canonical group structure. Another way to put this would be to say that the Manifold $M$ has a $G$-fibre attached to all its points, wherein the identity element is forgotten. Also, notice here that the projection map is insensitive to the $G$-action.

\ex{3.1.1} The simplest example of a principle bundle is $(G\times M, M,\pi)$ with the right action given by $(x,g_0)g := (x,g_0g)$.

\ex{3.1.2} An important example of a principle $G$-bundle is the so called {\it frame bundle} $LM$ over an $n$-dimensional manifold $M$, which is a collection of basis-frames
\begin{equation}
    L_mM\ :=\ \{ (b_1,\ldots,b_n) ~|~ \<b_1,\ldots,b_n\> = T_mM \} \cong GL(n,\R)
\end{equation}
corresponding to the tangent bundle $TM$ attached at each point $m$ of the manifold, i.e.
\begin{equation}
    LM\ :=\ \bigcup\limits_{m\in M} L_mM\ .
\end{equation}
The free right action of any $g \in GL(n,\R)$ on a given $(b_1,\ldots,b_n) \in L_mM$ is given by
\begin{equation}
    (b_1,\ldots,b_n)g\ :=\ (b_i\,g^i_{\ 1},\ldots,b_i\,g^i_{\ n})\ .
\end{equation}

\ex{3.1.3} Another important example of a principle bundle is $(G,G/H,\pi)$ where $H$ acts freely from right on $G$ via group multiplication resulting into orbits of cosets $G/H$. A famous example of this kind is the {\it Hopf bundle}, represented diagrammatically as follows: 
\begin{tikzcd}
   U(1) \arrow[r] &SU(2) \arrow[d] \\ 
                  &SU(2)/U(1)
\end{tikzcd} 
$\equiv$
\begin{tikzcd}
   S^1 \arrow[r] &S^3 \arrow[d] \\
                 &S^2
\end{tikzcd}

\defn{3.1.2} A {\it principle morphism} between a pair of bundles $(P,M,\pi)$ and $(P',M',\pi')$ is a bundle morphism $(\varphi,\psi)$ that, additionally, satisfies 
\begin{equation}
    \varphi(pg) \= \varphi(p)g \quad\forall\quad p \in P \und g\in G\ .
\end{equation}

\rem{3.1.2} The notion of trivialization, both local $(U,\varphi)$ as well as global $(G\times M,M,\pi_1)$, follows similar to the general theory of fibre bundles that we saw before, albeit with this extra condition of principle morphism. In a similar way, the idea of transition functions $g_{ij}$ between any two such overlapping local trivializations $(U_i,\varphi_i)$ and $(U_j,\varphi_j)$ and the corresponding structure group carries over. One can also define smooth sections on a principle bundle $P$ just like before.

\rem{3.1.3} The set of all principal morphisms between a bundle $P$ to itself forms a group $Aut(P)$ called the {\it automorphism group} of the principle bundle $P$. In the case of a trivial bundle $P=G\times M$ we have that $Aut(P)\cong C^\infty (M,G)$, where the latter is the well-known group of {\it gauge transformations}. 

\thm{3.1.1} A principle $G$-bundle $(P,M,\pi)$ is trivial if and only if it possesses a smooth section $\sigma: M \rightarrow P$.

\rem{3.1.4} An illustrative counter-example of the above theorem is the fact that the frame bundle $LS^2$ over the $2$-sphere is not trivial because there does not exist nowhere vanishing smooth vector fields, and hence a basis, on $TS^2$ (look at the north or south poles). This result is famously summarized as ``sphere can not be combed".

\rem{3.1.5} The topological properties, such as twisting, of the base manifold $M$ is intrinsically linked with that of the principal bundle $P$ and also carries over to the below defined associated bundles.

\vspace{12pt}
%----------------------------------------------------------------------------------------
\section{Associated bundles}
\vspace{2pt}

\defn{3.2.1} The $G$-product, denoted $X\times_G Y$, of two spaces $X$ and $Y$, both admitting right action of $G$, is the space of orbits under this action on the Cartesian product $X\times Y$. In other words, it is defined via the following equivalence relation
\begin{equation}
    (x,y) \sim (x',y')\quad \mathrm{iff}\quad \exists\ g \in G :\ x' \= xg \und y' \= yg\ ,
\end{equation}
where the equivalence class is denoted as $[x,y]$.

\defn{3.2.2} For a given principle $G$-bundle $(P,M,\pi)$ and a manifold $F$, which admits a left $G$-action, one defines the {\it associated bundle} $P_F$ by\footnote{Notice how we have employed left $G$-action to define right action on $F$ (see Remark 2.3.7).}
\begin{equation}
    P_F\ :=\ P\times_G F \quad \mathrm{where} \quad (p,v)g\ :=\ (pg,g^{-1}v)
\end{equation}
and the projection $\pi_F$ by
\begin{equation}
    \pi_F :\ P_F \to M\ , \quad \pi_F([p,v])\ :=\ \pi(p)\ .
\end{equation}

\rem{3.2.1} It can be shown that the associated bundle $(P_F,M,\pi_F)$ has the structure of a fibre bundle with typical fibre $F$. 

\ex{3.2.1} An important example of a fibre associated with the frame bundle $LM$ with Lie group $GL(n,\R)$ is the tensor bundle $T^{p,q}M$ with fibres $F = (\R^n)^{\times p}\times (\R^{n*})^{\times q}$ where the left action of a given $g \in GL(n,R)$ on some $v \in F$ is given by the following representation $\rho$
\begin{equation}
    (\rho(g)v)^{i_1,\ldots,i_p}_{\qquad j_1,\ldots,j_q}\ :=\ v^{i_1',\ldots,i_p'}_{\qquad j_1',\ldots,j_q'}\,(g)^{i_1}_{\ i_1'}\ldots(g)^{i_p}_{\ i_p'}\,(g^{-1})^{j_1'}_{\ j_1}\ldots(g^{-1})^{j_q'}_{\ j_q}\ .
\end{equation}
Another very useful generalization of this is the so called {\it tensor of density $\omega$} that admits a left-action via the following representation
\begin{equation}
    (\rho(g)v)^{i_1,\ldots,i_p}_{\qquad j_1,\ldots,j_q}\ :=\ (\mathrm{det} g)^\omega\,v^{i_1',\ldots,i_p'}_{\qquad j_1',\ldots,j_q'}\,(g)^{i_1}_{\ i_1'}\ldots(g)^{i_p}_{\ i_p'}\,(g^{-1})^{j_1'}_{\ j_1}\ldots(g^{-1})^{j_q'}_{\ j_q}\ .
\end{equation}

\defn{3.2.3} Given a principle morphism $(\varphi,\psi)$ between principal $G$-bundles $(P,M,\pi)$ and $(P',M',\pi')$ one defines an {\it associated bundle morphism} $\varphi_F$ between associated bundles $P_F\times_G F$ and $P'\times_G F$ by
\begin{equation}
    \varphi_F([p,v])\ :=\ [\varphi(p),v]\ ,
\end{equation}
which is well-defined since
\begin{equation}
    \varphi_F([pg,g^{-1}v]) \= [\varphi(pg),g^{-1}v] \= [\varphi(p)g,g^{-1}v] \= [p,v]\ .
\end{equation}

\rem{3.2.2} An associated bundle $P_F$ is called {\it trivial} if the underlying principal bundle $P$ is trivial. An important point to note here is that a trivial associated bundle is a trivial fibre bundle, but the converse is not true.

\defn{3.2.4} Let $H \subset G$ be a closed subgroup of $G$ while $P$ respectively $P'$ are principal $G$- respectively $H$-bundle defined over the same base space. If there exist a principal morphism $\varphi$ with respect to $H$, i.e. 
\begin{equation}
    \varphi(ph) \= \varphi(p)h
\end{equation}
for all $h\in H$, then $P$ is called a $G$-extension of $H$ while $P'$ is called a $H$-restriction of $P$.

\rem{3.2.3} While there always exists an extension $P$ of a given $P'$ as defined above, the converse is not always true. This has important ramifications both in Riemannian geometry as well as in Yang--Mills theory; for the latter this is related to the important question of spontaneous breakdown of the internal symmetry group from $G$ down to $H$.

\thm{3.2.1} A principle $G$-bundle $(P,M,\pi)$ can be restricted to a closed subgroup $H\subset G$ iff the bundle $(P/H,M,\pi_0)$ admits a smooth section.

\rem{3.2.4} An important application of the above theorem is that one can always have a Riemannian metric defined on any $n$-diemensional manifold $M$ as $LM/SO(n)$, for the closed subgroup $SO(n)\subset GL(n,\R)$, always admits a smooth section; the same is not always true for a pseudo-Riemannian metric as, $LM/SO(1,n{-}1)$ (with closed subgroup $SO(1,n)\subset GL(n,\R)$), for instance, does not always admits a smooth section because of possible topological restrictions.

\thm{3.2.2} There exists a one-to-one {\it $G$-equivariant} correspondence between a section $\sigma \in \Gamma^\infty(P_F)$ of an associated bundle $(P_F,M,\pi_F)$ and map $\phi: P \to F$ satisfying
\begin{equation}
    \phi(pg) \= g^{-1}\phi(p) \quad\forall\quad p \in P \und g \in G
\end{equation}
where the section $\sigma_\phi$, corresponding to $\phi$, arises as
\begin{equation}
    \sigma_\phi(x)\ :=\ [p,\phi(p)] \for p\in \pi^{-1}(x)\ .
\end{equation}

\defn{3.2.5} Given two local trivializing sections\footnote{This refers to the existence of a local trivialization $(U_i,\varphi_i)$ such that $\varphi_i^{-1}(x,g) := \sigma_i(x)g$.} $\sigma_i: U_i\subset M \to P$ and $\sigma_j: U_j\subset M \to P$ on a given principal $G$-bundle $P$ with $U_i\cap U_j \neq 0$ there exists some local {\it gauge functions} $\Lambda_{ij}: U_i\cap U_j \to G$ such that 
\begin{equation}
    \sigma_j(x) \= \sigma_i(x)\Lambda_{ij}(x) \quad\forall\quad x\in U_i\cap U_j\ .
\end{equation}
One defines {\it local representatives} $s_i: U_i \to F$, for a section $s\in \Gamma^\infty(P_F)$ corresponding to such local sections $\sigma_i$, by 
\begin{equation}
    s_i(x)\ :=\ \phi_{s_i}(\sigma_i(x))\ .
\end{equation}

\rem{3.2.5} The local representatives also satisfy the same gauge transformation rule as above, i.e.
\begin{equation}
    s_j(x) \= s_i(x)\Lambda_{ij}(x)\ .
\end{equation}
It turns out that these gauge functions $\Lambda_{ij}$ are nothing but transitions functions $g_{ij}$ of local trivializations arising from $\sigma_i$ and $\sigma_j$. For this thesis, we will identify the gauge group with the structure group.

\vspace{12pt}
%----------------------------------------------------------------------------------------
\section{Connections}
\vspace{2pt}

\rem{3.3.1} Let $(P,M,\pi)$ be a principal $G$-bundle. There exists a Lie algebra homomorphism between $L(G)\cong T_eG$ and $\Gamma^\infty(TP)$ given by
\begin{equation}
    i :\ T_eG \to \Gamma^\infty(TP)\ ,\quad A \to X^A\ ,
\end{equation}
where the induced vector-field $X^A$ arises from the right action of $G$ on $P$ as follows
\begin{equation}
    X^A[f]\ :=\ f'(p\,\mathrm{exp}(tA))(t{=}0)\ .
\end{equation}

\defn{3.3.1} For a given principal $G$-bundle $(P,M,\pi)$ one defines the {\it vertical subspace} at $p\in P$, denoted by $V_pP$, as follows
\begin{equation}
    V_pP\ :=\ \{X \in T_pP ~|~ \pi_*(X) \= 0  \}
\end{equation}
The {\it horizontal subspace} $H_pP$ arises as the orthogonal complement of $V_pP$ in the tangent space $T_pP$:
\begin{equation}
    T_pP \= V_pP \oplus H_pP\ .
\end{equation}

\rem{3.3.2} It can be shown that
\begin{equation}
    X^A_p \in V_pP \quad\forall\quad p\in P\ .
\end{equation}
This means that the above map $i$ induces an isomorphism $i_p$ between $T_eG$ and $V_pP$.

\defn{3.3.2} A {\it connection} on a principle $G$-bundle $(P,M,\pi)$ is a smooth assignment of $H_pP$ to each point $p\in P$ such that
\begin{enumerate}[(a)]
    \item $T_pP \= V_pP \oplus H_pP$,
    \item $(r_g)_*(H_pP) \= H_{pg}P$ (see Remark 2.3.7) and
    \item every $X_p \in T_pP$ has a unique decomposition according to (a) as follows
    \begin{equation}
        X_p \= ver(X_p) + hor(X_p)\ ,
    \end{equation}
    where $ver(X_p) \in V_pP$ and $hor(X_p) \in H_pP$.
\end{enumerate}

\rem{3.3.3} A more technically convenient way to deal with connections is by associating them with a Lie-algebra valued one-form $\omega$ in the following way
\begin{equation}
    \omega_p(X) \= i_p^{-1}(ver(X))\ ,
\end{equation}
which, in turn, imposes following conditions on $\omega$:
\begin{enumerate}[(i)]
    \item $\omega_p(X^A) = A$ for all $p\in P$ and $A\in L(G)$,
    \item $(r_g)^*\omega = \mathrm{Ad}_{g^{-1}}\omega$, i.e.~$(r_g^*\omega)_p(X) = \mathrm{Ad}_{g{-1}}(\omega_p(X))$ for all $X\in T_pP$ and
    \item $X\in H_pP$ iff $\omega_p(X) = 0$.
\end{enumerate}

\defn{3.3.3} For a local trivializing section $\sigma: U\subset M \to P$ of a principal $G$-bundle $P$, its local {\it gauge} fields arise from the following local representative of a Lie-algebra valued one-form $\omega$:
\begin{equation}
    \omega^U\ :=\ \sigma^*\omega\ .
\end{equation}
We label the $1$-form components of such local  gauge fields in Yang--Mills theory as
\begin{equation}\label{localGaugeA}
    A_\mu \equiv (\omega^U)_\mu\ ,
\end{equation}
and in general relativity as 
\begin{equation}
    \Gamma_\mu \equiv (\omega^U)_\mu\ .
\end{equation}

\thm{3.3.1} An explicit form of the local gauge field $\omega^U$ for the local trivialization $\varphi: \pi^{-1}(U) \to U\times G$ arising from $\sigma$ with $(\alpha,\beta) \in T_{(x,g)}(U\times G) \cong T_xU \oplus T_gG$ is given by
\begin{equation}
    (\varphi^*\omega)_{(x,g)} \= \mathrm{Ad}_{g^{-1}} (\omega^U_x(\alpha)) + \< \Omega_l, \beta \>_g\ ,
\end{equation}
where $\Omega_l$ is the Maurer--Cartan one-form.

\ex{3.3.1} An important example of such a local representative $\omega^U$ is in the case of the frame bundle $LM$ for an $n$-dimensional manifold $M$ with a given chart $(U,\phi)$. For a given local section $\sigma: U \subset M \to LM$ defined by
\begin{equation}
    \sigma(m)\ :=\ ((\pa_1)_m,\ldots,(\pa_n)_m,) 
\end{equation}
the corresponding $\omega^U$ has following components (the three indices below are divided into two indices $\alpha,\beta$ for the Lie-algebra $L(GL(n,\R))$ and one $1$-form index $\mu$):
\begin{equation}
    ((\omega^U)_\mu)^\alpha_{\ \beta}\ =:\ \Gamma^\alpha_{\ \mu \beta} \with \alpha,\beta,\mu = 1,\dots,n
\end{equation}
where $\Gamma^\alpha_{\ \mu \beta}$ is the famous {\it Christoffel symbol} of this {\it Levi-Civita or affine connection}\footnote{This relies on a choice of the natural metric on $M$, which is akin to choosing a basis $G^{\ a}_b$ of the Lie-algebra $L(GL(n,\R))$ with components $(G^{\ a}_b)^c_{\ d} := \delta^c_b\delta^a_d$.} that is widely used in Riemannian geometry and general relativity.

\thm{3.3.2} Let $(P,M,\pi)$ be a principal $G$-bundle with $U_i,U_j \subset M$ such that $U_i\cap U_j \neq 0$. Further, let $A_\mu^{(i)}$ and $A_\mu^{(j)}$ be local gauge functions arising from given local trivializing sections $\sigma_i: U_i \to P$ and $\sigma_j: U_j \to G$ respectively. The transformation of these fields under the action of gauge functions $\Lambda_{ij}: U_i\cap U_j \to G$ is, for any $m\in M$, given by 
\begin{equation}
    A_\mu^{(j)}(m) \= \mathrm{Ad}_{\Lambda_{ij}(m)^{-1}}\left(A_\mu^{(i)}(m)\right) + (\Lambda_{ij}^*\Omega_l)_\mu(m)\ .
\end{equation}

\rem{3.3.4} We notice that for matrix Lie groups the above transformation rule takes the following simple form:
\begin{equation}
    A_\mu^{(j)}(m) \= \Lambda_{ij}(m)^{-1}\left(A_\mu^{(i)}(m)\right)\Lambda_{ij}(m) + \Lambda_{ij}(m)^{-1}\pa_\mu \Lambda_{ij}(m)\ .
\end{equation}

\rem{3.3.5} This gauge transformation behaviour is what prevents a gauge field $\omega^U$ from being global, i.e.~$A ~\mathrm{or}~ \Gamma \not\in \Omega^1(M)$. In particular, this explains why the Christoffel symbol is not a tensor! This is because of the following transformation rule between any two given charts $(U,\phi)$ and $(U',\phi')$ on a $4$-dimensional manifold $M\ni m$ with $\phi(m) = \{ x^0,\ldots,x^4 \}$ and $\phi'(m) = \{ x'^{0},\ldots,x'^{4} \}$ where the indices, such as $\alpha$ below, takes temporal $\alpha {=} 0$ as well as spatial $\alpha {=} 1,2,3$ values:
\begin{equation}
    \Gamma^{'\alpha}_{\ \mu\beta} \= \frac{\pa x'^{\alpha}}{\pa x^{\tilde{\alpha}}}\,\frac{\pa x^{\tilde{\mu}}}{\pa x'^{\mu}}\,\frac{\pa x^{\tilde{\beta}}}{\pa x'^{\beta}}\,\Gamma^{\tilde{\alpha}}_{\ \tilde{\mu}\tilde{\beta}} + \frac{\pa x'^{\alpha}}{\pa x^{\lambda}}\,\frac{\pa x^\lambda}{\pa x'^{\mu}\pa x'^{\beta}}\ .
\end{equation}
An explicit expression for $\Gamma^{\alpha}_{\ \mu\beta}$ for a given basis of vector fields $\{ \pa_0,\ldots,\pa_4 \}$ on $M$ equipped with a pseudo-Riemannian metric $g$ with components $g(\pa_\mu,\pa_\nu) = g_{\mu\nu}$ is given by
\begin{equation}
    \Gamma^{\alpha}_{\ \mu\beta} \= \frac{1}{2}\,g^{\alpha\lambda} \left( g_{\mu\lambda,\beta} + g_{\lambda\beta,\mu} - g_{\mu\beta,\lambda} \right)\ ,
\end{equation}
where $g_{{\mu\beta,\lambda}} := \pa_\lambda[g_{\mu\beta}]$.

\vspace{12pt}
%----------------------------------------------------------------------------------------
\section{Parallel transport}
\vspace{2pt}

\defn{3.4.1} Owing to the fact that $\pi_*: H_pP \to T_{\pi(p)}M$ is an isomorphism, there exists the notion of a unique vector field for a given $X\in \vecF{M}$ known as the {\it horizontal lift} of $X$ and is denoted as $X^\uparrow$. This satisfies, for all $p\in P$, the following conditions
\begin{enumerate}[(i)]
    \item $ver(X_p^\uparrow) = 0$ and
    \item $\pi_*(X_p^\uparrow) = X_{\pi(p)}$.
\end{enumerate}

\rem{3.4.1} The act of horizontal lifting is $G$-equivariant i.e. $(r_g)_*(X_p^\uparrow) = X_{pg}^\uparrow$.

\defn{3.4.2} A {\it horizontal lift} of a smooth path $\gamma: [a,b]\subset \R \to M$ is another path $\gamma^\uparrow: [a,b] \to P$ which is horizontal, i.e.~$ver(\gamma^\uparrow) = 0$ such that $\pi(\gamma^\uparrow(t)) = \gamma(t)$ for all $t\in [a,b]$.

\thm{3.4.1} For each point $p\in \pi^{-1}(\{\gamma(a)\})$ there exist a unique horizontal lift $\gamma^\uparrow: [a,b] \to P$ of $\gamma$ such that $\gamma^\uparrow(a)=p$.

\rem{3.4.2} Given a path $\gamma: [a,b] \to M$ and another path $\beta: [a,b] \to P$ which projects down to $\gamma$, i.e.~$\pi(\beta(t)) = \gamma(t)$ for all $t\in [a,b]$, there exits some unique function $g: [a,b]\to G$ such that 
\begin{equation}
    \gamma^\uparrow(t) \= \beta(t)\,g(t) \quad\forall\quad t\in [a,b]\ .
\end{equation}

\thm{3.4.2} The unique path $g: [a,b] \to G$ defined above satisfies the following first order ODE in terms of a Lie-algebra valued $1$-form $\omega$:
\begin{equation}
    0 \= \mathrm{Ad}_{g(t)^{-1}*}\left(\omega_\beta(t)(X_{\beta,\beta(t)}) \right) + \< \Omega_l,X_g\>_{g(t)}\ ,
\end{equation}
where $X_{\beta,\beta(t)} \equiv \dot{\beta}(t)$ is the tangent vector to the curve $\beta$ at point $\beta(t)$ (see Definition 2.1.6).

\rem{3.4.3} For a matrix Lie group $G$ and a local gauge field $A_\mu$ \eqref{localGaugeA}, the above ODE takes the following simple form
\begin{equation}
    \dot{g}(t) \= - A_\mu(\gamma(t))\,\dot{\gamma}^\mu(t)\,g(t)\ ,
\end{equation}
where the components of the curve $\gamma$ in a local chart has been denoted as $\gamma^\mu$. The solution to this ODE, for an initial condition $g(0)=g_0$, is obtained as a path-ordered exponential in the following way
\begin{equation}
 \begin{aligned}
    g(t) &\= \left({\bf P}\,\mathrm{exp}\left( -\int\limits_a^t A_\mu(\gamma(s))\,\dot{\gamma}^\mu(s)\diff{s} \right) \right)g_0 \\
    &\= g_0 - \left(\int\limits_a^t A_\mu(\gamma(s))\,\dot{\gamma}^\mu(s)\diff{s}\right)g_0 \\
    &\qquad\qquad + \left(\int\limits_a^t\diff{s_1}\int_a^{s_1}\diff{s_2}\, A_{\mu_1}(\gamma(s_1))\,A_{\mu_2}(\gamma(s_2))\,\dot{\gamma}^{\mu_1}(s_1)\,\dot{\gamma}^{\mu_2}(s_2)\right)g_0 - \ldots\ .
 \end{aligned}
\end{equation}

\rem{3.4.4} From the above result we observe that the local expression for the horizontal lift $\gamma^\uparrow$, for the path $\gamma: [a,b] \to U \subset M$, is given by
\begin{equation}
    \gamma^\uparrow(t) \= \sigma(\gamma(t))\left({\bf P}\,\mathrm{exp}\left( -\int\limits_a^t A_\mu(\gamma(s))\,\dot{\gamma}^\mu(s)\diff{s} \right) \right)g_0\ .
\end{equation}

\defn{3.4.3} The {\it parallel transport} along a path $\gamma: [a,b] \to M$ is defined by the following map
\begin{equation}
    T_\gamma :\ \pi^{-1}(\gamma(a)) \to \pi^{-1}(\gamma(b))\ ,\quad p \mapsto \gamma^\uparrow_p(b)\ ,
\end{equation}
where $\gamma^\uparrow$ is the unique horizontal lift of $\gamma$ passing through $p\in \pi^{-1}(\gamma(a))$.

\rem{3.4.5} We observe that $T_\gamma$ is a bijection on fibres and thus on $G$. An interesting thing happens when $\gamma: [a,b] \to M$ is a loop i.e. $\gamma(a) = \gamma(b)$; one obtains a natural map from loops based at $\gamma(a) \in M$ to elements of $G$. The subgroup of all elements of $G$ that can be obtained in this way is called the {\it holonomy} group of the principle bundle $P$. This plays an important role in understanding the relation between certain topological properties of $M$ with the connection $A_\mu$.

\defn{3.4.4} Let $(P,M,\pi)$ be a principal $G$-bundle equipped with a connection $1$-form $\omega$. Furthermore, let $(P_F,M,\pi_F)$ be associated with $P$ via the left action of $G$ on $F$. A {\it vertical subspace} of $T_{[p,v]}P_F$ is defined, analogous to that of principal bundle, as
\begin{equation}
    V_{[p,v]}P_F\ :=\ \{ X \in T_{[p,v]}P_F ~|~ \pi_{F*}X = 0 \}\ .
\end{equation}
Similarly, the {\it horizontal subspace} $H_{[p,v]}P_F$ can be defined as the orthogonal complement:
\begin{equation}
    T_{[p,v]}P_F \= V_{[p,v]}P_F \oplus H_{[p,v]}P_F\ .
\end{equation}

\defn{3.4.5} The {\it horizontal lift} of a path $\gamma: [a,b] \to M$ to the associated bundle $P_F$ and passing through $[p,v] \in \pi_F^{-1}(\gamma(a))$ is defined as
\begin{equation}
    \gamma_F^\uparrow(t) \ :=\ [\gamma^\uparrow(t),v]
\end{equation}
where $\gamma^\uparrow(a) = p$.

\rem{3.4.6} One can define {\it parallel transport} $T_\gamma$ along a path $\gamma: [a,b] \to M$ on the associated bundle $P_F$ analogous to the principal bundle case using the above definition of the horizontal lifting. For associated vector bundles $P_V$ with the vector space $V$ admitting a linear representation of $G$, this notion then facilitates the following definition of the covariant derivative. 

\subsection{Covariant derivative}

\defn{3.4.6} The covariant derivative of a section $\psi: M \to P_V$ of an associated vector bundle $P_V$ along a path $\gamma: [0,\varepsilon] \to M$ with $\varepsilon>0$ at $m_0=:\gamma(0)$ is defined by
\begin{equation}
    D_{X_{\gamma,\gamma(0)}}\psi\ :=\ \lim_{t\to 0}\left( \frac{T_\gamma(\psi(\gamma(t))) - \psi(m_0)}{t} \right)\in \pi_V^{-1}(m_0)\ .
\end{equation}

\rem{3.4.7} The covariant derivative $D_X$, for $X\in T_mM$, has the following algebraic properties for any section $\psi,\widetilde{\psi}\in \Gamma^\infty(P_V)$:
\begin{enumerate}[(i)]
    \item $D_{fX+Y}\psi \= f\,D_X\psi + D_Y\psi$ for all $f\in C^\infty(M)$ and $X,Y\in T_mM$,
    \item $D_X(\psi + \widetilde{\psi}) \= D_X\psi + D_X\widetilde{\psi}$ for all $X\in T_mM$ and
    \item $D_X(f\psi) \= X[f]\,\psi + f\,D_X\psi$ for all $f\in C^\infty(M)$ and $X\in T_mM$.
\end{enumerate}
This following more generic notion of a covariant derivative \eqref{genericD} is related to this one, as we will see below.

\defn{3.4.7} The {\it exterior covariant derivative} of a $k$-form $\omega \in \Omega^k(P)$ on a principal bundle $P$ is a horizontal $(k{+}1)$-form defined by
\begin{equation}
    D\omega(X_1,\ldots,X_{k+1})\ :=\ \diff{\omega}(hor(X_1),\ldots,hor(X_{k+1}))
\end{equation}
for a given set of vector fields $X_1,\ldots,X_{k+1} \in \vecF{P}$.

\rem{3.4.8} This notion of covariant derivative can be extended to an associated vector bundle $P_V$ discussed before with the aid of a given $G$-equivariant map $\phi: P \to F$ that has an associated section $\sigma_\phi \in \Gamma^\infty(P_V)$ (see Theorem 3.2.2) as follows
\begin{equation}\label{genericD}
    D\phi\ :=\ \diff{\phi}\circ hor\ .
\end{equation}
One can show, for any $X\in \vecF{P}$ and a given connection $\omega$ on $P$, that
\begin{equation}
    F\ni D\phi(X) \equiv D_X\phi\ :=\ \diff{\phi}(X) + \omega(X)\phi\ ,
\end{equation}
where we notice that $\phi$ are $F$-valued functions on $P$. Furthermore, one can pull this definition back to $M$ using any local trivializing map $\sigma: U\subset M\to P$ and noticing the fact that the pull-back operation is natural:
\begin{equation}
    \sigma^*(D\phi)(X)\ :=\ \diff{\sigma^*\phi}(X) + \sigma^*(\omega)(X)(\sigma^*\phi) \quad\forall\quad X \in \vecF{P}\ .
\end{equation}

\ex{3.4.1} For a given local orthonormal coframe of vector fields $\{\pa_\mu \}$ with $\mu=0,1,2,3$ on a $4$-dimensional Lorentzian manifold $M$, e.g. the Minkowski space $\R^{1,3}$, the covariant derivative $D_{\pa_\mu}=:D_\mu$ of the above-mentioned section $\phi$ (or $\sigma_\phi$ to be precise) is given in terms of local gauge fields $A_\mu$ as
\begin{equation}
    D_\mu\phi\ :=\ \pa_\mu\phi + A_\mu\phi\ .
\end{equation}

\ex{3.4.2} The covariant derivative $D_\mu =: \nabla_\mu$ of the tensor bundle $T^{p,q}(M)$ associated with the frame bundle $LM$ on $M$ is given in terms of the Levi--Civita connection and can be expressed in terms of the Chritoffel symbols. For example, covariant derivative of $T \in T^{1,2}(M)$ with components $T^\mu_{\alpha\beta} := T(\diff{x}^\mu,\pa_\alpha,\pa_\beta)$ is given by
\begin{equation}
    D_\rho\,T^\mu_{\alpha\beta} \= \pa_\rho T^\mu_{\alpha\beta} + \Gamma^\mu_{\rho\lambda}T^\lambda_{\alpha\beta} - \Gamma^\lambda_{\rho\alpha}T^\mu_{\lambda\beta} - \Gamma^\lambda_{\rho\beta}T^\mu_{\lambda\alpha}\ .
\end{equation}

\rem{3.4.9} The Levi--Civita connection $\Gamma$ is metric compatible:
\begin{equation}
    \nabla g = 0 \nabla_\mu\quad \textrm{or}\quad g_{\alpha\beta} \= 0\ .
\end{equation}
Moreover this connection is torsion-free, i.e.
\begin{equation}
    0 \= T(X,Y)\ :=\ \nabla_XY - \nabla_Y X - [X,Y]
\end{equation}
for all $X,Y \in \vecF{M}$. Choosing vector fields $X=\pa_\alpha$ and $Y=\pa_\beta$, the torsion-free condition ensures that the  Christoffel symbol is symmetric in the subscript indices:
\begin{equation}
    \Gamma^\mu_{\alpha\beta} \= \Gamma^\mu_{\beta\alpha}\ .
\end{equation}

\subsection{Curvature}

\defn{3.4.8} If $\omega$ is a connection $1$-form on a principal $G$-bundle $P$ then $D\omega=:\Omega$ is the Lie-algebra valued {\it curvature} $2$-form of $\omega$.

\rem{3.4.10} It can be shown that the curvature $2$-form $\Omega$ satisfies the Bianchi identity:
\begin{equation}
    D\Omega \= 0\ .
\end{equation}

\thm{3.4.3} For arbitrary pair of vector fields $X,Y \in \vecF{P}$ the curvature $2$-form $D\omega$ satisfies the following Cartan structure equation
\begin{equation}
    D\omega(X,Y)\ :=\ \diff{\omega}(X,Y) + [\omega(X),\omega(Y)]\ ,
\end{equation}
where we have employed the Lie bracket on $L(G)$.

\ex{3.4.3} For the Levi--Civita connection $\Gamma$ the curvature $2$-form $D\Gamma=:R$---valued in $L(GL(n,\R))$ for an $n$-dimensional pseudo-Riemannian manifold $M$---is known as the Riemann curvature and is given by
\begin{equation}
    R \= \diff{\Gamma} + \Gamma\wedge \Gamma\ .
\end{equation}
This turns out to be a $(1,3)$-tensor, i.e.~$R \in T^{1,3}(M)$ and, in local coframe $\{ \pa_\mu \}$ and frame $\{ \diff{x}^\mu \}$ with $\mu =0,1,2,3,$ of $M$, admits the following expression in terms of Christoffel symbol
\begin{equation}
    R^\rho_{\ \sigma\mu\nu} \= \pa_\mu \Gamma^\rho_{\nu\sigma} - \pa_\nu \Gamma^\rho_{\mu\sigma} + \Gamma^\rho_{\mu\lambda}\,\Gamma^\lambda_{\nu\sigma} - \Gamma^\rho_{\nu\lambda}\,\Gamma^\lambda_{\mu\sigma}\ .
\end{equation}

\rem{4.3.11} Riemann curvature tensor gives rise to the so called Ricci tensor $R_{\mu\nu} := R^\rho_{\ \mu\rho\nu} \in T^{0,2}(M)$ through index contraction, which yields the scalar curvature $R:= g^{\mu\nu}R_{\mu\nu}$. With these tools and the insight of {\it equivalence principle}: \\ ``For any point $m\in M$ there always exists a local chart $(U,\phi)$ with $m\in U$ admitting a smooth local orthonormal frame $\sigma \in \Gamma^\infty(LU)$ of the local frame bundle $LU$; in other words, every (spacetime) manifold is locally flat in a Minkowski sense" \\ Albert Einstein was able to construct the following field equation relating the curvature of spacetime with its matter content
\begin{equation}
    G_{\mu\nu}\ :=\ R_{\mu\nu} - \ \frac{1}{2}g_{\mu\nu}R \= \frac{8\pi G}{c^4} T_{\mu\nu}\ ,
\end{equation}
where $G$ is the Newton constant, $c$ is the speed of light and $T_{\mu\nu}$ is the stress-energy tensor arising from the variation of the matter action $S_m$ with respect to the matric:
\begin{equation}
    T_{\mu\nu} \= -\frac{2}{\sqrt{-\mathrm{det}\,g}} \frac{\delta S_m}{\delta g^{\mu\nu}}\ .
\end{equation}

\ex{3.4.4} A trivial solution of the vacuum Einstein equation is the Minkowski metric $\eta_{\mu\nu}$ which is globally flat. Another very important solution is the FLRW metric for homogeneous and isotropic universe, where the evolution of the {\it scale factor} $a(t)$ is governed by an appropriate stress-energy tensor $T_{\mu\nu}$, is given by
\begin{equation}
    g \= -c^2\diff{t}^2 + a(t)^2\left(\frac{\diff{r}^2}{1-\kappa r^2} + r^2\diff{\theta}^2 + r^2\sin^2\theta\diff{\phi}^2 \right)\ ,
\end{equation}
where $\{r,\theta,\phi \}$ are coordinates on celestial spheres with respect to an observer (like us) and the parameter $\kappa = -1$, $0$ or $1$ denotes the topology of the $3$-dimensional Euclidean space as being open, flat or closed respectively. 

\ex{3.4.5} For local Yang--Mills field $A:=\sigma^*\omega$ the curvature $2$-form $D\omega=:F$ is given by
\begin{equation}
    F \= \diff{A} + A\wedge A\ .
\end{equation}

\vspace{12pt}
%----------------------------------------------------------------------------------------
\section{Yang-Mills equation}
\vspace{2pt}

\rem{3.5.1} A local expression for the Yang--Mills curvature $F$ on a Lorentzian manifold $M$ with local orthonormal coframe $\{\pa_\mu \}$ is given by
\begin{equation}
    F_{\mu\nu} \= \pa_\mu A_\nu - \pa_\nu A_\mu - [A_\mu,A_\nu]\ .
\end{equation}

\rem{3.5.2} The Yang--Mills curvature transforms under a local gauge transformation $\Lambda_{ij}(m)$ (see Definition 3.2.5) in the following way
\begin{equation}
    F^{(j)}_{\mu\nu}(m) \= \Lambda_{ij}(m)^{-1}\,F^{(i)}_{\mu\nu}(m)\,\Lambda_{ij}(m)\ ,
\end{equation}
and is thus global, i.e.~$F\in \Omega^2(M)$.

\rem{3.5.3} The Yang--Mills equation on a Lorentzian manifold $M$ (with $D$ as in \ref{genericD}) is given by
\begin{equation}
    *D(*F) \= J\ ,
\end{equation}
where the $1$-form $J \in \Omega^1(M)$ is the {\it current} that arises from the presence of any source ``charge" on the manifold.

\rem{3.5.4} The Yang--Mills action (coupling constant $g$)
\begin{equation}
    S_{YM} \= \frac{1}{2g^2}\int_M \mathrm{Tr}(F\wedge *F)
\end{equation}
together with the following action for the source $J$:
\begin{equation}
    S_J \= \frac{1}{g^2} \mathrm{Tr}(A \wedge *J)
\end{equation}
gives rise to the above Yang--Mills equation by variational principle.

\rem{3.5.5} Maxwell's theory of electromagnetism that we saw in the last chapter is a special case of Yang--Mills theory where the gauge group is $U(1)$.

%% file: Chapters/Chapter4.tex
%auto-ignore
\chapter{Yang--Mills equations on $\diff{S}_4$}
\label{Chapter4}
\justifying

Here we review the conformal relation of the $4$-dimensional de Sitter space with a finite Lorentzian cylinder with $S^3$-slicing, study calculus on $S^3$ and present the Yang--Mills field equations on the cylinder. The content of this chapter is partially taken from \cite{KL20,KPH21,KLP21}\footnote{The role of $L_a$ and $R_a$ in \cite{KL20} is interchanged as compared to this thesis.}.

\vspace{12pt}
%----------------------------------------------------------------------------------------
\vspace{2pt}
\section{The de Sitter-Minkowski correspondence via $S^3$-cylinder}
\noindent
The de Sitter space in four dimensions $\diff S_4$ has a natural embedding as a single-sheeted hyperboloid in five-dimensional Minkowski space $\R^{1,4}$ with coordinates $(q_0, q_{_A}); A=1,2,3,4$ and global length scale $\ell$ and is given by
\begin{equation} \label{dS4}
   -q_0^2\, + q_1^2\, + q_2^2\, + q_3^2\, + q_4^2\, \= \ell^2 .
\end{equation}
One can use the flat metric on $\R^{1,4}$:
\begin{equation} \label{flat_metric}
   \diff s_{(1,4)}^2 \= -\diff q_0^2\, + \diff q_1^2\, + \diff q_2^2\, + \diff q_3^2\, + \diff q_4^2
\end{equation}
to construct the corresponding metric on $\diff S_4$ using the coordinate constraint \eqref{dS4}. The metric thus obtained is conformally equivalent to the metric on a finite Lorentzian cylinder over the $3$-sphere $S^3$: $\mathcal{I}\times S^3$ with $\mathcal{I} = \left(0, \pi \right)$. To see this, we employ the following coordinates
\begin{equation}
   q_{_A} \= \ell\, \omega_{_A}\, \csc\tau \und q_{_5} \= -\ell\, \cot\tau \quad\with\quad \tau\, \in\, (0,\pi)\ ,
\end{equation}
where $\omega_{_A}$ are the natural embedding coordinates of $S^3\hookrightarrow\R^4$: $\omega_{_A}\omega_{_A} \!=\! 1$\footnote{Repeated indices are summed over.}. A natural hyperspherical parametrisation of $\omega_{_A}$ is given by
\begin{equation}\label{omegas}
   \omega_1 \= \sin\chi\,\sin\theta\,\cos\phi\ ,\quad \omega_2 \= \sin\chi\,\sin\theta\,\sin\phi\ ,\quad \omega_3 \= \sin\chi\,\cos\theta\ ,\quad \omega_4 \= \cos\chi\ ,
\end{equation}
with $0\leq \chi,\theta \leq \pi$ and $0\leq\phi\leq2\pi$. The modified metric has the following form:
\begin{equation}\label{metricdS}
   \diff s^2 \= \diff s_{(1,4)}^2|_{\diff S_4} \= \frac{\ell^2}{\sin^2\tau}\left( -\diff\tau^2 + \diff\Omega_3^2 \right); \quad \diff\Omega_3^{\ 2}\=\diff\omega_{A}\, \diff\omega_{A}|_{\diff S_4}\ .
\end{equation}
By gluing two copies of such Lorentzian cylinders at $\tau{=}0$ by taking $\tau \in \widetilde{\cal I} := \left(-\pi, \pi \right)$ one finds that half of the resultant cylinder $\widetilde{\cal I}\times S^3$ is conformally equivalent to the $4$-dimensional Minkowski space via following parametrization of $(t,x,y,z)\in \R^{1,3}$:
\begin{equation} \label{S3toMink}
   \cot\tau \= \frac{r^2 {-} t^2 {+} \ell^2}{2\,\ell\,t}\ ,\
   \omega_1 \= \gamma\,\frac{x}{\ell}\ ,\
   \omega_2 \= \gamma\,\frac{y}{\ell}\ ,\
   \omega_3 \= \gamma\,\frac{z}{\ell}\ ,\
   \omega_4 \= \gamma \frac{r^2 {-} t^2 {-} \ell^2}{2\,\ell^2}\ ,
\end{equation}
where we have abbreviated
\begin{figure}[!htbp]
\centering
\captionsetup{width=\linewidth}
\includegraphics[width = 0.65\paperwidth]{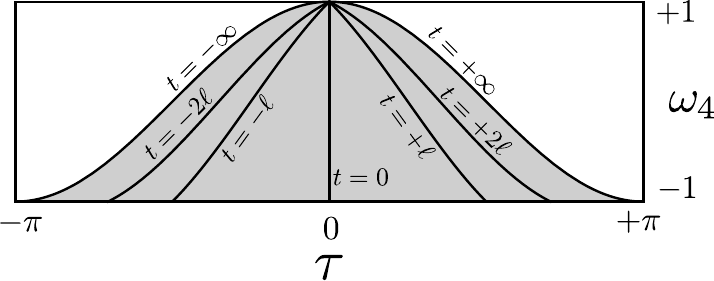}
\caption{An illustration of the map between a cylinder $2\mathcal{I}\times S^3$ and Minkowski space $R^{1,3}$.
The Minkowski coordinates cover the shaded area. The boundary of this area is given by the curve $\omega_4{=}\cos\chi{=}\cos\tau$. 
Each point is a two-sphere spanned by $\{\hat{\omega}_1,\hat{\omega}_2,\hat{\omega}_3\}$, which is mapped to a sphere of constant $r$ and $t$. }
\label{patching}
\end{figure}
\begin{equation}\label{gamma}
   r^2 = x^2+y^2+z^2\ \ \&\ \
   \gamma \= \frac{2\,\ell^2}{\sqrt{4\,\ell^2 t^2 + (r^2-t^2+\ell^2)^2}}
   =  \frac{2\,\ell^2}{\sqrt{4\,\ell^2 r^2 + (t^2-r^2+\ell^2)^2}}\ .
\end{equation}
A straightforward computation yields
\begin{equation}
   \cos\tau-\cos\chi\= \gamma\ >0\ ,
\end{equation}
which shows that only half of the cylinder---constrained by $\cos\tau <\cos\chi$---is allowed by the map \eqref{S3toMink}. Plugging this map back into the de Sitter metric \eqref{metricdS} we obtain
\begin{equation}\label{metricMink}
   \diff s^2 \=
   \frac{\ell^2}{t^2}\,\bigl(-\mathrm{d}t^2 +\diff x^2 +\diff y^2 +\diff z^2\bigr) 
   \with (x,y,z)\equiv(x^1,x^2,x^3)\in\R^3\ \&\ t\in\R \ ,
\end{equation}
which is the Minkowski metric up to a conformal factor\footnote{In fact, the map \eqref{S3toMink} covers only the positive half of the Minkowski space, i.e.~$t\in\R_+$ for the original cylinder ${\cal I}\times S^3$.}. A smooth gluing of the two cylinders across the time slice $t\!=\tau\!=0$ is nicely depicted in Figure \ref{patching}. Now, looking at the maps \eqref{omegas} and \eqref{S3toMink} we see a SO(3)-symmetry that we can exploit by writing
\begin{equation}\label{OmegaToHatOmega}
   \omega_a \= \sin\chi\; \hat{\omega}_a\ ~\textrm{and}~\ x^a \= r\,\hat{\omega}_a\ ~\textrm{for}~\ a\=1,2,3\ ~\textrm{and}~ r\= \sfrac\ell\gamma\,\sin\chi
\end{equation}
where the unit $S^2$ coordinates $\hat{\omega}_a$ are given by
\begin{equation}\label{S2coord}
   \hat{\omega}_1 \= \sin\theta\,\cos\phi\ , \quad \hat{\omega}_2 \= \sin\theta\,\sin\phi \und \hat{\omega}_3 \= \cos\theta\ .
\end{equation}
\begin{figure}[!ht]
\centering
\captionsetup{width=\linewidth}
\includegraphics[width = 0.4\linewidth]{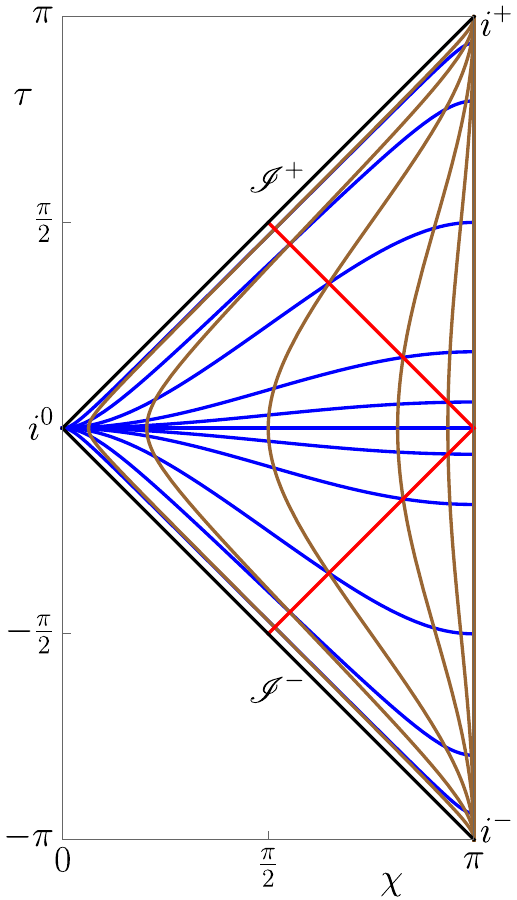}
\caption{
Penrose diagram of Minkowski space~$\R^{1,3}$. Each point hides a two-sphere $S^2{\ni}\{\theta,\phi\}$.
Blue curves indicate $t{=}\textrm{const}$ slices while brown curves depict the world volumes of $r{=}\textrm{const}$ spheres.
The lightcone of the Minkowski-space origin is drawn in red.
}
\label{CPplot}
\end{figure}
We can identify the unit $S^2$ between the Minkowski space and the de Sitter space using  (\ref{S3toMink}) and (\ref{OmegaToHatOmega}), so that, we have an effective map between coordinates $(t,r)$ and $(\tau,\chi)$:
\begin{equation} \label{tr2TauChi}
   \frac{t}{\ell} \= \frac{\sin\tau}{\cos\tau\,-\,\cos\chi} \quad\und\quad
   \frac{r}{\ell} \= \frac{\sin\chi}{\cos\tau\,-\,\cos\chi} \qquad\for\quad
   \chi>|\tau|\ ,
\end{equation}
which reveals that the triangular $(\tau,\chi)$ domain (where the points represent unit $S^2$) is nothing but the Penrose diagram of Minkowski space (See Figure \ref{CPplot}). The special lines and points in Figure \ref{CPplot} are given by
\begin{center}
\resizebox{\columnwidth}{!}{
 \begin{tabular}{|c|ccc|ccc|c|}
  \hline
   & \quad$\tau{=}0$ & south pole & boundary & \quad --- \quad{} & \!\!\!north pole\!\!\! & $\quad\tau{=}{\pm}\pi\quad$ & $\chi{\pm}\tau{=}\pi$ \\[2pt]
   $(\tau,\chi)$ & \quad$(0,\chi)$ & $(\tau,\pi)$ & $(\pm\chi,\chi)$ & $(0,\pi)$ & $(0,0)$ & $(\pm\pi,\pi)$ & $(\pm\pi{\mp}\chi,\chi)$ \\[4pt]
   $(t,r)$ & \quad$(0,r)$ & $(t,0)$ & $(\pm\infty,\infty)$ & $(0,0)$ & $(t,\infty)$ & $(\pm\infty,r)$ & $(\pm r,r)$ \\[2pt]
   & \quad$t{=}0$ & $r{=}0$ & $\mathscr{I}^\pm$ & \ \ origin \ {} & $i^0$ & $i^\pm$ & \ \  lightcone \ \ {} \\
  \hline
 \end{tabular}
 }
\end{center}
with the Minkowski spatial and temporal infinity $i^0$ and $i^\pm$ corresponding to the edges while the Minkowski null infinity~$\mathscr{I}^\pm$ corresponding to the conformal boundary $\chi{=}|\tau|$ of the Penrose diagram.

Equation \eqref{tr2TauChi} can be used to obtain the Jacobian for the transformation between the coordinates $y^m\in\{\tau,\chi,\theta,\phi\}$ and $x^{\mu}\in\{t,r,\theta,\phi\}$:
\begin{equation} \label{Jacobian}
   \bigl(J^m_{\ \ \mu}\bigr) \ :=\ \frac{\pa(\tau,\chi,\theta,\phi)}{\pa(t,r,\theta,\phi)}
   \= \frac1\ell\biggl(\begin{matrix} p & -q \\[4pt] q & -p \end{matrix} \biggr) \oplus\mathds{1}_2\ ,
\end{equation}
where the polynomials $p,q$ are given by
\begin{equation}
    p\= \sfrac{\gamma^2}{\ell^2}\,(r^2{+}t^2{+}\ell^2)/2 \= 1{-}\cos\tau\cos\chi \und q\= \sfrac{\gamma^2}{\ell^2}\,t\,r \= \sin\tau\sin\chi\ .
\end{equation}

A more direct way of visualizing the conformal correspondence between Minkwoski space and the $S^3$-cylinder is via Carter-Penrose transformation that readily produces the lightcone picture (see Appendix \ref{appendCPtransf} for details). 

\subsection{Structure of $3$-sphere and its harmonics}
\noindent
The presence of $S^3$ in the metric \eqref{metricdS} is an added advantage which can be exploited to write an $SO(4)$-invariant gauge connection and obtain the corresponding fields by solving Yang--Mills equation on the Lorentzian cylinder. These quantities can be later exported to the Minkowski spacetime owing to the conformal invariance of the vacuum Yang--Mills equation in $4$-dimensions. To this end, we start with the group $SO(4)$---which is isomorphic to two copies of $SU(2)$ (up to a $\mathds{Z}_2$ grading). Each of these $SU(2)$ has a group action that generates a left (right) multiplication (aka translation) on $S^3$. This can easily be checked using the map
\begin{equation}\label{map}
   g:\; S^3 \rightarrow SU(2);~~ (\omega_1,\ \omega_2,\ \omega_3,\ \omega_4)\, \mapsto\, -i
     \begin{pmatrix}
     \beta & \alpha^* \\ \alpha & -\beta^*
     \end{pmatrix}\ ,
\end{equation} 
with $\alpha := \omega_1 + \im\omega_2\ ~\textrm{and}~\ \beta := \omega_3 +\im\omega_4 $. This parameterization of $g$ ensures that the identity element $e = \mathds{1}_2$ of the group $SU(2)$ can be obtained from $(0,0,0,1)$, i.e.~the North pole of $S^3$. It is well known that $S^3$ is the group manifold of $SU(2)$. Keeping this in mind, we consider the Maurer--Cartan one-form
\begin{equation} \label{Cartan1}
   \Omega_l(g)\, :=\, g^{-1}\,\diff g \= e^a\,T_a,~~\textrm{with}~~ T_a = -i\sigma_a
\end{equation} 
being the $SU(2)$ generators. The left-invariant one-form\footnote{Named so because they remain invariant under the dragging induced by the left SU(2) multiplication.} $e^a$ can, alternatively, be expressed using the so called self-dual 't Hooft symbol $\eta^a_{_{BC}}$:
\begin{equation}\label{one-form}
   e^a \= -\eta^a_{_{\ BC}}\, \omega_{_B}\, \diff \omega_{_C} \quad\textrm{where} \quad \eta^a_{\ bc} \= \varepsilon_{abc} \und \eta^a_{\ b4} \= -\eta^a_{\ 4b} \= \delta^a_b\ .
\end{equation}
They satisfy following useful identities
\begin{equation}\label{MaurerCartan}
\delta_{ab}\, e^a\, e^b \= \diff\Omega_3^2 \quad\und\quad \diff e^a + \varepsilon_{abc}\, e^b\wedge e^c \= 0.
\end{equation} 
The left-invariant vector fields $L_a$---generating the right translations---are dual to $e^a$ and are given by
\begin{equation}\label{leftVF}
   L_a \= -\eta^a_{_{\ BC}}\,\omega_{_B}\,\sfrac{\partial}{\partial \omega_{_C}} \qquad\Rightarrow\qquad \left[ L_a , L_b \right] \= 2\,\varepsilon_{abc}\,L_c\ .
\end{equation}
In a similar way, the right-invariant vector fields $R_a$ (generating the left translations) are given by
\begin{equation}\label{rightVF}
   R_a \= -\tilde{\eta}^a_{_{\ BC}}\,\omega_{_B}\,\sfrac{\partial}{\partial\omega_{_C}} \qquad\Rightarrow\qquad
   \left[ R_a , R_b \right] = 2\,\varepsilon_{abc}\,R_c\ ,
\end{equation}
where the anti self--dual 't Hooft symbols $\tilde{\eta}^a_{_{\ BC}}$ are obtained from \eqref{one-form} by flipping the sign in the previous defintion whenever $B,C\!=4$. Furthermore, the vector fields $L_a$ and $R_a$ act on the one-forms $e^a$ via their Lie derivative, which can be performed using Cartan formula (see Section \ref{covectorFields}):
\begin{equation}\label{LieActionL}
\begin{aligned}
    L_a\,e^b\ :=\ \mathcal{L}_{L_a}\,e^b &\= \diff\circ\iota_{L_a} e^b + \iota_{L_a}\circ\diff{e^b} \\
    &\= \diff{\left(e^b(L_a)\right)} - \varepsilon^b_{\ ij}\,\iota_{L_a}\left(e^i\wedge e^j\right) \\
    &\= \diff{\left(\delta^b_a\right)} - \varepsilon^b_{\ ij}\left( e^i(L_a) e^j - e^j(L_a) e^i \right) \\
    &\= 2\,\varepsilon_{abc}\,e^c\ ,
\end{aligned}
\end{equation}
where in the second line we have used \eqref{MaurerCartan}. A similar calculation for the action of $R_a$ yields
\begin{equation}\label{LieActionR}
    R_a\, e^b \= 0\ .
\end{equation}
We can now write differential $\diff$ of the functions $f\in C^{\infty}\left(\mathcal{I}\times S^3\right)$ using $L_a$\footnote{One can use $R_a$ and its dual $1$-form as well.} as
\begin{equation}
   \diff f \= \diff\tau\; \partial_{\tau} f\ +\ e^a\,L_a f\ .
\end{equation}

Functions on $S^3$ can be expanded in a basis of harmonics~$Y_j(\chi,\theta,\phi)$ with $2j\in\N_0$, 
which are eigenfunctions of the scalar Laplacian\footnote{
The $SO(4)$ spin of these functions is actually $2j$, but we label them with their half-spin for reasons to be clear below.}:
\begin{equation}
-\mathop{}\!\mathbin\bigtriangleup_3 Y_j \= 2j(2j{+}2)\,Y_j \= 4j(j{+}1)\,Y_j \=
-\sfrac12(L^2+R^2)\,Y_j \= -\sfrac14(\Dcal^2+\Pcal^2)\,Y_j
\end{equation}
where $L^2=L_a L_a$ and $R^2=R_a R_a$ are (minus four times) the Casimirs of $su(2)_L$ and $su(2)_R$, respectively,
\begin{equation}
-\sfrac14 L^2\,Y_j \= -\sfrac14 R^2\,Y_j \= -\sfrac14\mathop{}\!\mathbin\bigtriangleup_3 Y_j \= j(j{+}1)\,Y_j\ .
\end{equation}
We have also introduced $\Dcal^2=\Dcal_a \Dcal_a$ and $\Pcal^2=\Pcal_a \Pcal_a$ with
\begin{equation} \label{DPdef}
\Dcal_a\ :=\ R_a + L_a \= -2\,\varepsilon_a^{\ bc}\,\omega_b\,\pa_c \und 
\Pcal_a\ :=\ R_a - L_a \= 2\,\omega_{[a}\,\pa_{4]}
\end{equation}
with $\pa_{_A}\equiv\sfrac{\pa}{\pa\omega_A}$ so that
\begin{equation}
\left[ \Dcal_a , \Dcal_b \right] \= 2\,\varepsilon_{ab}^{\ \ c}\,\Dcal_c \ ,\qquad
\left[ \Dcal_a , \Pcal_b \right] \= 2\,\varepsilon_{ab}^{\ \ c}\,\Pcal_c \ ,\qquad
\left[ \Pcal_a , \Pcal_b \right] \= 2\,\varepsilon_{ab}^{\ \ c}\,\Dcal_c \ .
\end{equation}
Hence, $\{\Dcal_a\}$ spans the diagonal subalgebra $su(2)_D\subset so(4)$, which generates the stabilizer subgroup $SO(3)$ 
in the coset representation $S^3\cong SO(4)/SO(3)$. Therefore, $\Dcal^2$ is (minus four times) the Casimir of $su(2)_D$, with eigenvalues $l(l{+}1)$ for $l=0,1,\ldots,2j$,
and $\sfrac14\Dcal^2=\mathop{}\!\mathbin\bigtriangleup_2$ is the scalar Laplacian on the $S^2$ slices traced out in $S^3$ by the $SO(3)_D$ action.

To further characterize a complete basis of $S^3$ harmonics, there are two natural options,
corresponding to two different complete choices of mutually commuting operators to be diagonalized. First, the left-right (or toroidal) harmonics $Y_{j;m,n}$ are eigenfunctions of $L^2=R^2$, $L_3$ and $R_3$:
\begin{equation} \label{Y-action}
\sfrac{\im}{2}\,L_3\,Y_{j;m,n} \= n\,Y_{j;m,n} \quad\und\quad
\sfrac{\im}{2}\,R_3\,Y_{j;m,n} \= m\,Y_{j;m,n} \ ,
\end{equation}
and hence the corresponding ladder operators 
\begin{equation}
L_\pm \= (L_1\pm\im L_2)/\sqrt{2} \quad\und\quad R_\pm \= (R_1\pm\im R_2)/\sqrt{2}
\end{equation}
act, in their Hermitian avatar, as
\begin{equation}\label{JpmAction}
\begin{aligned}
  \sfrac{\im}{2}\,L_\pm\,Y_{j;m,n} &\= \sqrt{(j{\mp}n)(j{\pm}n{+}1)/2}\,Y_{j;m,n\pm1}\ ,\\[4pt]
  \sfrac{\im}{2}\,R_\pm\,Y_{j;m,n} &\= \sqrt{(j{\mp}m)(j{\pm}m{+}1)/2}\,Y_{j;m\pm1,n}\ .
\end{aligned}
\end{equation}
The normalized harmonics $Y_{j;m,n}$ can be expanded in terms of functions $\alpha,\beta$ and their complex conjugates as
\begin{equation}\label{S3Harmonics}
\begin{aligned}
    Y_{j;m,n}(\omega) &\= \sum_{k=0}^{2j}\,K_{j,m,n}(k)\,\alpha^{n+m+k}\, \bar{\alpha}^{k}\, \beta^{j-m-k}\, \bar{\beta}^{j-n-k} \qquad\with \\
    K_{j,m,n}(k) &\= (-1)^{m+n+k}\,\sqrt{\frac{2j+1}{2\pi^2}}\,\frac{\sqrt{(j+m)!(j-m)!(j+n)!(j-n)!}}{(n+m+k)!(j-n-k)!(j-m-k)!k!}\ .
\end{aligned}    
\end{equation}
They satisfy the orthonormality condition
\begin{equation}\label{orthonormality}
    \int \diff^3\Omega_3\, Y_{j;m,n}\, \bar{Y}_{j';m',n'} \= \delta_{j,j'}\, \delta_{m,m'}\, \delta_{n,n'} \with \diff^3\Omega_3 \= \sin^2\chi\sin\theta\, \diff\chi\, \diff\theta\, \diff\phi\ .
\end{equation}
Second, the adjoint (or hyperspherical) harmonics~$\tY_{j;l,M}$ are eigenfunctions of $L^2=R^2$, $\Dcal^2$ and $\Dcal_3$ (its Hermitian version to be precise):
\begin{equation} 
-\sfrac{1}{4}\,\Dcal^2\,\tY_{j;l,M} \= l(l{+}1)\,\tY_{j;l,M} \quad\und\quad
\sfrac{\im}{2}\,\Dcal_3\,\tY_{j;l,M} \= M\,\tY_{j;l,M} \ ,
\end{equation}
with the ladder-operator actions~\cite{wybourne}
\begin{equation}
\begin{aligned}
\sfrac{\im}{2}\,\Dcal_\pm\,\tY_{j;l,M} &\= \sqrt{(l{\mp}M)(l{\pm}M{+}1)/2}\,\tY_{j;l,M\pm1} \ ,\\[4pt]
\sfrac{\im}{2}\,\Pcal_\pm\,\tY_{j;l,M} &\= 
\mp\sqrt{(l{\mp}M{-}1)(l{\mp}M)/2}\,c_{j,l}\,\tY_{j;l-1,M\pm1} \\[2pt] 
&\qquad\pm\ \sqrt{(l{\pm}M{+}1)(l{\pm}M{+}2)/2}\,c_{j,l+1}\,\tY_{j;l+1,M\pm1} \ ,\\[4pt]
\sfrac{\im}{2}\,\Pcal_3\,\tY_{j;l,M} &\= 
\sqrt{l^2{-}M^2}\,c_{j,l}\,\tY_{j;l-1,M} \ +\ 
\sqrt{(l{+}1)^2{-}M^2}\,c_{j,l+1}\,\tY_{j;l+1,M} \ ,
\end{aligned}
\end{equation}
where
\begin{equation}
c_{j,l} \= \sqrt{\bigl((2j{+}1)^2-l^2\bigr)/\bigl((2l{-}1)(2l{+}1)\bigr)}\ .
\end{equation}
In this case, there exists a recursive construction for harmonics on $S^{k+1}$ from those on $S^k$:
\begin{equation}\label{split-Y}
\tY_{j;l,M}(\chi,\theta,\phi) \= R_{j,l}(\chi)\,Y_{l,M}(\theta,\phi)\ ;\
R_{j,l}(\chi) \= \im^{2j+l}\,\sqrt{\sfrac{2j+1}{\sin\chi}\sfrac{(2j+l+1)!}{(2j-l)!}}\, P_{2j+\frac12}^{-l-\frac12}(\cos\chi)\ ,
\end{equation}
where $Y_{l,M}$ are the standard $S^2$ spherical harmonics and $P_a^b$ denote associated Legendre polynomials of the first kind.\footnote{
With fractional indices, it is rather a Gegenbauer polynomial, but also a hypergeometric function (see eq.~(2.8) of~\cite{Higuchi,HiguchiErratum}).} 
The two bases of harmonics are related via the standard Clebsch--Gordan series for the angular momentum addition
$j\otimes j = 0\oplus 1\oplus\ldots\oplus 2j$,
\begin{equation}\label{Y-new}
Y_{j;m,n} \= \sum\limits_{l=0}^{2j}\sum\limits_{M=-l}^{l}\,C_{m,n}^{l,M}\,\tY_{j;l,M}\ ,
\qquad\textrm{with}\qquad C_{m,n}^{l,M} = \<2j;l,M|j,m;j,n\>
\end{equation}
being the Clebsch--Gordan coefficients enforcing $m{+}n{=}M$ and $l\in\{0,1,\ldots,2j\}$.

A somewhat cumbersome calculation involving \eqref{S3toMink}, \eqref{gamma} and \eqref{one-form} gives us the cylinder one--forms in Minkowski coordinates:
\begin{equation}\label{1formsExpanded}
\begin{aligned}
e^0\ &:=\ \diff{\tau} \= \sfrac{\gamma^2}{\ell^3}\bigl(
\sfrac12(t^2{+}r^2{+}\ell^2)\,\diff t - t\,x^k \diff x^k \bigr) \\[2pt]
&\= \sfrac{\gamma^2}{\ell^3}\bigl(
\sfrac12(t^2{+}r^2{+}\ell^2)\,\diff t - t\,r\,\diff r \bigr)
\qquad\und \\[4pt]
e^a &\= \sfrac{\gamma^2}{\ell^3}\bigl(
t\,x^a \diff t - \bigl[\sfrac12(t^2{-}r^2{+}\ell^2)\,\delta^{ak} +
x^a x^k + \ell\,\varepsilon^{ajk}x^j \bigr]\,\diff x^k \bigr) \\[2pt]
&\= \sfrac{\gamma^2}{\ell^3}\bigl(
\hat{x}^a \bigl[ r\,t\,\diff t -\sfrac12(t^2{+}r^2{+}\ell^2)\,\diff r\bigr]
- \sfrac12(t^2{-}r^2{+}\ell^2)\,r\,\diff\hat{x}^a -
\ell\,r^2\varepsilon^{ajk}\hat{x}^j\diff\hat{x}^k \bigr)\ .
\end{aligned}
\end{equation}
We further note down the expressions for the vector fields $L_a$ in terms of $S^3$-angles $(\chi,\theta,\phi)$, using \eqref{omegas} and \eqref{leftVF}, for later purposes
\begin{equation}\label{Lfields}
 \begin{aligned}
   L_1 &\= \sin\theta\cos\phi\, \pa_\chi\ +\ (\cot\chi\cos\theta\cos\phi + \sin\phi)\,\pa_\theta\ -\ (\cot\chi\csc\theta\sin\phi - \cot\theta\cos\phi)\, \pa_\phi\ , \\
   L_2 &\= \sin\theta\sin\phi\, \pa_\chi\ +\ (\cot\chi\cos\theta\sin\phi - \cos\phi)\,\pa_\theta\ +\ (\cot\chi\csc\theta\cos\phi + \cot\theta\sin\phi)\, \pa_\phi\ , \\
   L_3 &\= \cos\theta\, \pa_\chi\ -\ \cot\chi\sin\theta\, \pa_\theta\ -\ \pa_\phi\ .
 \end{aligned}
\end{equation}

%----------------------------------------------------------------------------------------

\section{Yang-Mills field equations}
%\noindent
One can make use of the left-invariant one forms $e^a$ to expand a generic Yang--Mills gauge potential $\mathcal{A}$ (in the temporal gauge $\mathcal{A}_\tau {=}0$) as
\begin{equation}\label{one-formAnsatz}
\mathcal{A} \= X_a(\tau,g)\,e^a,
\end{equation}
where the three $SU(2)$ matrices $X_a$ depends, in general, on both the cylinder parameter $\tau$ and internal $S^3$-coordinates $\omega$ via the map $g$ \eqref{map}. The corresponding field strength $\mathcal{F}$, for this gauge connection $\mathcal{A}$, is given by
\begin{equation} \label{FcalExpanded}
\mathcal{F} \= \diff{\Acal} + \mathcal{A}\wedge\mathcal{A} \= \dot{X}_a\,e^0\wedge e^a + \frac{1}{2} \left( L_{[b}\, X_{c]} - 2\,\epsilon_{abc}\, X_a + \left[ X_b, X_c \right] \right) e^b\wedge e^c\ ,
\end{equation}
where $e^0 := \diff{\tau}$ and the dot on $X_a$ refers to its derivative w.r.t. $\tau$. The Yang--Mills equation, for this gauge field $\Acal$,
\begin{equation}
    \diff*\mathcal{F} + \Acal\wedge *\mathcal{F} - *\mathcal{F}\wedge\mathcal{A} \= 0\ ,
\end{equation}
after a straightforward calculation, yields the constraint condition
\begin{equation}\label{constraint}
    -2\,\im\,J_a\,X_a + [X_a,\dot{X}_a] \= 0\ ,
\end{equation}
along-with the field equation
\begin{align} \label{ge_eq}
\begin{split}
\ddot{X}_a =& -4\left(J^2{+}1\right)X_a - 4\,\im\,\epsilon_{abc}\,J_b\,X_c + 4\,J_a\,J_b\,X_b + 3\,\epsilon_{abc}\,[X_b,X_c]\\
&  + 2\,\im\,[X_a,J_b\,X_b] + 2\,\im\,[X_b,J_a\,X_b] + 4\,\im\,[J_b\,X_a,X_b] - [X_b,[X_a,X_b]]\ ,
\end{split}
\end{align}
where the operators $J_a$ and $J^2$ are given by
\begin{equation}\label{opJ}
    J_a\ :=\ \frac{\im}{2}\,L_a \und J^2\ :=\ J_a\,J_a\ .
\end{equation}
Finding a general solution for this equation is a daunting task, but progress can be made in two limiting cases as follows. 

One of the limiting case of \eqref{ge_eq} is the Abelian one, where commutators vanish to yield
\begin{equation}\label{eq_abelian}
\ddot{X}_a = -4\left( J^2 {+} 1 \right) X_a + 2\,\im\,\epsilon_{abc}\,J_b\,X_c\ .
\end{equation}
Furthermore the constraint \eqref{constraint} in this case becomes
\begin{equation}\label{constraintAbelian}
    J_a\,X_a \= 0\ .
\end{equation}
One thing to note here is that the above constraint, along-with the temporal gauge $A_\tau\!=0$, is not the usual Coulomb gauge on Minkowski space. In fact, we can make use of the inverse Jacobian \eqref{Jacobian}, while promoting the gauge potential to the Minkowski space $\Acal \!= \Acal_a\, e^a \!= A_\mu\,dx^\mu \!=A $, to get
\begin{equation}\label{gaugeCond2}
  0 = \Acal_\tau = \Acal \left(\partial_\tau \right) = A \left( \sfrac{\ell^2}{\gamma
  ^2}\left(p\,\partial_t + q\,\partial_r \right) \right) \implies \left(r^2+t^2+\ell^2\right)A_t + 2\,r\,t\,A_r = 0\ .
\end{equation}
Solution to \eqref{eq_abelian} was obtained in \cite{LZ17} using hyperspherical harmonics $Y_{j;m,n}$ as we will see in the Chapter \ref{Chapter5}.

Another limiting case of \eqref{ge_eq} is when, a more symmetric, $SO(4)$-equivariant condition is imposed yielding \cite{Friedan,Luescher77}:
\begin{equation}\label{Aansatz}
X_a \= \frac{1}{2}\left( 1 + \psi(\tau) \right)T_a
\end{equation}
with some function $\psi : \mathcal{I} \rightarrow \R$ and $SU(2)$ generators $T_a$ satisfying the Lie algebra
\begin{equation}
   [T_a,T_b] = 2\,\epsilon_{abc}\,T_c\ .
\end{equation}
Moreover, we work in the adjoint representation for the Lie algebra genertors $T_a$ where $\tr{(T_a\,T_b)} {=} {-8}\,\delta_{ab}$. The constraint \eqref{constraint} for this symmetric ansatz, i.e.
\begin{equation}
    [X_a,\dot{X}_a] \= 0
\end{equation}
is automatically satisfied and the field-equation \eqref{ge_eq} becomes
\begin{equation}
\ddot{X}_a = -4\,X_a + 3\,\epsilon_{abc}\,[X_b,X_c] - [X_b,[X_a,X_b]]\ .
\end{equation}
Solution to this equation was obtained in \cite{ILP17jhep} as we will see in Chapter \ref{Chapter6}.

%----------------------------------------------------------------------------------------

%% file: Chapters/Chapter5.tex
%auto-ignore
\chapter{Abelian solutions: $U(1)$}
\label{Chapter5} 
\justifying

In this chapter we present the knotted electromagnetic fields arising via the ``de Sitter" method and study its various properties. The content of this chapter has generated three published works: \cite{KL20,KPH21,KLP22}. Section \ref{nullFields} on null solutions is due to Olaf Lechtenfeld with verification by Colin Becker and some clarifications from Harald Skarke. I have been involved at all stages of research works for rest of the sections in this chapter.

\vspace{12pt}
%----------------------------------------------------------------------------------------
\vspace{2pt}
\section{Family of "knot" solutions}
It was shown in \cite{LZ17}, that the general solution of \eqref{eq_abelian} decomposes into spin-$j$ representations of~$so(4)$ and are labelled with hyperspherical harmonics of $S^3$. We review the construction of these solutions below in two steps: (a) we first solve \eqref{eq_abelian} on the $S^3$-cylinder and (b) we then pull these solution back to Minkowski space using $(\tau,\chi)\rightarrow (t,r)$ coordinate transformation. 

\subsection{Generic solution on the $S^3$-cylinder}
\noindent
To solve \eqref{eq_abelian} we first write it down in terms of ladder operators $J_{\pm}$ and $J_3$ \eqref{opJ} together with the redefined functions
\begin{equation}
    X_{\pm}\ :=\ \frac{1}{\sqrt{2}}\left(X_1 \pm \im X_2 \right)
\end{equation}
as follows:
\begin{equation} \label{knotEquations}
 \begin{aligned}
    \pa_{\tau}^2\,X_+ &\= -4\,(J^2-J_3+1)X_+ - 4\,J_+\,X_3\ ,\\[4pt]
    \pa_{\tau}^2\,X_3 &\= -4\,(J^2+1)X_3 + 4\,J_+\,X_- - 4\,J_-\,X_+\ ,\\[4pt]
    \pa_{\tau}^2\,X_- &\= -4\,(J^2+J_3+1)X_- + 4\, J_-\,X_3\ .
 \end{aligned}
\end{equation}
Similarly, the constraint condition \eqref{constraintAbelian} takes becomes
\begin{equation} \label{constraintAbelian2}
    0 \= J_3\,X_3 + J_+\,X_- +J_-\,X_+\ .
\end{equation}
We can solve \eqref{knotEquations} by the following ansatz
\begin{equation}
\begin{aligned}
    X_+ &\= \sum\limits_{j,m,n} Z^{j;m,n}_+\,e^{i\Omega_{j,m,n}\tau}\ ;\quad Z^{j;m,n}_+ \= c_+\,Y_{j;m,n+1}\ ,\\
    X_3 &\= \sum\limits_{j,m,n} Z^{j;m,n}_3\,e^{i\Omega_{j,m,n}\tau}\ ; \quad Z^{j;m,n}_3 \= c_3\,Y_{j;m,n}\ , \und \\
    X_- &\= \sum\limits_{j,m,n} Z^{j;m,n}_-\,e^{i\Omega_{j,m,n}\tau}\ ;\quad Z^{j;m,n}_- \= c_-\,Y_{j;m,n-1}\ ,
\end{aligned}
\end{equation}
where $X_*(j,m,n)$ with $*\in \{ +,3,- \}$ has been expanded in terms of $S^3$-harmonics. Plugging this ansatz back in \eqref{knotEquations} and using \eqref{JpmAction} we find, for every mode $(j,m,n)$, an eigenvalue equation for the vector $(c_+\ c_3\ c_-)^T$:
\begin{equation}
\begin{aligned}
    &\qquad\qquad M\begin{pmatrix}
    c_+ \\ c_3 \\ c_-
    \end{pmatrix} \= \Omega(j,m,n)^2\begin{pmatrix}
    c_+ \\ c_3 \\ c_-
    \end{pmatrix}\ ,\ \with \\
    &M \= \begin{psmallmatrix}
    4(j^2+j-n) & 2\sqrt{2(j-n)(j+n+1)} & 0 \\
    2\sqrt{2(j-n)(j+n+1)} & 4(j^2+j+1) & -2\sqrt{2(j+n)(j-n+1)} \\
    0 & -2\sqrt{2(j+n)(j-n+1)} & 4(j^2+j+n)
    \end{psmallmatrix}\ ,
\end{aligned}
\end{equation}
which admits an eigensystem with $3$ distinct eigenvalues $\Omega_j^2$ (that turns out to be independent of $m$ and $n$) and their corresponding eigenvectors. One of these eigenvectors does not satisfy the constraint \eqref{constraintAbelian2} and is, therefore, discarded. We label the remaining two eigensystems as type I, with eigenfrequency $\Omega_j^2=4(j{+}1)^2$, and type II, with eigenfrequency $\Omega_j^2=4\,j^2$, as follows
\begin{itemize}
\addtolength{\itemsep}{-4pt}
\item type I : \quad
$j{\geq}0\ ,\quad m = -j,\ldots,+j\ ,\quad n = -j{-}1,\ldots,j{+}1\ ,\ \Omega_j=\pm 2\,(j{+}1)\ ,$
\\
\begin{equation} \label{type1}
\begin{aligned}
Z_{+\ \textrm{I}}^{(j;m,n)}(\omega) &\= \sqrt{(j{-}n)(j{-}n{+}1)/2} \ \ Y_{j;m,n+1}(\omega) \ ,\\
Z_{3\ \textrm{I}}^{(j;m,n)}(\omega)\,&\= \sqrt{(j{+}1)^2-n^2} \ \ Y_{j;m,n}(\omega) \ ,\\
Z_{-\ \textrm{I}}^{(j;m,n)}(\omega)   &\= -\sqrt{(j{+}n)(j{+}n{+}1)/2} \ \ Y_{j;m,n-1}(\omega) \ .
\end{aligned}
\end{equation}
\item type II :\quad
$j{\geq}1\ ,\quad m = -j,\ldots,+j\ ,\quad n = -j{+}1,\ldots,j{-}1\ ,\ \Omega_j=\pm2\,j\ ,$
\\
\begin{equation} \label{type2}
\begin{aligned}
Z_{+\ \textrm{II}}^{(j;m,n)}(\omega)  &\= -\sqrt{(j{+}n)(j{+}n{+}1)/2} \ \ Y_{j;m,n+1}(\omega)\ ,\\
Z_{3\ \textrm{II}}^{(j;m,n)}(\omega)\,&\= \sqrt{j^2-n^2} \ \ Y_{j;m,n}(\omega) \ ,\\
Z_{-\ \textrm{II}}^{(j;m,n)}(\omega)   &\= \sqrt{(j{-}n)(j{-}n{+}1)/2} \ \ Y_{j;m,n-1}(\omega) \ .
\end{aligned}
\end{equation}
\end{itemize}

We can take a linear combination of these basis solutions and write down a real-valued connection $1$-form as
\begin{equation}\label{Asep}
\Acal \=  
\Bigl\{ \sum_{2j=0}^\infty X_{a\ \textrm{I}}^j(\omega) \ \ep^{2(j+1)\im\tau} \ +\ \textrm{c.c.} \Bigr\}e^a \ +\ 
\Bigl\{ \sum_{2j=2}^\infty X_{a\ \textrm{II}}^j(\omega) \ \ep^{2j\,\im\tau} \ +\ \textrm{c.c.} \Bigr\}e^a
\end{equation}
where we have reorganized the complex angular functions~$X_a^j$ as
\begin{equation}\label{XZ}
X_1^j=\sfrac{1}{\sqrt{2}} \bigl(Z_+^j + Z_-^j \bigr)\ ,\qquad 
X_2^j=\sfrac{\im}{\sqrt{2}} \bigl( Z_-^j - Z_+^j \bigr)\ ,\qquad 
X_3^j= Z_3^j
\end{equation}
for both types and expanded the functions~$Z^j_\pm$ and $Z^j_3$ into the above spin-$j$ basis solutions of type I \eqref{type1} and type II \eqref{type2} (for $*\in\{+,3,-\}$),
\begin{equation}\label{basisZ}
\begin{aligned}
Z_{*\ \textrm{I}}^j(\omega) &\= \sum_{m=-j}^j \sum_{n=-j-1}^{j+1} \lambda_{j;m,n}^{\textrm{I}}\ Z_{*\ \textrm{I}}^{(j;m,n)}(\omega)\ , \\
Z_{*\ \textrm{II}}^j(\omega) &\= \sum_{m=-j}^j \sum_{n=-j+1}^{j-1} \lambda_{j;m,n}^{\textrm{II}}\ Z_{*\ \textrm{II}}^{(j;m,n)}(\omega)
\end{aligned}
\end{equation}
with $(2j{+}1)(2j{+}3)$ arbitrary complex coefficients $\lambda^{\textrm{I}}_{j;m,n}$
and $(2j{+}1)(2j{-}1)$ coefficients $\lambda^{\textrm{II}}_{j;m,n}$\footnote{Note that type-II solutions are absent for $j{=}0$ and $j{=}\sfrac12$.}.

Inserting \eqref{type1} and \eqref{type2} into \eqref{basisZ} and the resulting expression into \eqref{XZ} provides a harmonic expansion
\begin{equation}\label{XY}
X_a^j(\omega) \= \sum_{m=-j}^j \sum_{n=-j}^j X_a^{j;m,n}\ Y_{j;m,n}(\omega)
\end{equation}
for both types of angular functions in~\eqref{Asep}\footnote{Note the different range of~$n$ for $X_a^{j;m,n}$ and $Z_*^{j;m,n}$; they are not easily related as $X_a^j$ and $Z_*^j$ are in~\eqref{XZ}.}.

It is useful for later purposes to introduce here the ``sphere-frame" electric $\Ecal_a$ and magnetic $\Bcal_a$ fields,
\begin{equation}\label{2FormEM}
    \Fcal \= \Ecal_a\,e^a \we e^\tau + \sfrac12\,\Bcal_a\,\varepsilon^a_{\ bc}\,e^b\we e^c\ .
\end{equation}
For a fixed type (I or II) and spin~$j$, we may eliminate $L_{[b}A_{c]}$ in \eqref{FcalExpanded} (without the commutator term) by using~\eqref{eq_abelian} and employ
\begin{equation}
\pa_\tau^2 \Acal^{(j)} = -\Omega_j^2\,\Acal^{(j)} \quad\und\quad
L^2\,\Acal^{(j)} \= -4j(j{+}1)\,\Acal^{(j)}
\end{equation}
to obtain
\begin{equation}\label{cur-field}
\Ecal_a^{(j)} \= -\pa_\tau X_a^{(j)} \quad\und\quad \Bcal_a^{(j)} \= \mp\Omega_j\,X_a^{(j)}\ ,
\end{equation}
where the upper sign pertains to type~I and the lower one to type~II.
We note in passing that, due to the compactness of the Lorentzian cylinder, the sphere-frame energy and action are always finite.

Due to the linearity of Maxwell theory, the overall scale of any solution is arbitrary.
Furthermore, the parity transformation $L\leftrightarrow R$---that corresponds to $m\leftrightarrow n$---interchanges a spin-$j$ solution of type~I with a spin-$(j{+}1)$ solution of type~II.
Finally, electromagnetic duality at fixed $j$ is realized by shifting $\Omega_j\tau$ by $\sfrac{\pi}{2}$ for type~I
or by $-\sfrac{\pi}{2}$ for type~II, which maps $\Acal$ to a dual configuration~$\Acal_\textrm{D}$
and likewise $\Fcal$ to $\Fcal_\textrm{D}$.

\subsection{Pulling the solution back to Minkowski space}

\begin{figure}[h!]
\captionsetup{width=\linewidth}
\centering
   \includegraphics[width = 5cm, height = 5cm]{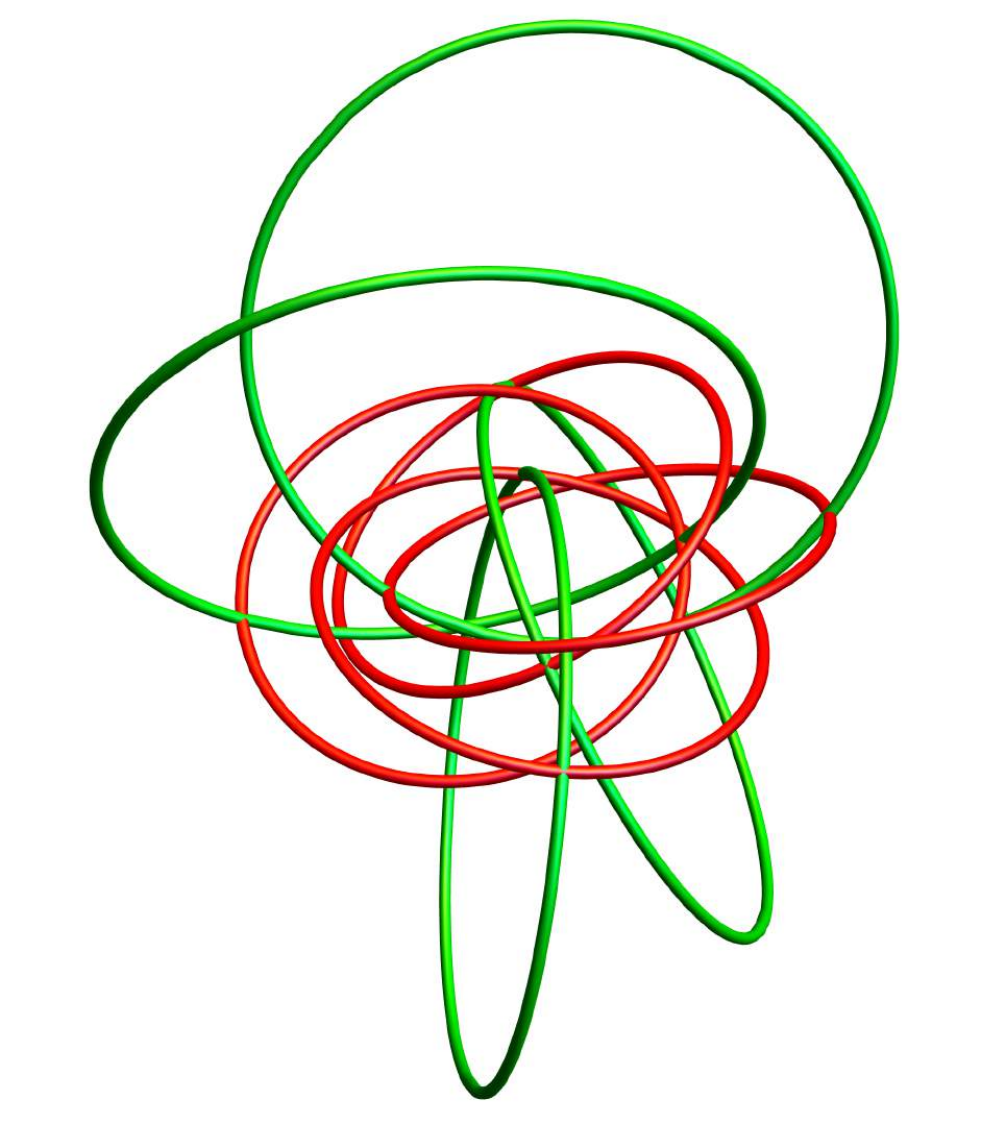}
   \includegraphics[width = 5cm, height = 5cm]{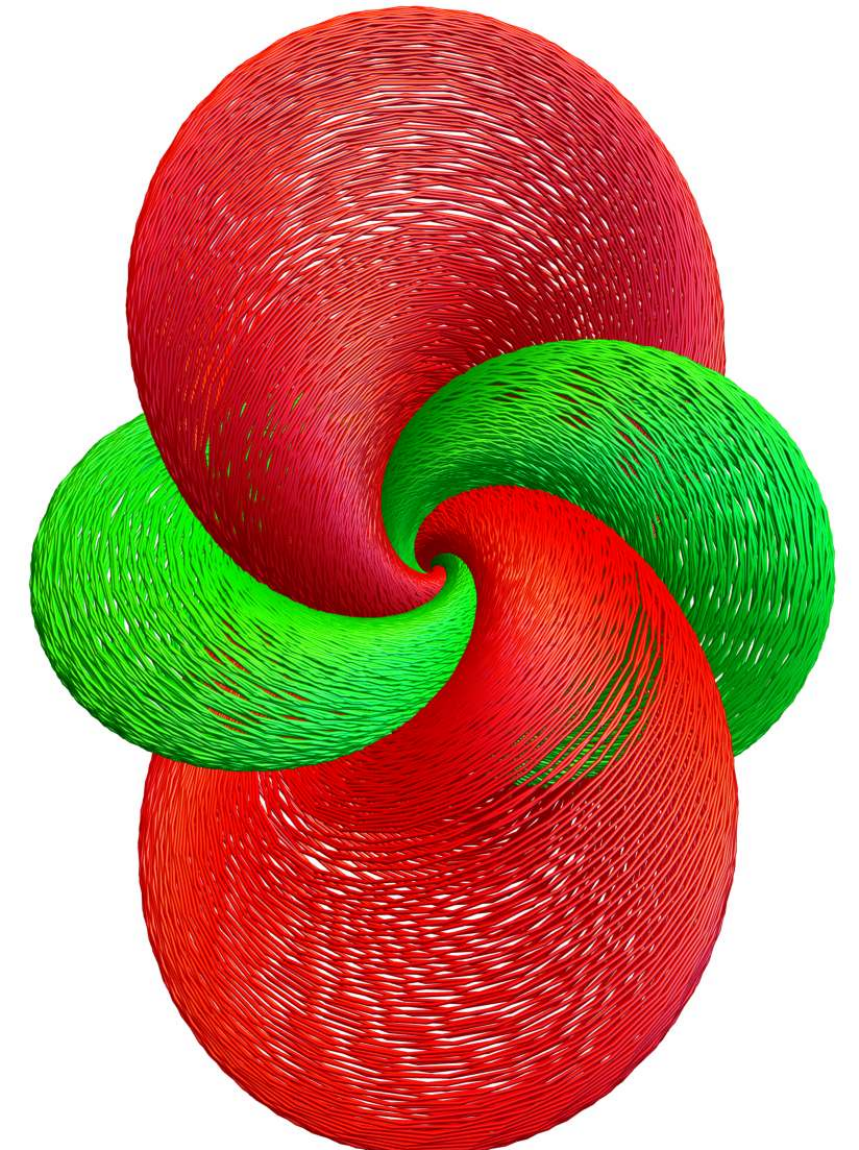}
   \includegraphics[width = 5cm, height = 5cm]{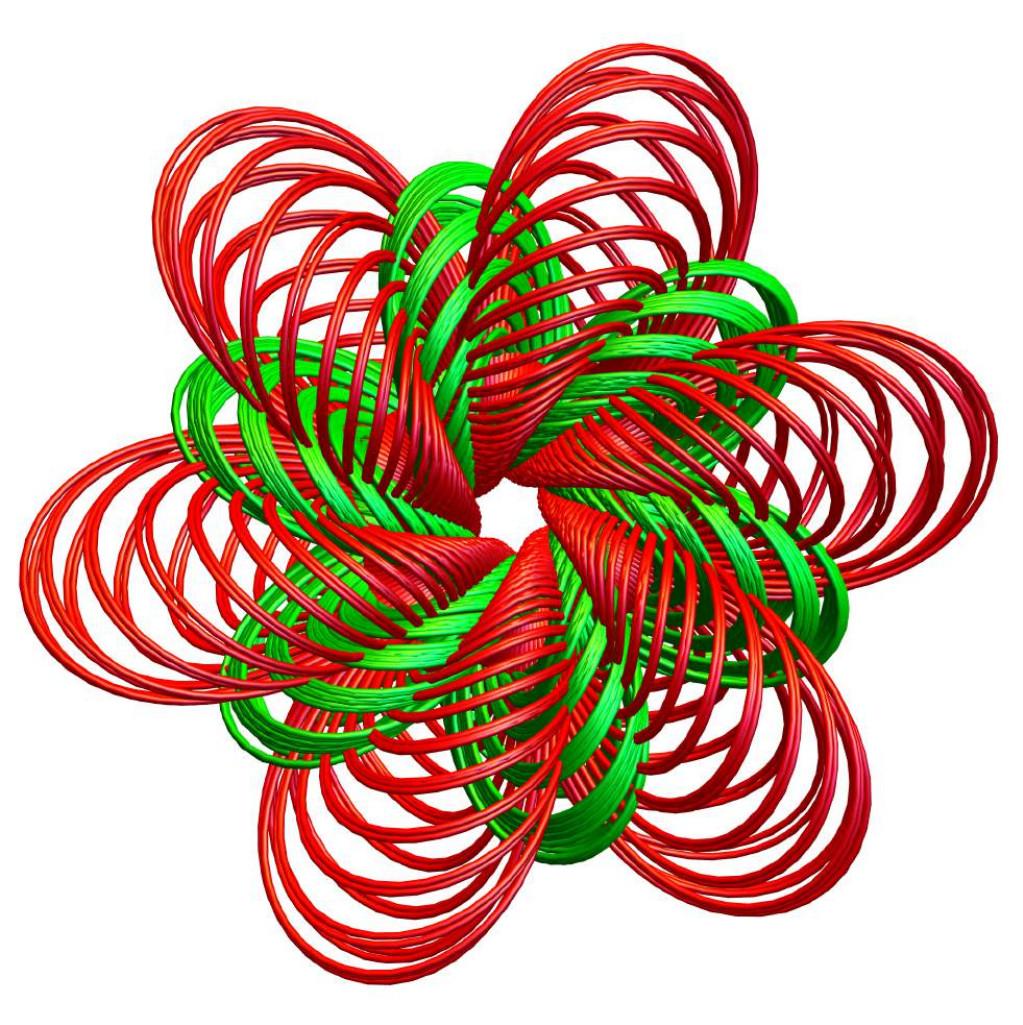}
 \caption{
 Electric (red) and magnetic (green) field lines at $t{=}0$ with 4 fixed seed points.
 Left: $(j;m,n)=(0;0,1)_I$ configuration, Center: $(j;m,n)=(\sfrac12;-\sfrac12,\sfrac32)_R$ configuration, Right: $(j;m,n)=(1;1,2)_I$ configuration. More self-knotted field lines start appearing with additional seed points in the simulation.}
\label{fieldLines}
\end{figure}

\noindent
We have completely solved the vacuum Maxwell equations on the Lorentzian cylinder ${\cal I}\times S^3$, which, by conformal invariance, carries over to any conformally equivalent spacetime including
de Sitter space dS$_4$ and Minkowski space $\R^{1,3}$. We can translate our Maxwell solutions from $\widetilde{\Ical}\times S^3$ to $\R^{1,3}$ simply by the coordinate change
\begin{equation}
\begin{aligned}
\tau=\tau(t,x,y,z) &\und \omega_{_A}=\omega_{_A}(t,x,y,z) \\[2pt]
\textrm{or}\qquad
\tau=\tau(t,r) &\und \chi=\chi(t,r)\ .
\end{aligned}
\end{equation}
In other words, abbreviating $x\equiv\{x^\mu\}$ and $y\equiv\{y^\rho\}$ and expanding
\begin{equation}
\begin{aligned}
\Acal &\= X_a\bigl(\tau(x),g(x)\bigr)\,e^a(x) 
\= A_\mu(x)\,\diff x^\mu
\= A_\rho(y)\,\diff y^\rho \quad\und \\[4pt]
\diff\Acal &\= \pa_\tau X_a\,e^0\we e^a + \bigl( L_b\,X_c - X_a\,\varepsilon^a_{\ bc}\bigr)\,e^b \we e^c \\
&\= \sfrac12 F_{\mu \nu}(x)\,\diff x^\mu \we \diff x^\nu 
\= \sfrac12 F_{\rho\lambda}(y)\,\diff y^\rho\we\diff y^\lambda
\end{aligned}
\end{equation}
using \eqref{1formsExpanded} we may read off $A_\mu$ (note that $A_t\neq0$, as discussed before) and 
$F_{\mu\nu}$ and thus the electric and magnetic fields
\begin{equation}\label{EMCartesian}
\Fcal \= F \= E_a\,\diff{x}^a\wedge\diff{t} + \frac{1}{2}B_a\,\varepsilon^a_{\ bc}\,\diff{x}^b\we \diff{x}^c
\end{equation}
in Cartesian or in spherical coordinates. In general, the knotted electromagnetic fields arising from the basis configurations \eqref{type1} are complex, and thus the basis configurations on Minkowski space will also be complex. Hence, they combine two physical solutions, namely the real and imaginary parts\footnote{The notation should not be confused with the type I configuration \eqref{type1}.}, which we denote as
\begin{equation*}
    (j;m,n)_R~ \textrm{configuration} \und (j;m,n)_I ~\textrm{configuration}\ ,
\end{equation*}
respectively. Such knotted electromagnetic fields \eqref{EMCartesian} for the basis configurations \eqref{type1} and \eqref{type2} increase in complexity with increasing $j$, as shown in Figure \ref{fieldLines}.

Furthermore, it comes in handy that $\Acal$, after computation, contains only even powers of~$\gamma$
and depends on~$\tau$ only through integral powers of
\begin{equation}
\exp (2\mathrm{i}\,\tau) \= \frac{[(\ell+\mathrm{i}t)^2+r^2]^2}{4\,\ell^2 t^2 + (r^2-t^2+\ell^2)^2}\ .
\end{equation}
Therefore, our Minkowski solutions have the remarkable property of being rational functions of~$(t,x,y,z)$.
More precisely, their electric and magnetic fields are of the form
\begin{equation}
\textrm{type I:} \quad \frac{P_{2(2j+1)}(x)}{Q_{2(2j+3)}(x)} \ ,\qquad
\textrm{type II:}\quad \frac{P_{2(2j-1)}(x)}{Q_{2(2j+1)}(x)} 
\end{equation}
where $P_r$ and $Q_r$ denote polynomials of degree~$r$.
Thus, as expected, their energy and action are finite.
Indeed, the fields fall off like $r^{-4}$ at spatial infinity for fixed time, but they decay
merely like $(t{\pm}r)^{-1}$ along the light-cone.
Hence, the asymptotic energy flow is concentrated on past and future null infinity~$\mathscr{I}^\pm$,
as it should be, but peaks on the light-cone of the spacetime origin.
Since our basis solutions (\ref{type1}) and~(\ref{type2}) form a complete set on de Sitter space, their Minkowski relatives are also complete on the space of finite-action configurations. 

For illustration, we display a type I basis solution with $(j;m,n)=(1;0,0)$
obtained from 
\begin{equation}\label{j1Prototype}
X_\pm \ \propto\ \sfrac{1}{\sqrt{2}}\,(\omega_1{\pm}\im\omega_2)(\omega_3{\pm}\im\omega_4)\,\cos 4\tau \und
X_3\ \propto\  (\omega_1^2{+}\omega_2^2{-}\omega_3^2{-}\omega_4^2)\,\cos 4\tau\ .
\end{equation}
It bodes well to combine these field configurations into the Riemann--Silberstein vector 
\begin{equation}\label{RSvector}
    {\bf S} \= {\bf E} + \im {\bf B}
\end{equation}
whose components, for \eqref{j1Prototype}, read (up to some overall scale)
{\small
\begin{equation}
\begin{aligned}
S_x &\= -\frac{2\im}{N}
\Bigl\{ 2y +3 \im t y -x z +2 t^2 y + 2\im t x z-8x^2 y-8y^3+4yz^2 + 4\im t^3 y \\%[4pt]
&\qquad\qquad  -6t^2 xz - 8\im t x^2 y-8\im t y^3+4\im t yz^2 +10x^3 z+10xy^2 z -2xz^3 \\%[4pt]
&\qquad\qquad\qquad\qquad + 2(\im t x z + yr^2)(-t^2{+}r^2)+(\im t y- x z) (-t^2{+}r^2)^2\Bigr\}
\ ,\quad{}
\end{aligned}
\end{equation}
\begin{equation}
\begin{aligned}
S_y &\= \frac{2\im}{N}
\Bigl\{ 2x +3 \im t x + y z+2 t^2 x-2\im t y z-8x^3-8xy^2 +4 xz^2 + 4 \im t^3 x \\%[4pt]
&\qquad\qquad + 6 t^2y z -8\im t x^3-8\im txy^2 +4 \im t x z^2-10 x^2 yz-10y^3 z +2y z^3 \\%[4pt]
&\qquad\qquad\qquad\qquad + 2 (-\im t y z {+} x r^2)(-t^2{+} r^2) +(\im t x+ y z)(-t^2{+}r^2)^2\Bigr\}
\ ,\qquad\&
\end{aligned}
\end{equation}
\begin{equation}
\begin{aligned}
S_z &\= \frac{\im}{N}
\Bigl\{1+2\im t+t^2-11 x^2- 11 y^2 +3 z^2 +4 \im t^3-16\im t x^2-16\im t y^2+4\im t z^2 - t^4 \\%[4pt]
&\qquad\qquad -2 t^2 x^2 -2t^2 y^2-2t^2 z^2+11 x^4+22x^2y^2-10 x^2 z^2+11y^4-10 y^2 z^2 \\%[4pt]
&\qquad\qquad\qquad\qquad\qquad +3z^4 + 2 \im t (t^2 {-} 3 r^2{+}2 z^2) (t^2{-}r^2)-(t^2{+}r^2)(-t^2{+}r^2)^2\Bigr\}
\end{aligned}
\end{equation}
}
with $N=\bigl((t{-}\im)^2-r^2\bigr)^5$. We can also obtain electromagnetic field configurations corresponding to the gauge field \eqref{Asep} that consists of complex linear combinations of these basis solutions. We demonstrate this for the famous Hopf--Ran\~{a}da field configuration, which was first discovered by Ran\~{a}da in 1989 \cite{Ranada89} using the Hopf fibration. The Riemann--Silberstein vector ${\bf S}$ for this field configuration is given by \cite{LZ17}
\begin{equation} \label{HRfields}
    {\bf S} \= \frac{1}{((t{-}\im)^2{-}r^2)^3} \begin{pmatrix}
        (x{-}\im\,y)^2 {-} (t{-}\im{-}z)^2 \\
        \im(x{-}\im\,y)^2 {+} \im(t{-}\im{-}z)^2 \\
        -2(x{-}\im\,y)(t{-}\im{-}z)
    \end{pmatrix}\ .
\end{equation}
We find that this solution, in our construction, is related to $(0,0,1)_I$ basis configurations and is obtained by using following coefficients in \eqref{basisZ}
\begin{equation}
    \lambda^I_{0;0,-1} \= 0\ ,\quad \lambda^I_{0;0,0} \= 0\ ,\und \lambda^I_{0;0,1} \= \frac{\pi}{4}\ .
\end{equation}
Moreover, we find that some of our basis configurations are related to the $(p,q)$-torus knots arising from Bateman's construction \cite{ABT17}. We illustrate this point in Figure \ref{fieldLines} where we find the following correspondences:
\begin{equation}
	\textrm{Hopfion} \leftrightarrow (1,1)\ ,\qquad (\sfrac12,-\sfrac12,\sfrac32)_R \leftrightarrow (2,1)\ ,\qquad (1,1,2)_I \leftrightarrow (1,3)\ . 
\end{equation}

\vspace{12pt}
%----------------------------------------------------------------------------------------
\section{Symmetry analysis}
\vspace{2pt}
\noindent
The main advantage of constructing Minkowski-space electromagnetic field configurations 
via the detour over de Sitter space is the enhanced manifest symmetry of our construction. 
The isometry group SO(1,4) of dS$_4$ is generated by 
($A,B=1,2,3,4$ and $a,b,c=1,2,3$, abbreviate $\frac{\partial}{\partial q_{_B}}\equiv\partial_{_B}$)
\begin{equation}
\{\Mcal_{_{AB}}{\equiv}\,{-}q_{[{\scriptscriptstyle A}}\partial_{{\scriptscriptstyle B}]}\,,\ 
\Mcal_{0{\scriptscriptstyle B}}{\equiv}\,q_{({\scriptstyle 0}}\partial_{{\scriptscriptstyle B})} \} \= 
\{ \Mcal_{ab}{=}\varepsilon_{abc}\Dcal_c\,,\ \Mcal_{4a}{=}\Pcal_a\,,\ \Mcal_{04}{=}\Pcal_0\,,\ \Mcal_{0b}{=}\Kcal_b\}\ ,
\end{equation}
which can be contracted (with $\ell{\to}\infty$) to the isometry group ISO(1,3) of $\R^{1,3}$ (the Poincar\'e group)
generated by ($\mu,\nu=0,1,2,3$ and $i,j,k=1,2,3$)
\begin{equation}
\{M_{\mu\nu}\,,\ P_\mu \} \=
\{ M_{ij}{=}\varepsilon_{ijk}D_k\,,\ P_i\,,\ P_0\,,\ M_{0j}{=}K_j\}\ ,
\end{equation}
where the two sets are ordered likewise, 
and we employ (as aleady earlier) calligraphic symbols for de Sitter quantities and straight symbols for Minkowskian ones.
Here, $D$ denotes spatial rotations, $P$ are translations, and $K$ stand for boosts in Minkowski space.

Since the two spaces are conformally equivalent already at $\ell{<}\infty$ via~\eqref{S3toMink},
the corresponding generators should be related. Indeed, the common SO(3) subgroup in
\begin{equation}\label{SO(3)subset}
SO(1,4) \supset SO(4) \supset SO(3) \quad\und\quad ISO(1,3) \supset SO(1,3) \supset SO(3)
\end{equation}
is identified, $\Dcal_i{=}D_i{=}{-}2\varepsilon_{ij}^{\ k}x^j\partial_k$. 
However, any other generator becomes nonlinearly realized when 
mapped to the other space via \eqref{tr2TauChi}.
For example, the would-be translation $\Pcal_3$ defined in~\eqref{DPdef} reads
\begin{equation}
\begin{aligned}
\Pcal_3 = L_3-R_3 &\= -2\,\cos\theta\,\pa_\chi + 2\,\cot\chi\sin\theta\,\pa_\theta \\[2pt]
&\= \sfrac1{\ell}\,\cos\theta\,\bigl( 2\,r\,t\,\pa_t + (t^2{+}r^2{+}\ell^2)\,\pa_r \bigr) 
- \sfrac1{\ell\,r} (t^2{-}r^2{+}\ell^2)\,\sin\theta\,\pa_\theta \\[2pt]
&\ \to\ 2\ell\,\bigl( \cos\theta\,\pa_r - \sfrac1r\,\sin\theta\,\pa_\theta \bigr)
\= 2\ell\,\pa_z \= \ell\,P_z \quad\for\ell\to\infty
\end{aligned}
\end{equation}
as it should be. Similarly, ${\cal P}_0\to \ell\,P_0$ and ${\cal K}_b\to K_j$ for $\ell{\to}\infty$ 
when expanded around $(t,r)=(\ell,0)$ corresponding to the $S^3$ south pole at $q_0{=}0$.
Nevertheless, the de Sitter construction enjoys an SO(4) covariance (generated by $\Dcal_a$ and $\Pcal_a$)
which extends the obvious SO(3) covariance in Minkowski space.
It allows us to connect all solutions of a given type (I or II) with a fixed value of the spin~$j$
by the action of SO(4) ladder operators $L_\pm$ and $R_\pm$ or $\Dcal_\pm$ and $\Pcal_\pm$, 
which is non-obvious on the Minkowski side. 
On the other hand, Minkowski boosts and translations have no simple realization on de Sitter space.

Actually, Maxwell theory on either space is also invariant under conformal transformations.
These may be generated by the isometry group together with a conformal inversion.
On the Minkowski side, the latter is
\begin{equation}
J : \quad x^\mu \ \mapsto\ \frac{x^\mu}{x\cdot x} \quad\with x\cdot x=r^2-t^2\ .
\end{equation}
We have to distinguish two cases:
\begin{equation}
\begin{aligned}
\textrm{spacelike:}\quad & t^2<r^2 \quad\Rightarrow\quad
J_> : \quad\bigl(t,r,\theta,\phi\bigr) \ \mapsto\ \bigl(\sfrac{t}{r^2-t^2}, \sfrac{r}{r^2-t^2},\theta,\phi\bigr)\ ,\\[2pt]
\textrm{timelike:} \quad & t^2>r^2 \quad\Rightarrow\quad
J_< : \quad\bigl(t,r,\theta,\phi\bigr) \ \mapsto\ \bigl(\sfrac{-t}{t^2-r^2}, \sfrac{r}{t^2-r^2},\pi{-}\theta,\phi{+}\pi\bigr)\ .
\end{aligned}
\end{equation}
On the de Sitter side, this is either (spacelike) a reflection on the $S^3$ equator~$\chi{=}\frac{\pi}{2}$
or (timelike) a $\pi$-shift in cylinder time~$\tau$ plus an $S^2$ antipodal flip,
\begin{equation}
\begin{aligned}
\textrm{spacelike:}\quad & |\tau|{+}\chi<\pi \quad\Rightarrow\quad
\Jcal_> :  \quad \bigl(\tau,\chi,\theta,\phi\bigr)\ \mapsto\ \bigl(\tau,\pi{-}\chi,\theta,\phi\bigr)\ , \\[2pt]
\textrm{timelike:} \quad & |\tau|{+}\chi>\pi \quad\Rightarrow\quad
\Jcal_< :  \quad \bigl(\tau,\chi,\theta,\phi\bigr)\ \mapsto\ \bigl(\tau{\pm}\pi,\chi,\pi{-}\theta,\phi{+}\pi\bigr)\ .
\end{aligned}
\end{equation}
In the spacelike case, merely the sign of $\omega_4{\equiv}\cos\chi$ gets flipped, 
which amounts to a parity flip $L\leftrightarrow R$. 
In the timelike case, both $\cos\tau$ and $\sin\tau$ change sign, 
which combines a time reversal with a reflection at $\tau{=}\frac{\pi}{2}$ or $\tau{=}{-}\frac{\pi}{2}$.
Note that it is different from the $S^3$ antipodal map, which
is not a reflection but a proper rotation, $\omega_{_A}\mapsto-\omega_{_A}$ or 
$(\chi,\theta,\phi)\mapsto(\pi{-}\chi,\pi{-}\theta,\phi{+}\pi)$.
The lightcone is singular under the inversion; it is mapped to the conformal boundary
$r{=}{\pm}t{=}\infty$ or $\chi{=}{\pm}\tau$.
We infer that the conformal inversion allows us to relate type-I and type-II solutions of the same spin.
It is easily checked that the spatial fall-off behavior of our rational solutions is not modified by the inversion.

Finally, one may consider dilatations in Minkowski space,
\begin{equation}
x^\mu \ \mapsto\ \lambda\,x^\mu \quad\for \lambda\in\R_+\ .
\end{equation}
However, this amounts to a trivial rescaling also achieved by changing the de Sitter radius,
$\ell\mapsto\lambda\,\ell$, as the scale~$\ell$ was removed on the Lorentzian cylinder. 

\vspace{12pt}
%----------------------------------------------------------------------------------------
\section{Conformal group and Noether charges}
\vspace{2pt}
\noindent
It is well known \cite{BR16} that free Maxwell theory on $\R^{1,3}$ arising from the action
\begin{equation}
    S\left[A_\mu\right] \= \int d^4x\, \mathcal{L}\ ;\qquad \mathcal{L} \= -\sfrac14 F^{\mu\nu}F_{\mu\nu}
\end{equation}
is invariant under the conformal group $SO(2,4)$. Furthermore, the above action is also invariant under the gauge transformations: $A_\mu(x)\rightarrow A_\mu(x)+\pa_\mu\lambda(x)$. The conformal group is generated by transformations $x^\mu \rightarrow x^\mu + \zeta^\mu(x)$, where the vector fields $\zeta^\mu$ obey the conformal Killing equations:
\begin{equation}\label{Killing}
    \zeta_{\mu,\nu}\, +\, \zeta_{\nu,\mu} \= \sfrac12\, \eta_{\mu\nu}\, \zeta^\alpha_{~,\alpha} \with \{\eta_{\mu\nu}\} \= \textrm{diag}(-1,1,1,1)\ .
\end{equation}
The conserved Noether current $J^\mu$ is obtained by equating the ``on-shell variation" of the action where the variations $\delta A_\mu$ are arbitrary and the fields $A_\mu$ satisfies the Euler-Langrange equations with its ``symmetry variation" where the fields $A_\mu$ are arbitrary but variations $\delta A_\mu$ satisfy the symmetry condition. The correct variation $\delta A_\mu$ is obtained by imposing the gauge invariance on the Lie derivative of $A_\mu$ w.r.t. the vector field $\zeta^\mu$:
\begin{equation}
    \mathcal{L}_{\zeta^\alpha}A_\mu := A'_\mu (x) - A_\mu (x) = -\zeta^\alpha\, \pa_\alpha A_\mu - A_\alpha\, \pa_\mu \zeta^\alpha ~\longrightarrow ~ \delta A_\mu \= F_{\mu\nu}\zeta^\nu\ .
\end{equation}
Finally, the conserved current is obtained as
\begin{equation}\label{currentJ}
    J^\mu \= \frac{\pa \mathcal{L}}{\pa A_{\rho,\mu}} \delta A_\rho + \zeta^\mu \mathcal{L} \= \zeta^\rho \left( F^{\mu\alpha}F_{\rho\alpha} - \sfrac14\delta^\mu_\rho F^2   \right) \with F^2 = F_{\beta \gamma}F^{\beta \gamma}\ ,
\end{equation}
which satisfies the continuity equation and gives the conserved (in time) charge $Q$\footnote{For source-free fields the current $\textbf{J}$ is assumed to vanish when the surface $\pa V$ is taken to infinity.}:
\begin{equation}
    \pa_\mu\, J^\mu = 0 \quad\implies\quad \frac{\diff Q}{\diff t} \= -\int_{\scriptsize\pa V} \diff^2{\bf s}\cdot{\bf J} \= 0\ ;\quad Q \= \int_{V} \diff^3x\, J^0\ .
\end{equation}
Before proceeding further, a couple of remarks pertaining to the subsequent calculations are in order:
\begin{itemize}
   \item All the charges $Q$ are computed at $t{=}0$ owing to the simple $J^0$ expressions on this time-slice. To that end, we record the following useful identities at $t{=}\tau{=}0$:
   \begin{equation}\label{t0Results}
      \begin{aligned}
       e^a_i &\= \sfrac1\ell \left( \gamma\, \omega_4\, \delta^a_i - \omega_a\, \omega_i + \epsilon_{aic}\, \gamma\, \omega_c \right)\ ,\quad  e^a_i\, e^b_i \= \sfrac{\gamma^2}{\ell^2} \delta^{ab} \\
       \gamma &\= 1-\omega_4\ ,~ \diff^3x \= \sfrac{\ell^3}{\gamma^3}\diff^3\Omega_3\ ~;~ \diff^3\Omega_3 := e^1\wedge e^2\wedge e^3\ .
      \end{aligned}
   \end{equation}
   Moreover, the electromagnetic fields at $t{=}0$ are given in terms of tetrads $e^\tau {=} e^\tau_\mu\, \diff x^\mu $ and $ e^a {=} e^a_\mu\, \diff x^\mu$ as follows:
   \begin{equation}\label{EMfields}
    E_i \= e^\tau_0\, e^a_i\, \mathcal{E}_a \quad\und\quad  B_i \= \sfrac12\, \epsilon_{ijk}\, \epsilon_{abc}\, e^b_j\, e^c_k\, \mathcal{B}_a\ 
   \end{equation}
   using the $2$-form \eqref{2FormEM} and its Minkowski counterpart \eqref{EMCartesian}.
   
   \item The charges are computed for type I \eqref{type1} solutions only (except for the energy and the related helicity) and we relabel the complex coefficients \eqref{basisZ} $\lambda^I_{j;m,n}$ as $\Lambda_{j;m,n}$ for convenience.
   
   \item  Furthermore, these charges $Q$ are computed for a fixed spin-$j$ and, thus, we will suppress the index $j$ from now onward, unless necessary. Note that the sphere-frame EM fields for fixed $j$ can be obtained by using the expansion \eqref{Asep} (for type I only) in \eqref{cur-field} as
   \begin{equation} \label{EBj}
     \Acal_a \= X_a(\omega)\,\ep^{\Omega\,\im\tau} + \bar{X}_a(\omega)\,\ep^{-\Omega\,\im\tau}
     \Rightarrow \biggl\{ \begin{array}{l}
     \Ecal_a \= -\im\,\Omega\,X_a\,\ep^{\Omega\,\im\tau} + \im\,\Omega\,\bar{X}_a\,\ep^{-\Omega\,\im\tau} \\[2pt]
     \Bcal_a \= \,-\,\Omega\,X_a\,\ep^{\Omega\,\im\tau}\,-\,\Omega\,\bar{X}_a\,\ep^{-\Omega\,\im\tau} \end{array} \biggr\}
   \end{equation}
   with $\bar{X}_a$ denoting the complex conjugate of $X_a$. Notice that for type II the overall sign in $\Bcal_a$ flips.
 
   \item  The simplifications below for the charge density $J^0$ are carried out using the harmonic expansion \eqref{XY} for the type I solution \eqref{type1}.
   
   \item We frequently use below the well known fact that an odd $\{\omega_{_A}\}$ integral over $S^3$ vanishes because of the opposite contributions coming from the antipodal points on the sphere. In particular, it can be checked that the following integral vanishes\footnote{Note that the power of $\omega_{_A}$ in $Y_{j;m,n}\bar{Y}_{j;m',n'}$ is always even. One way to check this is by employing the toroidal coordinates: $\omega_1 = \cos\eta\cos\kappa_1,\omega_2 = \cos\eta\sin\kappa_1, \omega_3 = \sin\eta\cos\kappa_2,\omega_4 = \sin\eta\sin\kappa_2$ with $\eta\in(0,\sfrac\pi2)$ and $\kappa_1,\kappa_2\in(0,2\pi)$ in \eqref{S3Harmonics}. The resultant selection rules coming from $\kappa_1,\kappa_2$ integral would yield $m-m'\in \sfrac{2k+1}{2}$ with $k\in\N_0$, which is not feasible for fixed $j$.}:
   \begin{equation}
      \int \diff^3\Omega_3\, (\omega_1)^a\, (\omega_2)^b\, (\omega_3)^c\, (\omega_4)^d\, Y_{j;m,n}\, \overline{Y}_{j;m',n'} \= 0 \quad\textrm{for}\quad a+b+c+d\ \in\ 2\N_0 + 1\ .
   \end{equation}
\end{itemize}
Having made these remarks, we now proceed to compute the charges $Q$ for various conformal transformations $\zeta^\mu$ obeying \eqref{Killing} in following four categories.

\subsection{Translations}
\noindent
An easily seen solution to \eqref{Killing} is the set of four constant translations
\begin{equation}
    \zeta^\mu = \epsilon^\mu\ ,
\end{equation}
which also partly generates the Poincar\'e group and give rise to the usual stress-energy tensor of electrodynamics
\begin{equation}\label{stress-energy}
    J^\mu \= T^\mu_{\ \; \nu} \=  F^{\mu\alpha}F_{\nu\alpha} - \sfrac14\delta^\mu_\nu F^2\ ,
\end{equation}
corresponding to the $\mu$-component of the translation for an arbitrary $\epsilon^\nu$. The corresponding charges are the energy $E$ and the momentum $\textbf{P}$.

\noindent
{\bf Energy.} The expression of the energy density $e:=T^{00}$ simplifies to  
\begin{equation}
    e\, :=\, \frac{1}{2} \left( E_i^2 + B_i^2 \right) \= \left(\frac{\gamma}{\ell}\right)^4\, \rho \quad\with\quad \rho \= \sfrac12 \left( \Ecal_a^2 + \Bcal_a^2 \right)\ ,
\end{equation}
which, in turn, simplifies the expression for the energy $E$ to
\begin{equation}
    E \= \int_V \diff^3x\;e \= \sfrac{1}{2\ell}\int_{S^3} \diff^3\Omega_3 \  (1-\cos\chi)\,\bigl({\cal E}_a^2 + {\cal B}_a^2\bigr)\ .
\end{equation}
Notice here that the orientation of the $S^3$ volume measure~$\diff^3\Omega_3$ is chosen to provide a positive result. From \eqref{EBj} we find that the ``sphere-frame'' energy density is time-independent and has similar expression for both solution types (with appropriate eigenfrequency $\Omega$):
\begin{equation}
\sfrac12 \bigl( \Ecal_a^2 + \Bcal_a^2 \bigr) \= 2\,\Omega^2\,X_a\bar{X}_a(\omega)\ .
\end{equation}

The resultant expression for $E$ in terms of complex parameters $\lambda_{m,n}$ (for both solution types) is given by
\begin{equation}
    E \= \frac{1}{\ell}\, (2j+1)\,\Omega^3 \sum_{m,n}\,|\lambda_{m,n}|^2\ .
\end{equation}
Notice again here that for type I solutions we would have $\Lambda_{m,n}$ in the above expression.

{\bf Helicity.} Although helicity is not a Noether charge of the conformal group, it is nevertheless a conserved quantity for the Maxwell system and turns out to be related to the energy. The expression for the helicity is metric-free and can thus be evaluated over any spatial slice. 
Choosing again $t{=}\tau{=}0$,
\begin{equation} \label{helicity}
h = h_{\textrm{mag}}+h_{\textrm{el}} \= \frac{1}{2} \int_{\R^3} \ \bigl( A\wedge F + A_D \wedge F_D \bigr)
\= -\frac{1}{2} \int_{S^3} \diff^3\Omega_3 \  (1-\cos\chi)\,\bigl(\Acal_a\Bcal_a + \Acal^D_a\Ecal_a\bigr)\ ,
\end{equation}
where the subscript/superscript $`D'$ refers to the dual fields. Once again, taking type~I (upper sign) or type~II (lower sign) and fixing the spin~$j$ we obtain
\begin{equation}
\Acal^D_a \= \pm\im\,X_a(\omega)\,\ep^{\Omega\,\im\tau} \mp\im\,\bar{X}_a(\omega)\,\ep^{-\Omega\,\im\tau}\ ,
\end{equation}
which yields a constant ``sphere-frame'' helicity density
\begin{equation}
-\sfrac12 \bigl( \Acal_a\Bcal_a + \Acal^D_a\Ecal_a \bigr) \= \pm 2\,\Omega\,X_a\bar{X}_a(\omega)\ .
\end{equation}
As a result, even before performing the $S^3$ integration, we find a linear helicity-energy relation 
\begin{equation}
\Omega\, h \= \pm \ell\, E \qquad\textrm{for fixed spin and type}\ .
\end{equation}

Since the helicity measure an average of the linking numbers of any two electric or magnetic field lines~\cite{moffatt17, moffatt69, berger},
the latter must be related to the value~$j$ of the spin. The individual linking number of two field lines, however, appears neither to be independent of the lines chosen nor constant in time, as our observations indicate.
An exception are the Ra\~nada--Hopf knots discussed before, 
which display a conserved linking number of unity between any pair of electric or magnetic field lines.

\noindent
{\bf Momentum.} For the momentum densities $p_i = T^{0i} = (\textbf{E}\times\textbf{B})_i$ we obtain an interesting correspondence relating the one-form $p := p_i\, \diff x^i$ on Minkowski space with a similar one on de Sitter space:
\begin{equation}\label{1formP}
    p \= (\sfrac\gamma\ell)^3\, \mathcal{P}_a\, e^a\ =:\ (\sfrac\gamma\ell)^3\, \mathcal{P} \quad\with\quad \mathcal{P}_a\ :=\ \varepsilon_{abc}\, \mathcal{E}_b\, \mathcal{B}_c\ .
\end{equation}
A straightforward calculation then yields the expression of momenta $P_i$:
\begin{equation}
    P_i \= \int_V \diff^3x\, p_i \= \int_{S^3} \diff^3\Omega_3\, \mathcal{P}_a\, e^a_i \= 2\im\ \Omega^2\ \varepsilon_{abc}\ \int_{S^3} \diff^3\Omega_3\, e^a_i\, X_b\, \bar{X}_c
\end{equation}
with $e^a_i$ given by \eqref{t0Results}. The results for $j=0$ are
\begin{equation}
  \begin{aligned}
    P_1^{(j=0)} &\= -\sfrac{\sqrt{2}}{\ell}\left(\left(\bar{\Lambda}_{0,-1} + \bar{\Lambda}_{0,1}\right) \Lambda_{0,0} + \bar{\Lambda}_{0,0}\left(\Lambda_{0,-1}+\Lambda_{0,1}\right)\right)\ , \\
    P_2^{(j=0)} &\=  \sfrac{\im\sqrt{2}}{\ell} \left(\left(-\bar{\Lambda}_{0,-1} + \bar{\Lambda}_{0,1}\right) \Lambda_{0,0} + \bar{\Lambda}_{0,0}\left(\Lambda_{0,-1}-\Lambda_{0,1}\right)\right)\ , \\
    P_3^{(j=0)} &\= \sfrac2\ell \left( |\Lambda_{0,-1}|^2 - |\Lambda_{0,1}|^2 \right)\ .
  \end{aligned}
\end{equation}
As a consistency requirement, we check that the vector charges $P_i$ are rotated according to the algebra of $\mathcal{D}_a$ \eqref{DPdef} (See appendix \ref{appendRotation}):
\begin{equation}\label{rotP}
    \mathcal{D}_a\, P_b \= 2\,\varepsilon_{abc}\, P_c\ .
\end{equation}
We also note down $P_3$ for $j=1/2\ ~\textrm{and}~ 1$ in table \ref{table1}.
\begin{table}[]
    \centering\setcellgapes{4pt}\makegapedcells \renewcommand\theadfont{\normalsize\bfseries}
% \begin{adjustbox}{width=\linewidth}
 \resizebox{\linewidth}{!}{
    \begin{tabular}{|c|P{6cm}|P{8cm}|}
        \hline
      & $j=1/2$ & $j=1$ \\ [0.5ex] \hline \hline
     $P_3$ & 
     $\begin{aligned}
         \sfrac{9}{\ell} \Big( &|\Lambda_{-1/2,-3/2}|^{2} - |\Lambda_{-1/2, 1/2}|^{2} \\
         &- 2 |\Lambda_{-1/2, 3/2}|^{2} + 2 |\Lambda_{1/2,-3/2}|^{2} \\
         &+ |\Lambda_{1/2,-1/2}|^{2} - |\Lambda_{1/2, 3/2}|^{2}\Big)
     \end{aligned}$ & 
     $\begin{aligned}
         \sfrac{24}{\ell}\Big( &|\Lambda_{-1,-2}|^{2} - |\Lambda_{-1,0}|^{2} - 2|\Lambda_{-1,1}|^{2} - 3|\Lambda_{-1,2}|^{2} \\
         &+ 2|\Lambda_{0,-2}|^{2} + |\Lambda_{0,-1}|^{2} - |\Lambda_{0, 1}|^{2} - 2|\Lambda_{0,2}|^{2} \\
         &+ 3|\Lambda_{1,-2}|^{2} + 2|\Lambda_{1,-1}|^{2} + |\Lambda_{1,0}|^{2} - |\Lambda_{1,2}|^{2} \Big)
     \end{aligned}$  \\ \hline
     $P_\phi$ & 
     $\begin{aligned}
         9 \Big( &2|\Lambda_{-1/2,-3/2}|^{2} + |\Lambda_{-1/2,-1/2}|^{2} \\
         &-|\Lambda_{-1/2,3/2}|^{2} + |\Lambda_{1/2,-3/2}|^{2} \\
         &- |\Lambda_{1/2,1/2}|^{2} - 2|\Lambda_{1/2, 3/2}|^{2}\Big)
     \end{aligned}$ & 
     $\begin{aligned}
         24 \Big( &3|\Lambda_{-1,-2}|^{2} + 2|\Lambda_{-1,-1}|^{2} + |\Lambda_{-1,0}|^{2} - |\Lambda_{-1,2}|^{2} \\
         &+ 2|\Lambda_{0,-2}|^{2} + |\Lambda_{0,-1}|^{2} - |\Lambda_{0,1}|^{2} - 2|\Lambda_{0,2}|^{2} \\
         &+ |\Lambda_{1,-2}|^{2} - |\Lambda_{1,0}|^{2} - 2|\Lambda_{1,1}|^{2} - 3|\Lambda_{1,2}|^{2} \Big)
     \end{aligned}$  \\ \hline
     $L_3$ & 
     $\begin{aligned}
         9\ell\, \Big( &-2|\Lambda_{-1/2,-3/2}|^{2} - |\Lambda_{-1/2,-1/2}|^{2} \\
         &+|\Lambda_{-1/2, 3/2}|^{2} - |\Lambda_{1/2,-3/2}|^{2} \\
         &+ |\Lambda_{1/2,1/2}|^{2} + 2|\Lambda_{1/2, 3/2}|^{2}\Big)
     \end{aligned}$ & 
     $\begin{aligned}
         -24\ell\, \Big( &3|\Lambda_{-1,-2}|^{2} + 2|\Lambda_{-1,-1}|^{2} + |\Lambda_{-1,0}|^{2} \\
         & - |\Lambda_{-1,2}|^{2} + 2|\Lambda_{0,-2}|^{2} + |\Lambda_{0,-1}|^{2} \\
         & - |\Lambda_{0, 1}|^{2} - 2|\Lambda_{0,2}|^{2} + |\Lambda_{1,-2}|^{2} \\
         &- |\Lambda_{1,0}|^{2} - 2|\Lambda_{1,1}|^{2} - 3|\Lambda_{1,2}|^{2} \Big)
     \end{aligned}$  \\ \hline
    \end{tabular}
}% \end{adjustbox}
    \caption{Expressions of $P_3$, $P_
    \phi$ and $L_3$ for $j=1/2 ~\textrm{and}~ 1$.}
    \label{table1}
\end{table}
One can compute the corresponding $P_1$ and $P_2$ for $j=1/2$ and $j=1$ by employing the action of an appropriate $\mathcal{D}_a$ of the table in Appendix \ref{appendRotation}.

We can additionally compute the spherical components of the momentum $(P_r,P_\theta,P_\phi)$ by letting the one-form $e^a$ in \eqref{1formP} act on the vector fields $(\pa_r,\pa_\theta, \pa_\phi)$. In practice, we first write $\pa_r = -\sfrac\gamma\ell \pa_\chi$ using \eqref{Jacobian} at $t{=}0$ and then invert the vector fields $(\pa_r,\pa_\theta, \pa_\phi)$ in terms of the left invariant vector fields $(L_1,L_2,L_3)$ using \eqref{Lfields}. Finally, using the duality relation $e^a(L_b)=\delta^a_b$ we obtain
\begin{equation}\label{sphericalP}
 \begin{aligned}
   P_r &\= -\frac1\ell\int_{S^3} \diff^3\Omega_3\,(1-\cos\chi)\, \left( \sin\theta\cos\phi\, \mathcal{P}_1 + \sin\theta\sin\phi\, \mathcal{P}_2 + \cos\theta\, \mathcal{P}_3 \right)\ , \\
   P_\theta &\= \int_{S^3} \diff^3\Omega_3\,\sin\chi\cos\chi \Big( (\cos\theta\cos\phi + \tan\chi\sin\phi)\, \mathcal{P}_1 \\
   &\qquad\qquad\qquad\qquad\qquad\qquad + (\cos\theta\sin\phi - \tan\chi\cos\phi)\mathcal{P}_2 - \sin\theta\, \mathcal{P}_3 \Big) ,\\
   P_\phi &\= \int_{S^3} \diff^3\Omega_3\, \sin^2\chi\sin\theta \Big( (\cos\theta\cos\phi - \cot\chi\sin\phi)\, \mathcal{P}_1 \\
   &\qquad\qquad\qquad\qquad\qquad\qquad + (\cos\theta\sin\phi + \cot\chi\cos\phi)\, \mathcal{P}_2 - \sin\theta\, \mathcal{P}_3 \Big) \ .
 \end{aligned}
\end{equation}
From the expression of $P_r$ we see that the integrand over $S^2$ i.e. $\hat{\omega}_a\mathcal{P}_a$ \eqref{OmegaToHatOmega} is an odd function\footnote{The terms of $\mathcal{P}_a$ using \eqref{split-Y} are proportional to  $Y_{l,M}Y_{l',M'}$, which is even in $\hat{\omega}_a$ for a fixed $j$.}, which makes $P_r$ vanish. We also find with explicit calculations (verified for up to $j=1$) that $P_\theta$ vanishes. For $j=0$ we find that $P_\phi$ is proportional to $P_3$:
\begin{equation}
    P_\phi^{(j=0)} \= \ell\, P_3^{(j=0)} \=  2\left(|\Lambda_{0,-1}|^{2}-|\Lambda_{0,1}|^{2}\right)\ .
\end{equation}
The expressions of $P_\phi$ for $j=1/2$ and $1$ has been recorded in the table \ref{table1}.
\subsection{Lorentz transformations}
\noindent
Another solution of \eqref{Killing} is given by
\begin{equation}
    \zeta^\mu(x) \= \epsilon^{\mu}_{\ \; \nu}\, x^\nu \quad\textrm{where}\quad \epsilon_{\mu\nu} \= -\epsilon_{\nu\mu}\ ,
\end{equation}
which correspond to the six generators of the Lorentz group $SO(1,3)$. These six together with the above four translations generates the full Poincar\'e group. The corresponding six charges are grouped into the boost ${\bf K}$ and the angular momentum ${\bf L}$.

\noindent
{\bf Boost.} The conserved charge densities arising from $J^0$ \eqref{currentJ} corresponding to $\epsilon_{0i}$ are the boost densities ${\bf k} = \rho\ {\bf x} - {\bf p}\ t$, which simplify for $t{=}0$ to
\begin{equation}
    k_i \= (\sfrac\gamma\ell)^3\, \mathcal{K}_i \quad\with\quad \mathcal{K}_i \= \rho\, \omega_i\ .
\end{equation}
The corresponding charges $K_i$ vanishes because of the odd integrand as discussed in an earlier remark:
\begin{equation}
K_i \= \int_V \diff^3x\, k_i \= \int_{S^3} \diff^3\Omega_3\, \mathcal{K}_i \= 2 \int_{S^3} \diff^3\Omega_3\, \omega_i\, X_a^j\, \bar{X}_a^j \= 0\ .
\end{equation}

\noindent
{\bf Angular momentum.} The other three conserved charge densities $J^0$ corresponding to $\epsilon_{ij}$ in \eqref{currentJ} are the angular momentum densities ${\bf l} = {\bf p}\times{\bf x}$, which takes a simple form just like in momentum \eqref{1formP}:
\begin{equation}
    l_i\, \diff x^i \= (\sfrac\gamma\ell)^2\, \mathcal{L}_a\, e^a \quad\with\quad \mathcal{L}_a \= \varepsilon_{abc}\, \mathcal{P}_b\, \omega_c\ .
\end{equation}
The expressions for the charges $L_i$ simplify to
\begin{equation}
    L_i \= \int_V \diff^3x\, l_i \= \int_{S^3} \diff^3\Omega_3\, (\sfrac\ell\gamma)\, \mathcal{L}_a\, e^a_i \= 2\im\ell\ \Omega^2\ \int_{S^3} \diff^3\Omega_3\, \sfrac{1}{\gamma}\, e^a_i \left( \bar{X}_a\, \omega_b X_b - X_a\, \omega_b\bar{X}_b \right)\ .
\end{equation}
Explicit calculation show that for $j{=}0$ the angular momenta $L_i$ are proportional to the momenta $P_i$:
\begin{equation}
    L_i^{(j=0)} \= -\ell^2\, P_i^{(j=0)}\ .
\end{equation}
This is, however, not true for higher spin $j$. The angular momenta $L_i$ has the same rotation behaviour as for the momenta $P_i$ \eqref{rotP}. We therefore note down the results of $L_3$ for $j=1/2$ and $1$ in table \ref{table1}, from which the corresponding expressions of $L_1$ and $L_2$ can be obtained using the table in Appendix \ref{appendRotation}.

We can again compute the spherical components of the angular momentum $(L_r,L_\theta,L_\phi)$ using the relations \eqref{sphericalP} by replacing $\mathcal{P}$ with $\mathcal{L}$ in it. We realize that the $S^2$ integrand for $L_r$ would only have terms like $\hat{\omega}_a\hat{\omega}_b\mathcal{P}_c$, which are all odd functions\footnote{Here again the terms of $\mathcal{P}_c$, which goes like $Y_{l,M}Y_{l',M'}$ \eqref{split-Y}, are all even functions of $\phi$ for a fixed $j$.} of $\phi$ and, therefore, the $\phi$ integral over the domain $(0,2\pi)$ would make $L_r$ vanish. We also find, with explicit computations, that the charges $L_\theta$ and $L_\phi$ for $j{=}0$ are proportional to $P_3$:
\begin{equation}
    L_\theta^{(j=0)} \= \sfrac43\, \ell^2\, P_3^{(j=0)} \quad\und\quad L_\phi^{(j=0)} \= -\sfrac13\, \ell^2\, P_3^{(j=0)}\ .
\end{equation}
As a non trivial example, we collect these charges for $j=1/2$ below:
\begin{equation}
  \begin{aligned}
     L_\theta^{\left(j=1/2\right)} \= \sfrac{12}{5}\ell\, \Big( &9|\Lambda_{-1/2,-3/2}|^{2} + 6|\Lambda_{-1/2,-1/2}|^{2} +|\Lambda_{-1/2, 1/2}|^{2}
     - 6|\Lambda_{-1/2,3/2}|^{2} \\ 
     & + 6|\Lambda_{1/2,-3/2}|^{2} - |\Lambda_{1/2, -1/2}|^{2} - 6|\Lambda_{1/2,1/2}|^{2} - 9|\Lambda_{1/2, 3/2}|^{2} \Big)\ ,
  \end{aligned}
\end{equation}
\begin{equation}
  \begin{aligned}
      L_\phi^{\left(j=1/2\right)} \= -\sfrac35 \ell\, \Big(  &6|\Lambda_{-1/2,-3/2}|^{2} - |\Lambda_{-1/2,-1/2}|^{2} -6|\Lambda_{-1/2, 1/2}|^{2} - 9|\Lambda_{-1/2,3/2}|^{2} \\
      &+ 9|\Lambda_{1/2,-3/2}|^{2} + 6|\Lambda_{1/2,-1/2}|^{2} + |\Lambda_{1/2,1/2}|^{2} - 6|\Lambda_{1/2, 3/2}|^{2}\\ 
      &+ \sqrt{3}\big( \bar{\Lambda}_{1/2,-3/2}\Lambda_{-1/2,-1/2} -\bar{\Lambda}_{1/2,1/2}\Lambda_{-1/2,3/2} \\
      &+ \bar{\Lambda}_{-1/2,-1/2}\Lambda_{1/2,-3/2} - \bar{\Lambda}_{-1/2,3/2}\Lambda_{1/2,1/2}\big) \Big)\ .
  \end{aligned}
\end{equation}

\subsection{Dilatation} 
It is easy to verify that a constant rescaling by $\lambda$:
\begin{equation}
    \zeta^\mu \= \lambda\, x^\mu
\end{equation}
is also a solution of \eqref{Killing}. The charge density corresponding to this single generator of the conformal group is ${\bf p}\cdot{\bf x}- e\,t$, which for $t{=}0$ simplifies to 
\begin{equation}
    p_i\, x_i \= (\sfrac\gamma\ell)^3\, \mathcal{P}_a\,\omega_a\ .
\end{equation}
The corresponding charge $D$ vanishes because of the odd integrand:
\begin{equation}
    D \= \int_V \diff^3x\, p_i\, x_i \= 2\im\,\Omega^2\, \varepsilon_{abc} \int_{S^3} \diff^3\Omega_3\, \omega_a\, X_b\, \bar{X}_c \= 0\ .
\end{equation}
\subsection{Special conformal transformations}
A fairly straightforward calculation shows that the following not so obvious transformation
\begin{equation}
    \zeta^\mu \= 2\, x^\mu\,b_\nu x^\nu\ -\ b^\mu\, x^\nu x_\nu 
\end{equation}
also satisfies \eqref{Killing}. The four generators corresponding to $b_\mu$ give rise to four different charges $V_0$ and ${\bf V}$.

\noindent
{\bf Scalar SCT.} The charge density $J^0$ corresponding to $b_0$ is 
\begin{equation}
    v_0 \= ({\bf x}^2+t^2)\, e - 2t\, {\bf p}\cdot{\bf x}\ ,
\end{equation}
which for $t{=}0$ simplifies to
\begin{equation}
    v_0 \= {\bf x}^2\, e \= (\sfrac\gamma\ell)^2 \rho\, \omega_a^2\ .
\end{equation}
The expression for the corresponding charge $V_0$ takes the following simple form:
\begin{equation}
    V_0 \= \int_V \diff^3x\, v_0 \= \int_{S^3} \diff^3\Omega_3\, (\sfrac\ell\gamma)\, (1-\omega_4^2)\, \rho \= 2\ell\, \Omega^2 \int_{S^3} \diff^3\Omega_3\, (1+\omega_4)\, X_a\, \bar{X}_a\ .
\end{equation}
Here again the $\omega_4$ term of the integral, being odd, vanishes and yields
\begin{equation}
    V_0 \= \ell^2\, E \= 8\,\ell\, (j+1)^3(2j+1)\, \sum_{m,n}\,|\Lambda_{m,n}|^2\ .
\end{equation}

\noindent
{\bf Vector SCT.} The charge densities $J^0$ that correspond to $b_i$ read:
\begin{equation}
    {\bf v} \= 2{\bf x}({\bf x}\cdot{\bf p}) - 2t\,{\bf x}\,e - ({\bf x}^2-t^2){\bf p}\ .
\end{equation}
This simplify at $t{=}0$ and takes a structure similar to the momentum densities $p_i$ \eqref{1formP}:
\begin{equation}\label{vectorSCT}
    v_i\, dx^i \= (\sfrac\gamma\ell)\, \mathcal{V}_a\, e^a \with \mathcal{V}_a \= 2\omega_a\, (\mathcal{P}_b\,\omega_b) - \omega_b^2\,\mathcal{P}_a\ . 
\end{equation}
The expressions for the charges $V_i$ then simplify to
\begin{equation}
    V_i \= \int_V \diff^3x\, v_i \= 2\im\,\ell^2\,\Omega^2 \int_{S^3} \diff^3\Omega_3\, \gamma^{-2} e^a_i \left( 2\varepsilon_{bcd}\,\omega_a\,\omega_b\,X_c\,\bar{X}_d - \varepsilon_{abc}\,(1-\omega_4^2)\,X_b\,\bar{X}_c \right)\ .
\end{equation}
With explicit computation we observe that the charges $V_i$ are proportional to the momenta $P_i$ (verified explicitly for up to $j{=}1$)
\begin{equation}
    V_i \= \ell^2\, P_i\ .
\end{equation}

As before, we can compute the spherical components $(V_r,V_\theta,V_\phi)$ by using the expressions \eqref{sphericalP} and replacing $\mathcal{P}$ with $\mathcal{V}$ in it. We notice that the charge $V_r$ vanishes owing to the odd $S^2$ integrand\footnote{Observe that the terms in $\mathcal{V}_a$ \eqref{vectorSCT} are all even in $\hat{\omega}_a$.} just like in the case of $P_r$. However, unlike $P_\theta$ here the charge $V_\theta$ is non-vanishing. Explicit calculations show that for $j=0$ the charges $V_\theta$ and $V_\phi$ are proportional to the momentum $P_3$:
\begin{equation}
    V_\theta^{(j=0)} \= -\sfrac43 \ell^3\, P_3^{(j=0)} \quad\und\quad V_\phi^{(j=0)} \= -\sfrac53 \ell^3\, P_3^{(j=0)}\ .
\end{equation}
Additionally, we record below the charges $V_\theta$ and $V_\phi$ for the non-trivial case of $j{=}1/2$
\begin{equation}
  \begin{aligned}
     V_\theta^{\left(j=1/2\right)} \= -\sfrac{6}{5}\ell^2\, \Big( &9|\Lambda_{-1/2,-3/2}|^{2} + |\Lambda_{-1/2,-1/2}|^{2} - 9|\Lambda_{-1/2,1/2}|^{2} - 21|\Lambda_{-1/2,3/2}|^{2} \\ 
     & + 21|\Lambda_{1/2,-3/2}|^{2} +
     9|\Lambda_{1/2,-1/2}|^{2} - |\Lambda_{1/2,1/2}|^{2} - 9|\Lambda_{1/2, 3/2}|^{2} \Big)\ ,
  \end{aligned}
\end{equation}
\begin{equation}
  \begin{aligned}
      V_\phi^{\left(j=1/2\right)} \= -\sfrac35 \ell^2\, \Big(  &42|\Lambda_{-1/2,-3/2}|^{2} + 33|\Lambda_{-1/2,-1/2}|^{2} + 8|\Lambda_{-1/2,1/2}|^{2} - 33|\Lambda_{-1/2,3/2}|^{2} \\
      &+ 33|\Lambda_{1/2,-3/2}|^{2} - 8|\Lambda_{1/2,-1/2}|^{2} - 33|\Lambda_{1/2,1/2}|^{2} - 42|\Lambda_{1/2,3/2}|^{2}\\ 
      &\qquad + 8\sqrt{3}\big( -\bar{\Lambda}_{1/2,-3/2}\Lambda_{-1/2,-1/2} +\bar{\Lambda}_{1/2,1/2}\Lambda_{-1/2,3/2} \\
      &\qquad\qquad\qquad - \bar{\Lambda}_{-1/2,-1/2}\Lambda_{1/2,-3/2} + \bar{\Lambda}_{-1/2,3/2}\Lambda_{1/2,1/2}\big) \Big)\ .
  \end{aligned}
\end{equation}
One can compute these charges for higher spin-$j$ by following the same strategy.

\subsection{Applications}
\label{sec4}
\noindent
The method of constructing rational electromagnetic fields presented in this paper has the added advantage that it produces a complete set labelled by $(j,m,n)$; any electromagnetic field configuration having finite energy can, in principle, be obtained from an expansion like in (\ref{Asep}-\ref{basisZ}), albeit with a varying $j$. The operational difficulty involved in this procedure has to do with the fact that this set is infinite as $j\in \sfrac{\N}{2}$. There are, however, many important cases where only a finite number of knot-basis solutions (sometimes only with a fixed $j$) need to be combined to get the desired EM field configuration. One such very important case is that of the Hopfian solution discussed before. Below we analyse two very interesting generalisations of the Hopfian solution presented in \cite{HSS15} in the context of present construction. It is imperative to note here that while the scope of construction of a new solution from the known ones as presented in \cite{HSS15} is limited, the same is not true for the method presented in this paper, which by design can produce arbitrary number of new field configurations. Some of these possible new field configurations obtained from the $j{=}0$ sector (possibly from $j{=}1/2~ \textrm{or}~ 1$ as well) could find experimental application with improved experimental techniques like in \cite{LSMetal18}. 

\begin{figure}[h!]
\captionsetup{width=\linewidth}
\centering
   \includegraphics[width = 5cm, height = 5cm]{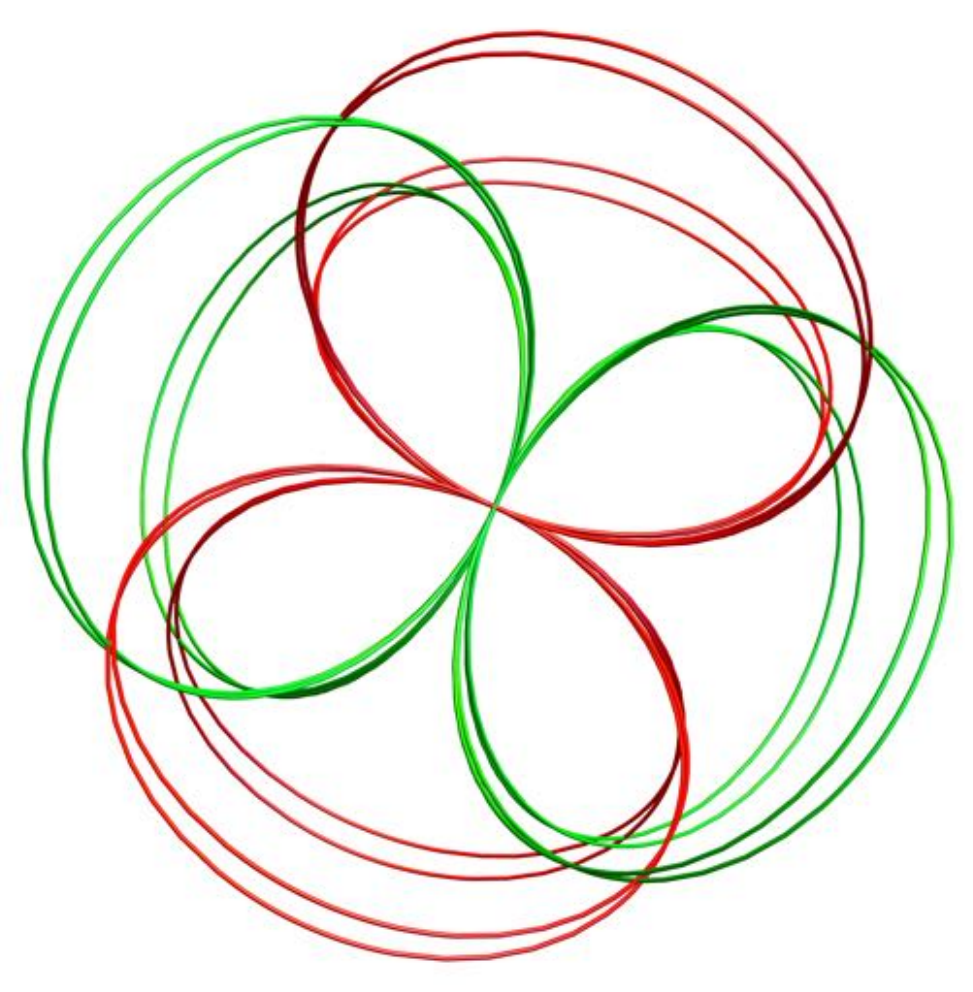}
   \hspace{2cm}
   \includegraphics[width = 5cm, height = 5cm, trim = {2cm 3cm 1.5cm 2cm},clip]{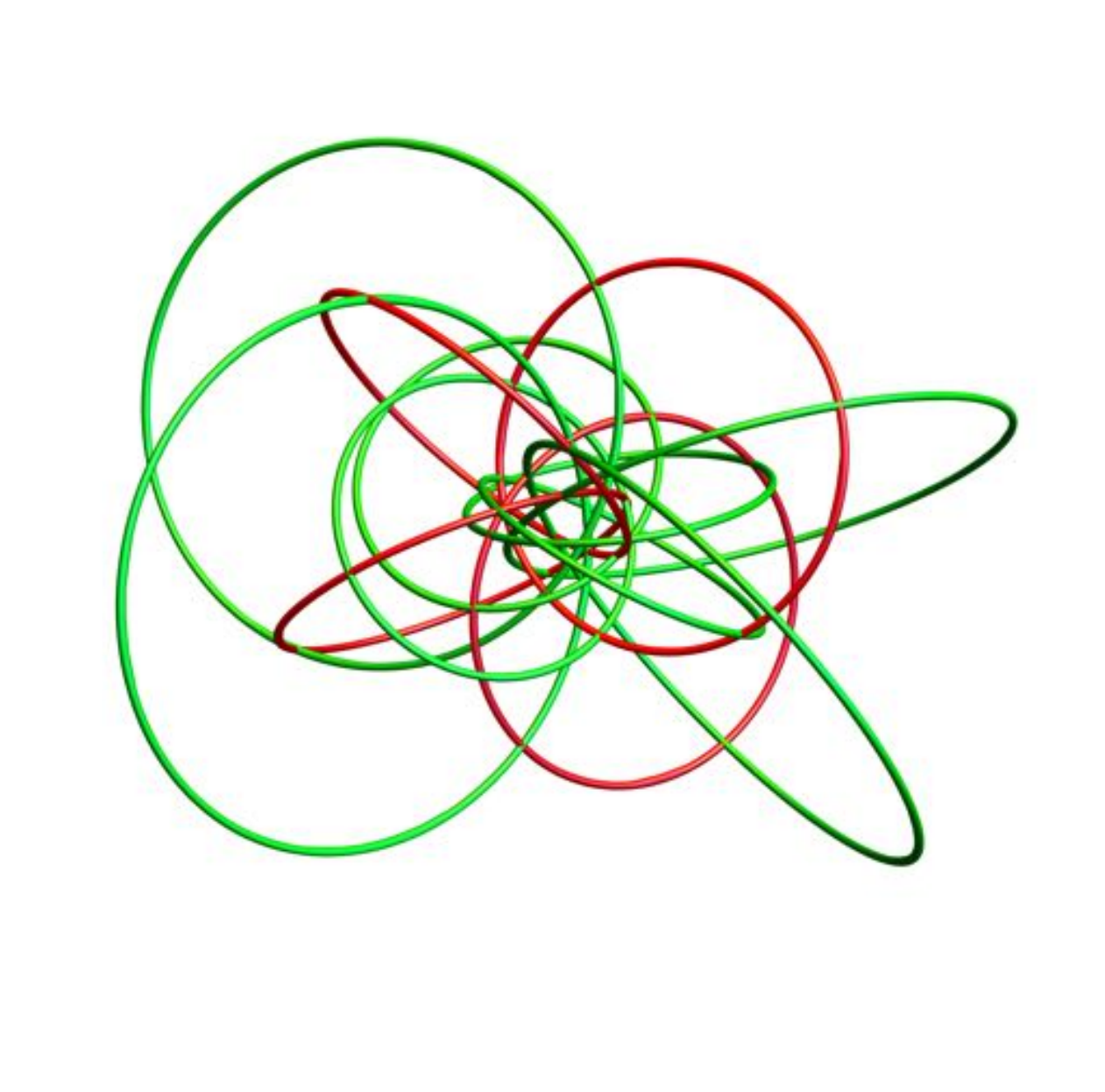}
 \caption{
 Sample electric (red) and magnetic (green) field lines at $t{=}0$.
 Left: time-translated Hopfian with $c=60$, Right: Rotated Hopfian with $\theta=1$.} 
\label{fig3}
\end{figure}
Bateman's construction, employed in \cite{HSS15}, hinges on a judicious ansatz for the Riemann-Silberstein vector \eqref{RSvector} satisfying Maxwell's equations:
\begin{equation}
    {\bf S} \= \nabla\alpha \times \nabla\beta \quad\implies\quad \nabla\cdot{\bf S}\=0 ~\ \& ~\ \im\, \partial_t{\bf S} \= \nabla\times{\bf S}\ ,
\end{equation}
using a pair of complex functions $(\alpha,\beta)$. An interesting generalization of the Hopfian solution is obtained in equations (3.16-3.17) of \cite{HSS15} using (complex) time-translation (TT) to obtain the following $(\alpha,\beta)$ pair (up to a normalization)
\begin{equation}
    (\alpha,\beta)_{TT} \= \left(\frac{A-1-\im z}{A+\im(t+\im c)}, \frac{x-\im y}{A+\im(t+\im c)} \right)\ ;\quad A\= \sfrac12 \left(x^2+y^2+z^2-(t+\im c)^2+1\right)\ ,
\end{equation}
where $c$ is a constant real parameter. The corresponding  EM field configuration is obtained, in our case, by choosing $\ell {=} 1{-}c$ and only the following $j{=}0$, and hence type I, complex coefficients in \eqref{basisZ}
\begin{equation}\label{j0TT}
    \left(\Lambda_{0;0,1}\ ,~\Lambda_{0;0,0}\ ,~\Lambda_{0;0,-1}\right)_{TT} \= \left( 0\ ,~0\ ,~-\im\frac{\pi}{2\ell^2} \right)\ .
\end{equation}
A sample electromagnetic knot configuration of this modified Hopfian is illustated in figure \ref{fig3}. Another interesting generalization of the Hopfian is constructed in equations (3.20-3.21) of \cite{HSS15} using a (complex) rotation (R) to get the following $(\alpha,\beta)$ pair (again, up to a normalization)
\begin{equation}
    (\alpha,\beta)_{R} \= \left(\frac{A-1+\im(z \cos{\im\theta}+x\sin{\im\theta})}{A+\im t}, \frac{x \cos{\im\theta}-z\sin{\im\theta}-\im y}{A+\im t} \right) \ ,
\end{equation}
where $A = \sfrac12 \left(x^2+y^2+z^2-t^2+1\right)$. To get this particular EM field configuration we need to set $\ell{=}1$ and use the following combination of only $j{=}0$ type I coefficients in \eqref{basisZ}
\begin{equation}\label{j0R}
    \left(\Lambda_{0;0,1}\ ,~\Lambda_{0;0,0}\ ,~\Lambda_{0;0,-1}\right)_{R} \= \left( \im\frac{\pi}{4}(\cosh{\theta}-1)\ ,~-\frac{\pi}{2\sqrt{2}}\sinh{\theta}\ ,~-\im\frac{\pi}{4}(\cosh\theta+1) \right)\ .
\end{equation}
We illustrate the EM field lines for a sample $\theta$ value of this modified Hopfian in figure \ref{fig3}. We note down the conformal charges corresponding to these solutions in table \ref{table3} by plugging the $j{=}0$ coefficients \eqref{j0TT} and \eqref{j0R} in the appropriate formulae of the previous section. The results matches with the ones given in \cite{HSS15} up to a rescaling of the energy, which can be achieved by an appropriate choice of normalization.
\begin{table}[]
    \centering\setcellgapes{4pt}\makegapedcells \renewcommand\theadfont{\normalsize\bfseries}
    \begin{tabular}{|c|P{5cm}|P{5cm}|}
        \hline
      & Time-translated Hopfian & Rotated Hopfian \\ [0.5ex] \hline \hline
     Energy (E) & 
     $\frac{2\pi^2}{(1-c)^5}=:E_{TT}$ & 
     $2\pi^2\cosh^2\theta =: E_{R}$  \\ \hline
     Momentum ($\textbf{P}$) & 
     $\left(0\ ,~0\ ,~\frac{1}{4}\right)E_{TT}$ & 
     $\left(0\ ,~-\frac{1}{4}\tanh{\theta}\ ,~\frac{1}{4}\sech\theta\right)E_{R}$ \\ \hline
     Boost ($\textbf{K}$) & 
     $\left(0\ ,~0\ ,~0\right)$ & 
     $\left(0\ ,~0\ ,~0\right)$ \\ \hline
     Ang. momentum ($\textbf{L}$) & 
     $\left(0\ ,~0\ ,~-\frac{1}{4}(1-c)^2\right)E_{TT}$ & 
     $\left(0\ ,~\frac{1}{4}\tanh{\theta}\ ,~-\frac{1}{4}\sech\theta\right)E_{R}$ \\ \hline
      Dilatation (D) & 
     $0$ & 
     $0$ \\ \hline
     Scalar SCT ($V_0$) & 
     $(1-c)^2E_{TT}$ & 
     $E_{R}$  \\ \hline
     Vector SCT ($\textbf{V}$) & 
     $\left(0\ ,~0\ ,~\frac{1}{4}(1-c)^2\right)E_{TT}$ & 
     $\left(0\ ,~-\frac{1}{4}\tanh{\theta}\ ,~\frac{1}{4}\sech\theta\right)E_{R}$ \\ \hline
    \end{tabular}
    \caption{Conformal charges for the time-translated and rotated Hopfian configurations.}
    \label{table3}
\end{table}

\vspace{12pt}
%----------------------------------------------------------------------------------------
\section{Null fields}
\label{nullFields}
\vspace{2pt}
\noindent
An interesting subset of vacuum electromagnetic fields are those with vanishing Lorentz invariants,
\begin{equation}
\vec{E}^2-\vec{B}^2 =0 \quad\und\quad \vec{E}\cdot\vec{B} =0 \qquad\Longleftrightarrow\qquad
\bigl(\vec{E}\pm\im\vec{B}\bigr)^2 =\ 0\ .
\end{equation}
As a scalar equation it must equally hold on the de Sitter side, and so
we can try to characterize such configurations with our SO(4) basis above. For a given type and spin, 
the expressions in~\eqref{EBj} immediately give the Riemann-Silberstein vector on the $S^3$~cylinder,
\begin{equation}
\Ecal_a \pm\im\,\Bcal_a \= -2\im\,\Omega\,X_a(\omega)\,\ep^{\Omega\,\im\tau}\ ,
\end{equation}
where the upper (lower) sign pertains to type~I (II).
Note that the negative-frequency part of this field has cancelled.
The vanishing of $(\Ecal_a{\pm}\im\Bcal_a)(\Ecal_a{\pm}\im\Bcal_a)$ is then equivalent to a condition on the angular functions,
\begin{equation} \label{nullcondition}
0 \= X_1(\omega)^2 + X_2(\omega)^2 + X_3(\omega)^2 \= 2\,Z_+(\omega) Z_-(\omega) + Z_3(\omega)^2\ .
\end{equation}
When expanding the angular functions~$Z^j_{*\ \textrm{I}}$ or $Z^j_{*\ \textrm{II}}$ into basis solutions with ~\eqref{basisZ},
one arrives at a system of homogeneous quadratic equations for the free coefficients~$\lambda_{j;m,n}^{\textrm{I/II}}$.

Let us analyze the situation for type~I and spin~$j$. 
The functions~$Z^j_*(\omega)$ transform under a $(j,j)$ representation of $su(2)_L\oplus su(2)_R$. 
The null condition~\eqref{nullcondition} then yields a representation content of $(0,0)\oplus(1,1)\oplus\ldots\oplus(2j,2j)$
and may thus be expanded into the corresponding harmonics. Furthermore, The independent vanishing of all coefficients produces
$\frac16(4j{+}1)(4j{+}2)(4j{+}3)$ equations for the $(2j{+}1)(2j{+}3)$ parameters~$\lambda_{j;m,n}$ (note the ranges of $m$ and~$n$ for type~I).
Clearly, this system is vastly overdetermined. However, it turns out that only $4j^2{+}6j{+}1$ equations are independent,
still leaving $2j{+}2$ free complex parameters for the solution space. The independent equations can be organized as (suppressing~$j$)
\begin{equation}
\begin{aligned}
\lambda_{m,n}^2 &\ \sim\ \lambda_{m,n-1}\,\lambda_{m,n+1} \qquad\ \for m,n=-j\,\ldots,j \ ,\\
\lambda_{m,j+1}\,\lambda_{m+1,-j-1} &\= \lambda_{m+1,j+1}\,\lambda_{m,-j-1} \quad\for m=-j,\ldots,j{-}1\ .
\end{aligned}
\end{equation}
We have checked for $j{\le}5$ that the upper equations are solved by~\footnote{These are the generic solutions. There exist also special solutions given by 
\eqref{extweights} and $\lambda_{m,n}=0$ for $|n|\neq j{+}1$, for arbitrarily selected choices of~$m\in\{-j,\ldots,j\}$.}
\begin{equation}
\lambda_{m,n}^{2j+2} \= {\textstyle\sqrt{\binom{2j+2}{j+1-n}}}\ \lambda_{m,-j-1}^{j+1-n}\,\lambda_{m,j+1}^{j+1+n}
\quad\for m=-j,\ldots,j \und n=-j{-}1,\ldots,j{+}1\ ,
\end{equation}
while the lower ones imply that the highest weights $n{=}j{+}1$ are proportional to the lowest weights $n{=}{-}j{-}1$ (independent of~$m$), 
\begin{equation} \label{extweights}
\lambda_{m,-j-1} \= w\,\lambda_{m,j+1} \quad\for w\in\mathds{C}^*\ .
\end{equation}
Therefore, the full (generic) solution reads
\begin{equation}
\lambda_{m,n} \= {\textstyle\sqrt{\binom{2j+2}{j+1-n}}}\ w^{\frac{j+1-n}{2j+2}}\ \ep^{2\pi\im k_m\frac{j+1-n}{2j+2}}\ z_m 
\qquad\with z_m\in\mathds{C} \und k_m\in\{0,1,\ldots,2j{+}1\}\ ,
\end{equation}
containing $2j{+}2$ complex parameters $z_m$ and~$q$ as well as $2j$ discrete choices~$\{k_m\}$ 
(one of them can be absorbed into~$z_m$). This completely specifies the type-I null fields for a given spin.
Type-II null fields are easily obtained by applying electromagnetic duality to type-I null fields.

In the simplest case of $j{=}0$, the single equation $\lambda_{0,0}^2=2\lambda_{0,-1}\lambda_{0,1}$ describes 
a generic rank-3 quadric in $\mathds{C} P^2$, or a cone over a sphere $\mathds{C} P^1$ inside the parameter space~$\mathds{C}^3$. For higher spin,
the moduli space of type-I null fields remains a complete-intersection projective variety of complex dimension~$2j{+}1$.

\begin{figure}[h!]
\captionsetup{width=\linewidth}
\centering
\includegraphics[width = 0.35\paperwidth]{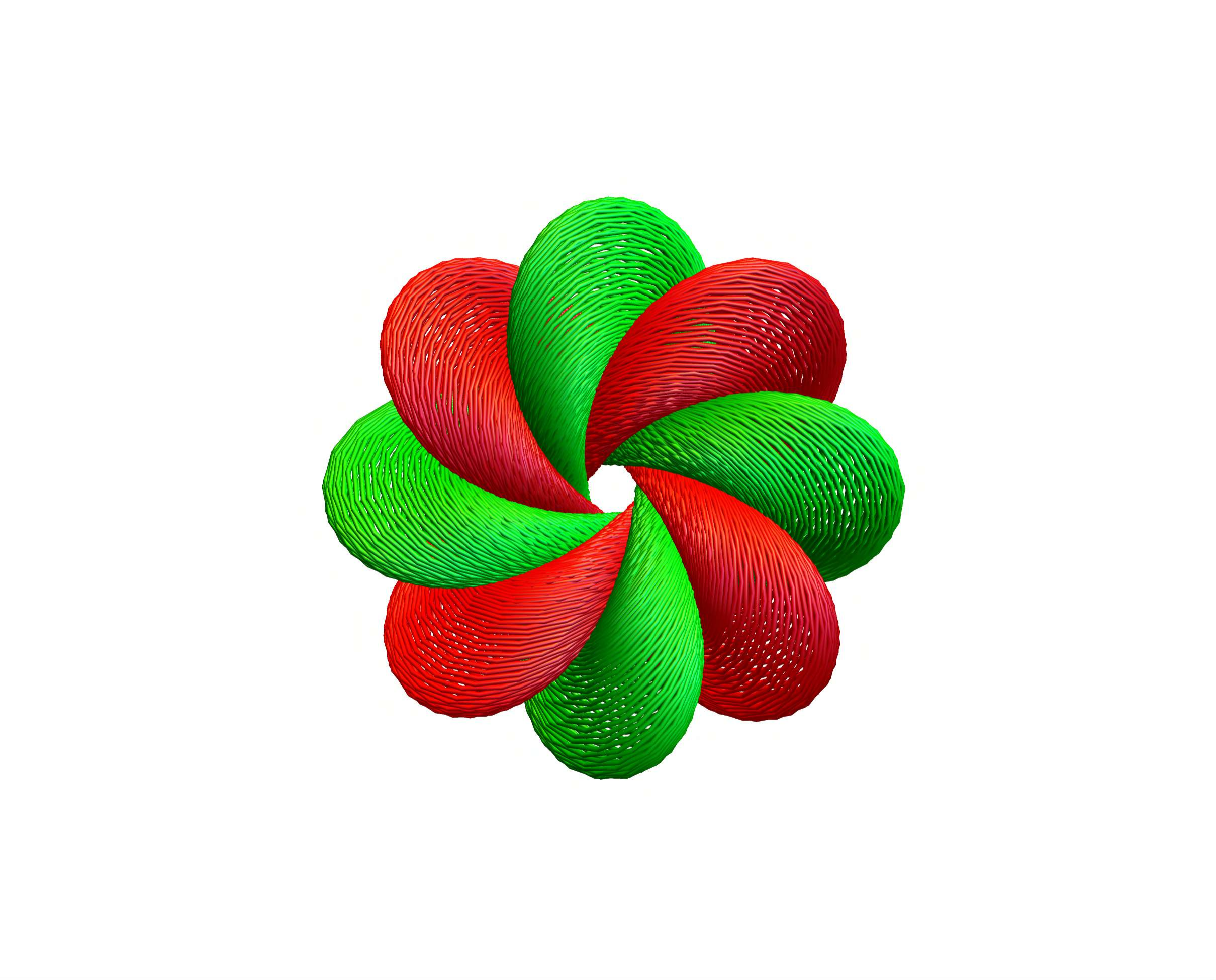}
\includegraphics[width = 0.35\paperwidth]{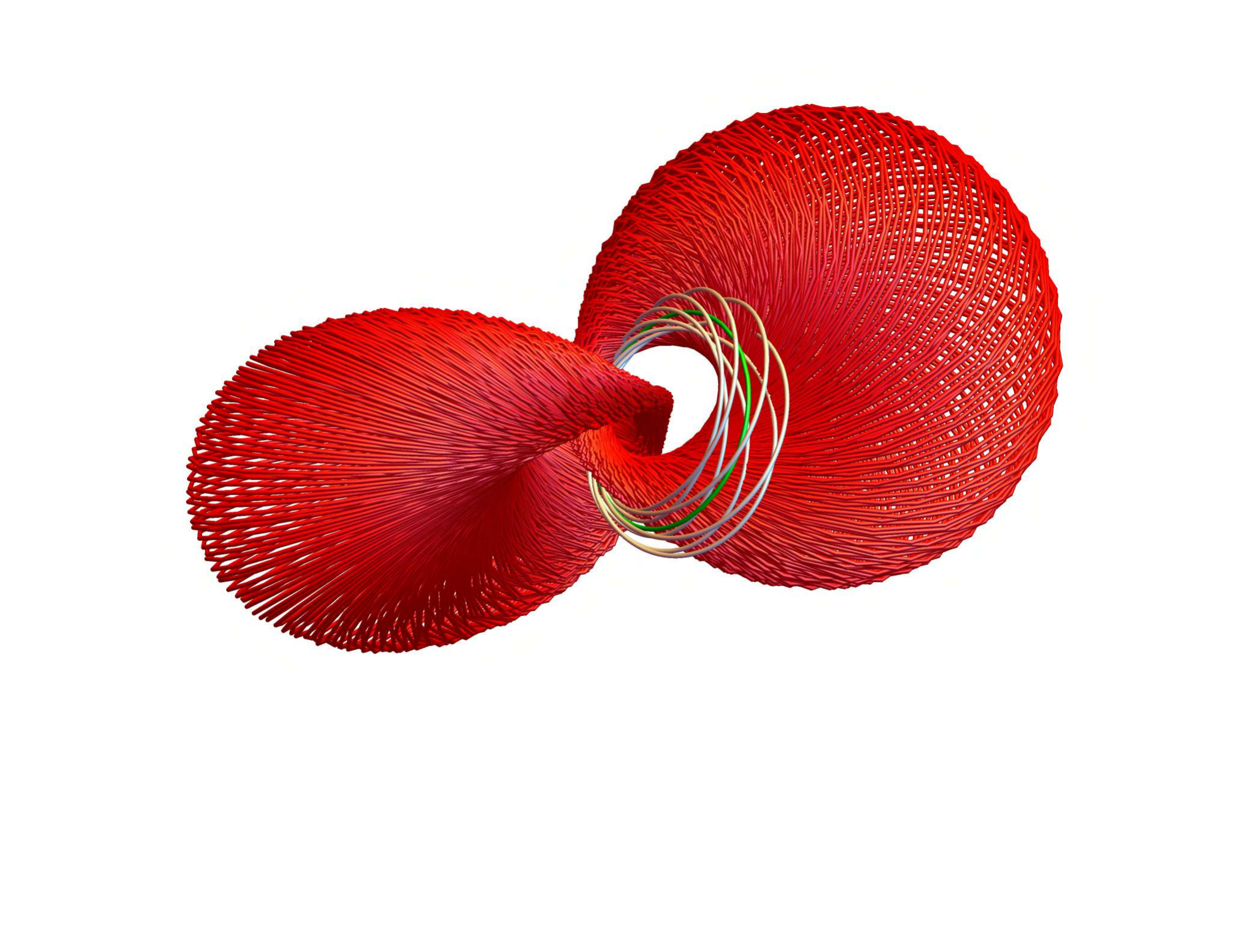}
\caption{
Sample electric (red) and magnetic (green) field lines at $t{=}0$.
Left: a pair of electric and a pair of magnetic field lines for the $(j;m,n) = \left(\sfrac12;\sfrac12,\sfrac32 \right)_R$ field configuration. Right: a pair of electric field lines, a magnetic field line
of self-linking one and a magnetic field line of self-linking seven for the $(j;m,n) = (1;0,2)_R$ field configuration.
}  
\label{nullFieldLines}
\end{figure}

We conclude the Section with a display of typical field lines (see Figure \ref{nullFieldLines}) for a type I $j{=}\sfrac12$ and $j{=}1$ null field at $t{=}0$. For $t{\neq}0$ the pictures get smoothly distorted.

\vspace{12pt}
%----------------------------------------------------------------------------------------
\section{Flux transport}
\vspace{2pt}
\noindent
We have seen that electromagnetic energy is radiated away along the light-cones. 
Let us try to quantify its amount over future null infinity~$\mathscr{I}^+$. Before proceeding further we note down the determinant of the Jacobian $J$ \eqref{Jacobian}
\begin{equation}
|\textrm{det}J| \= \frac{p^2{-}q^2}{\ell^2} \= \frac{\gamma^2}{\ell^2} 
\= \frac{\sin^2\!\tau}{t^2} \= \frac{\sin^2\!\chi}{r^2}
\quad\with \gamma^2=p^2{-}q^2
\end{equation}
and the spherical Minkowski components
\begin{equation}
A_t \= \Acal_\tau J^\tau_{\ t} + \Acal_\chi J^\chi_{\ t}  \ ,\qquad
A_r \= \Acal_\tau J^\tau_{\ r} + \Acal_\chi J^\chi_{\ r}  \ ,\qquad
A_\theta = \Acal_\theta \ ,\qquad A_\phi = \Acal_\phi\ ,
\end{equation}
or any other such tensor component arising due to \eqref{Jacobian}.
For later use, we also note here the transformation of the volume form
\begin{equation} \label{measure}
\diff^4x \= \diff t\,r^2\diff r\,\diff^2\Omega_2 \= r^2\,|\textrm{det}J|^{-1}\diff\tau\,\diff\chi\,\diff^2\Omega_2
\= \sin^2\!\chi\,|\textrm{det}J|^{-2}\diff\tau\,\diff\chi\,\diff^2\Omega_2
\=\frac{\ell^4}{\gamma^4}\,\diff\tau\,\diff^3\Omega_3\ .
\end{equation}

The energy flux at time~$t_0$ passing through a two-sphere of radius~$r_0$ centered at the spatial origin is given by
\begin{equation}
\Phi(t_0,r_0) \= \int_{S^2(r_0)}\!\!\diff^2\vec{\sigma}\cdot \bigl(\vec{E}\times\vec{B}\bigr)(t_0,r_0,\theta,\phi) 
\= \int_{S^2}r_0^2\,\diff^2\Omega_2\ T_{t\,r}^{\mathrm{(M)}}(t_0,r_0,\theta,\phi)\ ,
\end{equation}
where $\diff^2\Omega_2=\sin\theta\,\diff\theta\,\diff\phi$, 
and $T_{t\,r}^{\mathrm{(M)}}$ is the $(t,r)$ component of the Minkowski-space stress-energy tensor
\begin{equation}\label{stress-energyMink}
T_{\mu\,\nu}^{\mathrm{(M)}} \= F_{\mu\rho}F_{\nu\lambda}\,g^{\rho\lambda} - \sfrac14 g_{\mu\nu}F^2 
\quad\with (g_{\mu\nu}) = \textrm{diag}(-1,1,r^2,r^2\sin^2\!\theta)\ 
\end{equation}
for $\mu,\nu,\ldots\in\{t,r,\theta,\phi\}$. We carry out this computation in the $S^3$-cylinder frame by using the conformal relations
\begin{equation}
\ell^2\,T_{\mu\,\nu}^{\mathrm{(dS)}} \= t^2\,T_{\mu\,\nu}^{\mathrm{(M)}} \= \sin^2\!\tau\,T_{\mu\,\nu}^{\mathrm{(cyl)}} 
\= \sin^2\!\tau\,T_{m\,n}^{\mathrm{(cyl)}}\, J^m_{\ \ \mu}\,J^n_{\ \ \nu} 
\for m,n\in\{\tau,\chi,\theta,\phi\}
\end{equation}
with the Jacobian~\eqref{Jacobian} and the fact that \ $r\sin\tau=t\sin\chi$ \ so that 
\begin{equation}
\Phi(\tau_0,\chi_0) \= \int_{S^2}\sin^2\!\chi\,\diff^2\Omega_2\ T_{t\,r}^{\mathrm{(cyl)}}(\tau_0,\chi_0,\theta,\phi)
\=  \int_{S^2}\sin^2\!\chi\,\diff^2\Omega_2\ T_{m\,n}^{\mathrm{(cyl)}}\, J^m_{\ \ t}\,J^n_{\ \ r} \ .
\end{equation}
A straightforward computation using $(g_{mn})=\textrm{diag}(-1,1,\sin^2\!\chi,\sin^2\!\chi\sin^2\!\theta)$ then yields
\begin{equation}
\begin{aligned}
\Phi(\tau_0,\chi_0) &\= \frac{p\,q}{\ell^2}\int \diff^2\Omega_2\,\Bigl((\Fcal_{\tau\theta})^2 + (\Fcal_{\chi\theta})^2 
+ \sfrac{1}{\sin^2\!\theta}\bigl[(\Fcal_{\tau\phi})^2 + (\Fcal_{\chi\phi})^2\bigr]\Bigr) \\
&\quad +\ \frac{p^2{+}q^2}{\ell^2}\int \diff^2\Omega_2\,\Bigl( \Fcal_{\tau\theta}\,\Fcal_{\chi\theta} 
+ \sfrac{1}{\sin^2\!\theta}\Fcal_{\tau\phi}\,\Fcal_{\chi\phi}\Bigr)\ .
\end{aligned}
\end{equation}
The sphere-frame components $\Fcal_{mn}$ can be computed by expanding
$e^a=e^a_{\ m}\,\diff\xi^m$ in
\begin{equation}
\Fcal \=  \Ecal_a\,e^a \we e^\tau + \sfrac12\,\Bcal_a\,\varepsilon^a_{\ bc}\,e^b\we e^c \= \Fcal_{mn}\,\diff\xi^m\we\diff\xi^n
\quad\with\xi^n\in\{\tau,\chi,\theta,\phi\}\ .
\end{equation}
The expression for the flux in sphere-frame fields then becomes
\begin{equation}\label{flux}
\begin{aligned}
\ell^2\,\Phi &\= p\,q\,\sin^2\!\chi \int_{S^2}\diff^2\Omega_2\ \Bigl[
(\sin\phi\,\Ecal_1-\cos\phi\,\Ecal_2)^2 + (\cos\theta\cos\phi\,\Ecal_1+\cos\theta\sin\phi\,\Ecal_2 \\
&\qquad\qquad\qquad\qquad\qquad\qquad\qquad-\sin\theta\,\Ecal_3)^2 + (\sin\phi\,\Bcal_1-\cos\phi\,\Bcal_2)^2 \\
&\qquad\qquad\qquad\qquad\qquad\qquad+ (\cos\theta\cos\phi\,\Bcal_1+\cos\theta\sin\phi\,\Bcal_2-\sin\theta\,\Bcal_3)^2 \Bigr] \\
&\qquad +\ (p^2{+}q^2)\,\sin^2\!\chi \int_{S^2}\diff^2\Omega_2\ \Bigl[
(\sin\phi\,\Bcal_1-\cos\phi\,\Bcal_2) (\cos\theta\cos\phi\,\Ecal_1 \\
&\qquad\qquad\qquad\qquad\qquad\qquad\quad +\cos\theta\sin\phi\,\Ecal_2-\sin\theta\,\Ecal_3) - (\sin\phi\,\Ecal_1-\cos\phi\,\Ecal_2) \\
&\qquad\qquad\qquad\qquad\qquad\qquad\qquad\quad \cdot(\cos\theta\cos\phi\,\Bcal_1+\cos\theta\sin\phi\,\Bcal_2-\sin\theta\,\Bcal_3) \Bigr]\ .
\end{aligned}
\end{equation}

The total energy flux across future null infinity is obtained by evaluating this expression on $\mathscr{I}^+$ and integrating over it. Introducing cylinder light-cone coordinates
\begin{equation}
u=\tau{+}\chi \und v=\tau{-}\chi \qquad\textrm{so that}\qquad
t{+}r=-\ell\,\cot\sfrac{v}{2} \und t{-}r=-\ell\,\cot\sfrac{u}{2}
\end{equation}
we characterize $\mathscr{I}^+$ as
\begin{equation}
\biggl\{ \begin{array}{l} t{+}r\to\infty \\[4pt]  t{-}r \in\R \end{array} \biggr\} \quad\Leftrightarrow\quad
\biggl\{ \begin{array}{l} u\in(0,2\pi) \\[4pt]  v=0 \end{array} \biggr\} \qquad\Rightarrow\qquad
p=q=\sin^2\!\chi \und \gamma=0\ .
\end{equation}
Further noticing that
\begin{equation}
\diff(t{-}r) \= \frac{\ell\ \diff u}{p{+}q} \= \frac{\ell\ \diff u}{1-\cos u} \quad\und\quad
\sin^2\!\chi \= \sin^2\!\sfrac{u-v}{2} \= \sfrac12\bigl(1-\cos(u{-}v)\bigr)\ ,
\end{equation}
we may express this total flux as
\begin{equation}
\Phi_+ \= \int_{-\infty}^\infty \diff(t{-}r)\ \Phi\big|_{\mathscr{I}^+}
\= \int_0^{2\pi} \frac{\ell\ \diff u}{1-\cos u}\ \Phi(\sfrac{u}{2},\sfrac{u}{2})
\end{equation}
to obtain
\begin{equation} \label{totalflux}
\begin{aligned}
\Phi_+ &\=\frac{1}{8\ell} \int\!\diff u\,(1{-}\cos u)^2 \int\!\diff^2\Omega_2\ \Bigl[
\bigl\{ \cos\theta\cos\phi\,\Ecal_1+\cos\theta\sin\phi\,\Ecal_2-\sin\theta\,\Ecal_3+\sin\phi\,\Bcal_1 \\
&-\cos\phi\,\Bcal_2 \bigr\}^2 +\ \bigl\{ \cos\theta\cos\phi\,\Bcal_1+\cos\theta\sin\phi\,\Bcal_2-\sin\theta\,\Bcal_3-\sin\phi\,\Ecal_1+\cos\phi\,\Ecal_2 \bigr\}^2 \Bigr]
\end{aligned}
\end{equation}
The square bracket expression above can be further simplified for a fixed spin and type 
by employing~\eqref{EBj} along with \eqref{XZ}, \eqref{basisZ}, \eqref{type1} and~\eqref{type2} to get
\begin{equation}
\Phi_+^{(j)} = \frac{\Omega^2}{4\ell}\int\!\diff u\,(1{-}\cos u)^2 \int\!\diff^2\Omega_2\,
 \big| \pm Z_+^j e^{-\im\phi}(1\pm \cos\theta)\mp Z_-^j e^{\im\phi}(1\mp \cos\theta)-\sqrt{2}Z_3^j\sin\theta\, \big|^2\ ,
\end{equation}
where the upper (lower) sign corresponds to a type-I (type-II) solution. 
In the special case of $j=0$ $(\Omega{=}2)$ the contribution to the two-sphere integral only comes 
from the part which is independent of $(\theta,\phi)$, i.e.~$\frac{4}{3}\left(|Z^0_+|^2+|Z^0_-|^2+|Z^0_3|^2\right)$, 
so that the integration can easily be performed by passing to the adjoint harmonics 
$\tilde{Y}_{j;l,M}$ \eqref{Y-new} and using \eqref{split-Y}  to get
\begin{equation}
\Phi_+^{(0)} \= \frac{16}{3\,\ell}\int\limits_0^{2\pi}\!\diff u\ \sin^4\!\sfrac{u}{2}\,
\bigl| R_{0,0}(\sfrac{u}{2})\bigr|^2\sum\limits_{n=-1}^{1}|\lambda_{0,n}|^2 
\= \frac{8}{\ell}\sum\limits_{n=-1}^{1}|\lambda_{0,n}|^2 \= E^{(0)}\ .
\end{equation}
{\setstretch{2.0} The same equality $\Phi_+=E$ continues to hold true as we go up in spin $j$ 
(we verified it for $j{=}\sfrac12$ and $j{=}1$), 
thus validating the  energy conservation $\pa^\mu T_{\mu 0}=0$.}

\vspace{12pt}
%----------------------------------------------------------------------------------------
\section{Trajectories}
\vspace{2pt}
\noindent
Given a knotted electromagnetic field configuration, a natural question that arises is how do charged particles propagate in the background of such a field? We proceed to address this issue here by analyzing, with numerical simulations, the trajectories of several (identical) charged point particles for the family of knotted field configurations \eqref{EMCartesian} that we encountered before. We will consider type I \eqref{type1} basis field configurations (up to $j{=}1$ for simplicity) and the Hopf--Rañada field configuration \eqref{HRfields}. 

\begin{figure}[h!]
\captionsetup{width=\linewidth}
\centering
   \includegraphics[width = 5cm, height = 5cm]{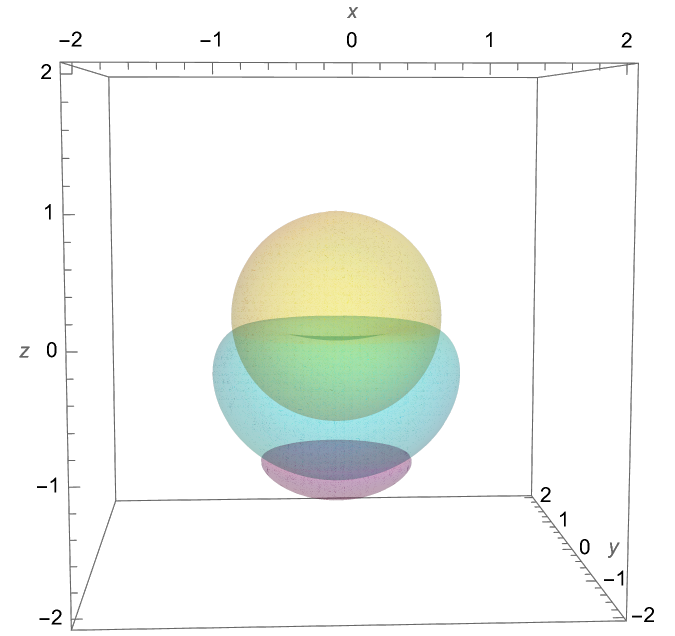}
   \includegraphics[width = 5cm, height = 5cm]{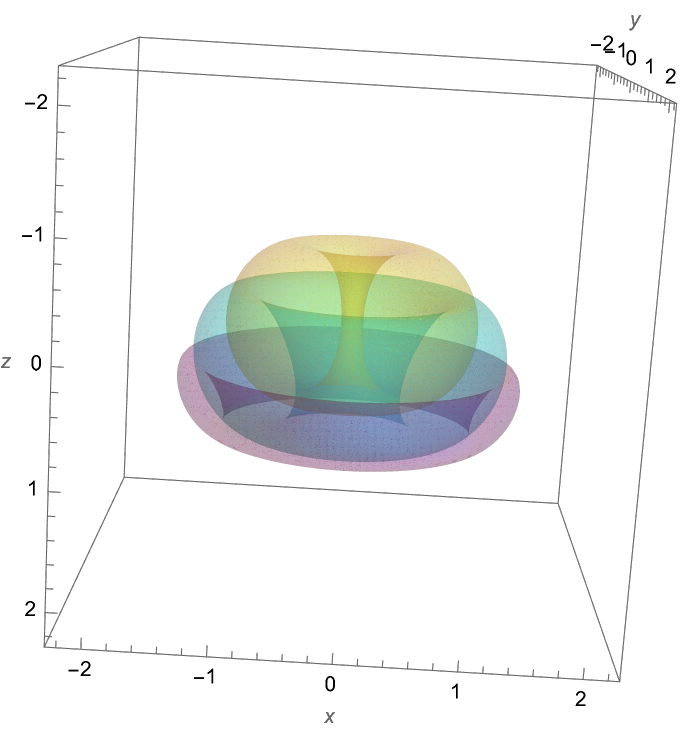}
   \includegraphics[width = 5cm, height = 5cm]{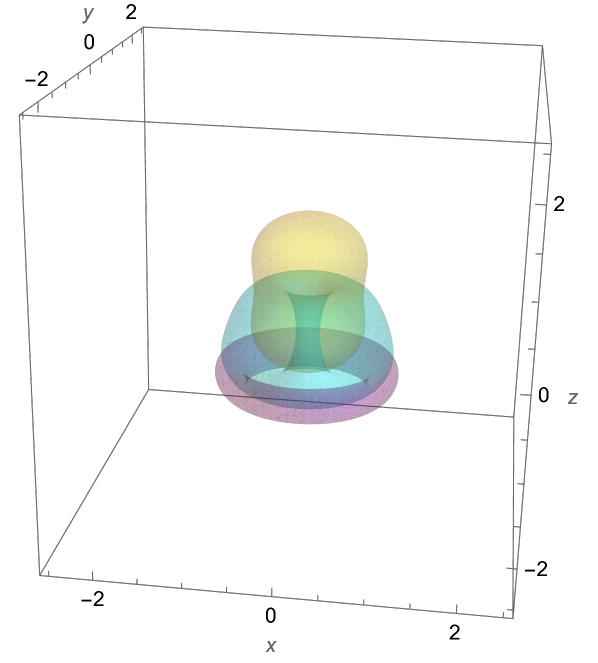}
 \caption{
 Contour plots for energy densities at $t{=}0$ (yellow), $t{=}1$ (cyan) and $t{=}1.5$ (purple) with contour value $0.9E_{max}(t{=}1.5)$.
 Left: Hopf--Ran\~{a}da configuration, Center: $(j;m,n)=(\sfrac12;-\sfrac12,-\sfrac32)_R$ configuration, Right: $(j;m,n)=(1;-1,1)_R$ configuration.}
\label{EnDen}
\end{figure}

In some of the simulations we employ the maximum of the energy density at time $t$, i.e. $E_{max}(t)$ (that occurs at several points ${\bf x}_{max}$ that are located symmetrically with respect to the origin), for different initial conditions and field configurations. In such cases we have employed a parameter $R_{max}(t)$ of ``maximal'' radius defined via
\begin{equation}\label{Rmax}
    E\left( t,{\bf x}_{max}(t) \right) \= E_{max}(t)\quad \implies \quad R_{max}(t)\ :=\ |{\bf x}_{max}(t)|\ .
\end{equation}
A characteristic feature of these basis knot electromagnetic fields is that they have a preferred $z$-axis direction due to our convention to diagonalize the $J_3$ action in \eqref{Y-action}; notice here that the SO(3) isometry subgroup, and hence its generators $J_a$, are identified on the cylinder and the Minkowski side as shown in \eqref{SO(3)subset}. This is clearly exemplified in Figure \ref{EnDen}, where the energy density $E := \sfrac12 ({\bf E}^2 + {\bf B}^2)$ decreases along the $z$-axis. As a result, the basis fields along the $z$-axis (i.e. ${\bf E}(t,x{=}0,y{=}0,z)$ and ${\bf B}(t,x{=}0,y{=}0,z)$) are either directed in the $xy$-plane or along the $z$-axis. In fact, for extreme field configurations $(j;\pm j,\pm (j{+}1))$, for any $j{>}0$, the fields along the $z$-axis vanish for all times.

The trajectories of these particles are governed by the relativistic Lorentz equation
\begin{equation}\label{eq:force1}
	\frac{\diff {\bf p}}{\diff t} = q({\bf E}_{\ell}+{\bf v} \times {\bf B}_{\ell})\ ,
\end{equation}
where $q$ is the charge of the particle, ${\bf p} = \widetilde{\gamma} m {\bf v}$ is the relativistic three-momentum, ${\bf v}$ is the usual three-velocity of the particle, $m$ is its mass, $\widetilde{\gamma} = (1-{\bf v}^{\,2})^{-1/2}$ is the Lorentz factor, and ${\bf E}_{\ell}$ and ${\bf B}_{\ell}$ are dimensionful electric and magnetic fields respectively. With the energy of the particle $E_{p} = \widetilde{\gamma} m$ and $\diff E_p/ \diff{t} = q\, {\bf v} \cdot {\bf E}$, one can rewrite (\ref{eq:force1}) in terms of the derivative of ${\bf v}$ \cite{landau2} as
\begin{equation}\label{eq:force2}
	\frac{\diff {\bf v}}{\diff t} = \frac{q}{\gamma m} \left( {\bf E}_{\ell} + {\bf v} \times {\bf B}_{\ell} - ({\bf v} \cdot {\bf E}_{\ell})\,{\bf v} \right)\ .
\end{equation}
Equations (\ref{eq:force1}) and (\ref{eq:force2}) are equivalent, and either one can be used for a simulation purpose; they only differ by the position of the nonlinearity in ${\bf v}$. In natural units $\hbar {=} c {=} \epsilon_0 {=} 1$, every dimensionful quantity can be written in terms of a length scale. We relate all dimensionful quantities to the de Sitter radius $\ell$ from equation \eqref{dS4} and work with the corresponding dimensionless ones as follows:
\begin{equation}
    T := \frac{t}{\ell}\ ,\quad {\bf X} := \frac{{\bf x}}{\ell}\ ,\quad {\bf V} := \frac{\diff X}{\diff T} \equiv {\bf v}\ ,\quad {\bf E} := \ell^2 {\bf E}_{\ell}, \und {\bf B} := \ell^2 {\bf B}_{\ell}\ .
\end{equation}
Moreover, the fields are solutions of the homogeneous (source-free) Maxwell equations, so they can be freely rescaled by any dimensionless constant factor $\lambda$. Combining the above considerations, one can rewrite (\ref{eq:force1}) (or analogously (\ref{eq:force2})) fully in terms of dimensionless quantities as
\begin{equation}
		\frac{\diff (\widetilde{\gamma}{\bf V})}{\diff T} = \kappa({\bf E} + {\bf V} \times {\bf B})\ ,
\end{equation}
where $\kappa = \sfrac{q\ell^3\lambda}{m}$ is a dimensionless parameter. One consequence of this parameter is that we can tune the values of each of the constants separately. In particular, we can make the charge as small as needed without changing $\kappa$ such that the effect of the backreaction on the trajectories becomes negligible. As for the initial conditions, we mostly work in the following two main scenarios:
\begin{enumerate}[label=(\arabic*)]
\item $N$ identical charged particles with ${\bf V}_0 {\equiv} {\bf V}(T{=}0) {=} 0$ located symmetrically (with respect to the origin), or

\item $N$ identical charged particles with $\,{\bf X}_0 {\equiv} {\bf X}(T{=}0) {=} 0\,$ with particle velocities directed radially outward in a symmetric fashion (with respect to the origin; shown in colored arrows),
\end{enumerate}
with the following 3 sub-cases for both of these conditions:
\begin{enumerate}[label=(\Alph*)]
    \item along a line,
    \item on a circle of radius $r$,
    \item on a sphere of radius $r$.
\end{enumerate}

\begin{figure}[H]
\captionsetup{width=\linewidth}
\centering
   \includegraphics[width = 7.5cm, height = 5cm]{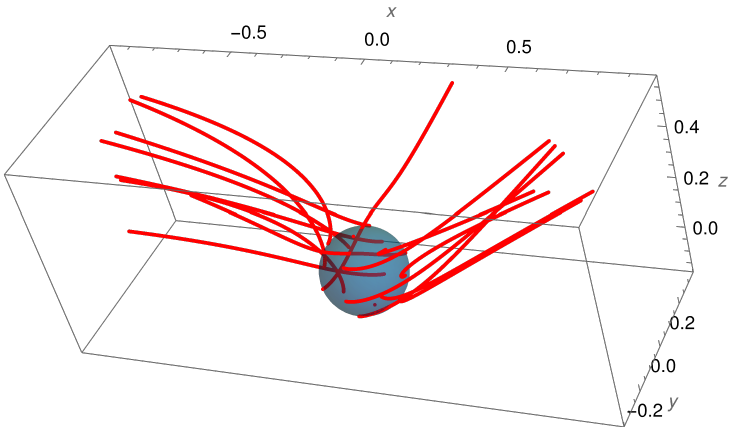}
   \includegraphics[width = 7.5cm, height = 5cm]{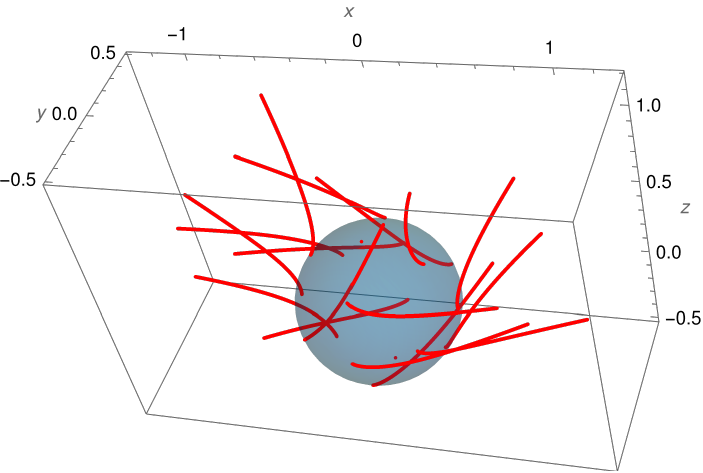}
 \caption{Simulation of $N{=}18$ particles in scenario (1C) for $(\sfrac12;-\sfrac12,-\sfrac32)_R$ with $\kappa{=}10$ and for $t\in[0,1]$. Left: $r{=}R_{max}(0){\approx}0.447$. Right: $r{=}R_{max}(0)/3$.}
\label{TrajRmaxneq0}
\end{figure}

We vary several parameters including the initial conditions with different directions of lines and planes for each configuration, the value of $\kappa$, and the simulation time in order to study the behavior of the trajectories. In several field configurations studied below, we find that $R_{max}(0)=0$, so we use a small radius $r$ for the initial condition of kind (1) to be able to probe the particles around a region of maximum energy of the field. In this scenario, the effect of the field on the trajectories of the particles is more prominent, as expected, and this helps us understand small perturbations of the trajectories as compared to a particle starting at rest from the origin. The effect of the fields on particles starting near the maximum of the energy density is also more prominent for $R_{max}(0){\neq}0$, as illustrated in Figure \ref{TrajRmaxneq0}. Moreover, for the initial condition of kind (2) we use the particle initial speeds in the range where it is (i) non-relativistic, (ii) relativistic (usually between $0.1$ and $0.9$), and (iii) ultrarelativistic (here, $0.99$ or higher).

\begin{figure}[H]
\captionsetup{width=\linewidth}
\centering
   \includegraphics[width = 5cm, height = 5cm]{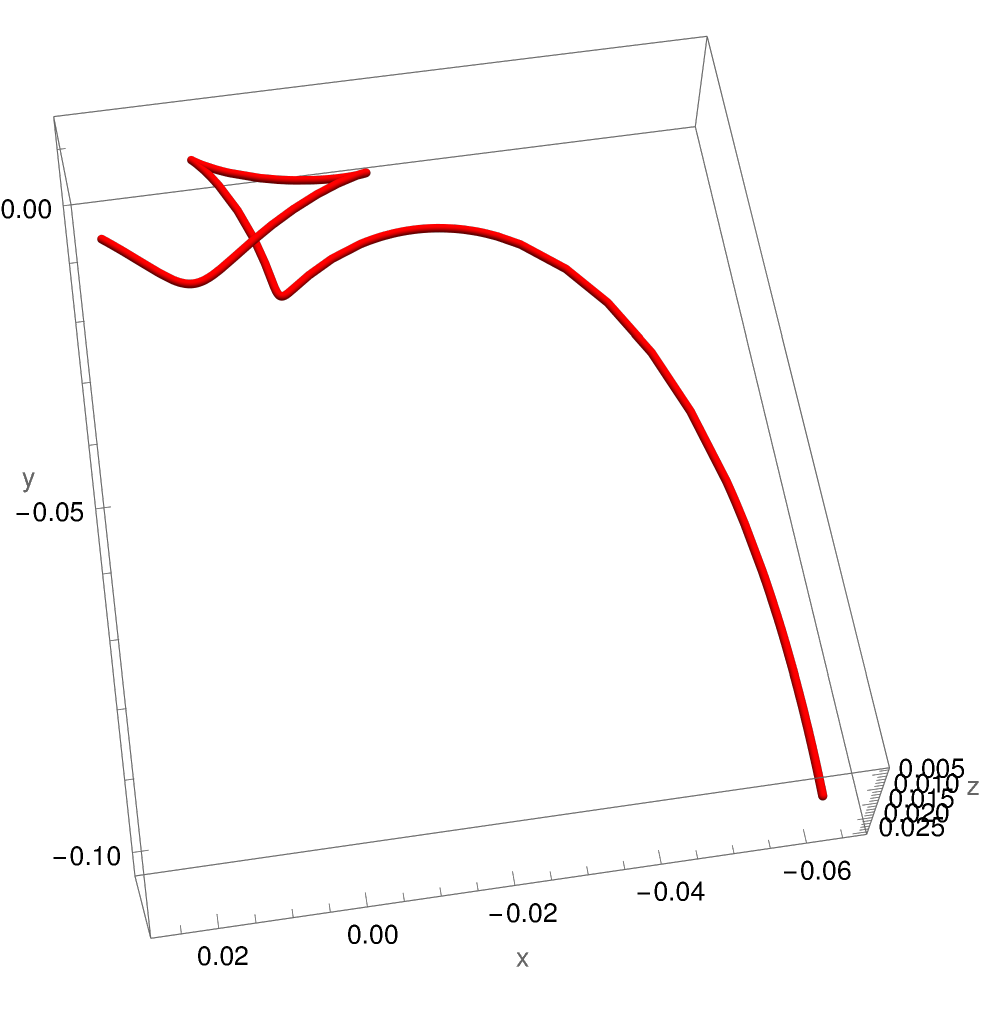}
   \includegraphics[width = 5cm, height = 5cm]{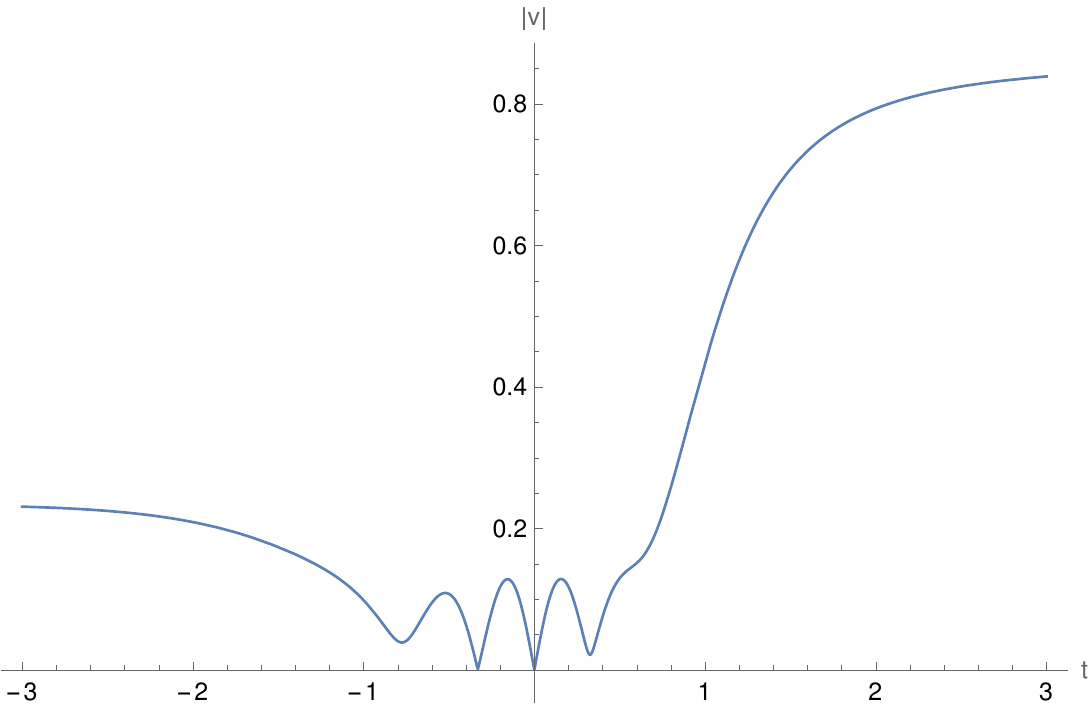}
   \includegraphics[width = 5cm, height = 5cm]{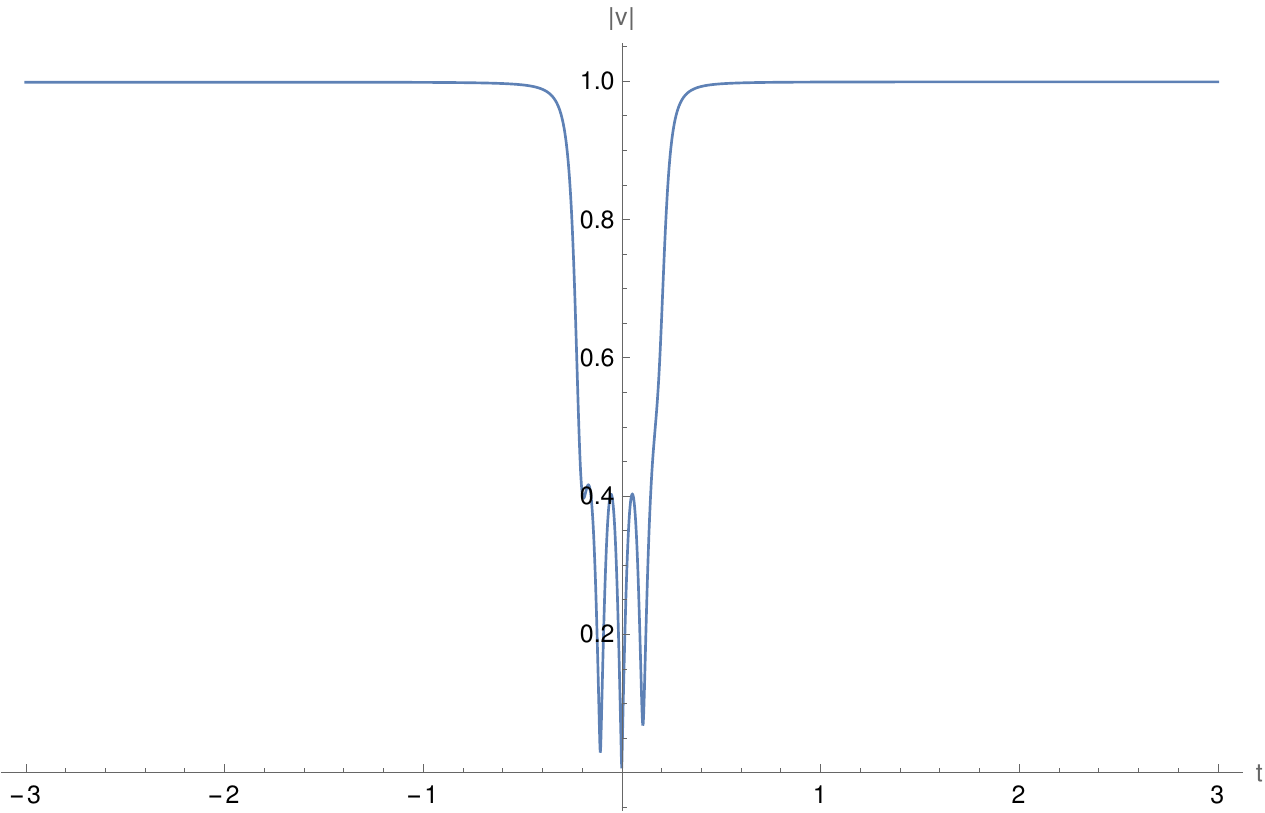}
 \caption{
 Trajectory of a charged particle for $(\sfrac12;-\sfrac12,-\sfrac32)_R$ configuration with initial conditions ${\bf X}_0 {=} (0.01,0.01,0.01)$ and ${\bf V}_0 {=} 0$ simulated for $t \in [-1,1]$.
 Left: Particle trajectory. Center: absolute velocity profile for $
\kappa {=} 10$. Right: absolute velocity profile for $\kappa {=} 100$.}
\label{singleTrajectory}
\end{figure}

\begin{figure}[H]
\captionsetup{width=\linewidth}
\centering
   \includegraphics[width = 7cm, height = 5cm]{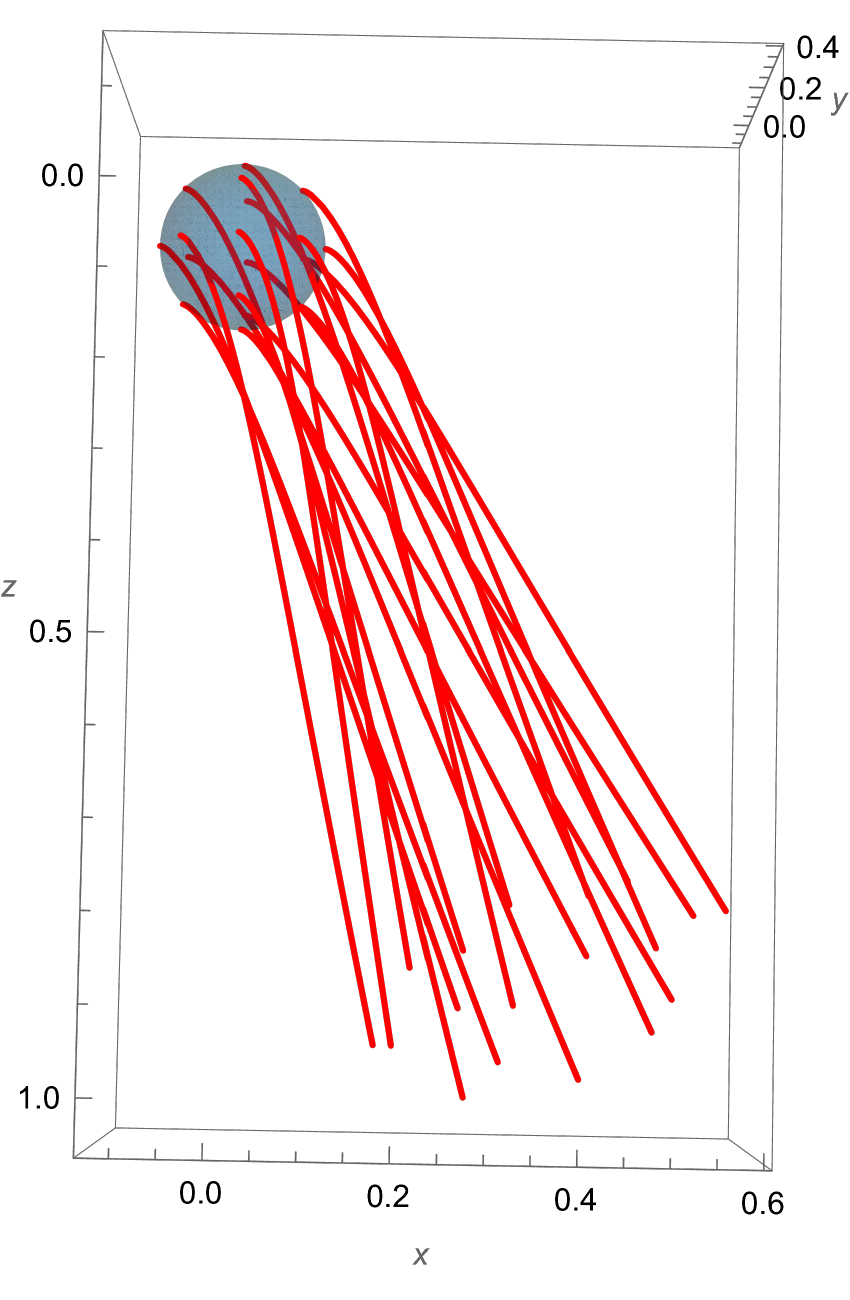}\hspace{1cm}
   \includegraphics[width = 7cm, height = 5cm]{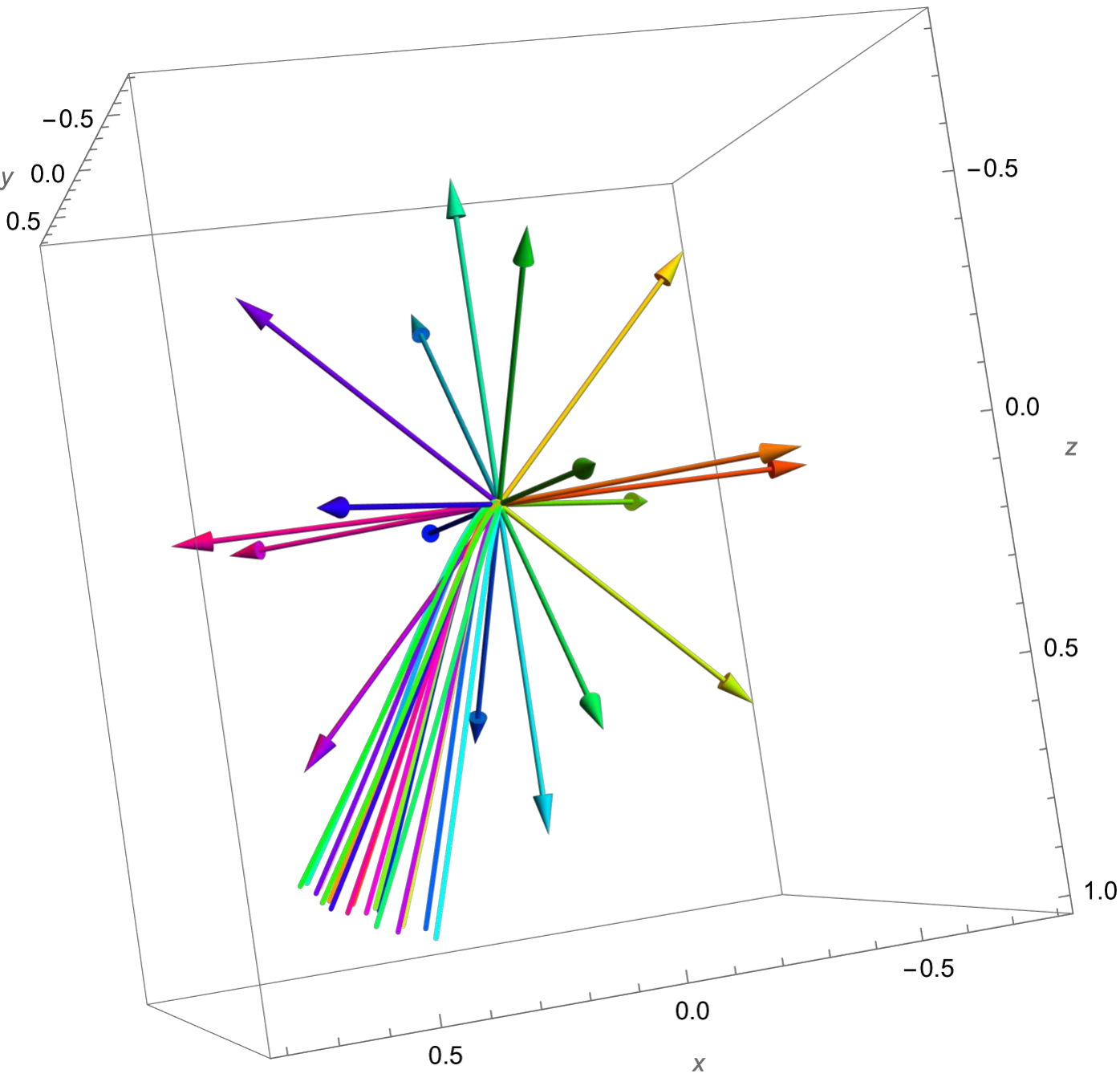}
 \caption{
 Simulation of $N{=}18$ particles for $(0,0,-1)_I$ configuration, with $\kappa {=} 100$ and $t\in[0,1]$. Left: scenario (1C) with $r{=}0.1$. Right: scenario (2C) with $r{=}0.75$.}
\label{j001plots}
\end{figure}

\begin{figure}[H]
\captionsetup{width=\linewidth}
\centering
   \includegraphics[width = 7cm, height = 5cm]{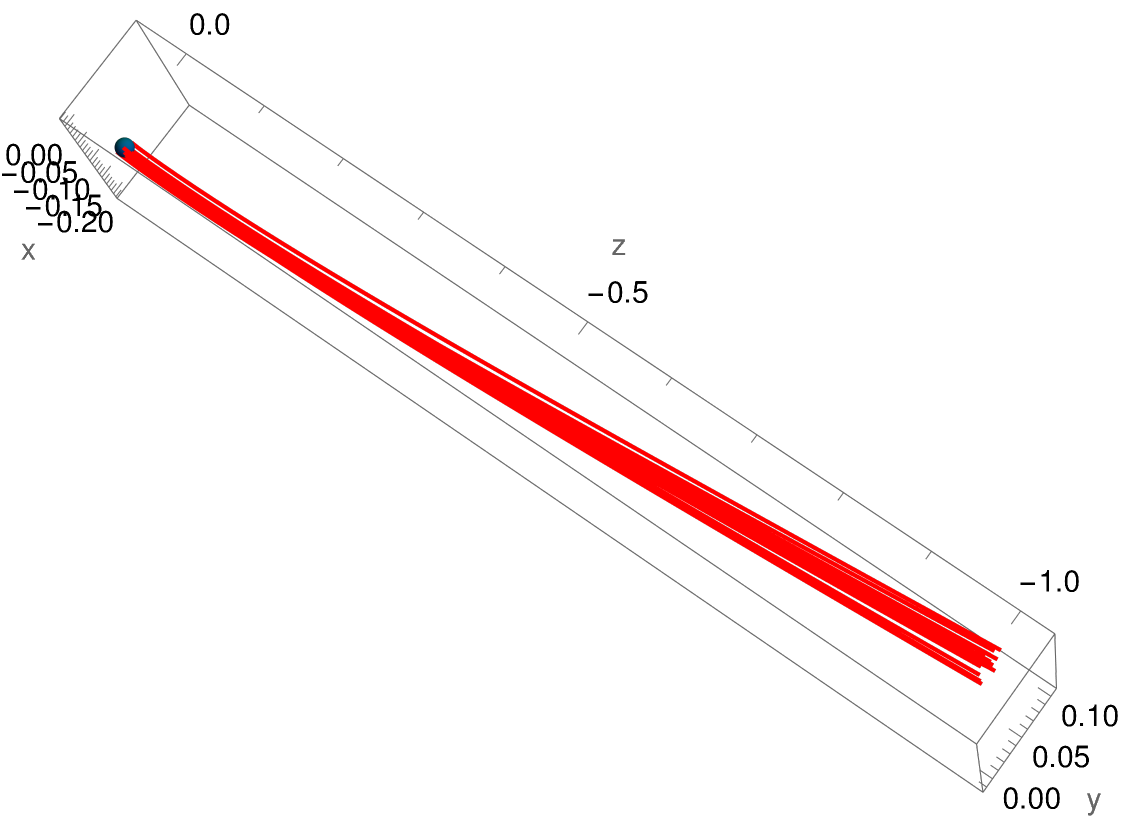}\hspace{1cm}
   \includegraphics[width = 7cm, height = 5cm]{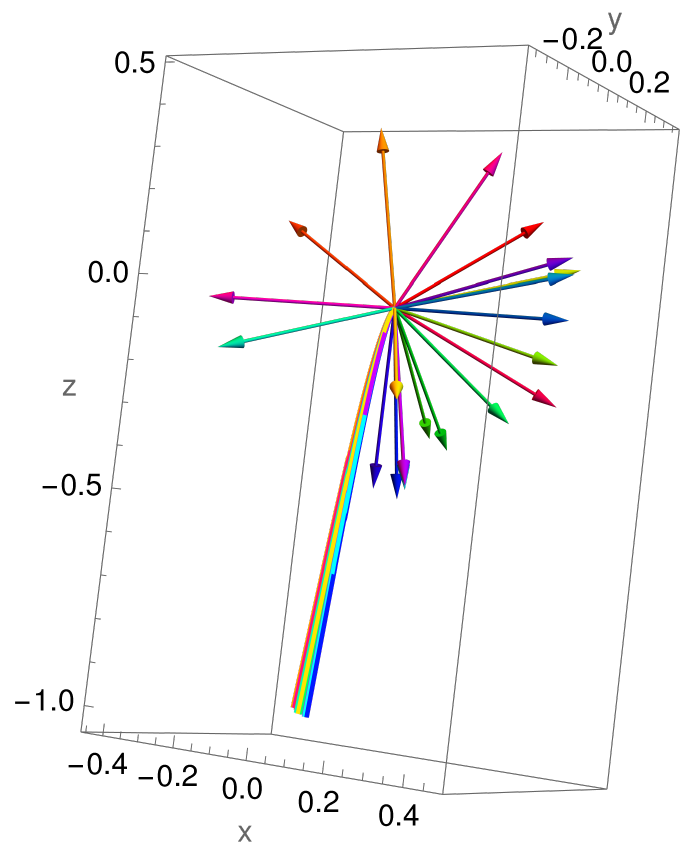}
 \caption{
    Simulation of $N{=}20$ particles for $(0,0,1)_I$ configuration, with $\kappa {=} 1000$ and $t\in[0,1]$. Left: particles starting from rest and located randomly inside a solid ball of radius $r{=}0.01$ ($R_{max}=0$). Right: particles located at origin and directed randomly (shown with colored arrows) with $|{\bf V}_0|{=}0.45$.}
\label{HRplots}
\end{figure}

\begin{figure}[H]
\captionsetup{width=\linewidth}
\centering
   \includegraphics[width = 7cm, height = 5cm]{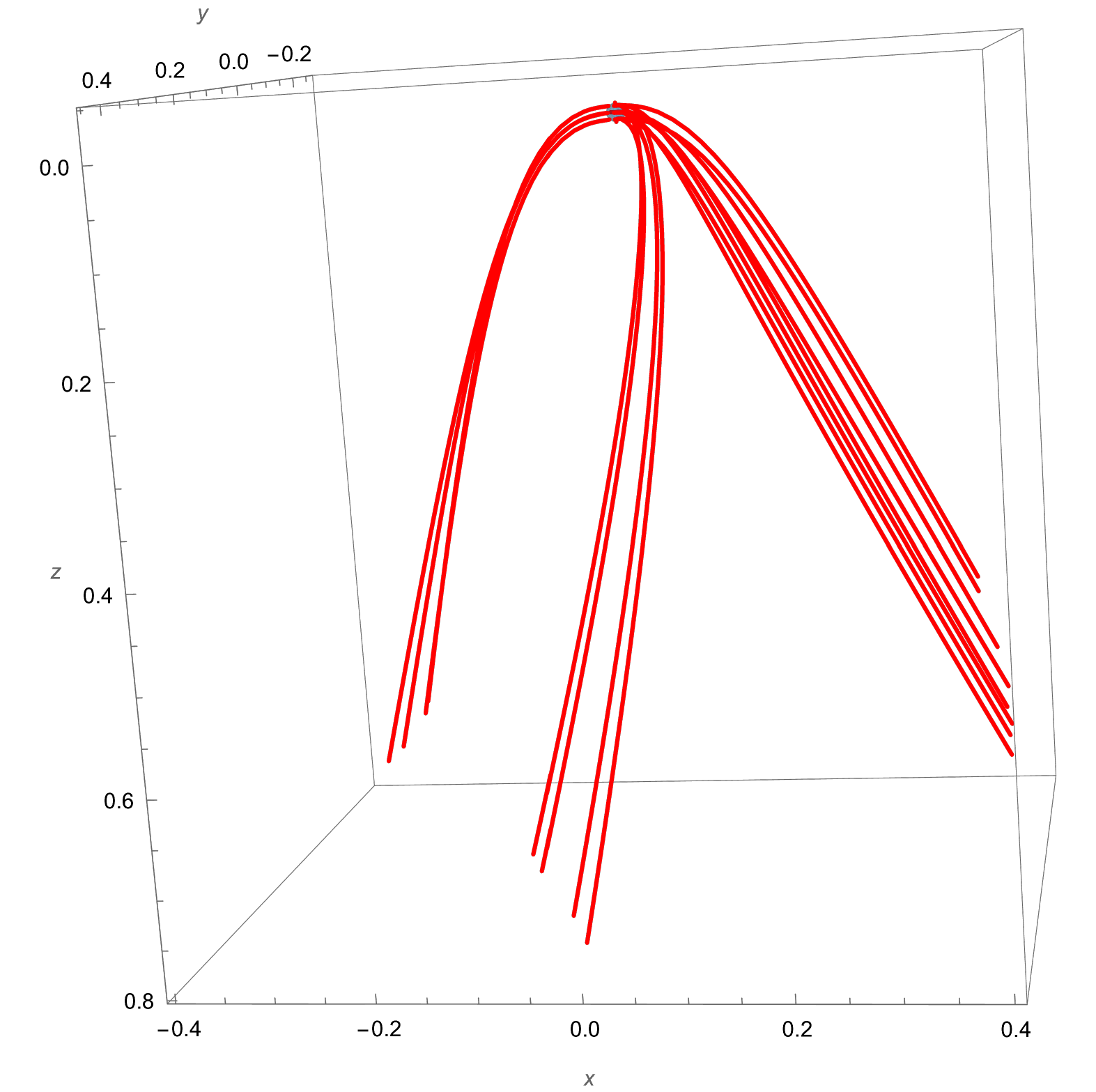}\hspace{1cm}
   \includegraphics[width = 7cm, height = 5cm]{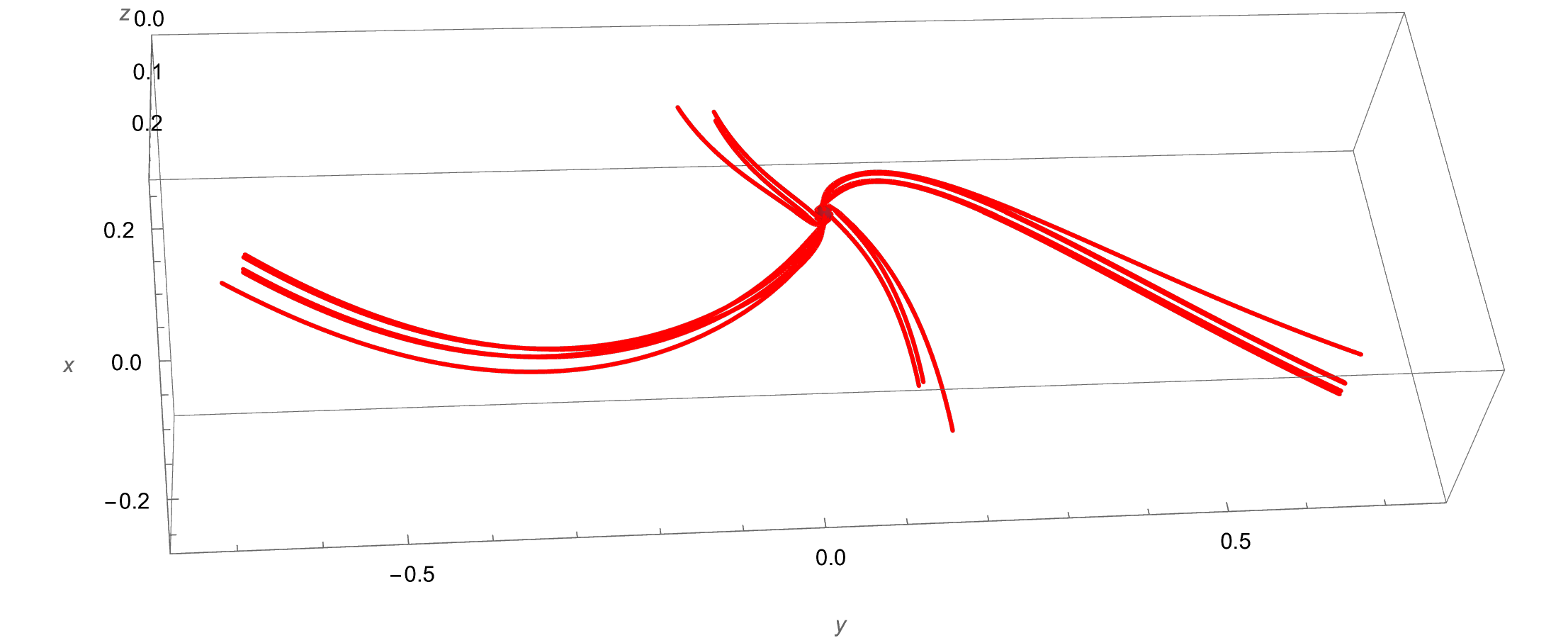}
 \caption{
 Simulation of $N{=}18$ particles in scenario (1C) with $r{=}0.01$ and for $t\in[0,3]$. Left: $(1,-1,-2)_R$ configuration with $\kappa {=} 500$. Right: $(1,-1,-1)_R$ configuration with $\kappa {=} 10$.}
\label{3-4furcation}
\end{figure}

We observe a variety of different behaviors for these trajectories, some of which we summarize below with the aid of figures. Firstly, it is worth noticing that, even with all fields decreasing as powers of both space and time coordinates, in most field configurations we observe particles getting accelerated from rest up to ultrarelativistic speeds. The limit of these ultrarelativistic speeds for higher times depend on the magnitude of the fields (see, for example, Figure \ref{singleTrajectory}).
With fixed initial conditions (of kind (1) or (2)) and for higher values of $\kappa$ one can expect, in general, that the initial conditions may become increasingly less relevant. For some fields configurations we indeed found that, with increasing $\kappa$, the particles get more focused and accumulate like a beam of charged particles along some specific region of space and move asymptotically for higher simulation times. This is exemplified below with two $j{=}0$ configurations: the $(0,0,-1)_I$ configuration in Figure \ref{j001plots}, and the HR configuration in Figure \ref{HRplots}. We have verified this feature not just with symmetric initial conditions of particles like that with initial conditions (1) and (2) (as in Figure \ref{j001plots}), but also in several initial conditions asymmetric with respect to the origin, like particles located randomly inside a sphere of fixed radius about the origin with zero initial velocity, and particles located at the origin but with different magnitudes of velocities. Figure \ref{HRplots} is an illustrative example for both of these latter scenarios of asymmetric initial conditions.
\vspace{-10pt}
This is not always the case though. For some $j{=}\tfrac12$ and $j{=}1$ configurations, and with initial particle positions in a sphere of very small radius about the origin, we are able to observe the splitting of particle trajectories (starting in some specific solid angle regions around the origin) into two, three or even four such asymptotic beams that converge along some particular regions of space (depending on the initial location of these particles in one of these solid angle regions). Trajectories generated by two such $j{=}1$ configurations have been illustrated in Figure \ref{3-4furcation}. 

\begin{figure}[H]
\captionsetup{width=\linewidth}
\centering
   \includegraphics[width = 7cm, height = 5cm]{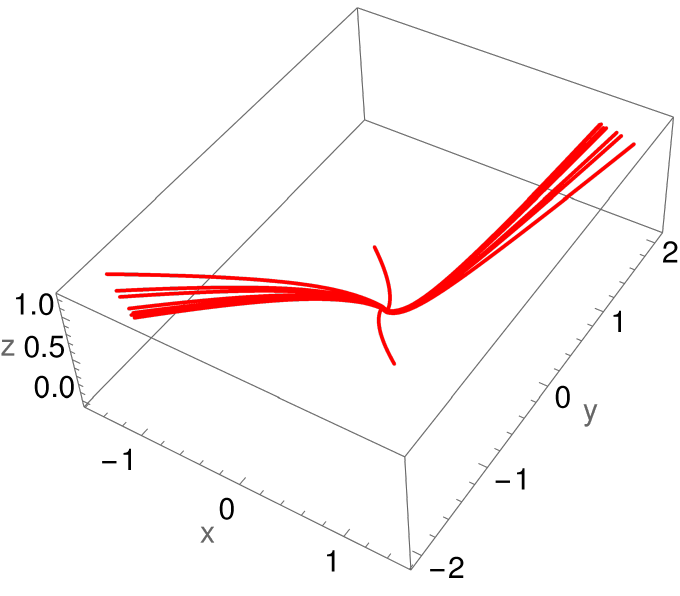}\hspace{1cm}
   \includegraphics[width = 7cm, height = 5cm]{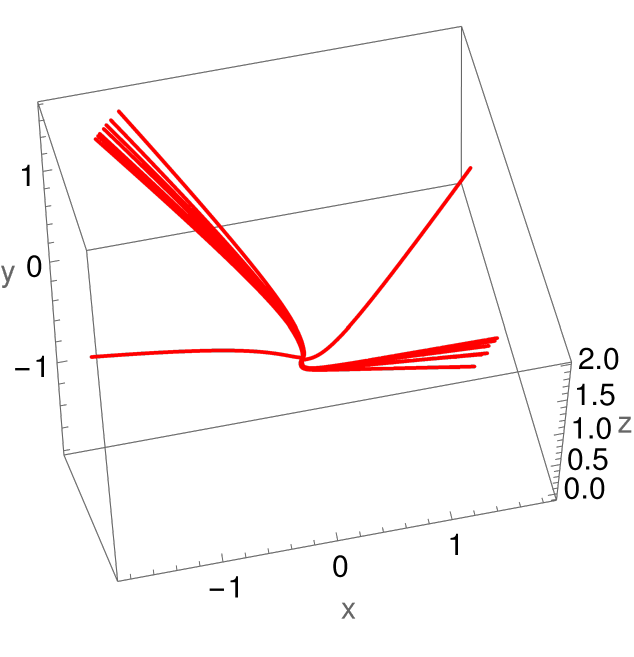}
 \caption{
 Simulation of $N{=}18$ particles in scenario (1C) with $\kappa{=}10$, $r{=}0.01$ and for $t\in[0,3]$. Left: $(\sfrac12;-\sfrac12,-\sfrac32)_R$ configuration. Right: $(1,0,-2)_I$ configuration.}
 \label{Bifurcation}
\end{figure}

Naturally, there are also regions of unstable trajectories for particles starting between these solid angle regions (see Figure \ref{Bifurcation}), which generally include the preferred $z$-axis, since in some cases trajectories that start at rest in the $z$-axis never leave it.

We employ the parameter $R_{max}$ \eqref{Rmax} in the the following Figures \ref{1Dpos}, \ref{1Dvel}, \ref{2Dvel}, \ref{3Dvel}, \ref{2Dpos}, and \ref{3Dpos} for both kinds of initial conditions viz.~(1) and (2) (it is especially relevant for the former) to understand the effect of field intensity on particle trajectories.
\begin{figure}[H]
\captionsetup{width=\linewidth}
\centering
   \includegraphics[width = 7cm, height = 5cm]{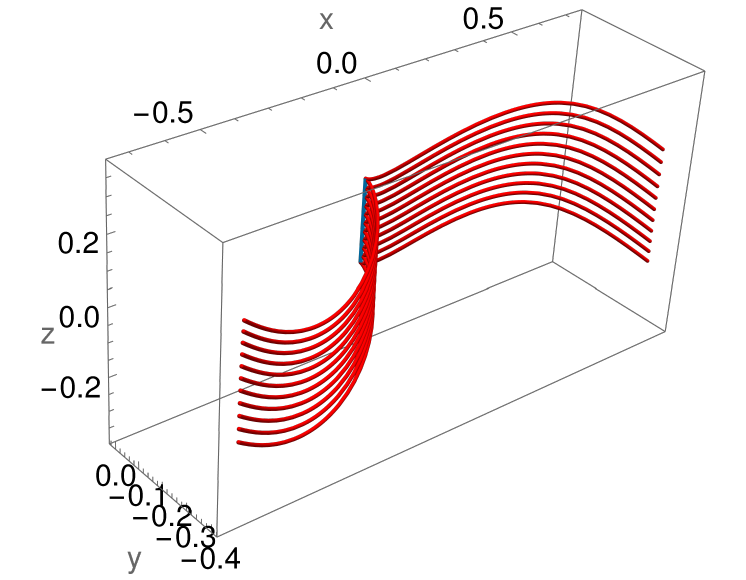}\hspace{1cm}
   \includegraphics[width = 7cm, height = 5cm]{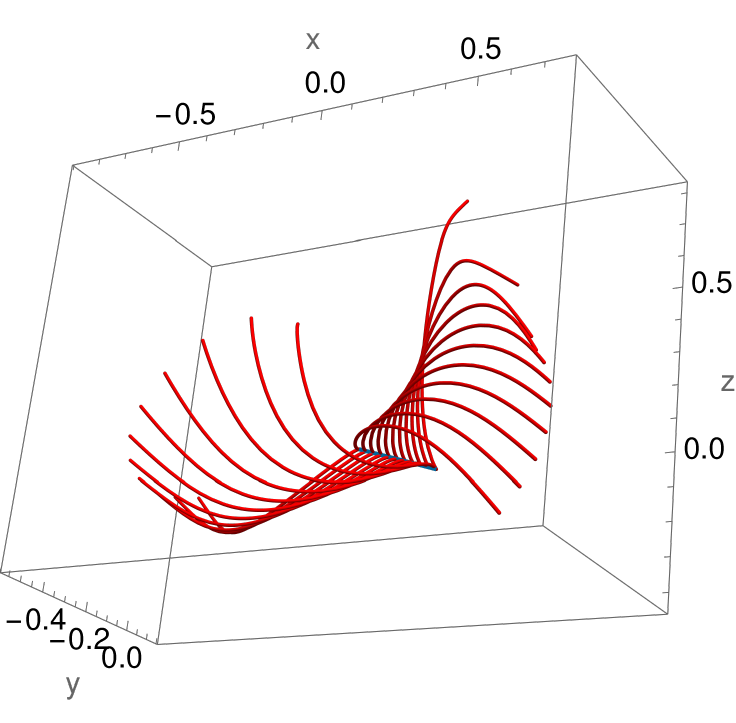}
 \caption{
Simulation of $N{=}11$ particles in scenario (1A) with $|{\bf X}_0| \propto 0.025$ (including one at the origin), for $(\sfrac12;\sfrac12,\sfrac12)_I$ configuration ($R_{max}=0$) with $\kappa{=}10$ and $t\in[-1,1]$. Left: particles initially located along $z$-axis (blue line). Right: particles initially located along some (blue) line in $xy$-plane.}
\label{1Dpos}
\end{figure}

One very interesting feature of trajectories for some of these field configurations is that they twist and turn in a coherent fashion owing to the symmetry of the background field. For particles with initial condition of kind (2), we see that their trajectories take sharp turns, up to two times, with mild twists before going off asymptotically. This has to do with the presence of strong background electromagnetic fields with knotted field lines. This is clearly demonstrated below in Figures \ref{1Dvel}, \ref{2Dvel}, and \ref{3Dvel}. It is worthwhile to notice in Figure \ref{1Dvel} that the particle which was initially at rest moves unperturbed along the $z$-axis; again, this has to do with the fact that these fields have preferred $z$-direction. This feature is even more pronounced in Figure \ref{2Dvel} and (the right subfigure of) Figure \ref{3Dvel} where we see that particles with ultrarelativistic initial speeds are forced to turn (almost vertically upwards) due to the strong electromagnetic field. These particles later take very interesting twists in a coherent manner. This twisting feature is much more refined for the case where initial particle velocities were directed along the $xy$-plane. Here also, we can safely attribute this behavior of the particle trajectories to the special field configurations, with preferred $z$-direction, that we are working with.

\begin{figure}[H]
\captionsetup{width=\linewidth}
\centering
   \includegraphics[width = 7cm, height = 5cm]{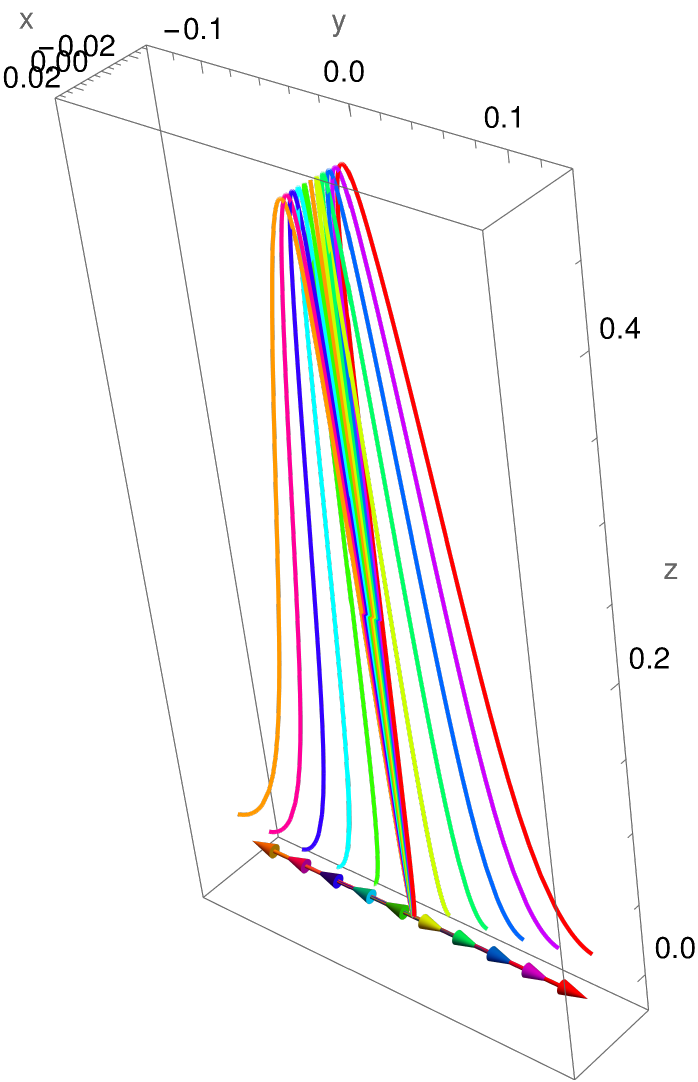}\hspace{1cm}
   \includegraphics[width = 7cm, height = 5cm]{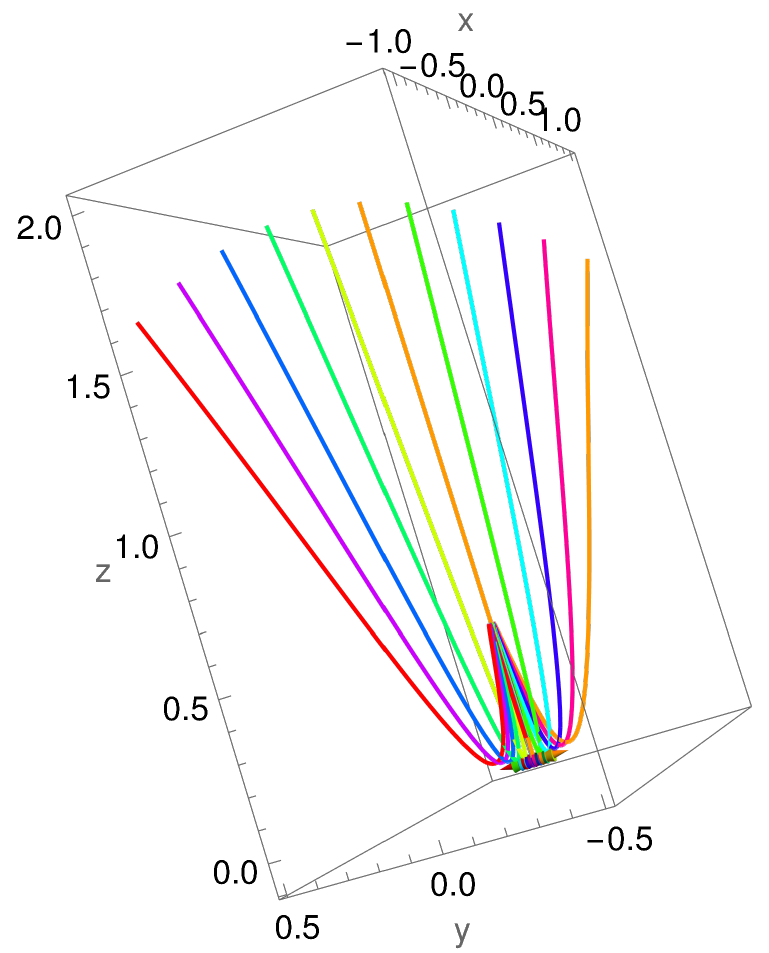}
 \caption{
 Simulation of $N{=}11$ particles in scenario (2A) with $|{\bf V}_0| \propto 0.025$ in the direction of $(0,1,0)$ (including one at rest), for $(1,0,0)_I$ configuration ($R_{max}=0$), with $\kappa{=}10$. Left: $t\in[0,1]$. Right: $t\in[0,3]$.}
\label{1Dvel}
\end{figure}
\begin{figure}[H]
\captionsetup{width=\linewidth}
\centering
   \includegraphics[width = 7cm, height = 5cm]{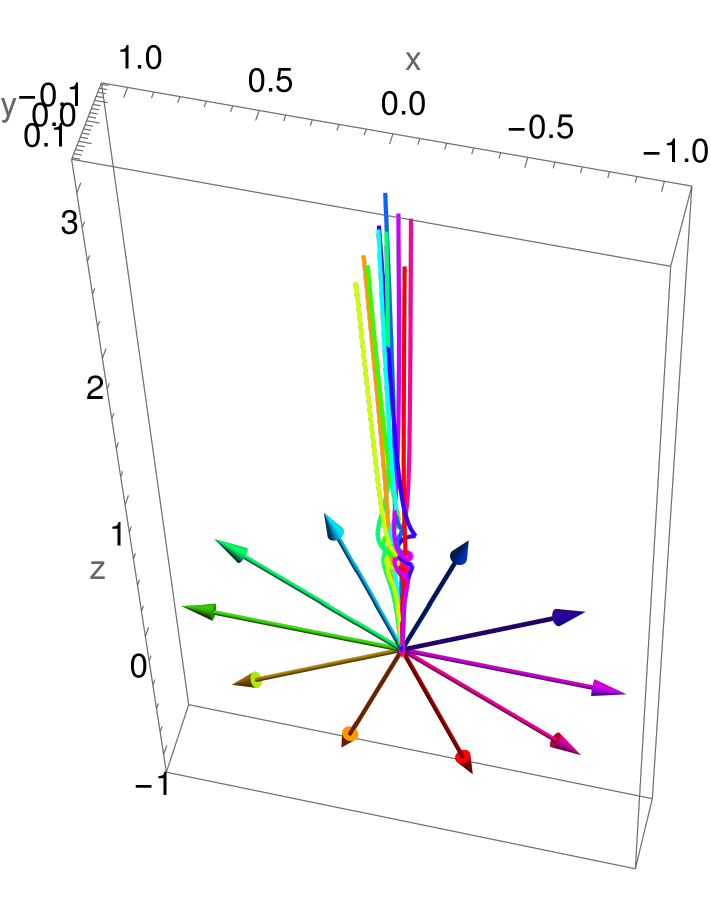}\hspace{1cm}
   \includegraphics[width = 7cm, height = 5cm]{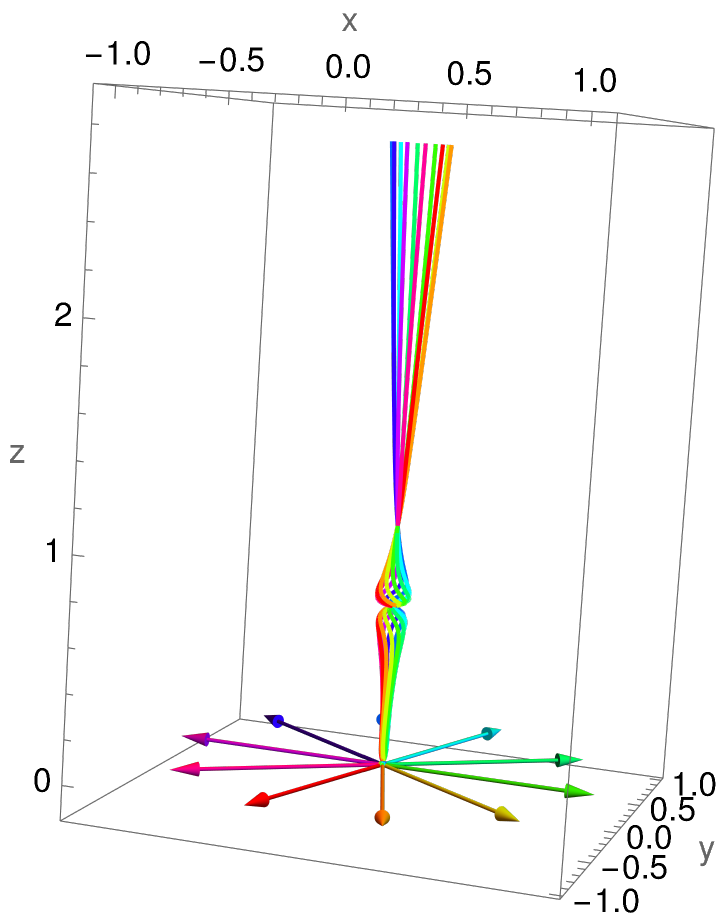}
 \caption{
 Simulation of $N{=}10$ particles in scenario (2B) with $r{=}0.99$ for $(1,-1,1)_R$ configuration with $\kappa{=}100$ and $t \in [0,3]$.
 Left: normal direction is $y$-axis. Right: normal direction is $z$-axis.}
\label{2Dvel}
\end{figure}
\begin{figure}[H]
\captionsetup{width=\linewidth}
\centering
   \includegraphics[width = 5cm, height = 5cm]{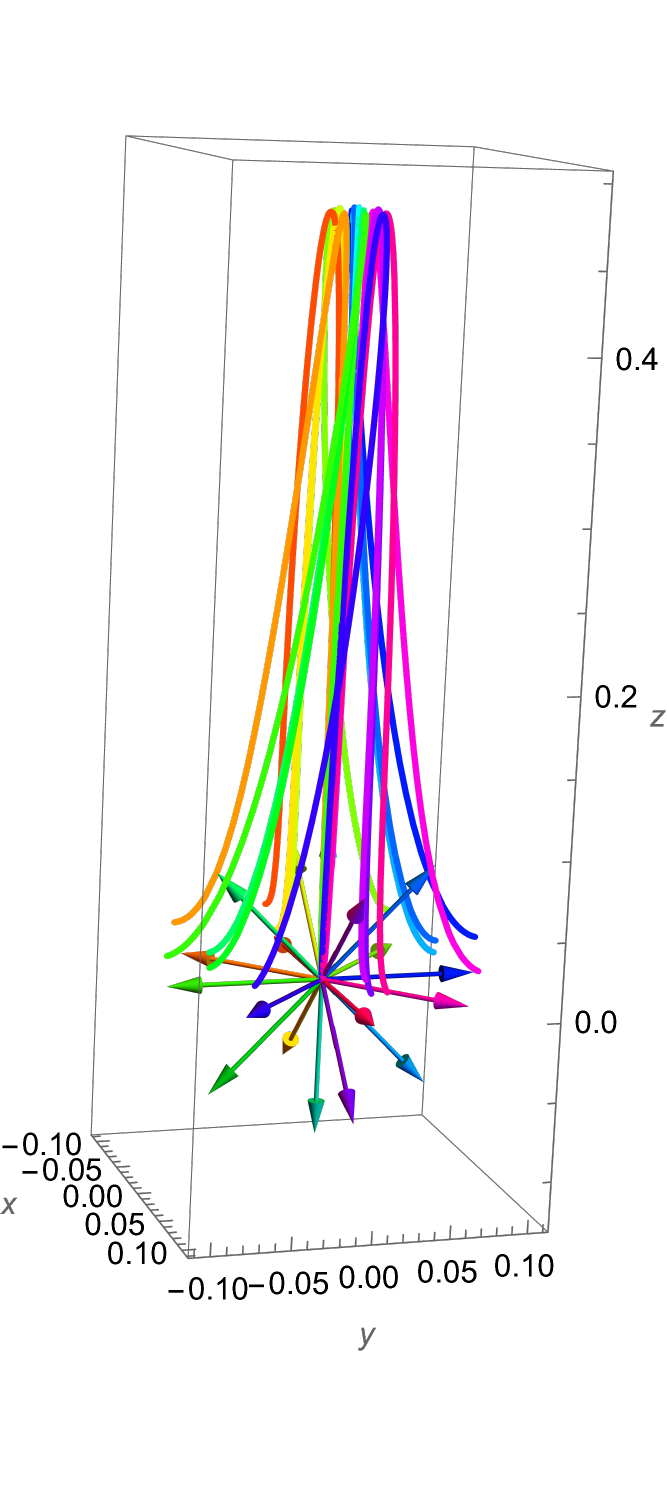}\hspace{2cm}
   \includegraphics[width = 7cm, height = 5cm]{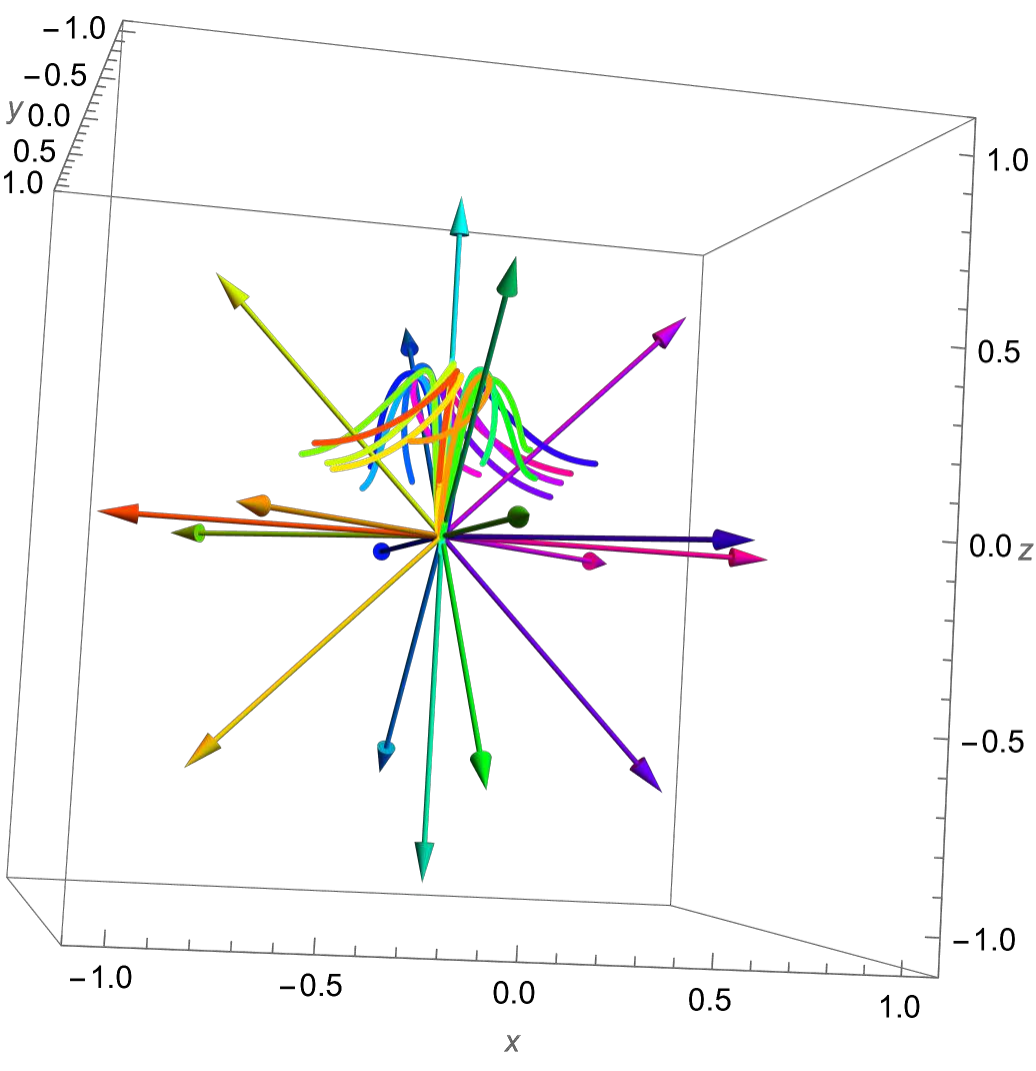}
 \caption{
Simulation of $N{=}18$ particles in scenario (2C) for $(1,0,0)_I$ configuration and $t \in [0,2]$. Left: $r{=}0.1$ and $\kappa{=}10$. Right: $r{=}0.99$ and $\kappa{=}100$.}
\label{3Dvel}
\end{figure}

We see in Figure \ref{2Dpos} that the trajectories of particles that were initially located on a circle whose normal is along the $z$-axis flow quite smoothly with mild twists for some time before they all turn symmetrically in a coherent way and go off asymptotically. Comparing this with the other case in Figure \ref{2Dpos}, where particles split into two asymptotic beams, we realize that this is yet another instance of the preferred choice of direction for the electromagnetic fields influencing the trajectories of particles.
\begin{figure}[H]
\captionsetup{width=\linewidth}
\centering
   \includegraphics[width = 7cm, height = 5cm]{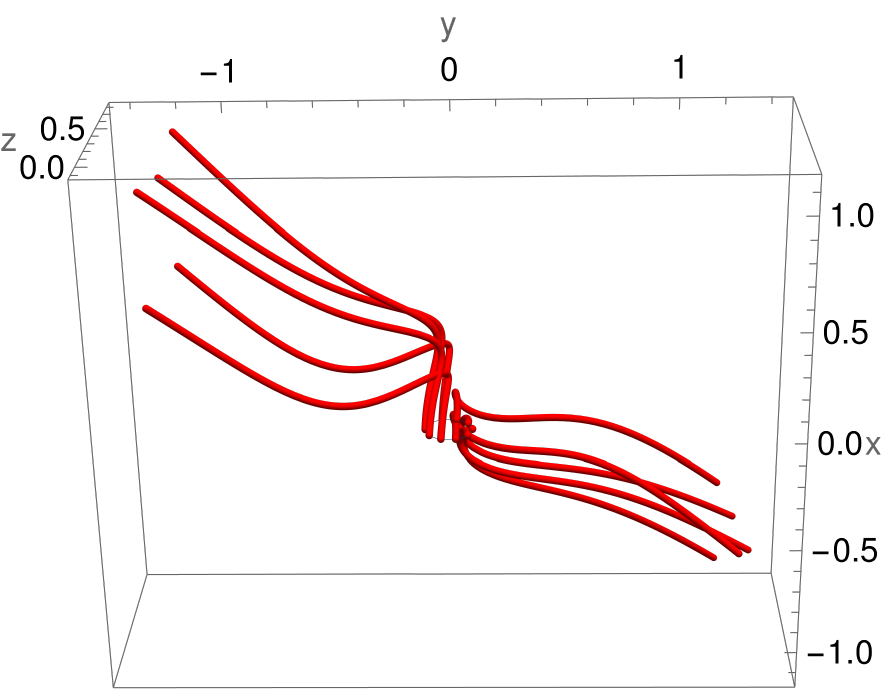}\hspace{1cm}
   \includegraphics[width = 7cm, height = 5cm]{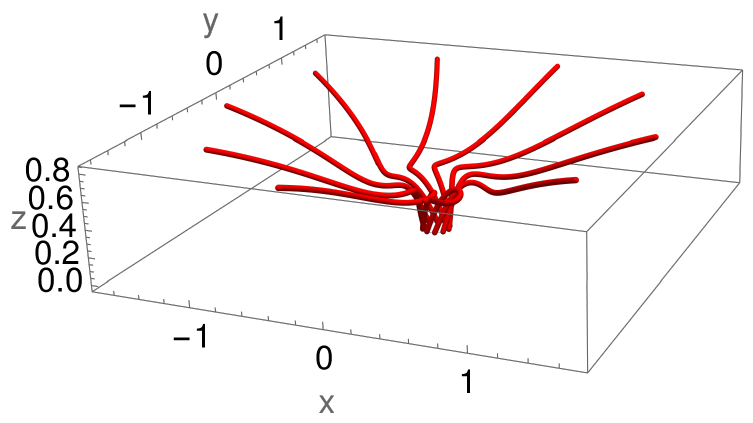}
 \caption{
 Simulation of $N{=}10$ particles in scenario (1B) with $r{=}0.1$ for $(1,0,0)_I$ configuration ($R_{max}=0$) with $\kappa {=} 10$ and $t \in [0,2]$.
 Left: normal direction is $x$-axis. Right: normal direction is $z$-axis.}
\label{2Dpos}
\end{figure}
\vspace{-10pt}
In Figures \ref{3Dpos} and \ref{2Dvel} we find examples of kind (1) and (2) respectively where both twisting as well as turning of trajectories is prominant. We see in Figure \ref{3Dpos} that the particles that start very close to the origin take a longer time to show twists as compared to the ones that start off on a sphere of radius $R_{max}$. This is due to the fact that the field is maximal at $R_{max}$ and hence its effect on particles is prominent, as discussed before. We also notice here that the particles sitting along the $z$-axis at $T{=}0$ (either on the north pole or on the south pole of this sphere) keep moving along the $z$-axis without any twists or turns. This exemplifies again the fact that these background fields have a preferred direction.
\begin{figure}[H]
\captionsetup{width=\linewidth}
\centering
   \includegraphics[width = 3cm, height = 7cm]{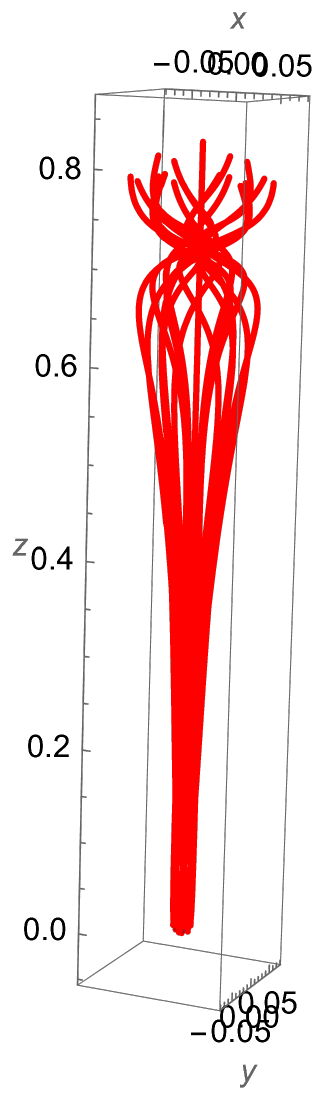}\hspace{2cm}
   \includegraphics[width = 8cm, height = 7cm]{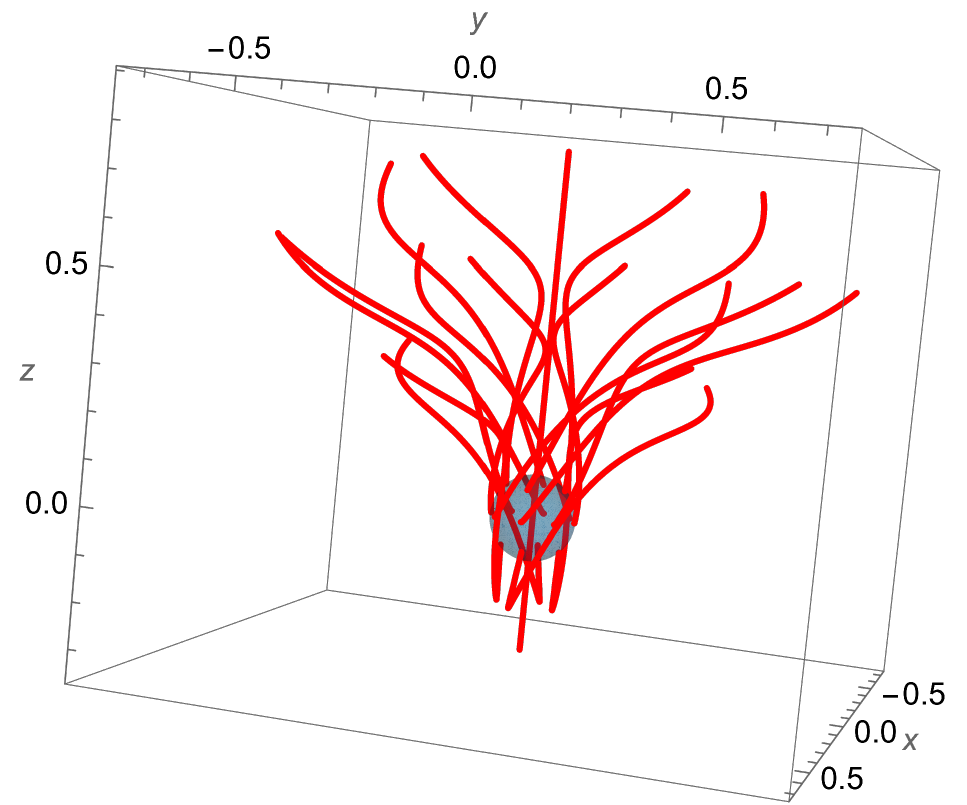}
 \caption{
 Simulation of $N{=}18$ particles in scenario (1C) for $(1,-1,1)_R$ configuration with $t \in [0,1]$. Left: $r{=}0.01$ and $\kappa{=}10$. Right: $r{=}0.1$ and $\kappa{=}100$.}
\label{3Dpos}
\end{figure}
\vspace{-10pt}
For higher-spin configurations the maximum of the energy density increases but it gets localized into an increasing number of lobes centered around the origin, due to the presence of higher-spin harmonics. Thus, only particles located very close to the tip of these lobes of maximum energy density get accelerated to ultrarelativistic speeds, while particles located outside (which effectively means most of the space) remain unaffected.

%% file: Chapters/Chapter6.tex
%auto-ignore
\chapter{Non-Abelian solutions: $SU(2)$}
\label{Chapter6} 
\justifying

Here we present cosmic Yang--Mills solution with $SU(2)$ gauge group and study their stability behaviour. The contents of this chapter, in large parts, are taken from the published work \cite{KLP21}. All the graphics in this chapter is due to Gabriel Pican\c{c}o Costa. I have been thoroughly involved at all stages of this research project.

\vspace{12pt}
%----------------------------------------------------------------------------------------
\section{$SO(4)$-symmetric cosmic Yang--Mills solution}
\vspace{2pt}
The Yang--Mills action on this ansatz simplifies to
\begin{equation}\label{YMaction}
   S \= \frac{-1}{4g^2}\int_{\Ical\times S^3} \!\!\! \tr\ \Fcal\wedge *\Fcal \=
   \frac{6\pi^2}{g^2} \int_\Ical \!\diff\tau\ \bigl[ \sfrac12\dot\psi^2 - V(\psi) \bigr]
   \with V(\psi)\=\sfrac12(\psi^2{-}1)^2\ ,
\end{equation}
where $g$ here denotes the gauge coupling.
Due to the principle of symmetric criticality~\cite{Palais}, solutions to the mechanical problem 
\begin{equation} \label{Newton}
   \ddot{\psi} + V'(\psi) \= 0
\end{equation}
will, via (\ref{Aansatz}), provide Yang--Mills configurations which extremize the action.
Conservation of energy implies that
\begin{equation} \label{energylaw}
   \sfrac12\dot{\psi}^2 +V(\psi) \= E \= \textrm{constant}\ ,
\end{equation}
and the generic solution in the double-well potential~$V$ is periodic in~$\tau$ with a period~$T(E)$.

Hence, fixing a value for~$E$ and employing time translation invariance to set $\dot{\psi}(0)=0$
uniquely determines the classical solution~$\psi(\tau)$ up to half-period shifts.
Its explicit form is
\begin{equation} \label{backgrounds}
   \psi(\tau) \= 
 \begin{cases}
   \ \sfrac{k}{\epsilon}\,\mathrm{cn}\bigl(\sfrac{\tau}{\epsilon},k\bigr)  
   \qquad\qquad\ \ \,\textrm{with}\quad T=4\,\epsilon\,K(k) 
   & \textrm{for}\quad \sfrac12<E<\infty \\[4pt]
   \ 0 \qquad\qquad\qquad\qquad\quad\! \textrm{with}\quad T=\infty 
   & \textrm{for}\quad E=\sfrac12 \\[4pt]
   \ \pm\sqrt{2}\,\mathrm{sech}\bigl(\sqrt{2}\,\tau\bigr) 
   \qquad\ \textrm{with}\quad T=\infty 
   & \textrm{for}\quad E=\sfrac12 \\[4pt]
   \ \pm\sfrac{k}{\epsilon}\,\mathrm{dn}\bigl(\sfrac{k\,\tau}{\epsilon},\sfrac1k\bigr)  
   \qquad\quad\ \textrm{with}\quad T=2\,\sfrac{\epsilon}{k}\,K(\sfrac1k) \quad
   & \textrm{for}\quad 0<E< \sfrac12 \\[4pt]
   \ \pm 1 \qquad\qquad\qquad\qquad \textrm{with}\quad T=\pi
   &  \textrm{for}\quad E=0 
 \end{cases}\ ,
\end{equation}
where cn and dn denote Jacobi elliptic functions, $K$ is the complete elliptic integral of the first kind, and
\begin{equation}
   2\,\epsilon^2 \= 2k^2{-}1 \= 1/\sqrt{2E} \qquad\textrm{with}\quad
   k=\sfrac{1}{\sqrt{2}},1,\infty \quad\Leftrightarrow\quad E=\infty,\sfrac12,0\ .
\end{equation}
For $E{\gg}\frac12$, we have $k^2{\to}\frac12$, and the solution is well approximated by 
$\frac{2}{\epsilon}\cos\bigl(\frac{2\sqrt{\pi^3}}{\Gamma(1/4)^2}\frac{\tau}{\epsilon}\bigr)$.
At the critical value of $E{=}\sfrac12$ ($k{=}1$), the unstable constant solution coexists 
with the celebrated bounce solution, and below it the solution bifurcates into oscillations 
in the left or right well of the double-well potential, which halfens the oscillation period. 
The two constant minima $\psi=\pm1$ correspond to the vacua $\Acal=0$ and $\Acal=g^{-1}\diff g$.
Actually, the time translation freedom is broken by the finite range of~$\Ical$,
so that time-shifted solutions differ in their boundary values 
and also in their values for the total energy and action.
\begin{figure}[h!]
\centering
\captionsetup{width=0.9\linewidth}
\includegraphics[width = 0.35\paperwidth]{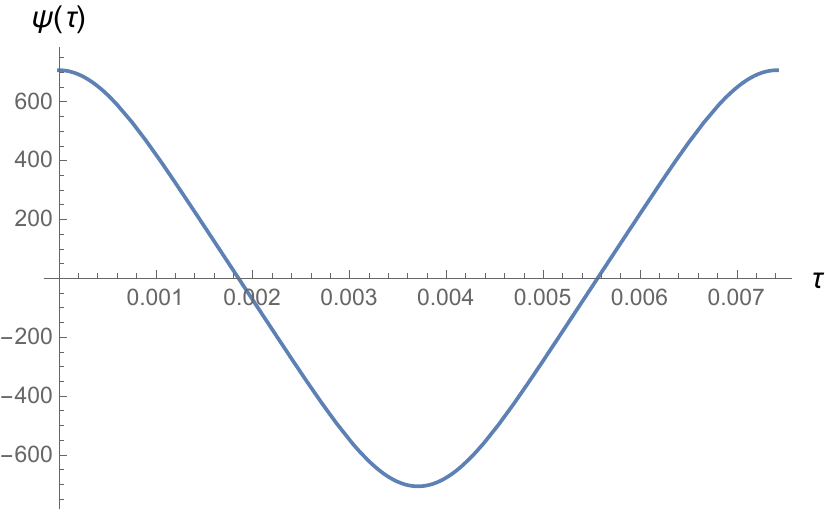} \quad
\includegraphics[width = 0.35\paperwidth]{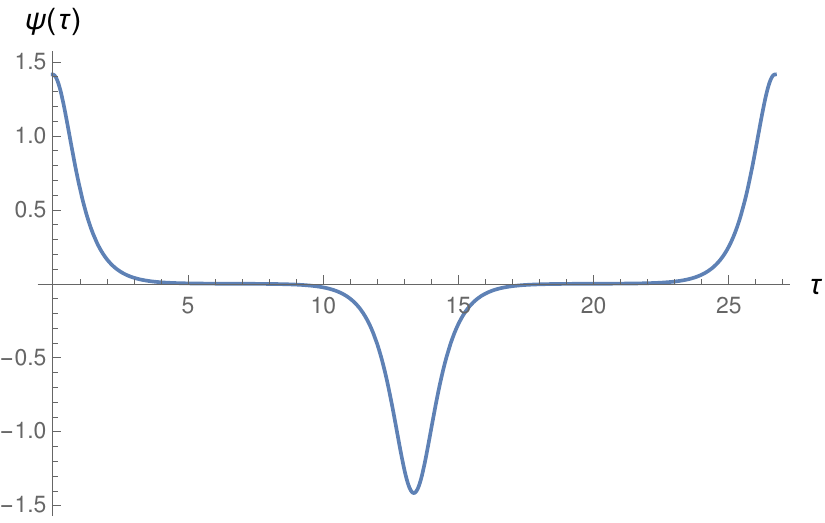} \\[12pt]
\includegraphics[width = 0.35\paperwidth]{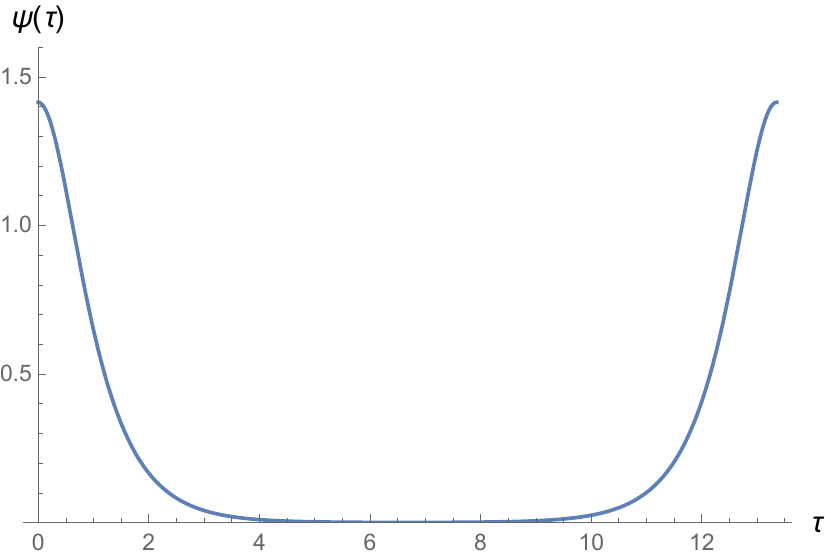} \quad
\includegraphics[width = 0.35\paperwidth]{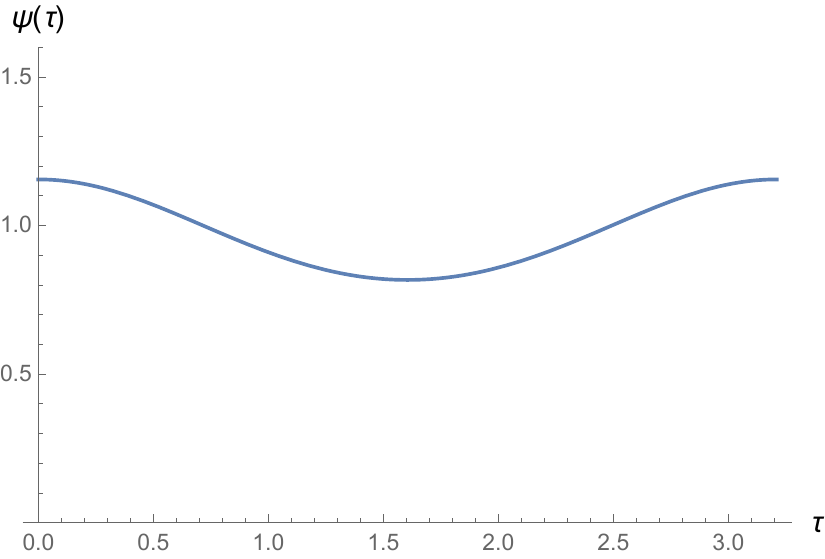}
\caption{Plots of $\psi(\tau)$ over one period, for different values of $k^2$:
$0.500001$ (top left), $0.9999999$ (top right), $1.0000001$ (bottom left) and $2$ (bottom right).}
\end{figure}

The corresponding color-electric and -magnetic field strengths read
\begin{equation}
   \Ecal_a \= \Fcal_{0a} \= \sfrac12\dot\psi\,T_a \quad\und\quad
   \Bcal_a \= \sfrac12\varepsilon_a^{\ bc}\Fcal_{bc} \= \sfrac12(\psi^2{-}1)\,T_a\ ,
\end{equation}
which yields a finite total energy (on the cylinder) of $6\pi^2 E/g^2$ and a finite action~\cite{ILP17prl,ILP17jhep}
\begin{equation}
   g^2 S[\psi] \= 6\pi^2\int_\Ical \!\diff\tau\ \bigl[E-(\psi^2{-}1)^2\bigr]
   \=6\pi^2\int_\Ical \!\diff\tau\ \bigl[\dot{\psi}^2-E\bigr]\ \ge\ -3\pi^2|\Ical|\ .
\end{equation}
The energy-momentum tensor of our SO(4)-symmetric Yang--Mills solutions is readily found as
\begin{equation}
   T \= \frac{3\,E}{g^2 a^2} \bigl( \diff\tau^2 + \sfrac13\,\diff\Omega_3^2 \bigr)\ ,
\end{equation}
which is traceless as expected.

The Einstein equations for a closed FLRW universe with cosmological constant~$\Lambda$
reduce to two independent relations, which can be taken to be its trace and its time-time component.
In conformal time one gets, respectively,
\begin{equation} \label{friedmann}
   \left.\begin{cases}
   \quad -R+4\,\Lambda \= 0 \\[4pt]
   \quad R_{\tau\tau}+\sfrac12R\,a^2-\Lambda\,a^2\= \kappa\,T_{\tau\tau}
   \end{cases} \right\}
   \quad\Leftrightarrow\quad
   \left.\begin{cases}
   \quad \ddot{a}+W'(a)\=0 \\[4pt]  \quad \sfrac12\dot{a}^2+W(a) \= \frac{\kappa}{2g^2}E \ =:\ E'
   \end{cases} \right\}
\end{equation}
with a gravitational coupling $\kappa=8\pi G$, 
a gravitational energy $E'$
and a cosmological potential 
\begin{equation} \label{gravpot}
   W(a) \= \sfrac12 a^2 - \sfrac{\Lambda}{6} a^4\ .
\end{equation}
The two anharmonic oscillators, with potential~$V$ for the gauge field and potential~$W$ for gravity,
are coupled only via the balance of their conserved energies,
\begin{equation}
   \frac1\kappa \bigl[ \sfrac12\dot{a}^2+W(a)\bigr] \= \frac1{2g^2} \bigl[ \sfrac12\dot{\psi}^2+V(\psi)\bigr]\ ,
\end{equation}
which is nothing but the Wheeler--DeWitt constraint~$H=0$.
\begin{figure}[h!]
\captionsetup{width=\linewidth}
\centering
\includegraphics[width = 0.35\paperwidth]{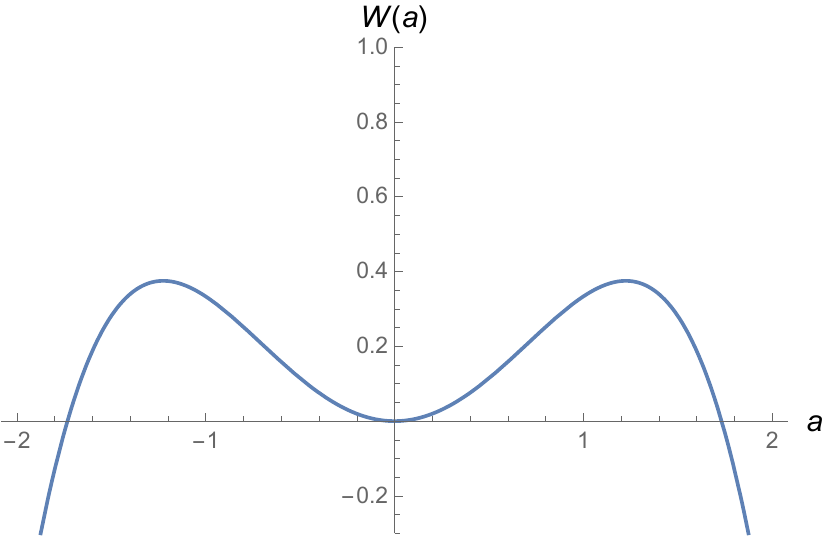} \quad
\includegraphics[width = 0.35\paperwidth]{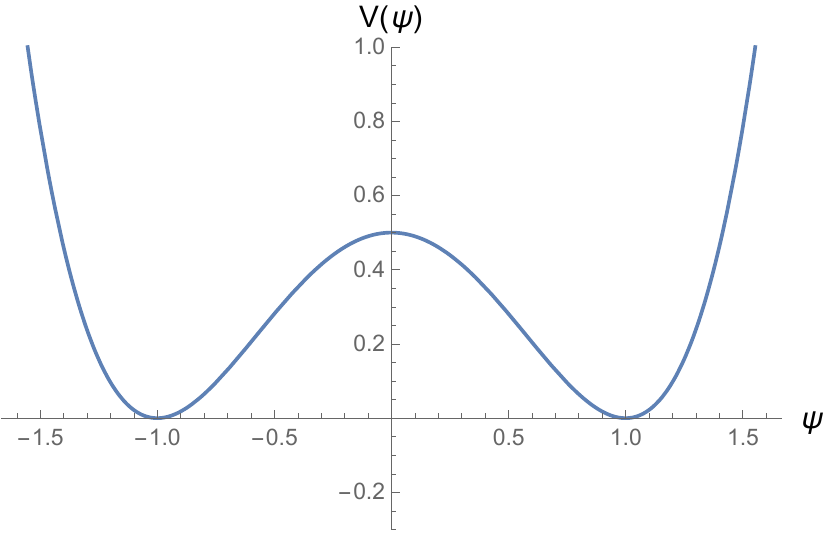} 
\caption{Plots of the cosmological potential $W(a)$ for $\lambda{=}1$ and the double-well potential $V$.}
\end{figure}

The Friedmann equation~(\ref{friedmann}), being a mechanical system with an inverted anharmonic
potential~(\ref{gravpot}), is again easily solved analytically,
\begin{equation} \label{universalscale}
   a(\tau) \= \begin{cases}
   \ \sqrt{\sfrac{3}{\Lambda}}\,\sfrac{1}{2\epsilon'}\,
   \sqrt{\frac{1-\mathrm{cn}\bigl(\sfrac{\tau}{\epsilon'},k'\bigr)}{1+\mathrm{cn}\bigl(\sfrac{\tau}{\epsilon'},k'\bigr)}}
   \qquad\qquad  \textrm{with}\quad T'=2\,\epsilon'K(k') 
   & \textrm{for}\quad \sfrac{3}{8\Lambda}<E'<\infty \\[4pt]
   \ \sqrt{\sfrac{3}{2\Lambda}}\,\tanh\bigl(\tau/\sqrt{2}\bigr) \qquad\qquad\quad\  \textrm{with}\quad T'=\infty 
   & \textrm{for}\quad E'=\sfrac{3}{8\Lambda} \\[4pt]
   \ \sqrt{\sfrac{3}{\Lambda}}\,\sfrac{1}{2\epsilon'}\,
   \sqrt{\frac{1-\mathrm{dn}\bigl(\sfrac{k'\tau}{\epsilon'},\sfrac{1}{k'}\bigr)}
   {1+\mathrm{dn}\bigl(\sfrac{k'\tau}{\epsilon'},\sfrac{1}{k'}\bigr)}}
   \qquad\quad \textrm{with}\quad T'=2\,\sfrac{\epsilon'}{k'}K(\sfrac{1}{k'}) \quad
   & \textrm{for}\quad 0<E'< \sfrac{3}{8\Lambda} \\[10pt]
   \ 0 \qquad\qquad\qquad\qquad\qquad\qquad\ \; \textrm{with}\quad T'=\pi
   &  \textrm{for}\quad E'=0 
   \end{cases}\ ,
\end{equation}
where we abbreviated~\footnote{
Our ${k'}^2$ should not be confused with the dual modulus $1{-}k^2$, which is often denoted this way.}
\begin{equation}
   2\,{\epsilon'}^2 \= 2{k'}^2{-}1 \= 1/\sqrt{\smash{\sfrac{8\Lambda}{3}}\, E'}
   \qquad\textrm{so that}\quad
   k'=\sfrac{1}{\sqrt{2}},1,\infty \quad\Leftrightarrow\quad E'=\infty,\sfrac{3}{8\Lambda},0\ .
\end{equation}
\begin{figure}[h!]
\centering
\captionsetup{width=\linewidth}
\includegraphics[width = 0.35\paperwidth]{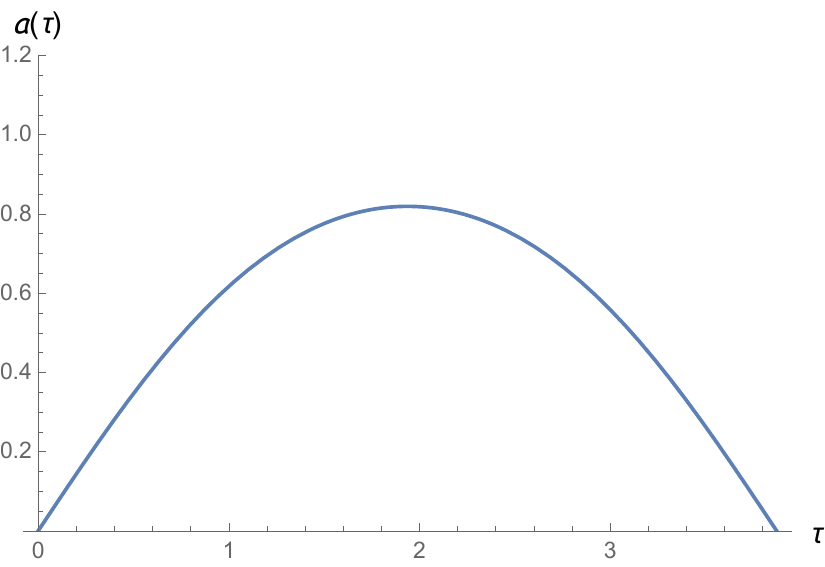} \quad
\includegraphics[width = 0.35\paperwidth]{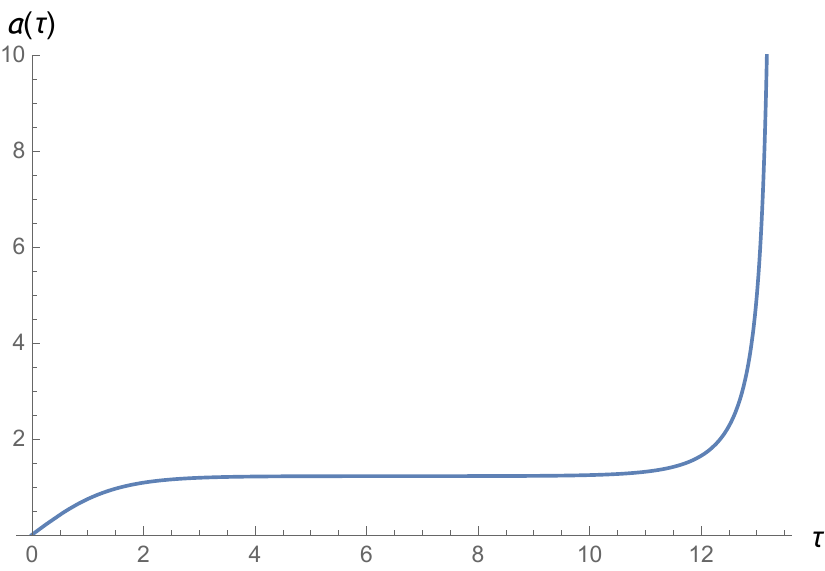} \\[12pt]
\includegraphics[width = 0.35\paperwidth]{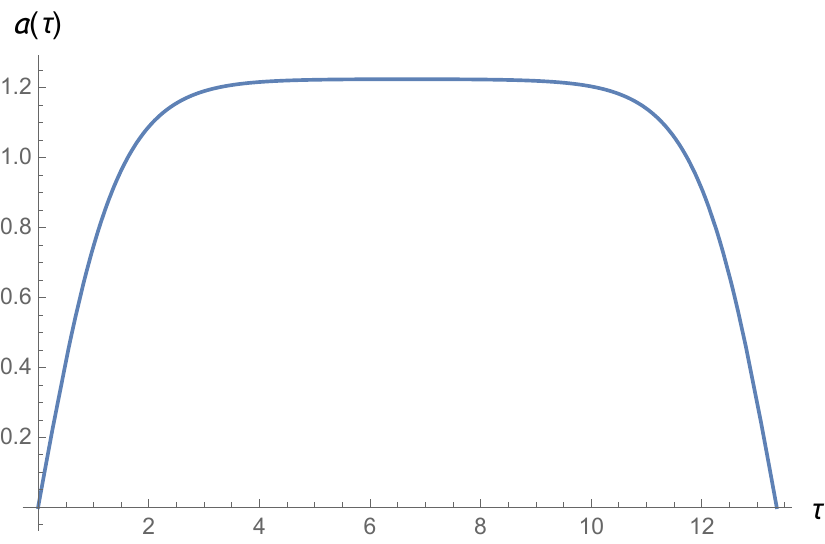} \quad
\includegraphics[width = 0.35\paperwidth]{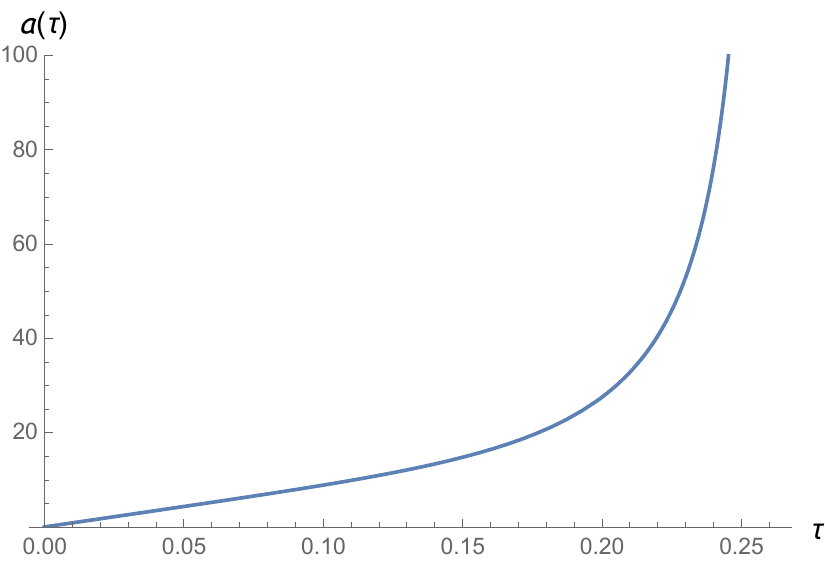}
\caption{Plots of $a(\tau)$ over one lifetime, for $\Lambda{=}1$ and different values of ${k'}^2$:
$0.505$ (top left), $0.9999999$ (top right), $1.0000001$ (bottom left) and $1.1$ (bottom right).}
\end{figure}
For $E'{\gg}\frac{3}{8\Lambda}$, we have ${k'}^2{\to}\frac12$, and the solution is well approximated by
$\sqrt{\frac{3}{\Lambda}}\frac{1}{2\epsilon'}\tan\bigl(\frac{\sqrt{\pi^3}}{\Gamma(1/4)^2}\frac{\tau}{\epsilon'}\bigr)$.
We only listed solutions with initial value $a(0)=0$ (big bang). 
There exist also (for $E'<\sfrac{3}{8\Lambda}$) bouncing solutions,
where the universe attains a minimal radius 
$a_{\textrm{min}}=\bigl[\frac{3}{2\Lambda}(1+\sqrt{1-\smash{\sfrac{8\Lambda}{3}}E'})\bigr]^{1/2}$
between infinite extension in the far past ($t{=}{-}\infty\leftrightarrow\tau{=}0$) and the
far future ($t{=}{+}\infty\leftrightarrow\tau{=}T'$). For $E'{>}0$ they are obtained by 
sending $\mathrm{dn}\to-\mathrm{dn}$ in~(\ref{universalscale}) above.
The quantity~$T'$ listed there is the (conformal) lifetime of the universe, 
from the big bang until either the big rip (for $E'>\frac{3}{8\Lambda}$) 
or the big crunch of an oscillating universe (for $E'<\frac{3}{8\Lambda}$).
The solution relevant to our Einstein--Yang--Mills system is entirely determined by the Newtonian energy~$E$
characterizing the cosmic Yang--Mills field: above the critical value of
\begin{equation}
   E_{\textrm{crit}} \= \frac{2\,g^2}{\kappa}\,\frac{3}{8\,\Lambda}
\end{equation}
the universe expands forever (until $t_{\textrm{max}}{=}\infty$), 
while below this value it recollapses (at $t_{\textrm{max}}{=}\int_0^{T'}\!\diff\tau\,a(\tau)$).
It demonstrates the necessity of a cosmological constant (whose role may be played
by the Higgs expectation value) as well as the nonperturbative nature of the cosmic Yang--Mills field,
whose contribution to the energy-momentum tensor is of~$O(g^{-2})$.

\vspace{12pt}
%----------------------------------------------------------------------------------------
\section{Natural perturbation frequencies}
\vspace{2pt}
\noindent
Our main task in this paper is an investigation of the stability of the cosmic Yang--Mills solutions
reviewed in the previous section. For this, we should distinguish between global and local stability.
The former is difficult to assess in a nonlinear dynamics but clear from the outset in case of a compact
phase space. The latter refers to short-time behavior induced by linear perturbations around
the reference configuration. We shall look at this firstly, in the present section and the following one. 
Here, we set out to diagonalize the fluctuation operator for our time-dependent Yang--Mills backgrounds
and find the natural frequencies. 

Even though our cosmic gauge-field configurations are SO(4)-invariant, we must allow for all kinds of
fluctuations on top of it, SO(4)-symmetric perturbations being a very special subclass of them.
A generic gauge potential ``nearby'' a classical solution~$\Acal$ on $\Ical\times S^3$ can be expanded as
\begin{equation}
   \Acal+\Phi \= \Acal(\tau,g)\ +\ \sum_{p=1}^3 \Phi_0^p(\tau,g)\,T_p\,\diff\tau\
   +\ \sum_{a=1}^3\sum_{p=1}^3 \Phi_a^p(\tau,g)\,T_p\,e^a(g)
\end{equation}
with, using $(\mu)=(0,a)$,
\begin{equation}
   \Phi_\mu^p(\tau,g) \= \sum_{j,m,n} \Phi_{\mu|j;m,n}^p(\tau)\,Y_{j;m,n}(g)\ ,
\end{equation}
on which we notice the following actions (suppressing the $\tau$ and $g$ arguments),
\begin{equation}
\begin{aligned}
   &(L_a \Phi_\mu^p)_{j;m,n} \= \Phi_{\mu|j;m,n'}^p \bigl(L_a)^{n'}_{\ n}\ ,\qquad
   (S_a \Phi)_0^p \= 0\ ,\\
   &(S_a \Phi)_b^p \= -2\varepsilon_{abc}\,\Phi_c^p \ ,\qquad
   (T_a \Phi)_\mu^p \= -2\varepsilon_{apq}\,\Phi_\mu^q \ ,
\end{aligned}
\end{equation}
where $L_a$ matrix elements are determined from \eqref{Y-action} and \eqref{JpmAction},
while $S_a$ are spin operator components. The (metric and gauge) background-covariant derivative reads
\begin{equation}
   D_\tau \Phi \= \partial_\tau \Phi \und
   D_a \Phi \= L_a \Phi + [ \Acal_a,\Phi ] \with
   \Acal_a \= \sfrac12\bigl(1+\psi(\tau)\bigr)\,T_a\ ,
\end{equation}
which is equivalent to
\begin{equation}
   D_a\Phi_b^p \= L_a\Phi_b^p - \varepsilon_{abc}\Phi_c^p + [ \Acal_a,\Phi_b]^p
   \quad\textrm{since}\quad 
   D_a\,e^b \= L_a\,e^b -  \varepsilon_{abc}\,e^c \= \varepsilon_{abc}\,e^c \ .
\end{equation}

The background $\Acal$ obeys the Coulomb gauge condition,
\begin{equation}
   \Acal_\tau =0 \qquad\und\qquad L_a\Acal_a = 0\ ,
\end{equation}
but we cannot enforce these equations on the fluctuation~$\Phi$. 
However, we may impose the Lorenz gauge condition,
\begin{equation} \label{gauge}
   D^\mu \Phi^p_\mu \= 0 \qquad\Rightarrow\qquad
   \partial_\tau \Phi_0^p - L_a\Phi^p_a - \sfrac12(1{+}\psi)(T_a\Phi_a)^p \= 0\ ,
\end{equation}
which is seen to couple the temporal and spatial components of~$\Phi$ in general.
We then linearize the Yang--Mills equations around $\Acal$ and obtain
\begin{equation}
   D^\nu D_\nu \Phi_\mu - R_{\mu\nu}\Phi^\nu + 2[\Fcal_{\mu\nu},\Phi^\nu] \= 0
\end{equation}
with the Ricci tensor
\begin{equation}
   R_{\mu 0} \= 0 \qquad\und\qquad R_{ab} \= 2\delta_{ab}\ .
\end{equation}
After a careful evaluation, the $\mu{=}0$ equation yields
\begin{equation} \label{temporal}
   \bigl[\partial_\tau^2 - L_b L_b + 2(1{+}\psi)^2\bigr] \Phi_0^p 
   - (1{+}\psi)L_b(T_b\Phi_0)^p - \dot{\psi}(T_b\Phi_b)^p \= 0\ ,
\end{equation}
while the $\mu{=}a$ equations read
\begin{equation} \label{spatial}
\begin{aligned}
   \bigl[\partial_\tau^2 - L_b L_b + 2(1{+}\psi)^2{+}4\bigr]\Phi_a^p 
   &-(1{+}\psi)L_b(T_b\Phi)_a^p - L_b(S_b\Phi)_a^p \\
   &- \sfrac12(1{+}\psi)(2{-}\psi)(S_bT_b\Phi)_a^p
   - \dot{\psi}(T_a\Phi_0)^p \= 0\ .
\end{aligned}
\end{equation}
It is convenient to package the orbital, spin, isospin, and fluctuation triplets into formal vectors,
\begin{equation}
   \vec{L}=(L_a)\ ,\qquad \vec{S}=(S_a)\ ,\qquad \vec{T}=(T_a)\ ,\qquad \vec{\Phi}=(\Phi_a)\ ,
\end{equation}
respectively, but they act in different spaces, hence on different indices,
such that $\vec{S}^2=\vec{T}^2=-8$ on~$\Phi$. 
In this notation, (\ref{gauge}), (\ref{temporal}) and~(\ref{spatial}) take the compact form
(suppressing the color index~$p$)
\begin{fleqn}[\parindent]
\begin{equation}
   \partial_\tau\Phi_0 - \vec{L}{\cdot}\vec{\Phi} - \sfrac12(1{+}\psi)\vec{T}{\cdot}\vec{\Phi} \= 0\ ,
   \label{gauge2}
\end{equation}
\begin{equation}
   \bigl[\partial_\tau^2{-}\vec{L}^2+2(1{+}\psi)^2\bigr] {\Phi}_0 
   - (1{+}\psi)\vec{L}{\cdot}\vec{T}\,\Phi_0 - \dot{\psi}\,\vec{T}{\cdot}\vec{\Phi} \= 0 \ ,
   \label{mu=0}
\end{equation}
\begin{equation}
\begin{split}
   \bigl[\partial_\tau^2{-}\vec{L}^2{-}\sfrac12\vec{S}^2{+}2(1{+}\psi)^2\bigr] {\Phi}_a
   &- (1{+}\psi)\vec{L}{\cdot}\vec{T}\,\Phi_a - \vec{L}{\cdot}(\vec{S}\,\Phi)_a \\
   &- \sfrac12(1{+}\psi)(2{-}\psi)\vec{T}{\cdot}(\vec{S}\,\Phi)_a - \dot{\psi}\,T_a\Phi_0 \= 0\ .
\end{split}
\label{mu=a}
\end{equation}
\end{fleqn}

A few remarks are in order. 
First, except for the last term, (\ref{mu=0}) is obtained from (\ref{mu=a}) by setting $\vec{S}=0$,
since $\Phi_0$ carries no spin index. 
Second, both equations can be recast as
\begin{fleqn}
\begin{equation} \label{2spins}
   \bigl[ \partial_\tau^2 - \sfrac{1{-}\psi}{2}\vec{L}^2 - \sfrac{1{+}\psi}{2}(\vec{L}{+}\vec{T})^2 
   - 2(1{+}\psi)(1{-}\psi) \bigr] \Phi_0 \= \dot{\psi}\,\vec{T}\cdot\vec{\Phi}\ ,  
\end{equation}
\begin{equation} \label{3spins}
\begin{split}
   \bigl[ \partial_\tau^2 - \sfrac{(1{-}\psi)(2{+}\psi)}{4}\vec{L}^2 - \sfrac{\psi(1{+}\psi)}{4}(\vec{L}{+}\vec{T})^2
   + \sfrac{\psi(1{-}\psi)}{4}(\vec{L}&{+}\vec{S})^2 - \sfrac{(1{+}\psi)(2{-}\psi)}{4}(\vec{L}{+}\vec{T}{+}\vec{S})^2 \\
   &- 2(1{+}\psi)(1{-}\psi) \bigr] \vec{\Phi} = \dot{\psi}\vec{T}\Phi_0\ ,
\end{split}
\end{equation}
\end{fleqn}
which reveals a problem of addition of three spins and a corresponding symmetry under
\begin{equation} \label{symmetry}
   \psi\ \leftrightarrow\ -\psi\ ,\qquad
   \vec{L}\ \leftrightarrow\ \vec{L}{+}\vec{T}{+}\vec{S} \quad\und\quad
   \vec{L}{+}\vec{S}\ \leftrightarrow\ \vec{L}{+}\vec{T}\ .
\end{equation}
Third, for constant backgrounds $(\dot{\psi}{=}0)$ the temporal fluctuation $\Phi_0$ decouples 
and may be gauged away. 
Still, the fluctuation operator in~(\ref{3spins}) is easily diagonalized only when 
the coefficient of one of the first three spin-squares vanishes, i.e.~for $\vec{L}{=}0$ ($j{=}0$),
for the two vacua $\psi{=}{\pm}1$, or for the ``meron'' $\psi{=}0$. 
The latter case has been analyzed by Hosotani~\cite{Hosotani}.

Let us decompose the fluctuation problem (\ref{gauge2})--(\ref{mu=a}) into finite-dimensional blocks
according to a fixed value of the spin~$j\in\sfrac12\N$,
\begin{equation}
   \vec{L}^2\, \Phi^p_{\mu|j} \= -4\,j(j{+}1)\,\Phi^p_{\mu|j}
\end{equation}
and suppress the $j$ subscript. We employ the following coupling scheme,\footnote{
Another (less convenient) scheme couples $\vec{L}{+}\vec{S}$, then $(\vec{L}{+}\vec{S}){+}\vec{T}=:\vec{V}$.}
\begin{equation}
   \vec{L}{+}\vec{T}=:\vec{U} \qquad\textrm{then}\qquad \vec{U}{+}\vec{S}=(\vec{L}{+}\vec{T}){+}\vec{S}=:\vec{V}\ .
\end{equation}
Clearly, $\vec{U}$ and $\vec{V}$ act on $\vec{\Phi}$ in $su(2)$ representations 
$j\otimes 1$ and $j\otimes 1\otimes 1$, respectively. On $\Phi_0$, we must put~$\vec{S}{=}0$
and have just $\vec{V}{=}\vec{U}$ act in a $j\otimes 1$ representation.
Combining the coupled equations (\ref{mu=0}) and (\ref{mu=a}) to a single linear system for
$(\Phi^p_\mu)=(\Phi^p_0,\Phi^p_a)$, we get a $12(2j{+}1)\times 12(2j{+}1)$ fluctuation matrix~$\Omega^2_{(j)}$,
\begin{equation} \label{fluctop}
   \bigl[ \delta_{\mu\nu}^{pq}\,\partial_\tau^2\ +\ (\Omega^2_{(j)})_{\mu\nu}^{pq} \bigr]\,\Phi_\nu^q \= 0\ .
\end{equation}
Actually, there is an additional overall $(2j{+}1)$-fold degeneracy present due to the trivial action of
the $su(2)_{\textrm{R}}$ generators~$R_a$, which plays no role here and will be suppressed.
Roughly speaking, the $3(2j{+}1)$ modes of $\Phi_0$ are related to gauge modes,\footnote{
Strictly, they are gauge modes only when $\dot\psi{=}0$. 
Otherwise, the gauge modes are mixtures with the $\Phi_a$ modes.}
and we still must impose the gauge condition~(\ref{gauge2}), which also has $3(2j{+}1)$ components.
Therefore, a subspace of dimension $6(2j{+}1)$ inside the space of all fluctuations will 
represent the physical gauge-equivalence classes in the end.

Our goal is to diagonalize the fluctuation operator~(\ref{fluctop}) for a given fixed value of~$j$.
It has a block structure,
\begin{equation} \label{block}
   \Omega^2_{(j)} \= \begin{pmatrix} \bar{N} & -\dot{\psi}\,T^\top \\[6pt] -\dot{\psi}\,T & N \end{pmatrix}\ ,
\end{equation}
where $\bar{N}$ and $N$ are given by the left-hand sides of (\ref{2spins}) and~(\ref{3spins}), respectively.
We introduce a basis where $\vec{U}^2$, $\vec{V}^2$ and $V_3$ are diagonal, i.e.
\begin{equation}
   \vec{U}^2\,|uvm\> \= -4\,u(u{+}1)\,|uvm\> \quad\und\quad 
   \vec{V}^2\,|uvm\> \= -4\,v(v{+}1)\,|uvm\>\ ,
\end{equation}
with $m=-v,\ldots,v$ and denote the irreducible $su(2)_v$ representations with those quantum numbers as $\rep{v}{u}$.
On the $\Phi_0$ subspace, $u$ is redundant since $u{=}v$ as $\vec{S}{=}0$.
Working out the tensor products, we encounter the values
\begin{equation} \label{uvreps}
\begin{aligned}
   \rep{v}{u} &\= \rep{j{-}2}{j{-}1}\ ;\!\!\!&\rep{j{-}1}{j{-}1}&\ ,\ \rep{j{-}1}{j}\ ;\ \rep{j}{j{-}1}\ ,
   \!\!\!&\rep{j}{j}&\ ,\ \rep{j}{j{+}1}\ ;\ \rep{j{+}1}{j}\ ,\!\!\!&\rep{j{+}1}{j{+}1}&\ ;\ \rep{j{+}2}{j{+}1} 
   &\quad\textrm{on}\ \vec{\Phi}\ ,& \\
   \rep{v}{u} &\= &\rep{j{-}1}{j{-}1}&\ ; &\rep{j}{j}&\ ; &\rep{j{+}1}{j{+}1}&
   &\quad\textrm{on} \ \Phi_0&\ ,
\end{aligned}
\end{equation}
with some representations obviously missing for $j{<}2$.

Let us treat the $\dot{\psi}\,T$ term in~(\ref{block}) as a perturbation and momentarily put it to zero,
so that $\Omega^2_{(j)}$ is block-diagonal for the time being.
Then, it is easy to see from (\ref{2spins}) and (\ref{3spins}) that $[\vec{V},\bar{N}]=[\vec{U},\bar{N}]=0$
and $[\vec{V},N]=0$, even though $[\vec{U},N]\neq0$ because $(\vec{L}{+}\vec{S})^2$ 
is not diagonal in our basis. Therefore, we have a degeneracy in~$m$.
Furthermore, both $\bar{N}$ and $N$ decompose into at most three respectively five blocks 
with fixed values of~$v$ ranging from $j{-}2$ to~$j{+}2$ and separated by semicolons in~(\ref{uvreps}). 
Moreover, the $\bar{N}$~blocks are irreducible and trivially also carry a value of~$u{=}v$. 
In contrast, $N$ is not {\it simply\/} reducible; its $\vec{V}$ representations have multiplicity one, two or three. 
Only the $N$~blocks with extremal $v$~values in~(\ref{uvreps}) are irreducible. 
The other ones are reducible and contain more than one $\vec{U}$~representation, 
hence the $u$-spin distinguishes between their (two or three) irreducible $v$~subblocks.
The only non-diagonal term in~$N$ is the $(\vec{L}{+}\vec{S})^2$ contribution, which couples different copies
of the same $v$-spin to each other, but of course not to any $u{=}v$ block of~$\bar{N}$, 
and does not lift the $V_3{=}m$~degeneracy. 
As a consequence, the unperturbed fluctuation equations 
for $\Phi_0{=}\Phi_{(\bar{v})}$ and $\vec{\Phi}{=}\Phi_{(v,\alpha)}$ take the form 
(suppressing the $m$ index)
\begin{fleqn}
\begin{equation}
\begin{split}
   \unity_{(\bar{v})}\bigl[ \partial_\tau^2\ +\ \bar\omega^2_{(\bar{v})}\bigr]\,\Phi_{(\bar{v})} \= 0 \qquad\und\qquad
   \unity_{(v)}\bigl[ \partial_\tau^2\ +\ \omega^2_{(v,\alpha)}\bigr]\,\Phi_{(v,\alpha)} \= 0 \\[4pt]
   \for \dot{\psi}=0 \with \bar{v} \in \{ j{-}1,\ j,\ j{+}1\} \und
   v \in \{ j{-}2,\ j{-}1,\ j,\ j{+}1,\ j{+}2 \} \ ,
\end{split}
\end{equation}
\end{fleqn}
where $\unity_{(v)}$ denotes a unit matrix of size~$2v{+}1$, and $\alpha$ counts the multiplicity 
of the $v$-spin representation in~$N$ (between one and three). 
According to~(\ref{2spins}) the unperturbed frequency-squares for $\bar{N}$ are the eigenvalues
\begin{equation}
   \bar{\omega}^2_{(\bar{v})} \= 
   2(1{-}\psi)\,j(j{+}1) + 2(1{+}\psi)\,\bar{v}(\bar{v}{+}1) - 2(1{+}\psi)(1{-}\psi)
\end{equation}
with multiplicity  $2\bar{v}{+}1$, hence we get
\begin{equation}
\begin{aligned}
   \bar{\omega}^2_{(j-1)} &\= 2\,\psi^2-4j\,\psi+2(2j^2{-}1)\ ,\\
   \bar{\omega}^2_{(j)} &\= 2\,\psi^2+2(2j^2{+}2j{-}1)\,\\
   \bar{\omega}^2_{(j+1)} &\= 2\,\psi^2+4(j{+}1)\,\psi +2(2j^2{+}4j{+}1)\ .
\end{aligned}
\end{equation}
Considering $N$ in (\ref{3spins}), we can read off the eigenvalues at $v=j{\pm}2$
because in these two extremal cases $(\vec{L}{+}\vec{S})^2=\vec{U}^2$ 
is already diagonal in the $\bigl\{ |uvm\>\bigr\}$ basis. For the other $v$-values
we must diagonalize a $2{\times}2$ or $3{\times}3$ matrix to find
\begin{equation}
\begin{aligned}
   \omega^2_{(j-2)} &\= \textrm{root of}\ Q_{j-2}(\lambda)
   \= -2(2j{-}1)\,\psi + 2(2j^2{-}2j{+}1)\ ,\\
   \omega^2_{(j-1,\alpha)} &\= \textrm{two roots of}\ Q_{j-1}(\lambda)\ ,\\
   \omega^2_{(j,\alpha)} &\= \textrm{three roots of}\ Q_j(\lambda)\ ,\\
   \omega^2_{(j+1,\alpha)} &\= \textrm{two roots of}\ Q_{j+1}(\lambda)\ ,\\
   \omega^2_{(j+2)} &\= \textrm{root of}\ Q_{j+2}(\lambda)
   \= 2(2j{+}3)\,\psi + 2(2j^2{+}6j{+}5)\ ,
\end{aligned}
\end{equation}
each with multiplicity $2v{+}1$, where $Q_v$ denotes a linear, quadratic or cubic polynomial.\footnote{
For $j{<}2$ some obvious modifications occur due to the missing of $v{<}0$ representations.}

Let us now turn on the perturbation $\dot{\psi}\,T$, which couples $N$ with~$\bar{N}$,
and consider the characteristic polynomial~${\cal P}_j(\lambda)$ of our fluctuation problem,
\begin{equation}
\begin{aligned}
   {\cal P}_j(\lambda) &\ :=\
   \det\,\Bigl(\begin{smallmatrix} \bar{N}{-}\lambda & -\dot{\psi}T^\top \\[4pt] 
   -\dot{\psi}T & N{-}\lambda \end{smallmatrix} \Bigr)
   \= \det (N{-}\lambda) \cdot \det\bigl[ (\bar{N}{-}\lambda) - \dot{\psi}^2\,T^\top (N{-}\lambda)^{-1} T \bigr] \\[4pt]
   &\= \bigl[{\textstyle\prod}_v \det (N_{(v)}{-}\lambda)\bigr] \cdot 
   \det\bigl[ (\bar{N}{-}\lambda) - \dot{\psi}^2\, T^\top \{{\textstyle\bigoplus}_{v} (N_{(v)}{-}\lambda)^{-1}\}\,T\bigr] \ ,
\end{aligned}
\end{equation}
where we made use of
\begin{equation}
\< u\,v\,m|\,N\,|u'v'm'> \= \bigl(N_{(v)}\bigr)_{uu'}\,\delta_{vv'}\delta_{mm'}\ .
\end{equation}
Since $T$ furnishes an $su(2)$ representation (and not an intertwiner) 
it must be represented by square matrices and thus cannot connect different $v$ representations.
Hence the perturbation does not couple different $v$ sectors but only links $N$ and $\bar{N}$
in a common $\bar{v}{=}v$~sector. Therefore, it does not affect the extremal sectors $v=j{\pm}2$.
Moreover, switching to a diagonal basis $\{|\alpha vm\>\}$ for $N$ we can simplify to
\begin{equation}
\begin{aligned}
   T^\top \{{\textstyle\bigoplus}_{v} (N_{(v)}{-}\lambda)^{-1}\}\,T\bigr]
   &\= {\textstyle\bigoplus}_{\bar{v}}\{T^\top (N{-}\lambda)^{-1} T\}_{(\bar{v})} \\
   &\={\textstyle\bigoplus}_{\bar{v}}\Bigl\{ {\textstyle\sum}_\alpha 
   (\omega^2_{(\bar{v},\alpha)}{-}\lambda)^{-1} \bigl(T^\top|\alpha\>\!\<\alpha|\,T\bigr)_{(\bar{v})}\Bigr\}\ .
\end{aligned}
\end{equation}
Observing that 
$\bigl(T^\top|\alpha\>\!\<\alpha|\,T\bigr)_{(\bar{v})}=
-t_{\bar{v},\alpha}\bigl(\vec{T}^2\bigr)_{(\bar{v})}=8\,t_{\bar{v},\alpha}\unity_{(\bar{v})}$
with some coefficient functions $t_{\bar{v},\alpha}(\psi)$, with $\sum_\alpha t_{\bar{v},\alpha}=1$,
we learn that the $V_3$ degeneracy remains intact and arrive at ($\bar{v}\in\{j{-}1,j,j{+}1\}$)
\begin{equation} \label{charP}
\begin{aligned}
   {\cal P}_j(\lambda) &\= \bigl[{\textstyle\prod}_v Q_v(\lambda)^{2v+1}\bigr] \cdot
   {\textstyle\prod}_{\bar{v}} \bigl\{(\bar{\omega}^2_{(\bar{v})}{-}\lambda) 
   - 8\dot{\psi}^2 {\textstyle\sum}_\alpha t_{\bar{v},\alpha} 
   (\omega^2_{(\bar{v},\alpha)}{-}\lambda)^{-1} \bigr\}^{2\bar{v}+1}\\[4pt]
   &\= (\omega^2_{(j-2)}{-}\lambda)^{2j-3}\cdot(\omega^2_{(j+2)}{-}\lambda)^{2j+5}\cdot
   {\textstyle\prod}_{\bar{v}} \bigl\{ (\bar{\omega}^2_{(\bar{v})}{-}\lambda)\,Q_{\bar{v}}(\lambda) - 8\dot{\psi}^2 P_{\bar{v}}(\lambda) \bigr\}^{2\bar{v}+1} \\[4pt]
   &\= (\omega^2_{(j-2)}{-}\lambda)^{2j-3}\cdot(\omega^2_{(j+2)}{-}\lambda)^{2j+5}\cdot
   {\textstyle\prod}_{\bar{v}} R_{\bar{v}}(\lambda)^{2\bar{v}+1}\ ,
\end{aligned}
\end{equation}
where $P_{\bar{v}}=Q_{\bar{v}}\sum_\alpha t_{\bar{v},\alpha} (\omega^2_{(\bar{v},\alpha)}{-}\lambda)^{-1}$
is a polynomial of degree one less than $Q_{\bar{v}}$ since all poles cancel, and $R_{\bar{v}}$ is a
polynomial of one degree more.
We list the polynomials $Q_v$, $P_{\bar{v}}$ and $R_{\bar{v}}$ for $j{\le}2$ in the Appendix.

To summarize, by a successive basis change ($m'=-j,\ldots,j$ and $m=-v,\ldots,v$)
\begin{equation}
   \bigl\{ |\mu\,p\,m'\> \bigr\} \quad\Rightarrow\quad 
   \bigl\{ |\bar{v} m\>, |u v m\> \bigr\} \quad\Rightarrow\quad
   \bigl\{ |\bar{v} m\>, |\alpha v m\> \bigr\} \quad\Rightarrow\quad
   \bigl\{ |\beta v m\> \bigr\}
\end{equation}
we have diagonalized~(\ref{fluctop}) to
\begin{equation} \label{diagonalized}
   \bigl[ \partial_\tau^2 + \Omega^2_{(j,v,\beta)} \bigr] \, \Phi_{(v,\beta)} \= 0
   \qquad\textrm{with}\quad v\in\{j{-}2,j{-}1,j,j{+}1,j{+}2\} \ ,
\end{equation}
where $\Omega^2_{(j,v,\beta)}$ are the distinct roots of the characteristic polynomial~${\cal P}_j$ in~(\ref{charP}),
and (for $j{\ge}2$) the multiplicity $\beta$ takes $1,3,4,3,1$ values, respectively:
\begin{equation}
   \Omega^2_{(j,j\pm2)} = \omega^2_{(j\pm2)}\ ,\quad
   \Omega^2_{(j,j\pm1,\beta)} = \textrm{three roots of}\ R_{j\pm1}(\lambda)\ ,\quad
   \Omega^2_{(j,j,\beta)} = \textrm{four roots of}\ R_{j}(\lambda)\ .
\end{equation}
The reflection symmetry~(\ref{symmetry}) implies that 
$\Omega^2_{(v,j,\cdot)}(\psi)=\Omega^2_{(j,v,\cdot)}(-\psi)$.
For $j{<}2$, obvious modifications occur due to the absence of some $v$~representations.

We still have to discuss the gauge condition~(\ref{gauge2}), which can be cast into the form
\begin{equation}
   0 \= \pa_\tau\Phi_0 - \bigl[ \sfrac12(1{-}\psi)\vec{L}+\sfrac12(1{+}\psi)\vec{U}\bigr]\cdot\vec\Phi
   \= \pa_\tau\Phi_{(\bar{v},\bar{m})} - K_{\bar{v},\bar{m}}^{\ v,m,\alpha}(\psi)\,\Phi_{(v,m,\alpha)}
\end{equation}
with a $3(2j{+}1){\times}7(2j{+}1)$ linear (in~$\psi$) matrix function~$K$.\footnote{
We have to bring back the $m$ indices because the gauge condition is not diagonal in them.}
Here the $v$~sum runs over $(j{-}1,j,j{+}1)$ only,
since the gauge condition~(\ref{gauge2}) has components only in the middle three $v$~sectors,
like the gauge-mode equation~(\ref{mu=0}).
It does not restrict the extremal $v$~sectors~$v=j{\pm}2$, since these fluctuations
do not couple to the gauge sector~$\Phi_0$ and are entirely physical.
For the middle three $v$~sectors (labelled by~$\bar{v}$), the $\dot{\psi}\,T$ perturbation leads to
a mixing of the $N$ modes with the $\bar{N}$ gauge modes, so their levels will avoid crossing. 
Performing the corresponding final basis change, the gauge condition takes the form
\begin{equation} \label{gauge4}
   \bigl[ L_{\bar{v},\bar{m}}^{\ \bar{v}'\!,\bar{m}'\!,\beta}(\psi)\,\pa_\tau 
   - M_{\bar{v},\bar{m}}^{\ \bar{v}'\!,\bar{m}'\!,\beta}(\psi) \bigr]\,\Phi_{(\bar{v}'\!,\bar{m}'\!,\beta)}\=0
\end{equation}
with certain $3(2j{+}1){\times}10(2j{+}1)$ matrix functions $L$ and~$M$.
This linear equation represents conditions on the normal mode functions $\Phi_{(\bar{v},\bar{m},\beta)}$
and defines a $7(2j{+}1)$-dimensional subspace of physical fluctuations, 
which of course still contains a $3(2j{+}1)$-dimensional subspace of gauge modes. 
For $j{<}1$, these numbers are systematically smaller.
Together with the two extremal $v$~sectors, we end up with $(7-3+2)(2j{+}1)=6(2j{+}1)$ 
physical degrees of freedom for any given value of~$j({\ge}2)$, as advertized earlier.

We conclude this section with more details for the simplest examples,
which are constant backgrounds and $j{=}0$ backgrounds.
For the vacuum background, say $\psi=-1$, which is isospin degenerate, one gets
\begin{equation}
   \bigl( \partial_\tau^2 -\sfrac12\vec{L}^2 -\sfrac12(\vec{L}{+}\vec{S})^2 \bigr)\,\vec{\Phi} \=0, \qquad
   \vec{L}{\cdot}\vec{\Phi} = 0\ ,\qquad \Phi_0 =0\ .
\end{equation}
It yields the positive eigenfrequency-squares
\begin{equation}
   \omega^2_{(j,u')} \= 2j(j{+}1) + 2u'(u'{+}1) \=
   \begin{cases}
   \ 4j^2 \ \textrm{at}\ j{{\ge}1} & \for u'=j{-}1 \\ 
   \ 4j(j{+}1) & \for u'=j \\
   \ 4(j{+}1)^2 & \for u'=j{+}1 
\end{cases}
\end{equation}
for $j=0,\sfrac12,1,\ldots$,
but the $\vec{L}{\cdot}\vec{\Phi}=0$ constraint removes the $u'{=}j$ modes.
Clearly, all (constant) eigenfrequency-squares are positive, hence the vacuum is stable.

For the ``meron'' background, $\psi\equiv0$, one has
\begin{equation}
   \bigl( \partial_\tau^2 -\sfrac12\vec{L}^2 -\sfrac12(\vec{L}{+}\vec{T}{+}\vec{S})^2 -2 \bigr)\,\vec{\Phi} \=0, \qquad
   \bigl(\vec{L}+\sfrac12\vec{T}\bigr)\cdot\vec{\Phi} = 0\ ,\qquad \Phi_0 =0\ .
\end{equation}
In this case, we read off 
\begin{equation}
   \omega^2_{(j,v)} +2 \= 2j(j{+}1) + 2v(v{+}1) \=
   \begin{cases}
   \ 4(j^2{-}j{+}1)  & \for v=j{-}2 \qquad (0\ \textrm{to}\ 1\times) \\ 
   \ 4j^2 & \for v=j{-}1 \qquad (0\ \textrm{to}\ 2\times) \\
   \ 4j(j{+}1) & \for v=j \qquad\quad\  (1\ \textrm{to}\ 3\times) \\
   \ 4(j{+}1)^2 & \for v=j{+}1 \qquad (1\ \textrm{to}\ 2\times) \\
   \ 4(j^2{+}3j{+}3)  & \for v=j{+}2 \qquad (1\times)
\end{cases}\ ,
\end{equation}
but the constraint removes one copy from each of the three middle cases (and less when $j{<}1$).
We end up with a spectrum $\{\omega^2\}=\{-2,1,6,7,10,\ldots\}$ with certain degeneracies~\cite{Hosotani}.
The single non-degenerate negative mode $\omega^2_{(0,0)}{=}{-}2$ is a singlet, $\Phi_a^p=\delta_a^p\phi(\tau)$, 
and it corresponds to rolling down the local maximum of the double-well potential. 
The meron is stable against all other perturbations.

For a time-varying background, the natural frequencies $\Omega_{(j,v,\beta)}$
inherit a $\tau$ dependence from the background~$\psi(\tau)$.
Direct diagonalization is still possible for $j{=}0$, where we should solve
\begin{equation}
\begin{aligned}
   & \partial_\tau\Phi_0 - \sfrac12(1{+}\psi)\vec{T}{\cdot}\vec{\Phi} \= 0 \ ,\\[4pt]
   & \bigl[\partial_\tau^2+2(1{+}\psi)^2\bigr] {\Phi}_0 - \dot{\psi}\,\vec{T}{\cdot}\vec{\Phi} \= 0\ ,\\[4pt]
   & \bigl[ \partial_\tau^2 +2(3\psi^2{-}1) -\sfrac14(1{+}\psi)(2{-}\psi)(\vec{S}{+}\vec{T})^2 \bigr]\ 
   \vec\Phi - \dot{\psi}\,\vec{T}\,\Phi_0 \= 0\ ,
\end{aligned}
\end{equation}
with
\begin{equation}
   (\vec{S}{+}\vec{T})^2\=\vec{V}^2\=-4\,v(v{+}1)\=0,-8,-24 \qquad\textrm{for}\quad v=0,1,2\ .
\end{equation}
It implies the unperturbed frequencies (suppressing the $j$ index)
\begin{equation} \label{unperturbed}
\begin{aligned}
   \bar\omega_{(1)}^2 &= 2(\psi{+}1)^2\ \ (3\times)\ ,\quad
   \omega_{(0)}^2 = 2(3\psi^2{-}1)\ \ (1\times)\ ,\\
   \omega_{(1)}^2 &= 2(2\psi^2{+}\psi{+}1)\ \ (3\times) ,\quad
   \omega_{(2)}^2 = 2(3\psi{+}5)\ \ (5\times)
\end{aligned}
\end{equation}
for
\begin{equation}
\begin{aligned}
   (\Phi_0)^p &\equiv \bigl(\Phi_{(\bar{v}=1)}\bigr)^p\ =:\ \delta^{pb}\bar\phi_b \ ,\\
   (\vec\Phi)_a^p \ &\equiv \bigl(\Phi_{(0)}+\Phi_{(1)}+\Phi_{(2)}\bigr)^p_a\ =:\
   \phi\,\delta_a^p + \epsilon^p_{\ ab}\,\phi_b + (\phi_{(ab)}{-}\delta_{ab}\phi)\delta^{bp}\ ,
\end{aligned}
\end{equation}
as long as $\dot{\psi}$ is ignored.
There are no $v$-spin multiplicities (larger than one) here.
Turning on $\dot{\psi}$ and observing that $(\vec{T}{\cdot}\vec\Phi)^p\sim\delta^{pb}\phi_b$, 
the characteristic polynomial of the coupled $12{\times}12$ system in the $|uvm\>$ basis reads
\begin{equation}
   {\cal P}_0(\lambda) \= \det \begin{pmatrix}
   (\bar{\omega}_{(1)}^2{-}\lambda)\unity_3 & 
   0 & -\dot{\psi}\,T_{(1)}^\top & 0 \\[4pt]
   0 & (\omega_{(0)}^2{-}\lambda)\unity_1 & 0 & 0 \\[4pt]
   -\dot{\psi}\,T_{(1)} & 0 & (\omega_{(1)}^2{-}\lambda)\unity_3 & 0 \\[4pt]
   0 & 0 & 0 & (\omega_{(2)}^2{-}\lambda)\unity_5 
\end{pmatrix}\ .
\end{equation}
Specializing the general discussion above to $j{=}0$, we find just $t_1{=}1$ so that $P_1{=}1$ and arrive at
\begin{equation}
\begin{aligned}
   {\cal P}_0(\lambda) &\= (\omega_{(0)}^2{-}\lambda)^1
   (\omega_{(1)}^2{-}\lambda)^3 (\omega_{(2)}^2{-}\lambda)^5
   \bigl[(\bar{\omega}_{(1)}^2{-}\lambda) - 8\dot{\psi}^2(\omega_{(1)}^2{-}\lambda)^{-1} \bigr]^3 \\
   &\= (\omega_{(0)}^2{-}\lambda) (\omega_{(2)}^2{-}\lambda)^5
   \bigl\{ (\bar{\omega}_{(\bar{1})}^2{-}\lambda)(\omega_{(1)}^2{-}\lambda) - 8\dot{\psi}^2 \bigr\}^3 \ .
\end{aligned}
\end{equation}
We see that the frequencies $\Omega_{(0)}^2{=}\omega_{(0)}^2$ and $\Omega_{(2)}^2{=}\omega_{(2)}^2$ 
are unchanged and given by~(\ref{unperturbed}), while the gauge mode $\bar{\omega}_{(\bar{1})}^2$ gets
entangled with the (unphysical) $v{=}1$ mode to produce the pair
\begin{equation}
\begin{aligned}
   \Omega^2_{(1,\pm)} &\= \sfrac12(\bar\omega_{(\bar{1})}^2{+}\omega_{(1)}^2)
   \,\pm\sqrt{\sfrac14(\bar\omega_{(\bar{1})}^2{+}\omega_{(1)}^2)^2-\bar\omega_{(\bar{1})}^2\omega_{(1)}^2+8\dot{\psi}^2}  \\
   &\= 3\psi^2{+}3\psi{+}2\, \pm\sqrt{\psi^2(\psi{-}1)^2+8\dot{\psi}^2}
\end{aligned}
\end{equation}
with a triple degeneracy. There are avoided crossings at $\psi{=}0$ and $\psi{=}1$.
Removing the unphysical and gauge modes in pairs, we remain with the singlet mode
$\Omega_{(0,0)}^2$ and the fivefold-degenerate $\Omega_{(0,2)}^2$. 
For all higher spins $j{>}0$, analytic expressions for the natural frequencies $\Omega_{(j,v,\beta)}$
now require merely solving a few polynomial equations of order four at worst.
We have done so up to $j{=}2$ and list them in the Appendix 
but refrain from giving further explicit examples here.
\begin{figure}[h!]
\centering
\captionsetup{width=0.9\linewidth}
\includegraphics[width = 0.35\paperwidth]{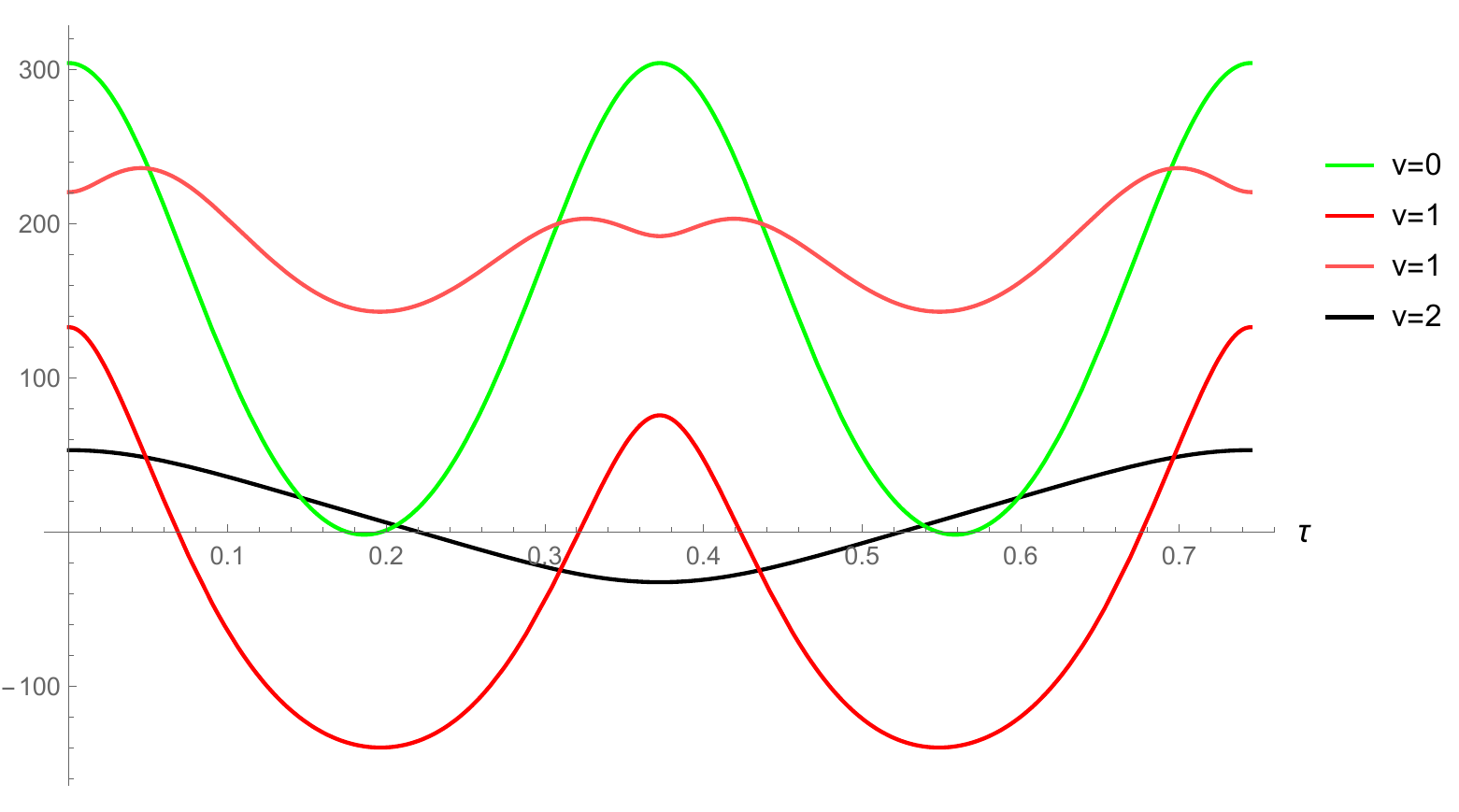} \quad
\includegraphics[width = 0.35\paperwidth]{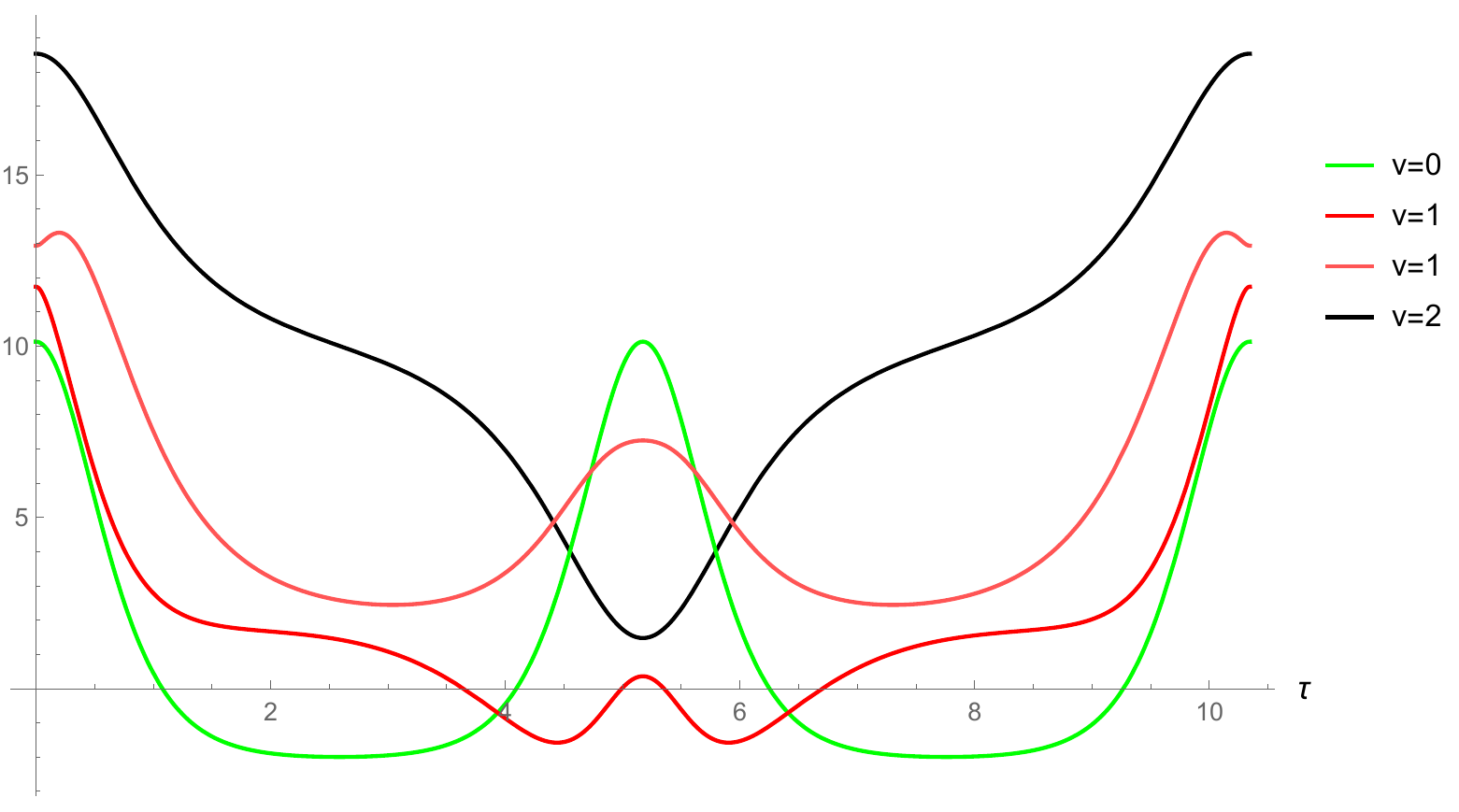} \\[12pt]
\includegraphics[width = 0.35\paperwidth]{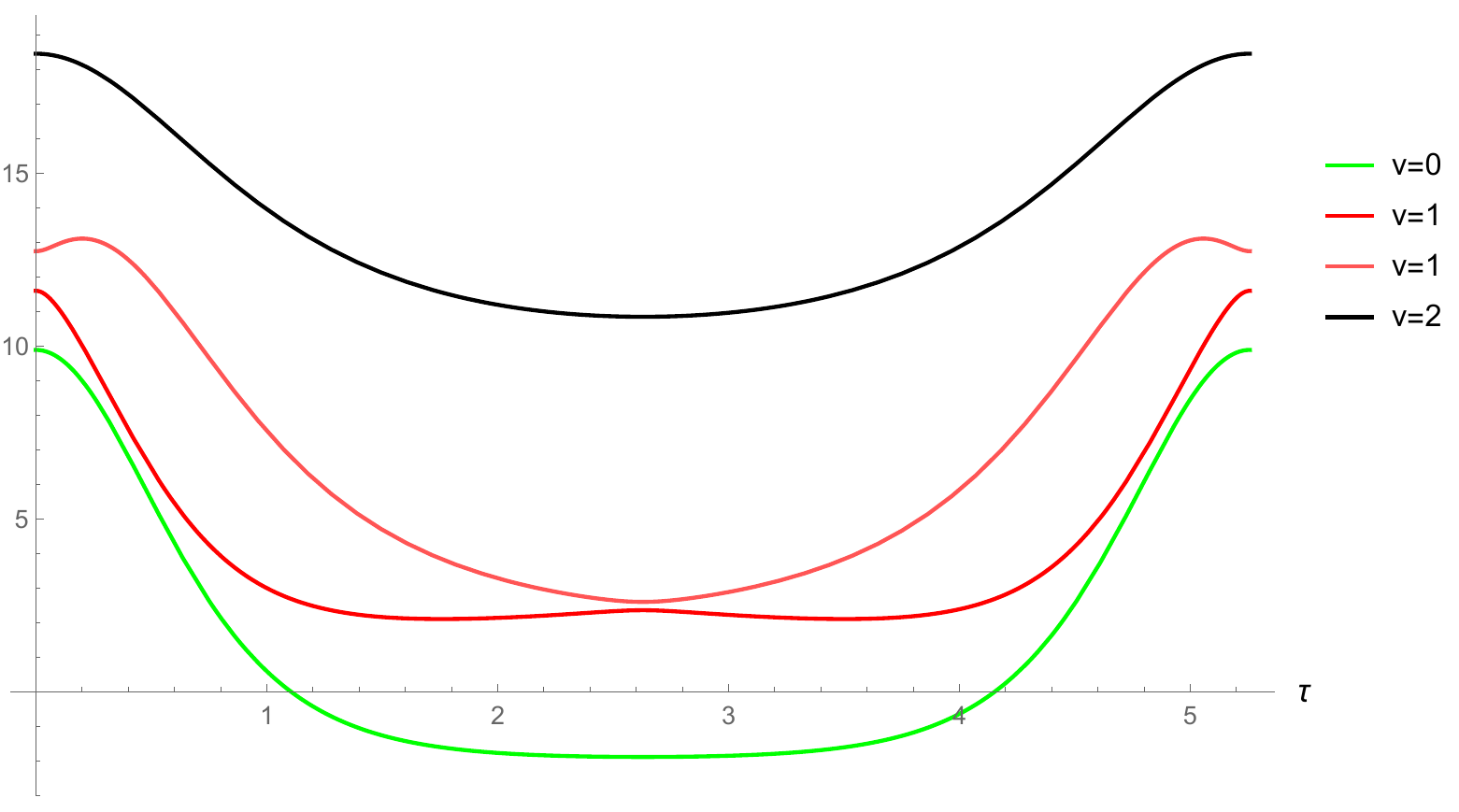} \quad
\includegraphics[width = 0.35\paperwidth]{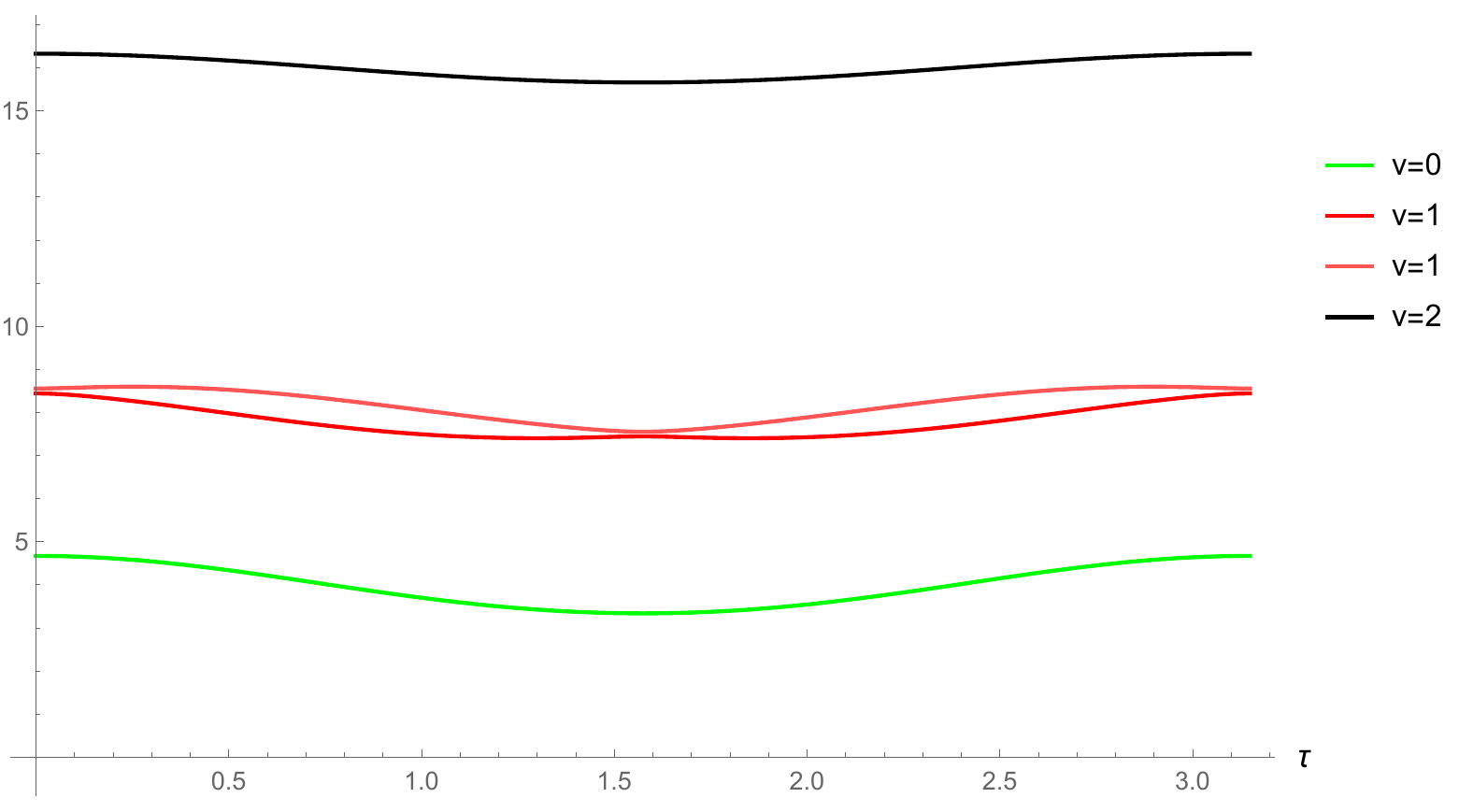}
\caption{Plots of $\Omega^2_{(0,v,\beta)}(\tau)$ over one period, for different values of $k^2$: 
$0.51$ (top left), $0.99$ (top right), $1.01$ (bottom left) and $5$ (bottom right).}
\end{figure}
\begin{figure}[h!]
\centering
\captionsetup{width=0.9\linewidth}
\includegraphics[width = 0.35\paperwidth]{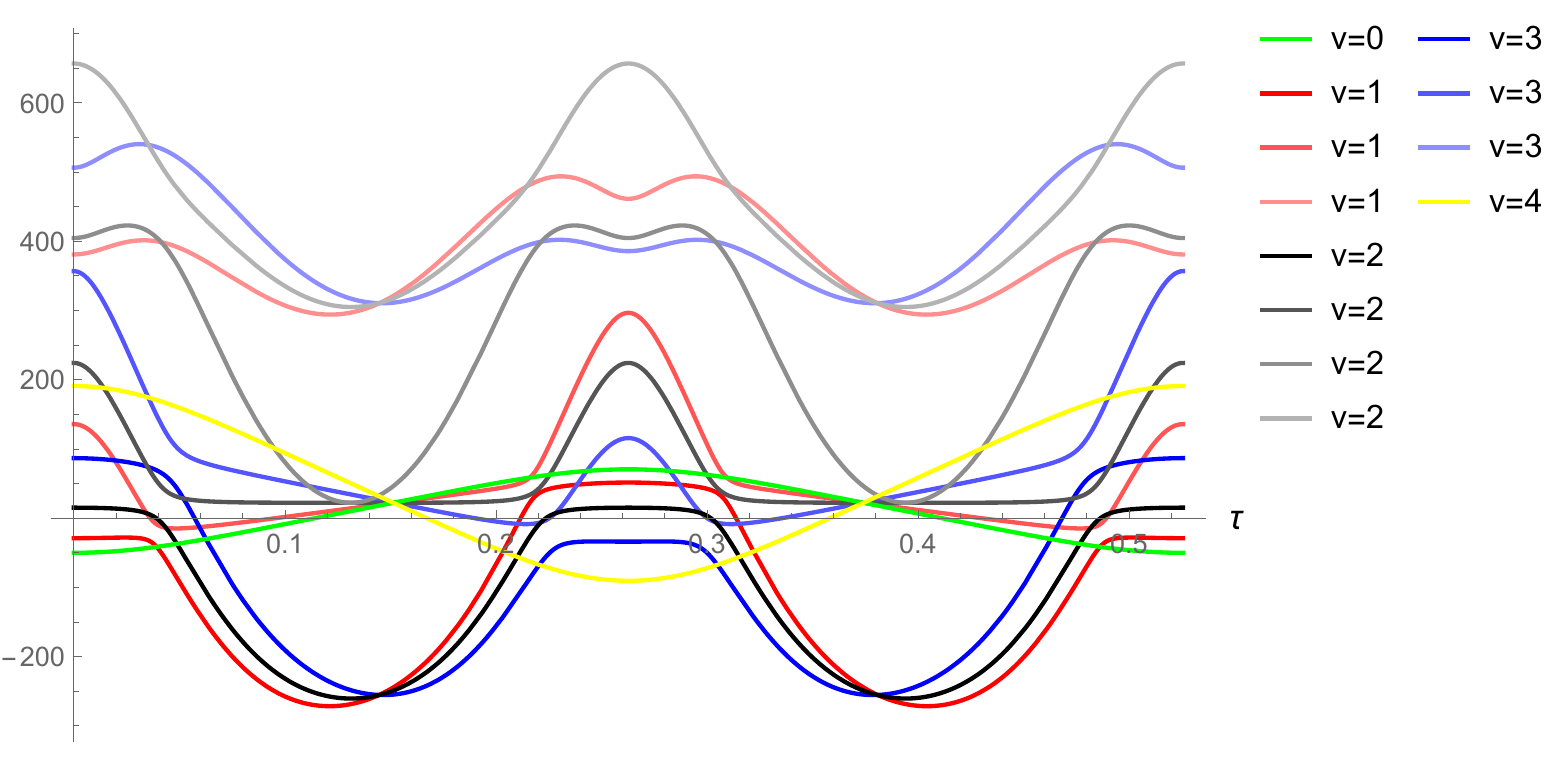} \quad
\includegraphics[width = 0.35\paperwidth]{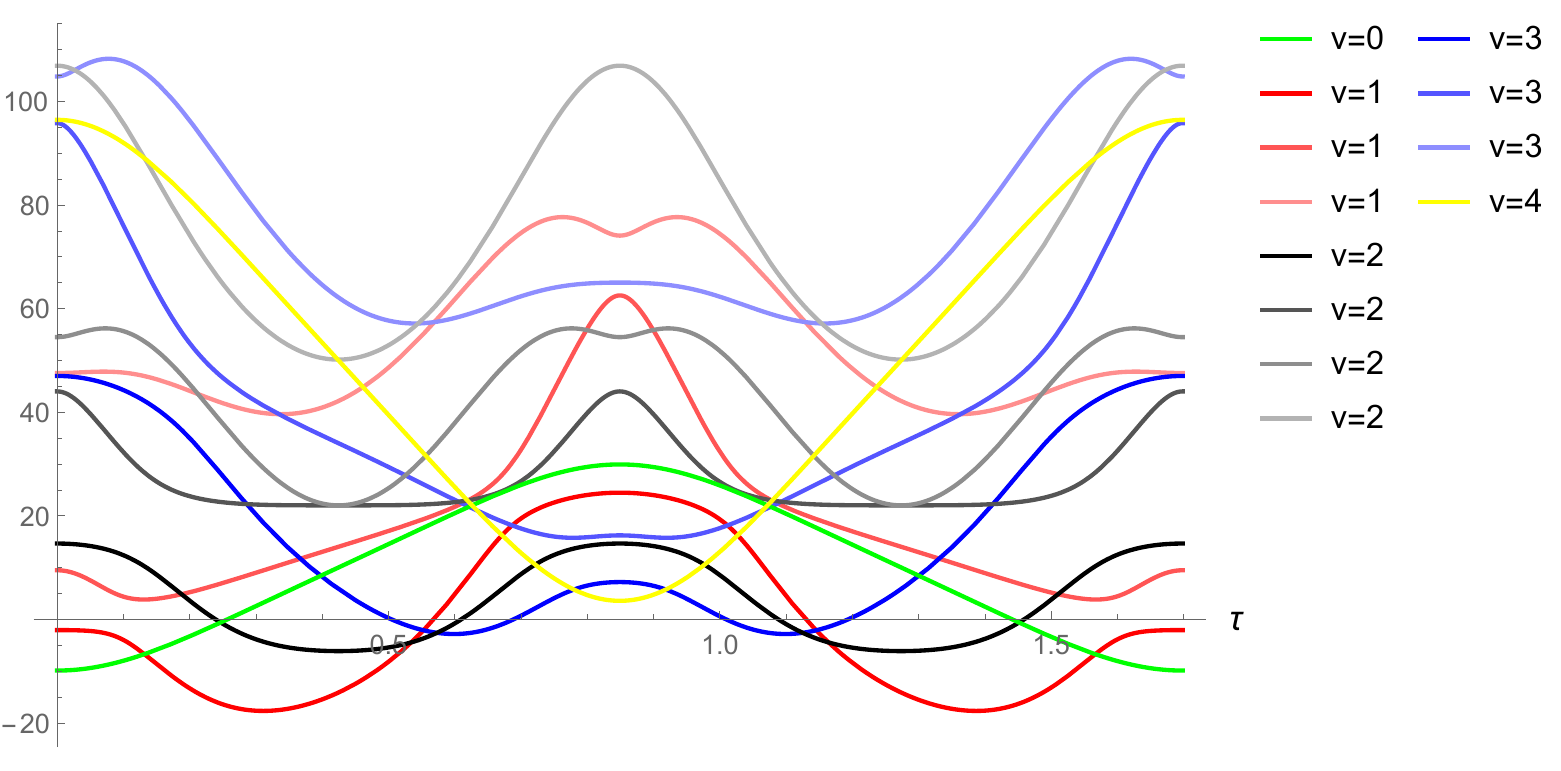} \\[12pt]
\includegraphics[width = 0.35\paperwidth]{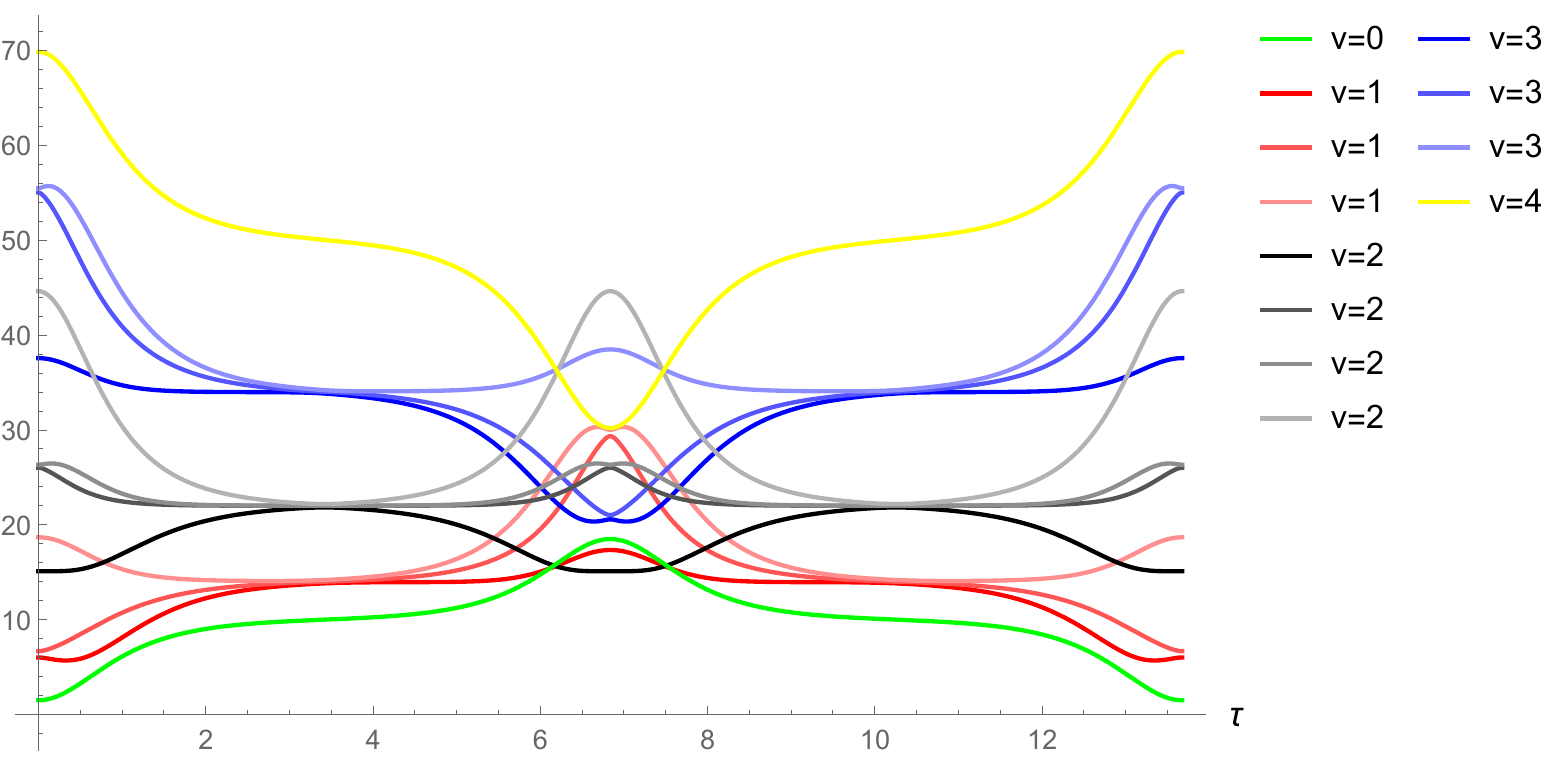} \quad
\includegraphics[width = 0.35\paperwidth]{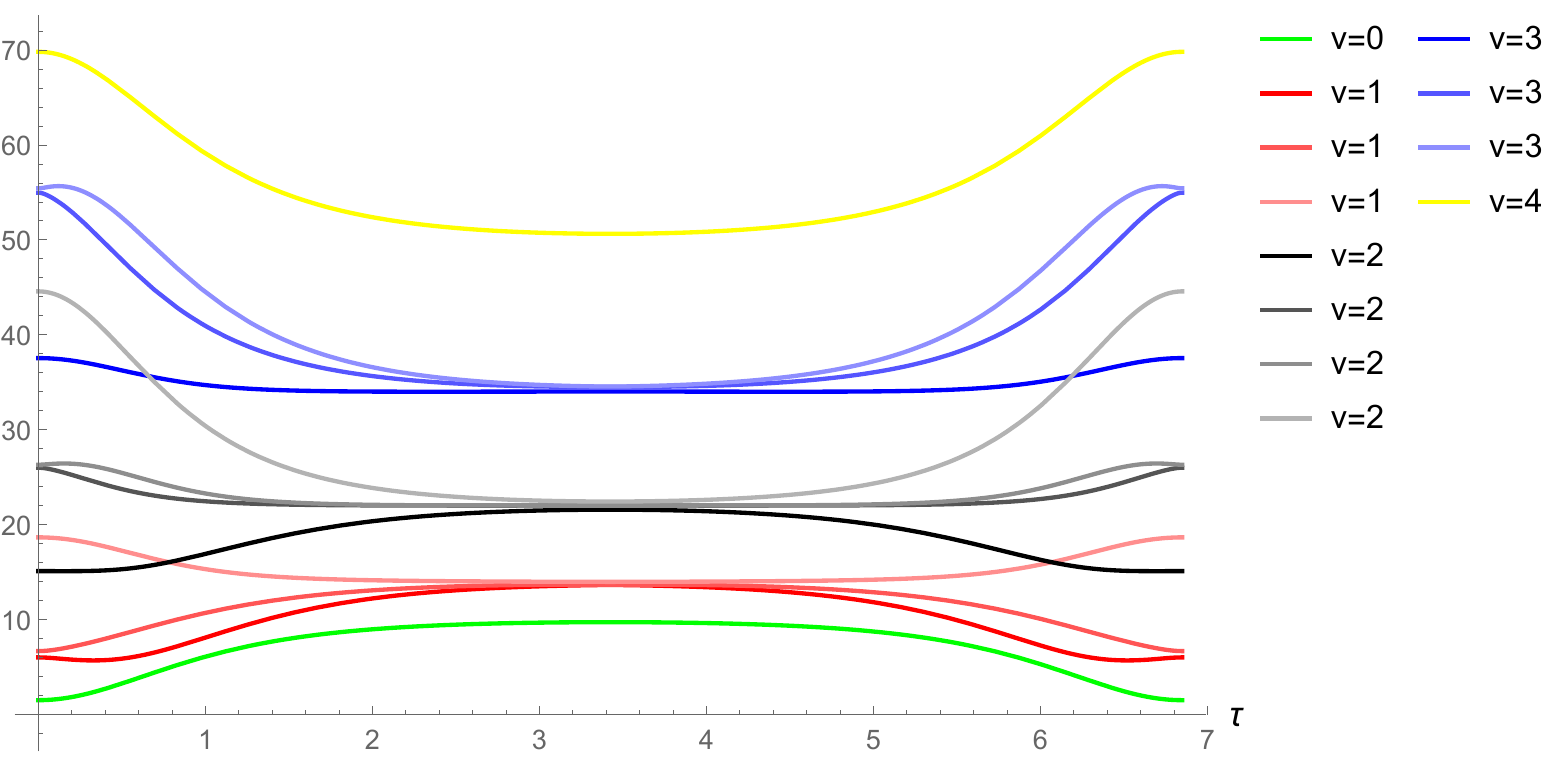}
\caption{Plots of $\Omega^2_{(2,v,\beta)}(\tau)$ over one period, for different values of $k^2$: 
$0.505$ (top left), $0.550$ (top right), $0.999$ (bottom left) and $1.001$ (bottom right).}
\end{figure}
From the cases of $j{=}0$ and $j{=}2$ displayed below one can see that some of the normal modes
dip into the negative regime, i.e.~their frequency-squares become negative, for a certain fraction of the time~$\tau$.
Because of this and, quite generally, due to the $\tau$ variability of the natural frequencies,
it is not easy to predict the long-term evolution of the fluctuation modes.
Clearly, the stability of the zero solution $\Phi{\equiv}0$, 
equivalent to the linear stability of the background Yang--Mills configuration, 
is not simply decided by the sign of the $\tau$-average of the corresponding frequency-square.

\vspace{12pt}
%----------------------------------------------------------------------------------------
\section{Stability analysis: stroboscopic map and Floquet theory}
\vspace{2pt}
\noindent
The diagonalized linear fluctuation equation~(\ref{diagonalized}) represents
a bunch of Hill's equations, where the frequency-squared is a root of a polynomial of order up to four
with coefficients given by a polynomial of twice that order in Jacobi elliptic functions.
A unique solution requires fixing two initial conditions, and so
for each fluctuation~$\Phi_{(j,v,\beta)}$ there is a two-dimensional solution space.
It is well known that Hill's equation, e.g.~in the limit of Mathieu's equation, displays parametric resonance phenomena,
which can stabilize otherwise unstable systems or destabilize otherwise stable ones.

For oscillating dynamical systems with periodically varying frequency, 
there exist some general tools to analyze linear stability.
Switching to a Hamiltonian picture and to phase space, 
it is convenient to transform the second-order differential equation into
a system of two coupled first-order equations (suppressing all quantum numbers),
\begin{equation} \label{first-order}
   \bigl[\partial^2_\tau+\Omega^2(\tau)\bigr] \Phi(\tau) \= 0 \quad\Leftrightarrow\quad
   \partial_\tau \begin{pmatrix} \Phi \\[4pt] \dot\Phi \end{pmatrix} \=
   \begin{pmatrix} 0 & 1 \\[4pt] -\Omega^2 & 0 \  \end{pmatrix}
   \begin{pmatrix} \Phi \\[4pt] \dot\Phi \end{pmatrix} \ =:\
   \im\,\widehat\Omega(\tau)\,\begin{pmatrix} \Phi \\[4pt] \dot\Phi \end{pmatrix}\ ,
\end{equation}
where the frequency~$\Omega(\tau)$ is $T$-periodic (sometimes $\frac{T}{2}$-periodic) in~$\tau$.
The solution to this first-order system is formally given by
\begin{equation}
   \begin{pmatrix} \Phi \\[4pt] \dot\Phi \end{pmatrix}(\tau) \=
   {\cal T} \exp\,\Bigl\{ \int_0^\tau \!\diff\tau'\ \im\,\widehat\Omega(\tau') \Bigr\} \,
   \begin{pmatrix} \Phi \\[4pt] \dot\Phi \end{pmatrix}(0) \ ,
\end{equation}
where ${\cal T}$ denotes time ordering.
Because of the time dependence of $\Omega$, the time evolution operator above
is not homogeneous thus does not constitute a one-parameter group, except when
the propagation interval is an integer multiple of the period~$T$. For $\tau{=}T$,
one speaks of the stroboscopic map \cite{Arnold}
\begin{equation}
   M\ :=\ {\cal T} \exp\,\Bigl\{ \int_0^T \!\diff\tau\ \im\,\widehat\Omega(\tau) \Bigr\} 
   \qquad\Rightarrow\qquad
   \begin{pmatrix} \Phi \\[4pt] \dot\Phi \end{pmatrix}(nT) \= M^n
   \begin{pmatrix} \Phi \\[4pt] \dot\Phi \end{pmatrix}(0)\ .
\end{equation}
The linear map~$M$ is a functional of the chosen background solution~$\psi$
and hence depends on its parameter~$E$ or~$k$.
This background is Lyapunov stable if the trivial solution $\Phi{\equiv}0$ is,
which is decided by the two eigenvalues $\mu_1$ and $\mu_2$ of~$M$. 
Since the system is Hamiltonian, $\det M{=}1$, we have three cases:
\begin{equation}
\begin{aligned}
   |\tr\,M| > 2 &\quad\Leftrightarrow\quad \mu_i\in\R  
   &\quad\;\Leftrightarrow\quad& \textrm{hyperbolic/boost} 
   &\quad\Leftrightarrow\quad& \textrm{strongly unstable} \ ,\\
   |\tr\,M| = 2  &\quad\Leftrightarrow\quad \mu_i=\pm1  
   &\quad\Leftrightarrow\quad& \textrm{parabolic/translation} 
   &\quad\Leftrightarrow\quad& \textrm{marginally stable}\ , \\
   |\tr\,M| < 2  &\quad\Leftrightarrow\quad \mu_i\in\textrm{U}(1)  
   &\quad\Leftrightarrow\quad& \textrm{elliptic/rotation} 
   &\quad\Leftrightarrow\quad&\textrm{strongly stable}\ .
\end{aligned}
\end{equation}
Clearly, $|\tr\,M|$ determines the linear stability of our classical solution.

Let us thus try to evaluate the trace of the stroboscopic map~$M$,
making use of the special form of the matrix~$\widehat\Omega$,
\begin{equation}
\begin{aligned}
   \tr\,M &\= \sum_{n=0}^\infty \im^n 
   \int_0^T\!\!\diff\tau_1\int_0^{\tau_1}\!\!\!\diff\tau_2\ \ldots\int_0^{\tau_{n-1}}\!\!\!\!\diff\tau_n\
   \tr\,\bigl[ \widehat\Omega(\tau_1)\,\widehat\Omega(\tau_2)\cdots\widehat\Omega(\tau_n)\bigr] \\[4pt]
   &\= 2 +  \sum_{n=1}^\infty(-1)^n 
   \int_0^T\!\!\diff\tau_1\int_0^{\tau_1}\!\!\!\diff\tau_2\ \ldots\int_0^{\tau_{n-1}}\!\!\!\!\diff\tau_n\
   H_n(\tau_1,\tau_2,\ldots,\tau_n)\Omega^2(\tau_1)\,\Omega^2(\tau_2)\,\cdots\Omega^2(\tau_n) \\[4pt]
   \textrm{with}& \quad H_n(\tau_1,\tau_2,\ldots,\tau_n) \= 
   (\tau_1{-}\tau_2)(\tau_2{-}\tau_3)\cdots(\tau_{n-1}{-}\tau_n)(\tau_n{-}\tau_1{+}1)
   \und H_1(\tau_1)=1\ .
\end{aligned}
\end{equation}
It is convenient to scale the time variable such as to normalize the period to unity,
\begin{equation}
   \tau = T\, x  \qquad\und\qquad \Omega^2(Tx) =: \omega^2(x)\ ,\quad H(\{Tx\})=:h(\{x\}) \ ,
\end{equation}
hence
\begin{equation} \label{Ansum}
\begin{aligned}
   \tr\,M &\= 2 +  \sum_{n=1}^\infty \bigl(-T^2\bigr)^n 
   \int_0^1\!\!\!\diff x_1\int_0^{x_1}\!\!\!\diff x_2 \ldots\int_0^{x_{n-1}}\!\!\!\!\diff x_n\,
   h_n(x_1,x_2,\ldots,x_n)\omega^2(x_1)\omega^2(x_2)\cdots\omega^2(x_n) \\[4pt]
   &\= \sum_{n=0}^\infty \frac{2}{(2n)!}\,M_n\,\bigl(-T^2\bigr)^n 
   \ =:\  2 - M_1 T^2 + \sfrac{1}{12}M_2 T^4 - \sfrac{1}{360}M_3 T^6 + \sfrac{1}{20160}M_4 T^8 - \ldots\ .
\end{aligned}
\end{equation}

It is impossible to evaluate the integrals $M_n$ without explicit knowledge of~$\omega^2(x)$. 
As a crude guess, we replace the weight function by its (constant) average value
\begin{equation}
   \<h_n\> \ :=\ \frac{1}{n!} \int_0^1\!\!\diff x_1\int_0^{x_1}\!\!\!\diff x_2\ \ldots\int_0^{x_{n-1}}\!\!\!\!\diff x_n\ 
   h_n(x_1,x_2,\ldots,x_n) \= \frac{2\,n!}{(2n)!}
\end{equation}
and obtain
\begin{equation}
   M_n \= \frac{(2n)!}{2}\,\<h_n\> 
   \int_0^1\!\!\diff x_1\int_0^{x_1}\!\!\!\diff x_2\ \ldots\int_0^{x_{n-1}}\!\!\!\!\diff x_n\ \prod_{i=1}^n \omega^2(x_i)
   \= \Bigl( \int_0^1\!\!\diff x\ \omega^2(x) \Bigr)^n \ =:\ \< \,\omega^2\>^n\ ,
\end{equation}
which yields
\begin{equation}
   \tr\,M \= 2\,\sum_{n=0}^\infty \frac{(-1)^n}{(2n)!}\,\<\,\omega^2\>^n\,T^{2n} 
   \= 2\,\cos\bigl(\sqrt{\<\,\omega^2\>}\,T\bigr)\ .
\end{equation}
This expression indicates stability as long as $\<\omega^2\>>0$.
However, the result for the $j{=}0$ singlet mode $\omega^2=\Omega^2_{(0,0)}$ in (\ref{j0average}) 
already showed that the averaged frequency-squared may turn negative in certain domains
thus changing the cos into a cosh there.

To do better, let us look at the individual terms~$M_n$ in~(\ref{Ansum}) for the simplest case
of the SO(4) singlet fluctuation, i.e.~$\Omega^2_{(0,0)}=6\psi^2{-}2$ in~(\ref{unperturbed}). 
Its average frequency-square is easily computed to be
\begin{equation} \label{j0average}
   \<\,\Omega_{(0,0)}^2 \> \= \frac{1}{\epsilon^2}\Bigl(6 \frac{E(k)}{K(k)}+4k^2-5\Bigr)\ , 
\end{equation}
where $E(k)$ and $K(k)$ denote the second and first complete elliptic integrals, respectively.
Plotting this expression as a function of the modulus~$k$, we see that it
becomes negative only in a very narrow range around $k{=}1$, namely for $|k{-}1|\lesssim0.00005$.
\begin{figure}[h!]
\centering
\includegraphics[width = 0.35\paperwidth]{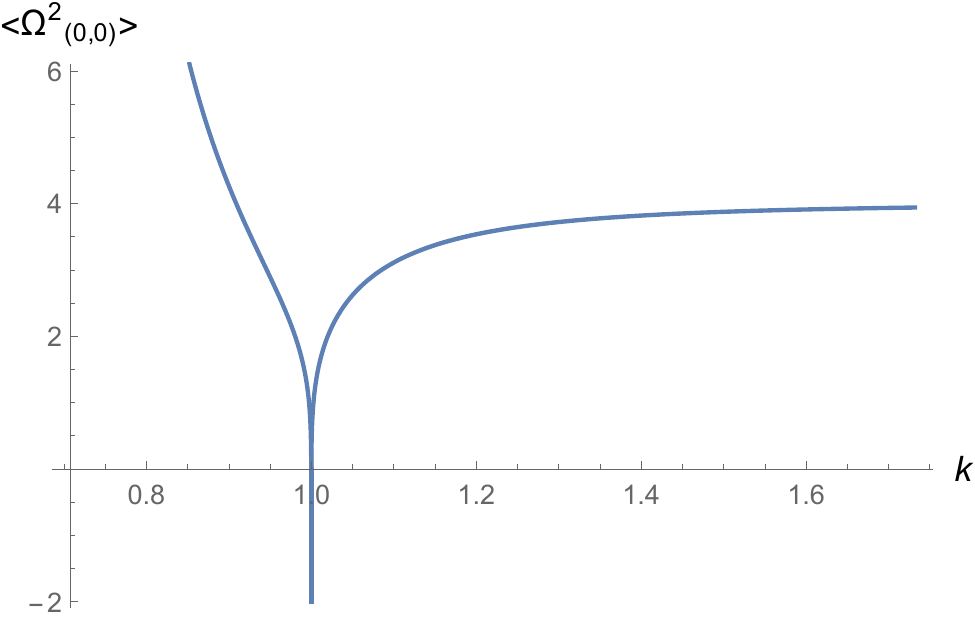} \quad
\includegraphics[width = 0.35\paperwidth]{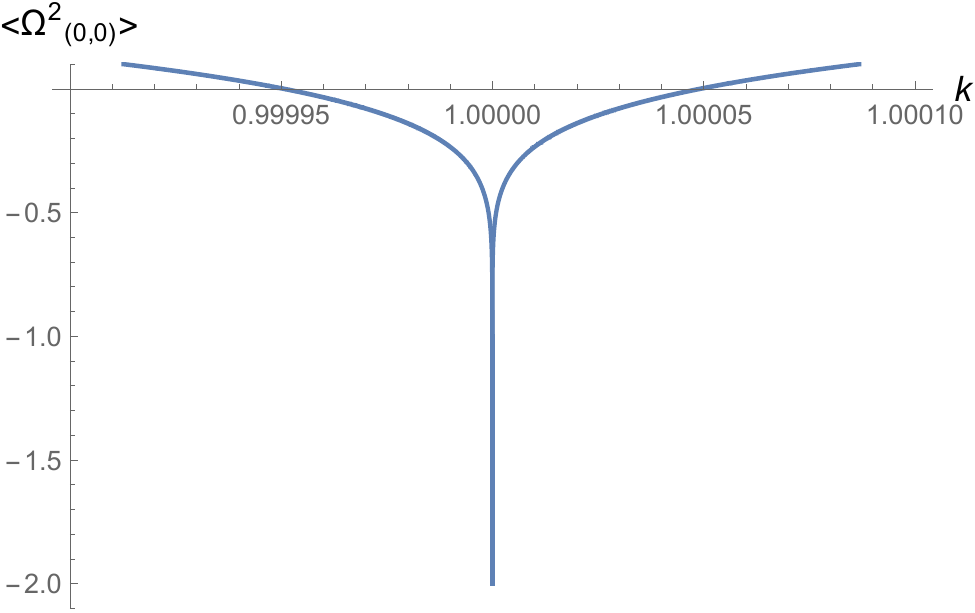}
\caption{Plot of $\<\Omega^2_{(0,0)}\>$ as a function of~$k$, with detail on the right.}
\end{figure}
We have only been able to analytically evaluate (with $k{<}1$ for simplicity)
\begin{equation}
   M_1 \= \<\,\Omega_{(0,0)}^2 \>  \und
   M_2 \= \<\,\Omega_{(0,0)}^2 \> ^2\ -\ \frac{1}{\epsilon^4}\Bigl(
   9\frac{2{-}k^2}{K(k)^2} - 27\frac{E(k)}{K(k)^3} + \frac{9\,\pi^2}{4\,K(k)^4} \Bigr)\ ,
\end{equation}
which does not suffice to rule out instability.
Indeed, numerical studies show that $M_n$ as a function of~$k$ looses its positivity in a range
around $k{=}1$ which increases with~$n$, where the series~(\ref{Ansum}) ceases to be alternating.
Moreover, even in the limit of a very large background amplitude, $k^2\to\frac12$, we find that
\begin{equation}
   \<\,\Omega_{(0,0)}^2 \>\ \to\ \frac{24\,\pi^2}{\epsilon^2\,\Gamma(\frac14)^4}\ \approx\ \frac{1.37}{\epsilon^2}
   \qquad\Rightarrow\qquad \sqrt{\<\,\Omega^2\>}\,T\ \to\ \sqrt{24\,\pi}\ \approx\ 8.68\ ,
\end{equation}
implying that we must push the series in~(\ref{Ansum}) at least to $O(M_{10}T^{20})$,
even though it turns out that $M_n<\<\,\Omega^2_{(0,0)}\>^n$ at $k^2=\frac12$ for $n>1$.

For a more complete analysis of linear stability in an oscillating system with time-dependent frequency
we can take recourse to Floquet theory.
It tells us that a general fundamental matrix solution
\begin{equation}
   \widehat\Phi(\tau) \= \begin{pmatrix} \Phi_1 & \Phi_2 \\[4pt] \dot\Phi_1 & \dot\Phi_2 \end{pmatrix}(\tau)
   \qquad\Rightarrow\qquad \partial_\tau \widehat\Phi(\tau) \= \im\,\widehat\Omega(\tau)\,\widehat\Phi(\tau)
\end{equation}
of our system~(\ref{first-order}) with some initial condition $\widehat\Phi(0)=\widehat\Phi_0$ 
can be expressed in so-called Floquet normal form as
\begin{equation}
   \widehat\Phi(\tau) \= Q(\tau)\;\ep^{\tau R} 
   \qquad\textrm{with}\qquad Q(\tau{+}2T) \= Q(\tau)\ ,
\end{equation}
where $Q(\tau)$ and $R$ are real $2{\times}2$ matrices,
so that the time dependence of the frequency can be transformed away by a change of coordinates,
\begin{equation}
   \Psi(\tau) \ :=\ Q(\tau)^{-1} \widehat\Phi(\tau) \qquad\Rightarrow\qquad
   \partial_\tau \Psi(\tau) \= R\,\Psi(\tau)\ .
\end{equation}
Due to the identity
\begin{equation}
   \widehat\Phi(\tau{+}T) \= \widehat\Phi(\tau)\,\widehat\Phi(0)^{-1}\,\widehat\Phi(T) 
   \= \widehat\Phi(T)\,\widehat\Phi(0)^{-1}\,\widehat\Phi(\tau) \= M\,\widehat\Phi(\tau)
\end{equation}
we see that our stroboscopic map~$M$ is nothing but the monodromy, and
\begin{equation}
   M^2 \= \widehat\Phi(2T)\,\widehat\Phi(0)^{-1} 
   \= Q(0)\,\widehat\Phi(0)^{-1}\,\widehat\Phi(2T)\,Q(0)^{-1}
   \= Q(0)\,\ep^{2 R T}\,Q(0)^{-1}\ ,
\end{equation}
so that its eigenvalues (or characteristic multipliers)
\begin{equation}
   \mu_i = \ep^{\rho_i T} \qquad\textrm{for}\quad i=1,2
\end{equation}
define a pair of (complex) Floquet exponents $\rho_i$ 
whose real parts are the Lyapunov exponents.
Since $\mu_1\mu_2=1$ implies that $\rho_1{+}\rho_2=0$,
our system is linearly stable if and only if both eigenvalues~$\rho_i$ of~$R$
are purely imaginary (or zero).

Generally it is impossible to find analytically the monodromy pertaining to
a normal mode~$\Phi_{(j,v,\beta)}$.\footnote{
An exception is the SO(4) singlet perturbation~$\Phi_{(0,0)}$, to be treated in the following section.}
However, we can get a qualitative understanding by looking numerically at some examples. 
Before numerically integrating Hill's equation, however, let us estimate at which energies~$E$
or, rather, moduli~$k$, possible resonance frequencies might occur.
To this end, we determine the period-average of the natural frequency $\Omega_{(j,v,\beta)}$ 
and compare it to its modulation frequency~$\sfrac{2\pi}{T}$. If we model 
\begin{equation}
   \Omega^2(\tau) \= \<\,\Omega^2\> \,\bigl(1+h(\tau)\bigr)  \with
   \<\,\Omega^2\> = \sfrac{1}{T}\smallint_0^T \!\diff\tau\ \Omega^2(\tau) \und
   h(\tau) \ \propto\ \cos(2\pi\tau/T) \ ,
\end{equation}
where $T=4\,\epsilon\,K(k)$, then the resonance condition is met for
\begin{equation}
   \sqrt{\<\Omega^2\>} \= \ell\,\frac{\pi}{T} \qquad\Rightarrow\qquad
   k=k_\ell(j,v,\beta)\qquad\textrm{for}\quad \ell=1,2,3,\ldots\ .
\end{equation}
Since this model reproduces only the rough features of $\Omega^2(\tau)$, we expect potential instability
due to parametric resonance effects in a band around or near the values~$k_\ell$.

Below we display, together with the would-be resonant values~$k_\ell$,
the function $\tr M(k)$ for the sample cases of $(j,v)=(2,0)$ and $(2,2)$.
\begin{figure}[h!]
\centering
\captionsetup{width=\linewidth}
\includegraphics[width = 0.35\paperwidth]{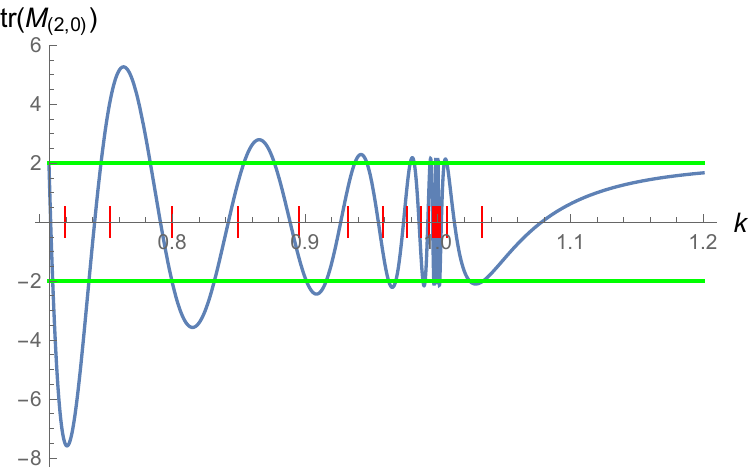} \quad
\includegraphics[width = 0.35\paperwidth]{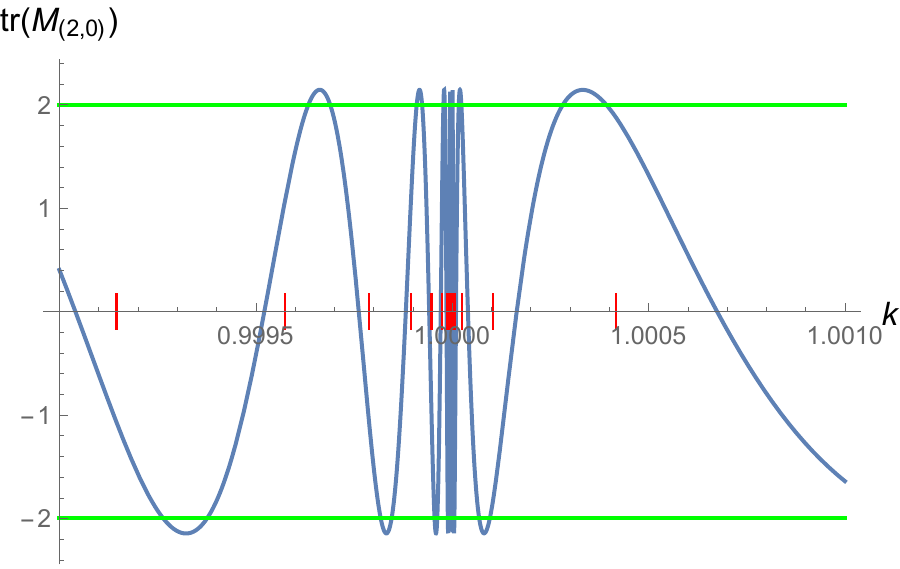} 
\caption{Plot of $\tr\,M(k)$ for $(j,v)=(2,0)$, with detail on the right. Would-be resonances marked in red.}
\end{figure}
\begin{figure}[h!]
\centering
\includegraphics[width = 0.35\paperwidth]{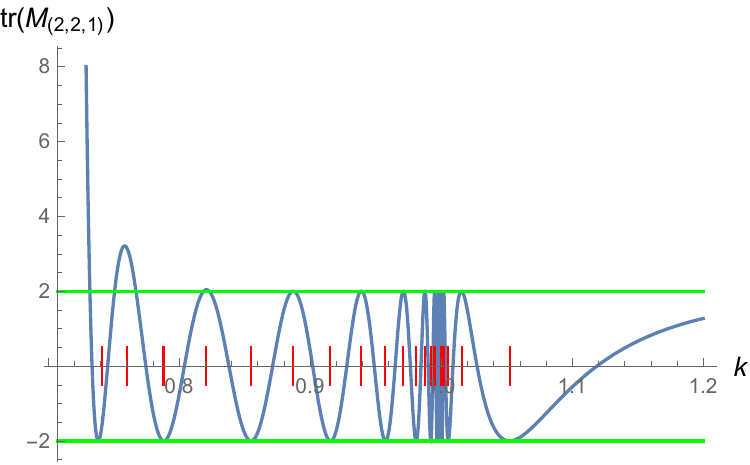} \quad
\includegraphics[width = 0.35\paperwidth]{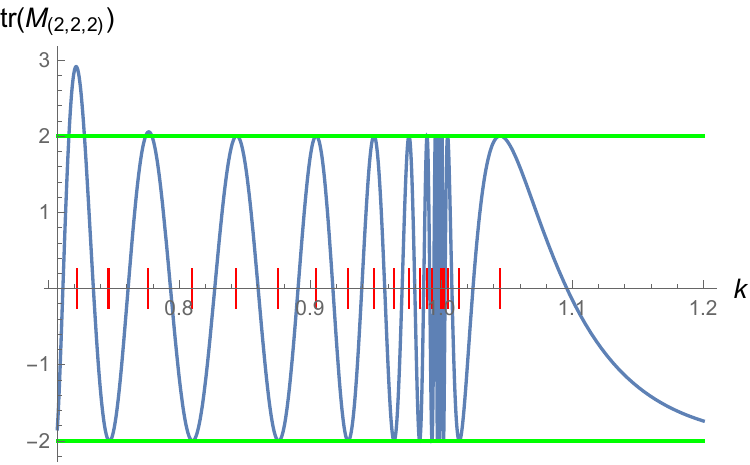} \\[8pt]
\includegraphics[width = 0.35\paperwidth]{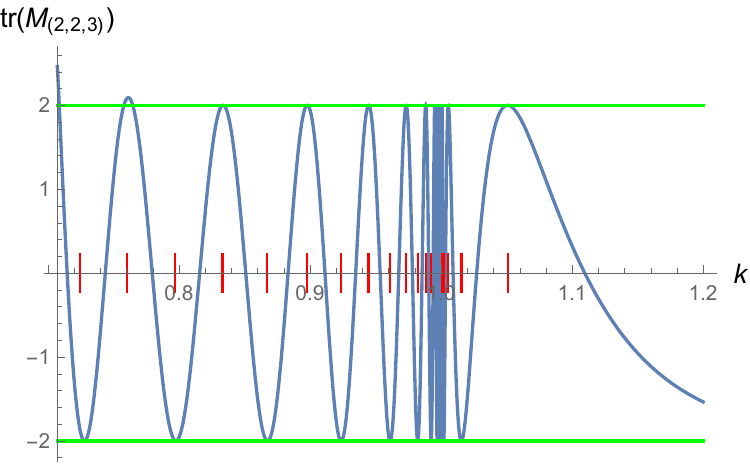} \quad
\includegraphics[width = 0.35\paperwidth]{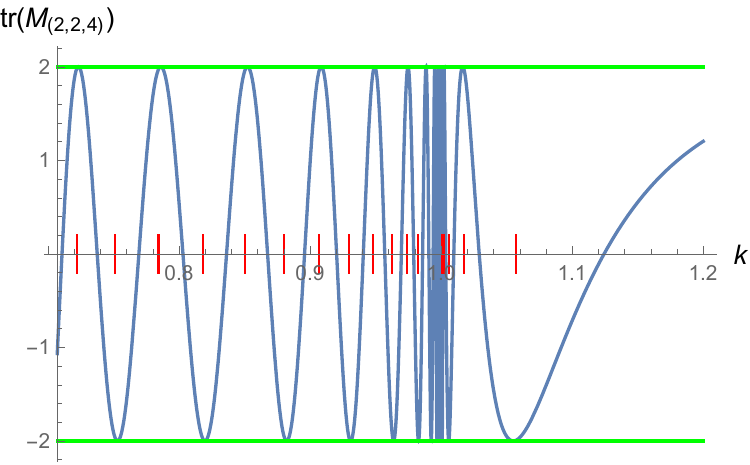} 
\caption{Plots of $\tr\,M(k)$ for $(j,v)=(2,2)$ and $\beta=1,2,3,4$. Would-be resonances marked in red.}
\end{figure}
One sees that, on both sides of the critical value of $E{=}\sfrac12$ (or $k{=}1$), 
corresponding to the double-well local maximum, the $k_\ell$ values accumulate at the critical point.
But while for $k{>}1$ (energy below the critical point) $\tr M(k)$ oscillates between values close to $2$ 
in magnitude and thus exponential growth is rare and mild, for $k{<}1$ (energy above the critical point)
the oscillatory behavior of $\tr M(k)$ comes with an amplitude exceeding~2 and growing with energy.
Hence, in this latter regime stable and unstable bands alternate.
This is supported by long-term numerical integration, as we demonstrate by plotting
$\Phi(\tau)$ for $(j,v,\beta)=(2,2,1)$ with initial values $\Phi(0){=}1$ and $\dot\Phi(0){=}0$ 
on both sides very close to the end of the first instability 
(at the highest value of~$E$ or the lowest value of~$k$).
\begin{figure}[h!]
\centering
\captionsetup{width=\linewidth}
\includegraphics[width = 0.35\paperwidth]{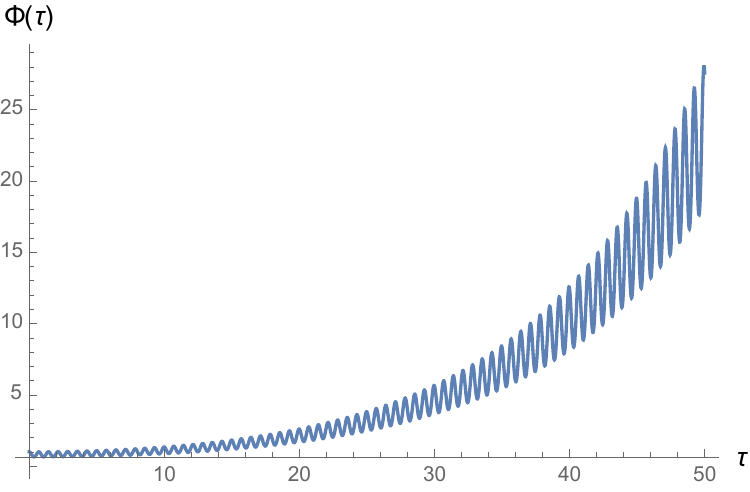} \quad
\includegraphics[width = 0.35\paperwidth]{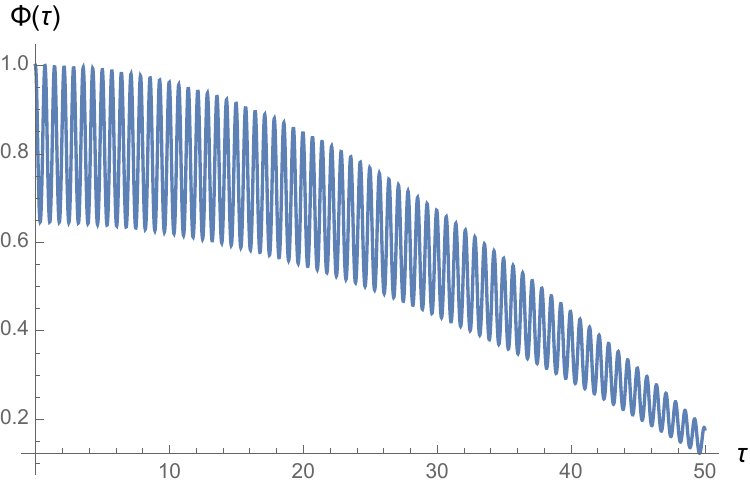} 
\caption{Plot of $\Phi(\tau)$ for $(j,v,\beta)=(2,2,1)$ and $k{=}0.73198$ (left) and $k{=}0.73199$ (right).}
\end{figure}

Most relevant for the cosmological application is the regime of very large energies, $E\to\infty$ (or $k\to 1/\sqrt{2}$).
In this limit, we observe the following universal behavior. 
Because the period~$T$ collapses with $\epsilon{=}\sqrt{k^2{-}1/2}$, we rescale
\begin{equation}
   \sfrac{\tau}{\epsilon} = z \in [0,4K(\sfrac12)]\ ,\quad
   \epsilon\,\psi = \tilde\psi\ ,\quad 
   \epsilon^2\dot\psi = \partial_z\tilde\psi\ ,\quad 
   \epsilon^2\Omega^2 = \tilde\Omega^2\ ,\quad 
   \epsilon^2\lambda = \tilde\lambda
\end{equation}
so that the tilded quantities remain finite in the limit,
and find, with $\bar\omega^2_{(\bar{v})}\to 2\psi^2$,\footnote{
For the cases $(j,v)=(0,1)$ and $(1,0)$, the factor $\tilde\lambda$ is missing; 
for $(j,v)=(0,0)$, one only has $R=Q\sim(\tilde\lambda{-}6\tilde\psi^2)$.}
\begin{equation}
\begin{aligned}
   &Q_{(v\pm2)}\ \sim\ \tilde\lambda\ , \\
   &Q_{(v\pm1)}\ \sim\ \tilde\lambda\,(\tilde\lambda-4\tilde\psi^2)\ ,\qquad\qquad R_{(v\pm1)}\ \sim\ \tilde\lambda\,\bigl[(\tilde\lambda-2\tilde\psi^2)(\tilde\lambda-4\tilde\psi^2)-8(\pa_z\tilde\psi)^2\bigr]\ ,\\
   &Q_{(v)}\ \sim\ \tilde\lambda\,(\tilde\lambda-4\tilde\psi^2)(\tilde\lambda-6\tilde\psi^2)\ ,\quad R_{(v)}\ \sim\ \tilde\lambda\,\bigl[(\tilde\lambda-2\tilde\psi^2)(\tilde\lambda-4\tilde\psi^2)-8(\pa_z\tilde\psi)^2\bigr](\tilde\lambda-6\tilde\psi^2)\ ,
\end{aligned}
\end{equation}
because all $j$-dependent terms in the polynomials are subleading and drop out in the limit.
Factorizing the $R$~polynomials, we find the four universal natural frequency-squares
\begin{equation}
   \tilde\Omega^2_1 = 0\ ,\quad
   \tilde\Omega^2_2 = 3\,\tilde\psi^2-\sqrt{\tilde\psi^4+8(\pa_z\tilde\psi)^2}\ ,\quad
   \tilde\Omega^2_3 = 3\,\tilde\psi^2+\sqrt{\tilde\psi^4+8(\pa_z\tilde\psi)^2}\ ,\quad
   \tilde\Omega^2_4 = 6\,\tilde\psi^2\ .
\end{equation}
One must pay attention, however, to the fact that the avoided crossings disappear in the $\epsilon\to0$ limit.
Therefore, the correct limiting frequencies to input into
\begin{equation}
   \bigl[ \pa_z^2 + \tilde\Omega^2_{(j,v,\beta)} \bigr]\,\tilde\Phi_{(j,v,\beta)} \= 0
\end{equation}
are
\begin{equation}
\begin{aligned}
   &\tilde\Omega^2_{(j,j\pm2)}\ \ = 0\ ,\\
   &\tilde\Omega^2_{(j,j\pm1,\beta)} \in \bigl\{ 
   \textrm{min}(\tilde\Omega^2_1,\tilde\Omega^2_2),\ \textrm{max}(\tilde\Omega^2_1,\tilde\Omega^2_2),\
   \tilde\Omega^2_3 \bigr\}\ ,\\
   &\tilde\Omega^2_{(j,j,\beta)} \ \ \ \in \bigl\{ 
   \textrm{min}(\tilde\Omega^2_1,\tilde\Omega^2_2),\ \textrm{max}(\tilde\Omega^2_1,\tilde\Omega^2_2),\ 
   \textrm{min}(\tilde\Omega^2_3,\tilde\Omega^2_4),\ \textrm{max}(\tilde\Omega^2_3,\tilde\Omega^2_4) \bigr\}\ ,
\end{aligned}
\end{equation}
of which we show below the last list as a function of~$z$. 
\begin{figure}[h!]
\centering
\captionsetup{width=0.9\linewidth}
\includegraphics[width = 0.5\paperwidth]{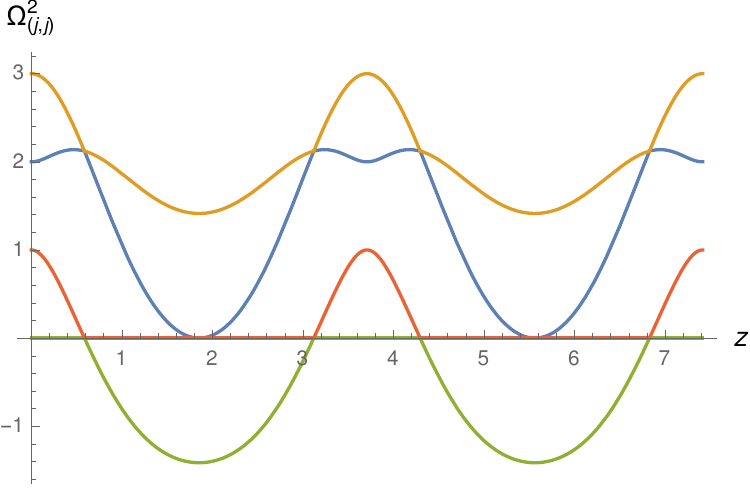} 
\caption{Plot of the universal limiting natural frequency-squares 
$\tilde\Omega^2_{(j,v,\beta)}$ for $v{=}j$ and $\beta=1,2,3,4$.}
\end{figure}
The monodromies are easily computed numerically,\footnote{
For the cases $(j,v)=(0,1)$ and $(1,0)$ one gets $\{56.769,\ -1.659\}$;
for $(j,v)=(0,0)$ we have $\tr\,M=2$.}
\begin{equation}
\begin{aligned}
   &\tr\,M_{(j,j\pm2)}(E{\to}\infty)\ \ \ = 2\ ,\\
   &\tr\,M_{(j,j\pm1,\beta)}(E{\to}\infty)\ \in \bigl\{ 306.704,\ -1.842,\ -1.659 \bigr\}\ ,\\
   &\tr\,M_{(j,j,\beta)}(E{\to}\infty) \ \ \ \ \in \bigl\{ 306.704,\ -1.842,\ 2.462,\ -1.067 \bigr\}\ ,
\end{aligned}
\end{equation}
in agreement with the figures above. In particular, the extremal $v$-values become marginally stable,
while part of the non-extremal cases are unstable for high energies. 

Of course, for each non-extremal value of~$v$ we still have to project out unphysical modes
by imposing the gauge condition~(\ref{gauge4}). However, in the $12(2j{+}1)$-dimensional fluctuation space
the gauge condition has rank~$3(2j{+}1)$ while we see that (for $j{\ge}2$) in total $4(2j{+}1)$ normal modes 
are unstable at high energy. Therefore, the projection to physical modes cannot remove all instabilities. 
We must conclude that, for sufficiently high energy~$E$, some fluctuations grow exponentially,
implying that the solution $\Phi{\equiv}0$ is linearly unstable, and thus is the Yang--Mills background.

\vspace{12pt}
%----------------------------------------------------------------------------------------
\section{Singlet perturbation: exact treatment}
\vspace{2pt}
\noindent
Even though the Floquet representation helped to reduce the long-time behavior of the perturbations
to the analysis of a single period~$T$, it normally does not give us an exact solution to Hill's equation.
However, for the SO(4) singlet fluctuation around $\psi(\tau)$, we can employ the fact that 
$\dot{\psi}$ trivially solves the fluctuation equation,
\begin{equation} \label{timeshift}
   (\dot\psi)^{\cdot\cdot} = (\ddot\psi)^{\cdot} = -\bigl(V'(\psi)\bigr)^{\cdot} 
   = -V''(\psi)\,\dot{\psi} \= -(6\psi^2{-}2)\,\dot\psi \= -\Omega^2_{(0,0)}(\tau)\,\dot\psi\ ,
\end{equation}
with a frequency function which is $\frac{T}{2}$-periodic.
This implies that all fluctuation modes are $T$-periodic.
With the knowledge of an explicit solution to the fluctuation equation 
we can reduce the latter to a first-order equation and solve that one to find a second solution.
The normalizations are arbitrary, so we choose
\begin{equation}
   \Phi_1(\tau)\=-\sfrac{\epsilon^3}{k}\,\dot{\psi}(\tau) \und
   \Phi_2(\tau) \= \Phi_1(\tau)\,\int^\tau\frac{\diff\sigma}{\Phi_1(\sigma)^2}
   \= -\sfrac{k}{\epsilon^3}\,\dot{\psi}(\tau)\,\int^\tau \frac{\diff\sigma}{\dot{\psi}^2(\sigma)}\ ,
\end{equation}
which are linearly independent since
\begin{equation}
   W(\Phi_1,\Phi_2)\ \equiv\ \Phi_1\dot\Phi_2-\Phi_2\dot\Phi_1 \= 1\ .
\end{equation}

For simplicity, we restrict ourselves to the energy range 
$\sfrac12{<}E{<}\infty$, i.e.~$1{>}k^2{>}\sfrac12$. 
Explicitly, we have
\begin{equation}
\begin{aligned}
   \!\!\!\Phi_1(\tau) &\= \epsilon\,\sn\,\dn \ , \\[4pt]
   \!\!\!\Phi_2(\tau) &\= \sfrac{1}{1-k^2}\,\cn
   \bigl[ (2k^2{-}1)\,\textrm{dn}^2\bigl(\sfrac{\tau}{\epsilon},k\bigr) -k^2 \bigr] \\
   &\qquad + \sn\,\dn \bigl[ \sfrac{\tau}{\epsilon} +
   \sfrac{2k^2{-}1}{1{-}k^2}\,E\bigl(\textrm{am}(\sfrac{\tau}{\epsilon},k),k\bigr)\bigr]\;, 
\end{aligned}
\end{equation}
where $\textrm{am}(z,k)$ denotes the Jacobi amplitude and $E(z,k)$ is the elliptic integral of the second kind.
\begin{figure}[h!]
\centering
\captionsetup{width=\linewidth}
\includegraphics[width = 0.35\paperwidth]{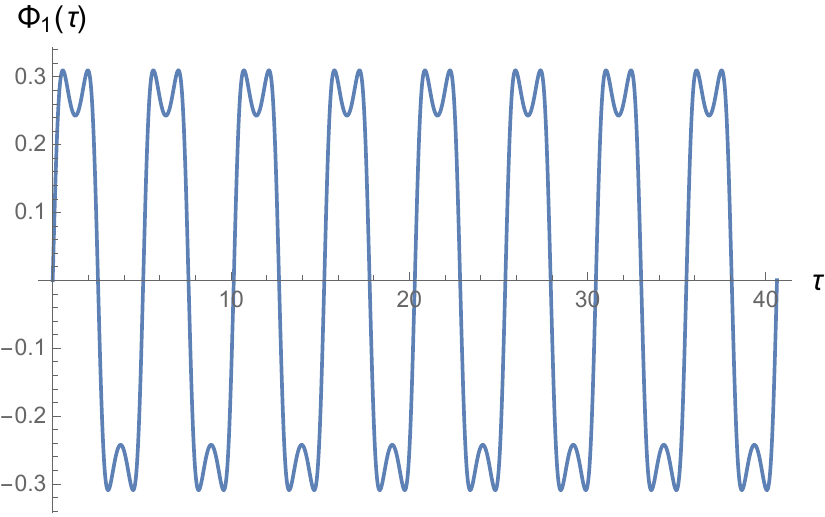} \quad
\includegraphics[width = 0.35\paperwidth]{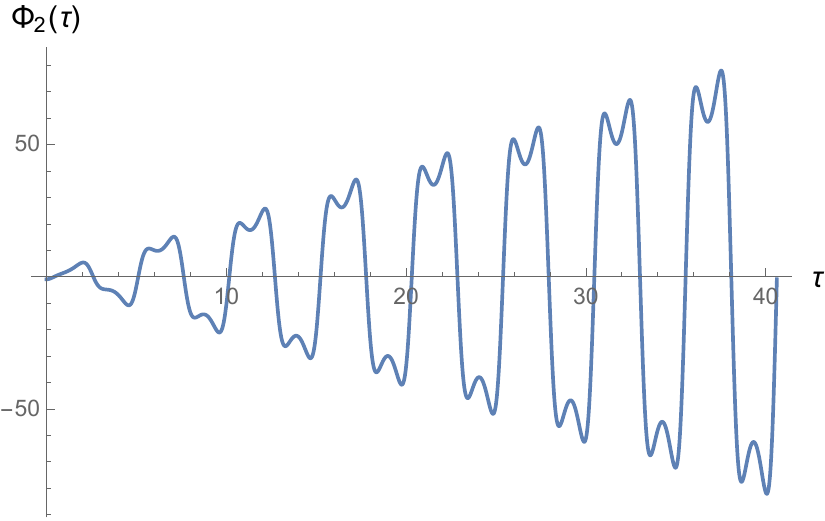} 
\caption{Plot of the SO(4) singlet fluctuation modes $\Phi_1$ and $\Phi_2$ over eight periods for $k^2{=}0.81$.}
\end{figure}
As can be checked, the initial conditions are
\begin{equation}
   \Phi_1(0)=0\ ,\quad \dot\Phi_1(0)=1 \qquad\und\qquad
   \Phi_2(0)=-1\ ,\quad \dot\Phi_2(0)=0\ ,
\end{equation}
which fixes the ambiguity of adding to $\Phi_2$ a piece proportional to~$\Phi_1$. Hence,
\begin{equation}
   \widehat\Phi(0) \= \Bigl(\begin{smallmatrix} 0 & \!{-}1 \\[4pt] 1 & 0 \end{smallmatrix} \Bigr)
   \qquad\Rightarrow\qquad
   M\= \widehat\Phi(T)\, \Bigl(\begin{smallmatrix} 0 & 1 \\[4pt] {-}1 & 0 \end{smallmatrix} \Bigr)\ .
\end{equation}
We know that $\Phi_1\sim\dot\psi$ is $T$-periodic, and so is $\dot\Phi_1$, 
but not the second solution,
\begin{equation}
   \Phi_2(\tau{+}T) \= \Phi_2(\tau) + \gamma\,T\,\Phi_1(\tau) \with \gamma \= \frac{1}{T}\int_0^T\frac{\diff\sigma}{\Phi_1(\sigma)^2}\bigg|_{\textrm{reg}}
   \ =:\ \sfrac{k^2}{\epsilon^6}\, \bigl\langle \dot\psi^{-2}\bigr\rangle_{\textrm{reg}}\ ,
\end{equation}
where the integral diverges at the turning points and must be regularized by subtracting
the Weierstra\ss\ $\wp$~function with the appropriate half-periods.
Since $\Phi_1$ has periodic zeros, $\Phi_2$ does return to~${-}1$ at integer multiples of~$T$.
It follows that the $\Phi_2$ oscillation linearly grows in amplitude with a rate (per period) of
\begin{equation}
   \gamma \= \frac{1}{\epsilon^2}\,\Bigl[\,1 + \frac{2k^2{-}1}{1{-}k^2}\,\frac{E(k)}{K(k)} \,\Bigr]\ ,
\end{equation}
which is always larger than 7.629, attained at $k\approx0.882$.
\begin{figure}[h!]
\centering
\includegraphics[width = 0.5\paperwidth]{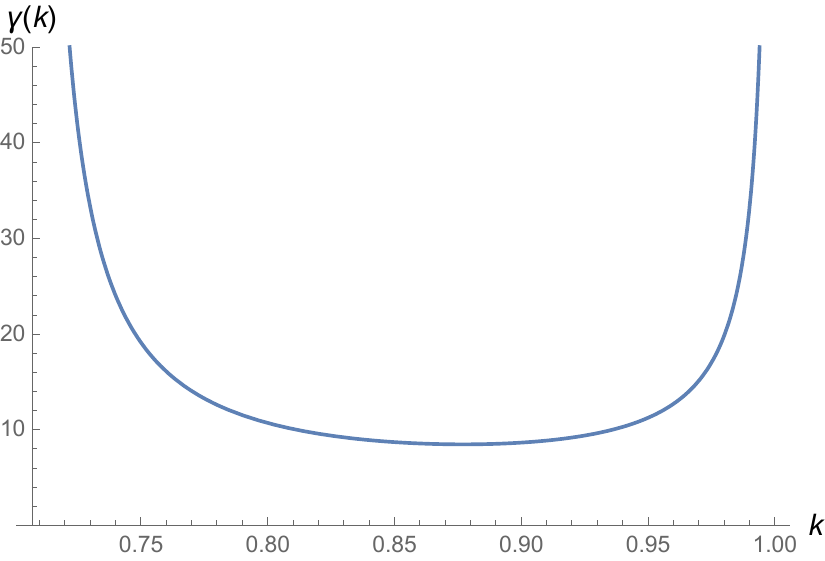} 
\caption{Plot of the linear growth rate~$\gamma$ as a function of~$k$.}
\end{figure}

In essence, we have managed to compute the monodromy
\begin{equation}
   M \= \begin{pmatrix} {-}\Phi_2(T) & \Phi_1(T) \\[4pt] {-}\dot\Phi_2(T) & \dot\Phi_1(T) \ \end{pmatrix} 
   \= \begin{pmatrix} \ 1 & 0 \\[4pt] \ \!\!{-}\gamma\,T & 1 \ \end{pmatrix} 
   \= \exp \Bigl\{ {-}\gamma\,T\,\bigl(\begin{smallmatrix} 0 & 0 \\[4pt] 1 & 0 \end{smallmatrix}\bigr) \Bigr\}
\end{equation}
and thus easily obtain the Floquet representation,
\begin{equation}
   R \= \begin{pmatrix} 0 & \gamma \\[4pt] 0 & 0 \end{pmatrix} \quad\Rightarrow\quad
   \ep^{\tau R} \= \begin{pmatrix} 1 & \gamma\,\tau \\[4pt] 0 & 1 \end{pmatrix}  \und
   Q(\tau)\ = \begin{pmatrix} \Phi_1 & \Phi_2{-}\Phi_1\gamma\,\tau \\[4pt] 
   \dot\Phi_1 & \dot\Phi_2{-}\dot\Phi_1\gamma\,\tau \end{pmatrix}\ .
\end{equation}
Obviously, we have encountered a marginally stable situation, since $M$ is of parabolic type. 
There is no exponential growth, and $\Phi_1$ is periodic thus bounded, but $\Phi_2$ 
grows without bound as long as one stays in the linear regime.
Note that we never made use of the form of our Newtonian potential.
In fact, this behavior is typical for a conservative mechanical system with oscillatory motion.

What to make of this linear growth? It can be (and actually is) easily overturned by nonlinear effects.
Going beyond the linear regime, though, requires expanding the Yang--Mills equation to higher orders
about our classical Yang--Mills solution~(\ref{Aansatz}).
While this is a formidable task in general, it can actually be done to all orders for the singlet perturbation!
The reason is that a singlet perturbation leaves us in the SO(4)-symmetric subsector,
thus connecting only to a neighboring ``cosmic background'', $\psi\to\tilde\psi$.
Since (\ref{backgrounds}) gives us analytic control over all solutions~$\psi(\tau)$,
the full effect of such a shift can be computed exactly. 
Splitting an exact solution~$\tilde{\psi}$ into a background part and its (full) deviation,
\begin{equation}
   \tilde{\psi}(\tau) \= \psi(\tau)\ +\ \eta(\tau)\ ,
\end{equation}
and inserting $\tilde{\psi}$ into the equation of motion~(\ref{Newton}), we obtain
\begin{equation} \label{nonlinear}
   0 \= \ddot\eta + V''(\psi)\,\eta + \sfrac12 V'''(\psi)\,\eta^2 + \sfrac16 V''''(\psi)\,\eta^3
   \= \ddot\eta + (6\psi^2{-}2)\,\eta + 6\psi\,\eta^2 + 2\,\eta^3\ ,
\end{equation}
extending the linear equation~(\ref{timeshift}) by two nonlinear contributions.
Perturbation theory introduces a small parameter~$\epsilon$ and formally expands
\begin{equation}
   \eta \= \epsilon\eta_{(1)} + \epsilon^2\eta_{(2)} + \epsilon^3\eta_{(3)} + \ldots\ ,
\end{equation}
which yields the infinite coupled system
\begin{equation}
\begin{aligned}
   & \bigl[\pa_\tau^2+(6\psi^2{-}2)\bigr]\,\eta_{(1)} \= 0 \ ,\\
   & \bigl[\pa_\tau^2+(6\psi^2{-}2)\bigr]\,\eta_{(2)} \= -6\psi\,\eta_{(1)}^2 \ ,\\
   & \bigl[\pa_\tau^2+(6\psi^2{-}2)\bigr]\,\eta_{(3)} \= -12\psi\,\eta_{(1)}\eta_{(2)} -2\,\eta_{(1)}^3\ ,\\
   & \ldots \ ,
\end{aligned}
\end{equation}
which could be iterated with a seed solution~$\eta_{(1)}$ of the linear system.

However, we know that the exact solutions to the full nonlinear equation~(\ref{nonlinear}) 
is simply given by the difference
\begin{equation}
   \eta(\tau) \= \tilde\psi(\tau) - \psi(\tau)
\end{equation}
of two analytically known backgrounds. The SO(4)-singlet background moduli space is parametrized by
two coordinates, e.g.~the energy~$E$ (or elliptic modulus~$k$) and the choice of an initial condition
which fixes the origin~$\tau{=}0$ of the time variable. In~(\ref{backgrounds}), we selected $\dot\psi(0)=0$,
but relaxing this we can reintroduce this collective coordinate by allowing shifts in~$\tau$. 
We may then parametrize the SO(4)-invariant Yang--Mills solutions as
\begin{equation}
   \psi_{k,\ell}(\tau) \= \psi(\tau{-}\ell) 
   \qquad\textrm{with}\qquad 2E=1/(2k^2{-}1)^2 \quad\und\quad \ell\in\R
\end{equation}
where $\psi$ is taken from~(\ref{backgrounds}). 
Note that $\dot\psi_{k,\ell}$ solves the background equation~(\ref{timeshift}) with a frequency-squared
$\omega_{k,\ell}^2=6\psi_{k,\ell}^2{-}2$.
Without loss of generality we assign $\psi=\psi_{k,0}$ and $\tilde\psi=\psi_{k{+}\delta k,\delta\ell}$,
hence
\begin{equation}
\begin{aligned}
   \eta(\tau) &\= \delta k\,\pa_k\psi(\tau) - \delta\ell\,\dot\psi(\tau) 
   + \sfrac12(\delta k)^2\,\pa_k^2\psi(\tau) - \delta k \delta\ell\,\pa_k\dot\psi(\tau) 
   + \sfrac12(\delta\ell)^2\,\ddot\psi(\tau) + \ldots \\[4pt]
   &\=  \delta k\,\pa_k\psi(\tau{-}\delta\ell) + \sfrac12(\delta k)^2\,\pa_k^2\psi(\tau{-}\delta\ell)
   + \sfrac16(\delta k)^3\,\pa_k^3\psi(\tau{-}\delta\ell) + \ldots\ ,
\end{aligned}
\end{equation}
because $\pa_\ell\psi=-\dot\psi$. 
Clearly, a shift in~$\ell$ only shifts the time dependence of the frequency and does not alter the energy~$E$,
which is not very interesting. Its linear part corresponds to the mode $\Phi_1\sim\dot\psi$ of the previous section.
A change in~$k$, in contract, will lead to a solution with an altered frequency and energy.
Its linear part is given by $\Phi_2$, which grows linearly in time. 
However, due to the boundedness of the full motion, the nonlinear corrections have to limit this growth and
ultimately must bring the fluctuation back close to zero. This is the familiar wave beat phenomenon:
the difference of two oscillating functions, $\tilde\psi$ and $\psi$, with slightly different frequencies,
will display an amplitude oscillation with a beat frequency given by the difference.
This is borne out in the following plots.
\begin{figure}[h!]
\centering
\captionsetup{width=\linewidth}
\includegraphics[width = 0.35\paperwidth]{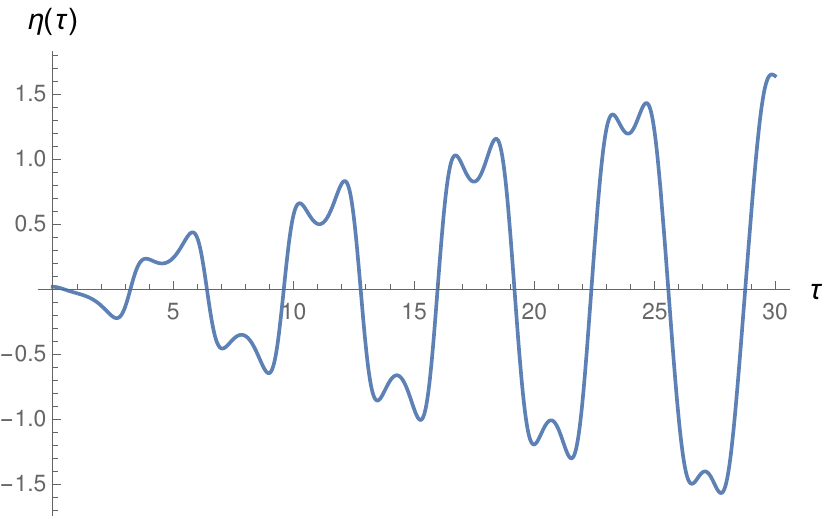} \quad
\includegraphics[width = 0.35\paperwidth]{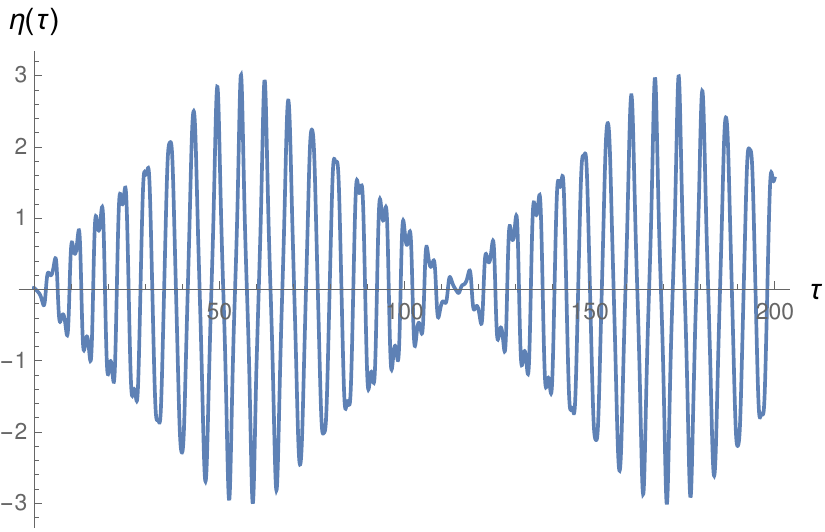}
\caption{Plots of the full perturbation $\eta$ at $k{=}0.95$ for $\eta(0){=}0.02$, $\dot{\eta}(0){=}0$, giving a beat ratio of ${\sim}19$.}
\end{figure}
As a result, we can assert a long-term stability of the cosmic Yang--Mills fields 
against the SO(4) singlet perturbation, even though on shorter time scales an excursion 
to a nearby solution is not met with a linear backreaction.

%% file: Chapters/Chapter7.tex
%auto-ignore
\chapter{Conclusion \& Outlook}
\label{Chapter7} 

We have studied stability behaviour of some well known solutions of $SU(2)$ Yang--Mills fields in $4$-dimensional de Sitter space $\diff{S}_4$ under generic gauge perturbation. These solutions could be of relevance to early time cosmology (before the electro-weak symmetry breaks down) in a scenario recently presented by Friedan \cite{Friedan}. An $SO(4)$ symmetric sector is analytically solvable and reduces to three coupled anharmonic oscillators (for the metric, an $SU(2)$ Yang–Mills field and the Higgs field, the latter being frozen to its vacuum state). We have presented a complete analysis of the linear gauge-field perturbations of the time-dependent Yang–Mills solution, by diagonalizing the fluctuation operator and studying the long-time behavior of the ensuing Hill’s equations using the stroboscopic map and Floquet theory. For parametrically large gauge-field energy (as is required in Friedan’s setup) the natural frequencies and monodromies become universal, and some unstable perturbation modes survive even in this limit. This provides strong evidence that such oscillating cosmic Yang–Mills fields are unstable against small perturbations, although we have not yet included metric fluctuations here. Their influence will be analyzed in
follow-up work.

We have also analyzed a family of electromagenetic knot configurations recently developed by making use of a conformal equivalence between $\diff{S}_4$, Minkwoski space $\R^{1,3}$ and a finite Lorentzian $S^3$-cylinder. These solutions are constructed on the cylinder in an $SO(4)$ covariant way and are then pulled back to the Minkowski space using the conformal map (that leaves the Maxwell's theory invariant). These ``basis-knot" solutions of Maxwell's equation are labelled with $S^3$ harmonics $Y_{j,m,n}$ and give rise to field configurations of knotted field lines when pulled back to the Minkowski space. We have analyzed the symmetry feature of these basis knotted electromagnetic field configurations with the isometry group $SO(1,4)$ of the de Sitter space. We have further studied, numerically, the effect of these basis configurations on the trajectories of multiple identical charged particles with different initial conditions. Various behaviors were obtained, including a separation of trajectories into different “solid angle regions” that converge asymptotically into a beam of charged particles along a few particular regions of space, an ultrarelativistic acceleration of particles and coherent twists/turns of the trajectories before they go off asymptotically. The results contribute to an effort to better understand the interactions between electromagnetic knots and charged particles \cite{AT10}. This becomes increasingly relevant as laboratory generation of knotted
fields progresses \cite{LSMetal18}. Further work in this direction could be to analyze a single Fourier mode of these
solutions to understand its experimental realization via monochromatic laser beams.

Furthermore, we have considered complex linear combinations of these basis-knot field configurations (that can model any finite-energy field configuration) and characterized the corresponding moduli space of null fields. We have also computed Noether charges of such a linearly combined configuration with fixed $j$ for the conformal group $SO(2,4)$, which is the largest symmetry group for the Maxwell theory. Here again the ``de Sitter" method proved to be advantageous in that the expressions of these charge densities simplifies immensely on the cylinder and, thus, can be computed with ease. We found that many of these charges vanish owing to the orthogonality of the harmonics and that the energy and momentum are the only independent charged in many cases. A nice geometric structure of $1$-forms facilitated the computation of spherical components of vector charges as well. We verified our results against the results for some modified Hopf--Ran{\~a}da field configurations of \cite{HSS15}. We would like to further check the validity of our results by comparing them with other solutions presented in \cite{HSS15}.

%% file: Appendices/AppendixA.tex
%auto-ignore
\chapter{Carter-Penrose transformation}
\label{appendCPtransf}
\noindent
The metric on the Minkowski space $\R^{1,3}$ in polar coordinates is given by
\begin{equation}
    \diff s_{Mink}^2 \= -\diff t^2 + \diff r^2 + r^2\,\diff\Omega_2^2\ ;\qquad \diff\Omega_2^2 \= \diff\theta^2 + \sin^2\theta\,\diff\phi^2 ,
\end{equation}
where $-\infty < t < \infty,\ 0 \leq r < \infty,\ 0 \leq \theta \leq \pi$, and $0 \leq \phi \leq 2\pi$. We first employ the light-cone coordinates $(u,v)$ to transform the metric in the following way
\begin{equation}\label{transf1}
\begin{aligned}
    &u\ :=\ t-r\ ,\quad v\ :=\ t+r\ ;\quad  -\infty < u \leq v < \infty\  \\[2pt]
    &\implies \diff s_{Mink}^2 \= -\diff v\, \diff u + \sfrac14 (v-u)^2\,\diff\Omega_2^2\ .
\end{aligned}
\end{equation}
In the second step we compactify the spacetime with the help of the coordinate (U,V) as follows:
\begin{equation}\label{transf2}
\begin{aligned}
   &U\ :=\ \arctan u\ ,\quad V\ :=\ \arctan v\ ;\quad  -\sfrac\pi2 < U \leq V < \sfrac\pi2\  \\[2pt]
    &\implies \diff s_{Mink}^2 \= \frac{1}{4\cos^2V\cos^2U}\left[-4\,\diff V\, \diff U +  \sin^2(V-U)\,\diff\Omega_2^2\right]\ .
\end{aligned}
\end{equation}
Finally, we rotate the coordinate system back using $(\tau,\chi)$ to obtain the desired form of the metric:
\begin{equation}\label{transf3}
\begin{aligned}
   &\tau\ :=\ V + U\ ,\quad \chi\ :=\ \pi + U - V\ ;\quad  0 < \chi \leq \pi\ , |\tau| < \chi\  \\[2pt]
    &\implies \diff s_{Mink}^2 \= \gamma^{-2}\left[-\diff \tau^2 + \diff\chi^2 +  \sin^2\chi\,\diff\Omega_2^2\right]\ ; \quad \gamma \= \cos\tau - \cos\chi\ .
\end{aligned}
\end{equation}
We realize that the Minkowski metric is conformally equivalent to a Lorentzian cylinder $\mathcal{I}\times S^3\ ;\ \mathcal{I}=(-\pi,\pi)$ with a conformal factor that can be recasted in terms of Minkowski coordinates using the above transformations (\ref{transf1}-\ref{transf3}):
\begin{equation}\label{metricCyl}
      \diff s_{cyl}^2\ :=\ -\diff\tau^2 + \diff\Omega_3^2 \= \gamma^2\, \diff s_{Mink}^2\ ;\quad \gamma = \frac{2\ell^2}{\sqrt{4\,t^2\,\ell^2 + (r^2-t^2+\ell^2)^2}}\ ,
\end{equation}
where we have made $\gamma$ dimensionless using the de Sitter radius $\ell$. The lightcone structure of the spacetime as presented in the Penrose diagram (see Figure \ref{CPplot}) is a direct consequence of \eqref{transf3}.  

%% file: Appendices/AppendixB.tex
%auto-ignore
\newgeometry{left=3cm,top=1.2cm,right=2.8cm,bottom=0.5cm}
\chapter{Rotation of indices}

\label{appendRotation}
\noindent
By construction the gauge potential $\Acal$ is $SO(4)$ invariant, which means it is also invariant under the action of $SO(3)$ generators $\mathcal{D}_a$ \eqref{DPdef}. For a complex-valued $\Acal = \Acal_a\,e^a$ expanded using \eqref{Asep} (for type I) and \eqref{XY} at a fixed $j$ this means 
\begin{equation}
\begin{aligned}
    0 &\= \mathcal{D}_a (\Acal) \\
    &\= \sum_{m,n,\tilde{n}} C^{n,\tilde{n}}_{j,b}\, \ep^{\im\Omega\tau}\, \left( \mathcal{D}_a(\Lambda_{m,\tilde{n}})\,e^b\,Y_{j;m,n} + \Lambda_{m,\tilde{n}}\,\mathcal{D}_a(e^b)\,Y_{j;m,n} + \Lambda_{m,\tilde{n}}\,e^b\,\mathcal{D}_a(Y_{j;m,n})  \right)
\end{aligned}    
\end{equation}
where $\mathcal{D}_a(e^b)$ are determined from \eqref{LieActionL} and \eqref{LieActionR} while $\mathcal{D}_a(Y_{j;m,n})$ are determined from (\ref{Y-action}-\ref{JpmAction}). By collecting the coefficients of various linearly independent $e^b$ and $Y_{j;m,n}$ terms in the above expansion for a fixed $\mathcal{D}_a$ one gets a set of coupled linear equations for $\mathcal{D}_a(\Lambda_{m,\tilde{n}})$, which can be easily solved. The action of the generators $\mathcal{D}_a$ on $\Lambda_{m,\tilde{n}}$ for $j=0,~ 1/2 ~\textrm{and}~ 1$ is given in the following table.

\begin{center}
\begin{adjustbox}{width=\linewidth}
%\begin{table}[H]
%  \resizebox{\linewidth}{6cm}{
%    \hspace{0cm}\setcellgapes{4pt}\makegapedcells \renewcommand\theadfont{\normalsize\bfseries}
    \begin{tabular}{|P{3.25cm}|P{6.3cm}|P{6.25cm}|P{1.8cm}|}
        \hline
      & $\mathcal{D}_1$ & $\mathcal{D}_2$ & $\mathcal{D}_3$ \\ [0.5ex] \hline \hline
     $\left.j=0: \right.                                                %j=0!
     \begin{aligned}
        &\Lambda_{0,-1} \mapsto \\
        &\Lambda_{0,0} \mapsto \\
        &\Lambda_{0,1} \mapsto
     \end{aligned}$ & 
     $\begin{aligned}                                               %D_1
         &\sqrt{2}\im \Lambda_{0,0} \\
         &\sqrt{2}\im (\Lambda_{0,1}+\Lambda_{0,-1}) \\
         &\sqrt{2}\im \Lambda_{0,0}
     \end{aligned}$ & 
     $\begin{aligned}                                               %D_2
         &-\sqrt{2}  \Lambda_{0,0} \\
         &\sqrt{2}  (\Lambda_{0,-1} - \Lambda_{0,1}) \\
         &\sqrt{2}  \Lambda_{0,0} 
     \end{aligned}$ &
     $\begin{aligned}                                               %D_3
         &-\sqrt{2}\im  \Lambda_{0,-1} \\
         &\quad 0 \\
         &\sqrt{2}\im  \Lambda_{0,1}
     \end{aligned}$ \\ \hline
     $\left.\begin{aligned}
        &\quad j=\sfrac12: \\                                                %j=1/2!
        &\underbrace{\textrm{Notation}}\\
        &\pm\sfrac12\equiv\pm \\
        &\pm\sfrac32\equiv\uparrow\downarrow
     \end{aligned}\right\}
     \begin{aligned}
        &\Lambda_{-,\downarrow} \mapsto \\[.2em]
        &\Lambda_{-,-} \mapsto \\[.2em]
        &\Lambda_{-,+} \mapsto \\[.2em]
        &\Lambda_{-,\uparrow} \mapsto \\[.2em]
        &\Lambda_{+,\downarrow} \mapsto \\[.2em]
        &\Lambda_{+,-} \mapsto \\[.2em]
        &\Lambda_{+,+} \mapsto \\[.2em]
        &\Lambda_{+,\uparrow} \mapsto
     \end{aligned}$ & 
     $\begin{aligned}                                               %D_1
         &\im(\sqrt{3}\Lambda_{-,-} + \Lambda_{+,\downarrow} ) \\
         &\im(\sqrt{3}\Lambda_{-,\downarrow} + 2\Lambda_{-,+} + \Lambda_{+,-} ) \\
         &\im(2\Lambda_{-,-} + \sqrt{3}\Lambda_{-,\uparrow} + \Lambda_{+,+} ) \\
         &\im(\sqrt{3}\Lambda_{-,+} + \Lambda_{+,\uparrow} ) \\
         &\im(\Lambda_{-,\downarrow} + \sqrt{3}\Lambda_{+,-} ) \\
         &\im(\Lambda_{-,-} + \sqrt{3}\Lambda_{+,\downarrow} + 2 \Lambda_{+,+} ) \\
         &\im(\Lambda_{-,+} + 2\Lambda_{+,-} + \sqrt{3}\Lambda_{+,\uparrow} ) \\
         &\im(\Lambda_{-,\uparrow} + \sqrt{3}\Lambda_{+,+} ) \\
     \end{aligned}$ &
     $\begin{aligned}                                               %D_2
         &-\sqrt{3}\Lambda_{-,-} - \Lambda_{+,\downarrow} \\
         &\sqrt{3}\Lambda_{-,\downarrow} - 2\Lambda_{-,+} - \Lambda_{+,-} \\
         &2\Lambda_{-,-} - \sqrt{3}\Lambda_{-,\uparrow} - \Lambda_{+,+} \\
         &\sqrt{3}\Lambda_{-,+} - \Lambda_{+,\uparrow} \\
         &\Lambda_{-,\downarrow} - \sqrt{3}\Lambda_{+,-} \\
         &\Lambda_{-,-} + \sqrt{3}\Lambda_{+,\downarrow} -2\Lambda_{+,+} \\
         &\Lambda_{-,+} +2\Lambda_{+,-} -\sqrt{3}\Lambda_{+,\uparrow} \\
         &\Lambda_{-,\uparrow} + \sqrt{3}\Lambda_{+,+} \\
     \end{aligned}$ &
     $\begin{aligned}                                               %D_3
         &-4\im\Lambda_{-,\downarrow} \\[.2em]
         &-2\im\Lambda_{-,-} \\[.2em]
         &0 \\[.2em]
         &2\im\Lambda_{-,\uparrow} \\[.2em]
         &-2\im\Lambda_{+,\downarrow} \\[.2em]
         &0 \\[.2em]
         &2\im\Lambda_{+,+} \\[.2em]
         &4\im\Lambda_{+,\uparrow}
     \end{aligned}$ \\ \hline
     $\left.\begin{aligned}
        &\quad j=1: \\                                                %j=1!
        &\underbrace{\textrm{Notation}}\\
        &\pm1\equiv\pm \\
        &\pm2\equiv\uparrow\downarrow
     \end{aligned}\right\}
     \begin{aligned}
        &\Lambda_{-,\downarrow} \mapsto \\[.2em]
        &\Lambda_{-,-} \mapsto \\[.2em]
        &\Lambda_{-,0} \mapsto \\[.2em]
        &\Lambda_{-,+} \mapsto \\[.2em]
        &\Lambda_{-,\uparrow} \mapsto \\[.2em]
        &\Lambda_{0,\downarrow} \mapsto \\[.2em]
        &\Lambda_{0,-} \mapsto \\[.2em]
        &\Lambda_{0,0} \mapsto \\[.2em]
        &\Lambda_{0,+} \mapsto \\[.2em]
        &\Lambda_{0,\uparrow} \mapsto \\[.2em]
        &\Lambda_{+,\downarrow} \mapsto \\[.2em]
        &\Lambda_{+,-} \mapsto \\[.2em]
        &\Lambda_{+,0} \mapsto \\[.2em]
        &\Lambda_{+,+} \mapsto \\[.2em]
        &\Lambda_{+,\uparrow} \mapsto
     \end{aligned}$ & 
     $\begin{aligned}                                               %D_1
         &\im(2\Lambda_{-,-} + \sqrt{2}\Lambda_{0,\downarrow} ) \\
         &\im(2\Lambda_{-,\downarrow} + \sqrt{6}\Lambda_{-,0} + \sqrt{2}\Lambda_{0,-} ) \\
         &\im(\sqrt{6}\Lambda_{-,-} + \sqrt{6}\Lambda_{-,+} + \sqrt{2}\Lambda_{0,0} ) \\
         &\im(\sqrt{6}\Lambda_{-,0} + 2\Lambda_{-,\uparrow} + \sqrt{2}\Lambda_{0,+} ) \\
         &\im(2\Lambda_{-,+} + \sqrt{2}\Lambda_{+,\uparrow} ) \\
         &\im(\sqrt{2}\Lambda_{-,\downarrow} + 2\Lambda_{0,-} + \sqrt{2}\Lambda_{+,\downarrow} ) \\
         &\im\sqrt{2}(\Lambda_{-,-} + \sqrt{2}\Lambda_{0,\downarrow} + \sqrt{3}\Lambda_{0,0} + \Lambda_{+,-} ) \\
         &\im\sqrt{2}(\Lambda_{-,0} + \sqrt{3}\Lambda_{0,-} + \sqrt{3}\Lambda_{0,+} + \Lambda_{+,0} ) \\
         &\im\sqrt{2}(\Lambda_{-,+} + \sqrt{3}\Lambda_{0,0} + \sqrt{2}\Lambda_{0,\uparrow} + \Lambda_{+,+} ) \\
         &\im(\sqrt{2}\Lambda_{-,\uparrow} + 2\Lambda_{0,+}  + \sqrt{2}\Lambda_{+,\uparrow} ) \\
         &\im(\sqrt{2}\Lambda_{0,\downarrow} + 2\Lambda_{+,-} ) \\
         &\im(\sqrt{2}\Lambda_{0,-} + 2\Lambda_{+,\downarrow} + \sqrt{6}\Lambda_{+,0} ) \\
         &\im(\sqrt{2}\Lambda_{0,0} + \sqrt{6}\Lambda_{+,-} + \sqrt{6}\Lambda_{+,+} ) \\
         &\im(\sqrt{2}\Lambda_{0,+} + \sqrt{6}\Lambda_{+,0} + 2\Lambda_{+,\uparrow} ) \\
         &\im(\sqrt{2}\Lambda_{0,\uparrow} + 2\Lambda_{+,+} ) \\        
     \end{aligned}$ &
     $\begin{aligned}                                               %D_2
         &-2\Lambda_{-,-} - \sqrt{2}\Lambda_{0,\downarrow} \\
         &2\Lambda_{-,\downarrow} - \sqrt{6}\Lambda_{-,0} - \sqrt{2}\Lambda_{0,-} \\
         &\sqrt{6}\Lambda_{-,-} - \sqrt{6}\Lambda_{-,+} - \sqrt{2}\Lambda_{0,0} \\
         &\sqrt{6}\Lambda_{-,0} - 2\Lambda_{-,\uparrow} - \sqrt{2}\Lambda_{0,+} \\
         &2\Lambda_{-,+} - \sqrt{2}\Lambda_{0,\uparrow} \\
         &\sqrt{2}\Lambda_{-,\downarrow} - 2\Lambda_{0,-} - \sqrt{2}\Lambda_{+,\downarrow} \\
         &\sqrt{2}(\Lambda_{-,-} + \sqrt{2}\Lambda_{0,\downarrow} - \sqrt{3}\Lambda_{0,0} - \Lambda_{+,-}) \\
         &\sqrt{2}(\Lambda_{-,0} + \sqrt{3}\Lambda_{0,-} - \sqrt{3}\Lambda_{0,+} - \Lambda_{+,0}) \\
         &\sqrt{2}(\Lambda_{-,+} + \sqrt{3}\Lambda_{0,0} - \sqrt{2}\Lambda_{0,\uparrow} - \Lambda_{+,+}) \\
         &\sqrt{2}\Lambda_{-,\uparrow} + 2\Lambda_{0,+} - \sqrt{2}\Lambda_{+,\uparrow} \\
         &\sqrt{2}\Lambda_{0,\downarrow} + -2\Lambda_{+,-} \\
         &\sqrt{2}\Lambda_{0,-} + 2\Lambda_{+,\downarrow} - \sqrt{6}\Lambda_{+,0} \\
         &\sqrt{2}\Lambda_{0,0} + \sqrt{6}\Lambda_{+,-} - \sqrt{6}\Lambda_{+,+} \\
         &\sqrt{2}\Lambda_{0,+} + \sqrt{6}\Lambda_{+,0} - 2\Lambda_{+,\uparrow} \\
         &\sqrt{2}\Lambda_{0,\uparrow} + 2\Lambda_{+,+} \\
     \end{aligned}$ &
     $\begin{aligned}                                               %D_3
         &-6\im\Lambda_{-,\downarrow} \\[.2em]
         &-4\im\Lambda_{-,-} \\[.2em]
         &-2\im\Lambda_{-,0} \\[.2em]
         &0 \\[.2em]
         &2\im\Lambda_{-,\uparrow} \\[.2em]
         &-4\im\Lambda_{0,\downarrow} \\[.2em]
         &-2\im\Lambda_{0,-} \\[.2em]
         &0 \\[.2em]
         &2\im\Lambda_{0,+} \\[.2em]
         &4\im\Lambda_{0,\uparrow} \\[.2em]
         &-2\im\Lambda_{+,\downarrow} \\[.2em]
         &0 \\[.2em]
         &2\im\Lambda_{+,0} \\[.2em]
         &4\im\Lambda_{+,+} \\[.2em]
         &6\im\Lambda_{+,\uparrow}
     \end{aligned}$ \\ \hline
    \end{tabular}
 % }
 %\caption{Action of $\mathcal{D}_a$ on $\Lambda_{m,\tilde{n}}$ for $j=0, ~1/2 ~\textrm{and}~ 1$.}
%\label{table4}
%\end{table}
\end{adjustbox} 
\end{center}

\restoregeometry

\nopagebreak[0]

%% file: Appendices/AppendixC.tex
%auto-ignore
\newgeometry{left=3cm,top=1.5cm,right=3cm,bottom=0.5cm}
\chapter{Polynomials in the characteristic equation}
\label{appendPolynomials}

\begin{center}
%\begin{adjustbox}{width=\linewidth}
\resizebox{\linewidth}{10cm}{
\begin{tabular}{| P{10mm} | P{25mm} | P{45mm} | P{80mm} |} 
\hline
$j\ ,\ v $ & P($\lambda$) & Q($\lambda$)  & R($\lambda$) \\ [0.5ex] 
 \hline\hline
$0\ ,\ 0$ \phantom{\Big|} & N/A & $(-2 + 6 \psi^2) - \lambda$ & $-(2 - 6 \psi^2) + \lambda$ \\
 \hline
$0\ ,\ 1$ \phantom{\Big|} & 1 & $(2 + 2 \psi + 4 \psi^2) - \lambda$ & $(4 + 12\psi + 20 \psi^2 + 20 \psi^3 + 8 \psi^4 - 8 \dot{\psi}^2) - (4 + 6 \psi + 6 \psi^2) \lambda + \lambda^2$ \\
 \hline
$0\ ,\ 2$ \phantom{\Big|} & N/A & $(10 + 6 \psi) - \lambda$ & N/A \\
 \hline
$\sfrac12\ ,\ \sfrac12$ \phantom{\Big|} & $-(1 + 6 \psi^2) + \lambda$ & $-(1 + 2 \psi^2 + 24 \psi^4) + (2 + 10 \psi^2) \lambda - \lambda^2$ & $-(1 + 4 \psi^2 + 28 \psi^4 + 48 \psi^6 - 8 \dot{\psi}^2 - 48 \psi^2 \dot{\psi}^2) + (3 + 16 \psi^2 + 44 \psi^4 - 8 \dot{\psi}^2) \lambda - (3 + 12 \psi^2) \lambda^2 + \lambda^3$ \\ 
 \hline
$\sfrac12\ ,\ \sfrac32$ \phantom{\Big|} & $-(7 + 3 \psi) + \lambda$ & $-(49 + 42 \psi + 32 \psi^2 + 12 \psi^3)   + (14 + 6 \psi + 4 \psi^2) \lambda - \lambda^2$ & $-(343 + 588 \psi + 574 \psi^2 + 360 \psi^3 + 136 \psi^4 + 24 \psi^5 - 56 \dot{\psi}^2 - 24 \psi \dot{\psi}^2) + (147 + 168 \psi + 124 \psi^2 + 48 \psi^3 + 8 \psi^4 - 8 \dot{\psi}^2) \lambda - (21 + 12 \psi + 6 \psi^2) \lambda^2 + \lambda^3$ \\ 
 \hline
$\sfrac12\ ,\ \sfrac52$ \phantom{\Big|} & N/A & $(17 + 8 \psi) - \lambda$ & N/A \\
 \hline
$1\ ,\ 0$ \phantom{\Big|} & $1$ & $(2 - 2 \psi + 4 \psi^2) - \lambda$ & $( 4 - 12 \psi + 20 \psi^2 - 20 \psi^3 + 8 \psi^4 - 8 \dot{\psi}^2) - (4 - 6 \psi + 6 \psi^2) \lambda + \lambda^2$ \\
 \hline
$1\ ,\ 1$ \phantom{\Big|} & $(36 + 36 \psi^2) - (12 + 6\psi^2) \lambda + \lambda^2$ & $(216 + 192 \psi^2 + 104 \psi^4) - (108 + 92 \psi^2 + 24 \psi^4) \lambda + (18 + 10 \psi^2) \lambda^2 - \lambda^3$ & $(1296 + 1584 \psi^2 + 1008 \psi^4 + 208 \psi^6 - 288 \dot{\psi}^2 - 288 \psi^2 \dot{\psi}^2) - (864 + 960 \psi^2 + 432 \psi^4 + 48 \psi^6 - 96 \dot{\psi}^2 - 48 \psi^2 \dot{\psi}^2) \lambda + (216 + 188 \psi^2 + 44 \psi^4 - 8 \dot{\psi}^2) \lambda^2 - (24 + 12\psi^2) \lambda^3 + \lambda^4$ \\
 \hline
$1\ ,\ 2$ \phantom{\Big|} & $-(14 + 4 \psi) + \lambda$ & $-(196 + 112 \psi + 60 \psi^2 + 16 \psi^3) + (28 + 8 \psi + 4 \psi^2) \lambda - \lambda^2$ & $-(2744 + 3136 \psi + 2128 \psi^2 + 928 \psi^3 + 248 \psi^4 + 32 \psi^5 - 112 \dot{\psi}^2 - 32 \psi \dot{\psi}^2) + (588 + 448 \psi + 236 \psi^2 + 64 \psi^3 + 8 \psi^4 - 8 \dot{\psi}^2) \lambda - (42 + 16 \psi + 6 \psi^2) \lambda^2 + \lambda^3$ \\
 \hline
$1\ ,\ 3$ \phantom{\Big|} & N/A & $(26 + 10 \psi) - \lambda$ & N/A \\ 
 \hline
$\sfrac32\ ,\ \sfrac12$ \phantom{\Big|} & $-(7 - 3 \psi ) + \lambda$ & $-(49 - 42 \psi + 32 \psi^2 - 12 \psi^3) + (14 - 6 \psi+ 4 \psi^2) \lambda - \lambda^2$ & $-(343 - 588 \psi + 574 \psi^2 - 360 \psi^3 + 136 \psi^4 - 24 \psi^5 - 56 \dot{\psi}^2 + 24 \psi \dot{\psi}^2) + (147 - 168 \psi + 124 \psi^2 - 48 \psi^3 + 8 \psi^4 - 8 \dot{\psi}^2) \lambda - (21 - 12 \psi + 6 \psi^2) \lambda^2 + \lambda^3$ \\ 
 \hline
$\sfrac32\ ,\ \sfrac32$ \phantom{\Big|} & $(169 + 78 \psi^2) - (26 + 6 \psi^2) \lambda + \lambda^2$ & $(2197 + 962 \psi^2 + 216 \psi^4) - (507 + 204 \psi^2 + 24 \psi^4) \lambda + (39 + 10 \psi^2) \lambda^2 - \lambda^3$ & $(28561 + 16900 \psi^2 + 4732 \psi^4 + 432 \psi^6 - 1352 \dot{\psi}^2 - 624 \psi^2 \dot{\psi}^2) + (-8788 - 4628 \psi^2 - 936 \psi^4 - 48 \psi^6 + 208 \dot{\psi}^2 + 48 \psi^2 \dot{\psi}^2) \lambda + (1014 + 412 \psi^2 + 44 \psi^4 - 8 \dot{\psi}^2) \lambda^2 - (52 + 12 \psi^2) \lambda^3 + \lambda^4$ \\
 \hline
$\sfrac32\ ,\ \sfrac52$ \phantom{\Big|} & $-(23 + 5 \psi) + \lambda$ & $-(529 + 230 \psi + 96 \psi^2 + 20 \psi^3) + (46 + 10 \psi + 4 \psi^2) \lambda - \lambda^2$ & $-(12167 + 10580 \psi + 5566 \psi^2 + 1880 \psi^3 + 392 \psi^4 + 40 \psi^5 - 184 \dot{\psi}^2 + 40 \psi\dot{\psi}^2) + (1587 + 920 \psi + 380 \psi^2 + 80 \psi^3 + 8 \psi^4 - 8 \dot{\psi}^2) \lambda + (-69 - 20 \psi - 6 \psi^2) \lambda^2 + \lambda^3$ \\
 \hline
$\sfrac32\ ,\ \sfrac72$ \phantom{\Big|} & N/A & $(37 + 12 \psi) - \lambda$ & N/A \\ 
 \hline
$2\ ,\ 0$ \phantom{\Big|} & N/A & $(10- 6 \psi) - \lambda$ & N/A \\
 \hline
$2\ ,\ 1$ \phantom{\Big|} & $-(14 - 4 \psi) + \lambda$ & $-(196 - 112 \psi + 60 \psi^2 - 16 \psi^3) + (28 - 8 \psi + 4 \psi^2) \lambda - \lambda^2$ & $-(2744 - 3136 \psi + 2128 \psi^2 - 928 \psi^3 + 248 \psi^4 - 32 \psi^5 - 112 \dot{\psi}^2 + 32 \psi \dot{\psi}^2) + (588 - 448 \psi + 236 \psi^2 - 64 \psi^3 + 8 \psi^4 - 8 \dot{\psi}^2) \lambda - (42 - 16 \psi + 6 \psi^2) \lambda^2 + \lambda^3$ \\
 \hline
$2\ ,\ 2$ \phantom{\Big|} & $(484 + 132 \psi^2) - (44 + 6 \psi^2) \lambda + \lambda^2$ & $(10648 + 2816 \psi^2 + 360 \psi^4) - (1452 + 348 \psi^2 + 24 \psi^4) \lambda + (66 + 10 \psi^2) \lambda^2 - \lambda^3$ & $(234256 + 83248 \psi^2 + 13552 \psi^4 + 720 \psi^6 - 3872 \dot{\psi}^2 - 1056 \psi^2 \dot{\psi}^2) - (42592 + 13376 \psi^2 + 1584 \psi^4 + 48 \psi^6 - 352 \dot{\psi}^2 - 48 \psi^2 \dot{\psi}^2) \lambda + (2904 + 700 \psi^2 + 44 \psi^4 - 8 \dot{\psi}^2) \lambda^2 - 
 (88 + 12 \psi^2) \lambda^3 + \lambda^4$ \\
 \hline
$2\ ,\ 3$ \phantom{\Big|} & $-(34 + 6 \psi) + \lambda$ & $-(1156 + 408 \psi + 140 \psi^2 + 24 \psi^3) + (68 + 12 \psi + 4 \psi^2) \lambda - \lambda^2$ & $-(39304 + 27744 \psi + 11968 \psi^2 + 3312 \psi^3 + 568 \psi^4 + 48 \psi^5 - 272 \dot{\psi}^2 - 48 \psi \dot{\psi}^2) + (3468 + 1632 \psi + 556 \psi^2 + 96 \psi^3 + 8 \psi^4 - 8 \dot{\psi}^2) \lambda - (102 + 24 \psi + 6\psi^2) \lambda^2 + \lambda^3$ \\ 
 \hline
$2\ ,\ 4$ \phantom{\Big|} & N/A & $(50 + 14 \psi) - \lambda$ & N/A \\ 
 \hline
\end{tabular}
}
%\end{adjustbox}
\end{center}

\restoregeometry